\documentclass[aps,pra,twocolumn,showpacs,superscriptaddress]{revtex4-1}

\usepackage{amsmath}  
\usepackage{amsthm}  
\usepackage{amssymb}  
\usepackage{color}
\usepackage{longtable}
\usepackage{multirow}
\usepackage{mathrsfs}
\usepackage{braket}
\usepackage{xfrac}
\usepackage[T1]{fontenc}

\let\savedegree\corresponds
\let\corresponds\relax
\usepackage{mathabx}
\let\corresponds\savedegree

\usepackage{float}
\usepackage{bm}
\usepackage[percent]{overpic}
\usepackage{mathrsfs}
\usepackage{mathcomp}
\usepackage[table,dvipsnames]{xcolor}

\usepackage[caption=false]{subfig}
\usepackage{graphicx}

\begin{document}
\title{Allowed region of the mean values of angular momentum observables and their uncertainty relations}

\author{Arun Sehrawat}
\email[Email: ]{arunsehrawat@hri.res.in}
\affiliation{Harish-Chandra Research Institute, HBNI, Chhatnag Road, Jhunsi, Allahabad 211019, India}

%\date{\today}

\begin{abstract}
The expectation values of operators drawn from a single quantum state cannot be outside of a particular region, called their allowed region or the joint numerical range of the operators. 
Basically, the allowed region is an image of the state space under the Born rule.
The maximum-eigenvalue-states---of every linear combination of the operators of interest---are sufficient to generate boundary of the allowed region.
In this way, we obtain the numerical range of certain Hermitian operators (observables) that are functions of the angular momentum operators.
Especially, we consider here three kinds of functions---combinations of powers of the ladder operators, powers of the angular momentum operators and their anticommutators---and discover  
the allowed regions of different shapes.
By defining some specific concave (and convex) functions on the joint numerical range, we also achieve tight uncertainty (and certainty) relations for the observables.
Overall, we demonstrate how the numerical range and uncertainty relations
change as the angular momentum quantum number grows.
Finally, we apply the quantum de Finetti theorem
by taking a multi-qubit system and attain the allowed regions and tight uncertainty relations in the limit where the quantum number goes to infinity.

\end{abstract}

\maketitle

%===========================================
\section{Introduction}\label{sec:Intro}

Can the numbers in ${\vec{r}=(x,y,\cdots,z)}$ be the expectation values---obtained from a single state $\rho$---of the operators in ${\vec{E}=(X,Y,\cdots,Z)}$? 
To provide a definite ``yes or no'' answer, we need to know the region $\mathcal{E}$
of all possible mean vectors 
$\vec{\boldsymbol{\varepsilon}}:=\langle \vec{E}\,\rangle_\rho$
for a given vector operator $\vec{E}$.
If and only if $\vec{r}\in\mathcal{E}$ then there exists at least one $\rho$ such that $\vec{r}=\vec{\boldsymbol{\varepsilon}}$, thus $\mathcal{E}$ is called \emph{allowed region} of the mean values.
The allowed region is also known as the joint numerical range \cite{Schwonnek17,Szymanski17}
and the quantum convex support \cite{Weis11} of $\vec{E}$.

Only pure states ${\rho=\rho^2}$ are usually considered to define the numerical range
\cite{Gustafson97,Keeler97,Szymanski18}, but here we take the whole state space that includes mixed states.
Hence, $\mathcal{E}$ is always a convex set.
Here, we are interested in $\mathcal{E}$ for
$n$ \emph{bounded} Hermitian operators (observables) 
of a $d$-level quantum system (qudit), where 
$n$ and $d$ are finite numbers, and we treat ${d\rightarrow\infty}$ as a limiting case.
Many important results about the numerical range of ${n=2}$ Hermitian operators are contained in Book \cite{Gustafson97}.

Although the numerical range is largely investigated
in mathematics, 
recently it is used to study the phase transitions \cite{Chen15,Chen16,Zauner16,Chen17,Chen17b,Szymanski17}.
The numerical range also plays a central role in \cite{Schwonnek17,Szymanski17,Szymanski18b}, where
numerical techniques are provided to get---an uncertainty relation (UR)---a state-independent
lower bound of the sum of squared standard deviations of two observables.
A UR based on such a sum is introduced in \cite{Maccone14}.
In \cite{Schwonnek18b}, the 
Wigner distribution is built on the numerical range of a set of observables, and its properties are analyzed.

In Sec.~\ref{sec:all-region}, we present a simple procedure
to obtain $\mathcal{E}$ for a general $\vec{E}$. It is---taken from \cite{Chen15,Schwonnek17,Szymanski17} and Chap.~5 in \cite{Gustafson97}---established on the fact that all the expectation values of a Hermitian operator 
lie between its extreme eigenvalues. If we consider operators ${\widehat{\eta}\cdot\vec{E}}$ for all real unit vectors ${\widehat{\eta}\in\mathbb{R}^{n}}$, then it will be sufficient to draw the boundary
$\partial\mathcal{E}$ of the permitted region by using the above fact.
Since $\mathcal{E}$ is a convex and bounded set in $\mathbb{R}^{n}$ in the case of bounded operators, it is completely specified by its boundary, and 
$\partial\mathcal{E}\subseteq \mathcal{E}$
for a finite $d$ \cite{Gustafson97}.
In \cite{Sehrawat17b} a different scheme---based on a result from \cite{Kimura03,Byrd03} about the positivity of $\rho$---is provided to get the allowed region $\mathcal{E}$.
There it is shown how 
to build uncertainty and certainty measures on $\mathcal{E}$ to achieve a tight UR and certainty relation (CR), which is briefly repeated
in Sec.~\ref{sec:all-region}.

For a qubit, ${d=2}$, $\partial\mathcal{E}$ is always an ellipsoid (possibly degenerate)
\cite{Kaniewski14,Meyer00,Abbott16,Sehrawat17b}, and
every tight UR and CR can be achieved by using it
\cite{Abbott16,Sehrawat17,Sehrawat17b}.
In the case of a qutrit, ${d=3}$,
all possible shapes of $\mathcal{E}$ are classified in \cite{Keeler97,Szymanski18}
for ${n=2,3}$ Hermitian operators.
In any dimension $d$,
the joint numerical range of a pair of projectors that have no common eigenvector is the convex hull of two ellipses \cite{Landau61,Lenard72}.
To specify $\mathcal{E}$ for an arbitrary $\vec{E}$ in a higher dimension 
is a difficult task, therefore we consider the
angular momentum operators in Sec.~\ref{sec:spin-ops}
and their polynomials in Secs.~\ref{sec:J+^2 and J-^2}--\ref{sec:N-qubit}
at the places of ${X,Y,\cdots,Z}$ and 
present $\mathcal{E}$, tight URs, and CRs
for the angular momentum quantum numbers ${j=\tfrac{d-1}{2}=\tfrac{1}{2},1,\tfrac{3}{2},2,\cdots.}$
There are known URs for the angular momentum operators 
\cite{Sanchez-Ruiz93,Hofmann03,Rivas08,Dammeier15,Riccardi17},
but our uncertainty measures and URs are different as described in Sec.~\ref{sec:all-region}.

Our main contribution lies in Secs.~\ref{sec:J+^2 and J-^2}--\ref{sec:N-qubit}
and Appendix~\ref{sec:Sup-material}, while Secs.~\ref{sec:all-region} and \ref{sec:spin-ops} are kept to introduce necessary notations for the paper.
Basically, we consider three types of operators:
combinations of powers of the ladder operators in Sec.~\ref{sec:J+^2 and J-^2},
powers of the angular momentum operators in Secs.~\ref{sec:Jx2 and Jy2}, \ref{sec:S-power}, and \ref{sec:N-qubit}
and their anticommutators in Secs.~\ref{sec:anticomm} and \ref{sec:N-qubit}.
The quadratic polynomials of the momentum operators in these sections are responsible for the spin squeezing \cite{Kitagawa93}.

We want to emphasize that the plots given in Sec.~\ref{sec:Jx2 and Jy2} for the boundary 
$\partial\mathcal{E}$ already appeared in \cite{Chen17}, where the authors used a multi-qubit system that inspires Sec.~\ref{sec:N-qubit} here.
In Sec.~\ref{sec:N-qubit}, we choose a multi-qubit system and use the 
quantum de Finetti theorem \cite{Stormer69,Hudson76} to obtain the allowed regions---for the operators investigated in the earlier sections---in the limit ${j\rightarrow\infty}$.
We summarize the main results and present an outlook in Sec.~\ref{sec:conclusion}.

%===========================================
\section{Allowed region and uncertainty measures on it}\label{sec:all-region}

With a given density operator $\rho$ on
a $d$-dimensional Hilbert space $\mathscr{H}_d$, one can compute the average values of all the $n$ Hermitian operators in ${\vec{E}=(X,Y,\cdots,Z)}$ by the Born rule. The rule can be perceived as a linear map 
\begin{equation}
\label{E}
\rho\longmapsto   \text{tr}(\rho \vec{E}\,)=\langle \vec{E}\,\rangle_\rho
=\vec{\boldsymbol{\varepsilon}}
\end{equation}
from the state space $\boldsymbol{\Omega}$ onto the allowed region,
also known as the joint numerical range \cite{Schwonnek17,Szymanski17}, 
\begin{equation}
\label{set-of-expt}
\mathcal{E}:=\big\{\,\vec{\boldsymbol{\varepsilon}}\ \big|\
\rho\in\boldsymbol{\Omega}\, \big\}\subset\mathbb{R}^{n}.
\end{equation}
Basically, $\boldsymbol{\Omega}$ is the collection of all positive semidefinite operators,
on $\mathscr{H}_d$, with the unit trace. 
We often write $\langle \ \rangle_{\rho}$ without the subscript as
all the mean values in $\vec{\boldsymbol{\varepsilon}}$ are drawn from a same state $\rho$.

For a single Hermitian operator ${X=X^\dagger}$, $\mathcal{E}$ is just the closed 
interval $[x_\text{m}, x_\textsc{m}]$ that includes all real numbers
between the minimum and maximum eigenvalues of $X$ \cite{Gustafson97}.
The numerical range
$\mathcal{E}$ of $\vec{E}$ is always a convex set because $\boldsymbol{\Omega}$ is. As long as the dimension $d$ is finite, $\mathcal{E}$ is also
a compact (closed and bounded) set \cite{Bengtsson06} in a hyperrectangle specified by the Cartesian product 
\begin{equation}
\label{hyperrectangle}
\mathcal{H}:=
[x_\text{m}, x_\textsc{m}]\times
[y_\text{m}, y_\textsc{m}]\times\cdots\times
[z_\text{m}, z_\textsc{m}]
\end{equation}
of closed intervals bounded by the extreme eigenvalues of the operators in $\vec{E}$.
$\mathcal{E}$ touches each facet of the hyperrectangle at some points that come from
the smallest- or largest-eigenvalue-states of one of the $n$ operators.
If ${X,Y,\cdots,Z}$ mutually commute---that is, ${XY=YX}$ and so on---then we can find their common eigenbasis ${\{|\textsf{e}_l\rangle \}_{l=1}^{d}}$. And, 
$\mathcal{E}$ will be a convex polytope
\begin{equation}
\label{conv-hull}
\text{Conv}\,
\big\{\, \langle \textsf{e}_l|\vec{E}\,|\textsf{e}_l\rangle
\;\big|\; l=1,\cdots,d\,
 \big\}
\end{equation}
in $\mathbb{R}^{n}$. The convex hull of a set ${\{\cdots\}}$
is denoted by ${\text{Conv}\{\cdots\}}$.
All the extreme points of \eqref{conv-hull} will be a subset of $\{ \langle \textsf{e}_l|\vec{E}\,|\textsf{e}_l\rangle\}_{l=1}^d$. An extreme point of $\mathcal{E}$ inevitably comes from a pure state---that is an extreme point of the state space $\boldsymbol{\Omega}$---but not every pure state provides an extreme point of $\mathcal{E}$ unless \eqref{E} is a bijective mapping.

The following procedure to obtain $\mathcal{E}$
is borrowed from \cite{Chen15,Schwonnek17,Szymanski17} and Chap.~5 in \cite{Gustafson97}, and a different method is given in \cite{Sehrawat17b}.
Let us take ${n=3}$ Hermitian operators ${X,Y,}$ and $Z$, and the procedure can be extended to any number of operators. 
First, we build a two-parameter family of operators
as
\begin{eqnarray}
\label{L}
\Lambda(\theta,\phi)&:=&
\widehat{\eta}(\theta,\phi)\cdot \vec{E}\,,
\quad \mbox{where}\quad
\vec{E}=(X,Y,Z)\,,
\\
\label{n-vector}
\widehat{\eta}(\theta,\phi)&:=&
(\sin\theta\cos\phi\,,\,\sin\theta\sin\phi\,,\,\cos\theta)\,,
\end{eqnarray}
${\theta\in[0,\pi]}$, and ${\phi\in[0,2\pi)}$. 
An eigenvalue and eigenket of $\Lambda$,
\begin{equation}
\label{L-eigen}
\Lambda(\theta,\phi)\,|\lambda(\theta,\phi)\rangle
=\lambda(\theta,\phi) \,|\lambda(\theta,\phi)\rangle\,,
\end{equation}
generally depend on the angles $\theta$ and $\phi$.
We call a density operator $\varrho$ an eigenstate associated with an eigenvalue $\lambda$ of $\Lambda$ if and only if 
${\Lambda\varrho=\lambda\varrho=\varrho \Lambda}$.

For every ${(\theta,\phi)}$, there are lower and upper bounds 
\begin{equation}
\label{L-exp}
\lambda_\text{m}(\theta,\phi)
\leq
\big\langle \Lambda(\theta,\phi)\big\rangle_\rho
\leq
\lambda_\textsc{m}(\theta,\phi)\,,
\end{equation}
where $\lambda_\text{m}$ and $\lambda_\textsc{m}$ are
the minimum and maximum eigenvalues of $\Lambda$.
It is sufficient to take only the maximum eigenvalues $\lambda_\textsc{m}$ because
${\Lambda(\pi-\theta,\pi+\phi)=-\Lambda(\theta,\phi)}$ and the smallest eigenvalue of ${\Lambda(\theta,\phi)}$ is the largest eigenvalue of ${\Lambda(\pi-\theta,\pi+\phi)}$.
Or, equivalently, one can keep both the eigenvalues and take ${\theta,\phi\in[0,\pi)}$.

Now let us define a supporting (tangent) hyperplane
\begin{equation}
\label{hyperplane}
\mathbb{H}(\theta,\phi):=\big\{\vec{r}=(x,y,z)\in\mathbb{R}^3\ |\ 
\widehat{\eta}(\theta,\phi)\cdot\vec{r}=\lambda_\textsc{m}(\theta,\phi)
\big\}
\end{equation}
of $\mathcal{E}$ with the normal vector $\widehat{\eta}(\theta,\phi)$.
According to the right-hand-side inequality in \eqref{L-exp}---that is,
\begin{equation}
\label{n.e<lambda}
\widehat{\eta}(\theta,\phi)\cdot\vec{\boldsymbol{\varepsilon}}\;
\leq\,\lambda_\textsc{m}(\theta,\phi)
\end{equation}
due to \eqref{L} and \eqref{n-vector}---the allowed region $\mathcal{E}$ is enclosed by these tangent planes for all $\theta$'s and $\phi$'s.
For each ${(\theta,\phi)}$, the intersection
\begin{equation}
\label{face}
\mathbb{H}(\theta,\phi)\cap\mathcal{E}=:\mathcal{F}(\theta,\phi)
\end{equation}
is called a proper \emph{face} of $\mathcal{E}$.
The boundary of $\mathcal{E}$ is the union of these faces:
\begin{equation}
\label{boundary}
\partial\mathcal{E}=\bigcup_{\theta,\phi}\,\mathcal{F}(\theta,\phi)\,.
\end{equation}
Since $\mathcal{E}$ is a compact and convex set, 
$\mathcal{F}$ is also a compact and convex set, and the set of all 
extreme points 
\begin{equation}
\label{ext-points}
\text{ext}(\mathcal{E})=\bigcup_{\theta,\phi}\,
\text{ext}\big(\mathcal{F}(\theta,\phi)\big)\,.
\end{equation}
For more details on the supporting hyperplane and the faces, we point to \cite{Grunbaum03}.

Corresponding to 
the biggest eigenvalue $\lambda_\textsc{m}(\theta,\phi)$ of 
$\Lambda(\theta,\phi)$,
let us denote the eigenspace by
${\mathscr{E}(\theta,\phi)\subseteq\mathscr{H}_d}$ and
an eigenket by ${|\theta,\phi \rangle_k}$.
Note that every ket in the whole paper represents a normalized vector.
At a particular $(\theta,\phi)$
if $\lambda_\textsc{m}$ is degenerate (that is, the dimension of $\mathscr{E}$ is more than 1) then we generally use a subscript $k$ to differentiate eigenkets.
All the maximum-eigenvalue-states---which are ${\{|\theta,\phi \rangle_k\langle \theta,\phi|\}}$
and all there convex combinations (mixtures)---saturate
inequality \eqref{n.e<lambda} and generate the whole face in \eqref{face}:
\begin{equation}
\label{E-para}
\mathcal{F}(\theta,\phi)=
\text{Conv}\,\big\{\langle \theta,\phi |\vec{E}\,| \theta,\phi \rangle
\ \big|\
|\theta,\phi \rangle\in\mathscr{E}(\theta,\phi)\big\}\,.
\end{equation}
If a face only has one point---that certainly occurs when $\lambda_\textsc{m}$ is nondegenerate and may even occur in a degenerate case [for example, see Sec.~\ref{sec:Jx2 and Jy2}]---then it is called an \emph{exposed}-extreme point.
Degeneracy is a necessary but not sufficient requirement for a face to have more than one distinct points.

In conclusion, we exploit \eqref{n.e<lambda} and \eqref{E-para} for all ${\theta\in[0,\pi]}$ and ${\phi\in[0,2\pi)}$ to completely specify---the allowed region through its boundary---$\partial\mathcal{E}$.
With $\mathcal{E}$ one can provide a definite yes/no answer to the question asked at the beginning of Introduction.
Note that \eqref{n.e<lambda} represents a single necessary condition, but
once we take \eqref{n.e<lambda} for all $\theta$'s and $\phi$'s then they will be sufficient.
Necessary and sufficient restrictions on 
the mean vector $\vec{\boldsymbol{\varepsilon}}$
are collectively called as the \emph{quantum contains} (QCs) in \cite{Sehrawat17b}.

\textit{A side remark:}
If each operator in ${\vec{E}'=(\Gamma_1,\cdots,\Gamma_\textsc{k})}$ is a real linear sum,
${\Gamma_k=T_{kx}X+T_{ky}Y+T_{kz}Z}$, of the operators in $\vec{E}$ then the allowed region $\mathcal{E}'$ for $\vec{E}'$ can be obtained directly from $\mathcal{E}$ by the 
${\textsc{k}\times 3}$ real matrix $T$.
As an example, through the orthogonal projection ${\vec{E}\mapsto(X,Y,0)=\vec{E}'}$
one can have the allowed region for $X$ and $Y$ from $\mathcal{E}$.
Another example, if $T$ is an orthogonal matrix on $\mathbb{R}^3$, then 
$\mathcal{E}$ can be turned into $\mathcal{E}'$ by a composition of a rotation and at most one reflection.

When an operator $X$ has more than two distinct outcomes (eigenvalues), then we need two \emph{independent} real numbers, ${\langle X\rangle}$ and ${\langle X^2\rangle}$, to characterize uncertainty, ${\Delta X=\sqrt{\langle X^2\rangle-\langle X\rangle^2}}$, about its measurement-outcomes. 
In the case of Shannon entropy 
${H(\vec{p}\,)=-\sum_{l=1}^{d}p_l\ln p_l}$, we require ${d-1}$ real numbers from 
${\vec{p}=(p_1,\cdots,p_d)}$, 
where $p_l$ is the probability of getting $l$th outcome in a measurement for $X$.
If $X$ has a degenerate eigenvalue such as in the following sections, then 
there exist infinitely-many distinct projective measurements for $X$, and thus there is no unique probability vector $\vec{p}$ and 
the Shannon entropy $H(\vec{p}\,)$.
However, the mean values ${\langle X\rangle}$, ${\langle X^2\rangle}$,
and thus the standard deviation $\Delta X$ are the same for every measurement for $X$.

Now suppose we only have the mean value ${\langle X\rangle}$, and we want to quantify
uncertainty---about the outcomes of $X$---using it.
In this situation, we can build uncertainty and certainty measures as described in \cite{Sehrawat17b}:
For an operator $X$, provided ${x_\text{m}\neq\,x_\textsc{m}}$, first we construct two positive semi-definite operators 
\begin{equation}
\label{Xdot}
\dot{X}:=\frac{x_\textsc{m}\,I-X}{x_\textsc{m}-x_\text{m}}
\quad\text{and}\quad
\mathring{X}:=\frac{X-x_\text{m}\,I}{x_\textsc{m}-x_\text{m}}
\end{equation}
such that ${\dot{X}+\mathring{X}}$ is the identity operator $I$.
Both ${\langle \dot{X}\rangle,\langle \mathring{X}\rangle\in[0,1]}$ are functions of 
${\langle X\rangle\in[x_\text{m}\,,\,x_\textsc{m}]}$ only.
Then we define certain concave and convex functions of ${\langle X\rangle}$
\begin{eqnarray}
\label{h(X)}
h(\langle X\rangle)&:=&-(\langle \dot{X}\rangle\ln\,\langle \dot{X}\rangle+\langle \mathring{X}\rangle\ln\,\langle \mathring{X}\rangle)\,,
\\
\label{u(X)}
u_\kappa(\langle X\rangle)&:=&{\langle \dot{X}\rangle}^\kappa+{\langle \mathring{X}\rangle}^\kappa\,,\quad 0<\kappa<\infty\,,
\quad\text{and}\qquad\\
\label{umax(X)}
u_\text{max}(\langle X\rangle)&:=&\max\,\{\,
\langle \dot{X}\rangle\,,\,\langle \mathring{X}\rangle\,\}\,.
\end{eqnarray}
The concave functions
${h\in[0,\ln2]}$ and ${u_\kappa\in[1,2^{1-\kappa}]}$ for ${0<\kappa<1}$ 
can be treated as uncertainty measures,
and 
the convex functions ${u_\kappa\in[2^{1-\kappa},1]}$ for ${1<\kappa<\infty}$ and ${u_\text{max}\in[\tfrac{1}{2},1]}$
will be certainty measures.

Note that, like the standard deviation $\Delta X$ and Shannon entropy $H(\vec{p}\,)$,
the above uncertainty measures do not reach their minimum values when the system is in an eigenstate of $X$ unless the state corresponds to an extreme eigenvalue $x_m$
or $x_\textsc{m}$. Naturally, if one can access only the mean value then she can be certain only in the two situations  
${\langle X\rangle=x_m}$ and 
${\langle X\rangle=x_\textsc{m}}$.
So, \eqref{h(X)}--\eqref{umax(X)} are (un)certainty measures about the outcomes 
of $X$ given only the mean value ${\langle X\rangle}$, not the state $\rho$, and they are
different from $\Delta X$ and $H(\vec{p}\,)$
particularly when $X$ has more than two distinct outcomes.

Since the addition of concave functions is also a concave
function,
for example,
\begin{eqnarray}
\label{u-half-sum}
u_{\sfrac{1}{2}}(\vec{\boldsymbol{\varepsilon}}\,):=
u_{\sfrac{1}{2}}(\langle X\rangle)+
u_{\sfrac{1}{2}}(\langle Y\rangle)+
u_{\sfrac{1}{2}}(\langle Z\rangle)
\end{eqnarray}
acts as a combined uncertainty measure on the compact and convex set $\mathcal{E}$.
Its global minimum will be at $\text{ext}(\mathcal{E})\subseteq\partial\mathcal{E}$ 
[see \eqref{boundary}--\eqref{E-para}].
Suppose ${u_{\sfrac{1}{2}}(\vec{\boldsymbol{\varepsilon}}\,)}$ reaches its 
absolute minimum, symbolized by $\mathfrak{u_{\sfrac{1}{2}}}$, at some $\theta$ and $\phi$, then the tight UR
${\mathfrak{u_{\sfrac{1}{2}}}\leq u_{\sfrac{1}{2}}(\vec{\boldsymbol{\varepsilon}}\,)}$
is saturated by a subset of kets in the eigenspace ${\mathscr{E}(\theta,\phi)}$.
For every UR based on the above formulation,
its minimum uncertainty kets always lie in 
\begin{equation}
\label{U-of-en-spaces}
\bigcup_{\theta,\phi}\,\mathscr{E}(\theta,\phi)\,.
\end{equation}
In the same way, one can reach the global maximum of a convex function (combined certainty measure) and then enjoy a tight CR with its maximum certainty kets in \eqref{U-of-en-spaces}.

By replacing the average vector 
${\vec{\boldsymbol{\varepsilon}}=(\langle X\rangle,\langle Y\rangle,
\langle Z\rangle)}$ with ${\vec{r}=(x,y,z)}$ in the tight UR
${\mathfrak{u_{\sfrac{1}{2}}}\leq u_{\sfrac{1}{2}}(\vec{\boldsymbol{\varepsilon}}\,)}$,
one can specify a region 
\begin{equation}
\label{R_u}
\mathcal{R}_{u_{\sfrac{1}{2}}}:=
\big\{\vec{r}\in\mathcal{H}
\ |\ 
\mathfrak{u_{\sfrac{1}{2}}}\leq u_{\sfrac{1}{2}}(\vec{r}\,)
\big\}\,,
\end{equation}
in hyperrectangle~\eqref{hyperrectangle}.
$\mathcal{R}$---bounded by a UR or CR---is a convex set in the space of expectation values.
Obviously $\mathcal{R}$ contains the allowed region $\mathcal{E}$, and
there exists no quantum state $\rho$ for any point outside $\mathcal{E}$, for instance, 
in the relative complement ${\mathcal{R}\backslash\mathcal{E}}$.
The boundary $\partial\mathcal{E}$ touches the boundary $\partial\mathcal{R}_{u_{\sfrac{1}{2}}}$ at those points that come from the minimum uncertainty states associated with the UR [for example, see Fig.~\ref{fig:regions}].

An uncertainty measure such as ${u_{\sfrac{1}{2}}(\vec{\boldsymbol{\varepsilon}}\,)}$ 
reaches its trivial lower bound $3$---which is just the sum of individual lower bounds of ${u_{\sfrac{1}{2}}(\langle X\rangle), u_{\sfrac{1}{2}}(\langle Y\rangle)},$ and ${u_{\sfrac{1}{2}}(\langle Z\rangle)}$---if and only if a vertex ${(x_{\text{m},\textsc{m}}\,,\,y_{\text{m},\textsc{m}}\,,\,z_{\text{m},\textsc{m}})}$
of $\mathcal{H}$ lies in the numerical range $\mathcal{E}$.
This will happen when there exist a common eigenket ${|\textsf{e}\rangle}$ \emph{that corresponds to an extreme eigenvalue of every operator} in ${\vec{E}=(X,Y,Z)}$.
It can also happen in a limiting case where $\mathcal{E}$ shares a few corners of $\mathcal{H}$ in a limit, say, ${j\rightarrow\infty}$ [see Secs.~\ref{sec:Jx2 and Jy2}, \ref{sec:S-power}, and \ref{sec:N-qubit}].
In such cases, all URs and CRs based on \eqref{Xdot}--\eqref{u-half-sum} will become trivial, and their regions ${\mathcal{R}=\mathcal{H}}$.

Since the standard deviation $\Delta$ and $h$ of \eqref{h(X)} are different functions,
URs based on them are difficult to compare.
Nevertheless, suppose both 
${\boldsymbol{\delta}\leq\Delta X+\Delta Y}$ and 
${\mathfrak{h}\leq 
h(\langle X\rangle)+h(\langle Y\rangle)}$ 
are \emph{tight} URs for the two non-commuting operators $X$ and $Y$,
where ${\boldsymbol{\delta}\geq 0}$ and ${\mathfrak{h}\geq 0}$ depend on the operators but not on the state $\rho$.
\emph{If and only if}
${\boldsymbol{\delta}=0}$ then there exists a state (common eigenstate) 
where both $X$ and $Y$ have no spread in their measurement-outcomes.
We cannot make such a statement regarding ${\mathfrak{h}}$ because
there can be a situation [see Sec.~\ref{sec:Jx2 and Jy2}] where $X$ and $Y$
share a common eigenket (${\boldsymbol{\delta}=0}$) but it does not correspond to their extreme eigenvalues (${\mathfrak{h}>0}$).
So ${\mathfrak{h}=0}$ implies ${\boldsymbol{\delta}=0}$ but not vice versa.
Clearly our URs and CRs based on \eqref{Xdot}--\eqref{u-half-sum} are different from the URs based on the standard deviation \cite{Robertson29,Hofmann03,Maccone14,Rivas08,Dammeier15,Schwonnek17} and on the Shannon entropy \cite{Maassen88,Sanchez-Ruiz93,Riccardi17}.

%===========================================
\section{Angular momentum operators}\label{sec:spin-ops}

One can describe the three 
angular momentum operators
\begin{equation}
\label{Jx,y,z}
J_x=\tfrac{1}{2}(J_++J_-)\,,\quad
J_y=\tfrac{1}{2\text{i}}(J_+-J_-)\,,
\ \mbox{and}\
J_z\quad
\end{equation}
as
\begin{eqnarray}
\label{Jpm}
J_\pm\,|m\rangle&=&
\sqrt{(j\mp m)(j\pm m+1)}\;
|m\pm 1\rangle
\quad\mbox{and}\qquad\\
\label{Jz}
J_z\,|m\rangle&=&m\,|m\rangle\,, 
\end{eqnarray}
where ${\text{i}=\sqrt{-1}\,}$,
\begin{equation}
\label{basis-Jz}
\mathfrak{B}_z=
\big\{\,|m\rangle \;\big|\; m=j\,,\,{j-1}\,,\,\cdots\,,\,{-j+1}\,,\,-j\,\big\}
\end{equation}
is an eigenbasis of $J_z$, and
$j$ can be ${\tfrac{1}{2}, 1,\tfrac{3}{2},2,\cdots\,}$.
In the paper, every operator is dimensionless, and each ket is represented in the basis $\mathfrak{B}_z$.
As the quantum number $m$ can only acquire ${2j+1}$ discrete values for a fixed $j$,
here our system of interest is of ${d=2j+1}$ levels.

To obtain the permissible region $\mathcal{E}$ for 
${\vec{J}=(J_x,J_y,J_z)}$,
we begin with
\begin{equation}
	\label{a1a2a3-v1v2v3}
	\begin{pmatrix}
	J'_x\\
	J'_y\\
	J'_z
	\end{pmatrix}
	=
	\begin{pmatrix}
\cos\theta\cos\phi & \cos\theta\sin\phi & -\sin\theta  \\
	-\sin\phi & \cos\phi & 0 \\
\sin\theta\cos\phi & \sin\theta\sin\phi & \cos\theta  
	\end{pmatrix}
	\begin{pmatrix}
	J_x\\
	J_y\\
	J_z
	\end{pmatrix}.
\end{equation}
One can show that the components of ${\vec{J}'=(J'_x,J'_y,J'_z )}$ follow the same commutation relations
that $J_x,J_y,$ and $J_z$ obey, for example, ${J_xJ_y-J_yJ_x=\text{i}J_z}$.
Hence, ${J'_x,J'_y,}$ and $J'_z$ are also angular momentum operators, and their spectrum ${j,\cdots,-j}$ is independent of $\theta$ and $\phi$.

In fact, $J'_z$ is our ${\Lambda(\theta,\phi)}$ here, and 
the angular momentum coherent state-vector \cite{Atkins71,Arecchi72}
\begin{equation}
\label{bloch-ket}
|\theta,\phi\rangle=\sum_{m=-j}^{j}
{\scriptstyle\sqrt{\tfrac{(2j)!}{(j+m)!\,(j-m)!}}
	\left(\cos\tfrac{\theta}{2}\right)^{j+m}
	\left(\sin\tfrac{\theta}{2}\right)^{j-m}
	e^{-\text{i}m\phi}}|m\rangle
\end{equation}
is its eigenket corresponding to ${\lambda_\textsc{m}=j}$. 
Furthermore,
\begin{equation}
\label{JxJyJz-para}
\langle\theta,\phi|\vec{J}\,|\theta,\phi\rangle= j\, \widehat{\eta}(\theta,\phi)
\end{equation}
generate the boundary of $\mathcal{E}$, which is a closed
ball classified by
\begin{equation}
\label{JxJyJz,QC}
{\langle J_x\rangle}^2+{\langle J_y\rangle}^2+{\langle J_z\rangle}^2\leq j^2\,.
\end{equation}
Here every face ${\mathcal{F}(\theta,\phi)}$ of $\mathcal{E}$ [see \eqref{E-para}] is made of a single exposed-extreme point \eqref{JxJyJz-para}.
For the three momentum observables, URs based on the standard deviation and entropy are achieved in \cite{Sanchez-Ruiz93,Hofmann03,Rivas08,Dammeier15,Riccardi17}.
While in \cite{Sehrawat17b}, tight URs and CRs based on \eqref{Xdot}--\eqref{u-half-sum} 
are derived by employing QC~\eqref{JxJyJz,QC}, where a maximum-certainty or minimum-uncertainty ket is a coherent ket \eqref{bloch-ket}.

%===========================================
\section{Powers of ladder operators}\label{sec:J+^2 and J-^2}

The ladder operators ${J_\pm}$ [see \eqref{Jpm}] are non-Hermitian operators
such that ${J_\pm^\dagger=J_\mp}$, whereas
\begin{eqnarray}
\label{Lgamma}
\Lambda_\gamma(\phi)&=&
e^{\text{i}\phi}J_+^{\,\gamma}+e^{-\text{i}\phi}J_-^{\,\gamma}
={\cos\phi\, X_\gamma+ \sin\phi\, Y_\gamma}\,,
\quad\ \\
\label{ABgamma}
X_\gamma&=&\big(J_+^{\,\gamma}+J_-^{\,\gamma}\big)\,,
\quad \mbox{and}\quad
Y_\gamma=\text{i}\,\big(J_+^{\,\gamma}-J_-^{\,\gamma}\big)
\end{eqnarray}
are Hermitian operators for every angle ${\phi\in[0,2\pi)}$
and the power ${\gamma=1,\cdots,d}$. For ${\gamma=1}$,
${X_\gamma=2J_x}$ and ${Y_\gamma=-2J_y}$, and they are the null operator
when ${\gamma=d}$.

It is known due to \cite{Byrd03,Kimura03} that the characteristic equation of an operator, say, $\Lambda$ on a $d$-dimensional space is 
\begin{equation}
\label{c-eq}
\sum_{l=0}^{d}(-1)^{l} S_l\,\lambda^{d-l}=0\,,
\end{equation}
where
\begin{eqnarray}
&&
\label{S1}
S_0=1\,,\quad
S_1=\text{tr}(\Lambda)\,,
\quad\text{and}\qquad\\
&&
\label{Sn}
S_l=\tfrac{1}{l}
\sum_{i=1}^{l}(-1)^{i-1}\,
\text{tr}(\Lambda^{i})\,S_{l-i}
\end{eqnarray}
are the symmetric functions of its eigenvalues.
By showing that $\text{tr}(\Lambda^{i})$ is independent of $\phi$ for every ${i=1,\cdots,d}$, one can certify that $S_l$ ${(l=1,\cdots,d)}$, characteristic equation \eqref{c-eq}, and all the
eigenvalues of ${\Lambda_\gamma(\phi)}$ are also independent of $\phi$.
As a result, inequalities \eqref{L-exp} and \eqref{n.e<lambda} here become
\begin{equation}
\label{ABgamma-plane}
-\lambda_\textsc{m}\leq
\cos\phi\,\langle X_\gamma\rangle +
\sin\phi\, \langle Y_\gamma\rangle 
\leq\lambda_\textsc{m}\,,
\end{equation}
where $\lambda_\textsc{m}$ is the largest eigenvalue of ${X_\gamma}$ (and of ${Y_\gamma}$) for a given $\gamma$; see Table~\ref{tab:lmax}.
Inequalities \eqref{ABgamma-plane} tell that the average value of
${\Lambda_\gamma(\phi)}$ must be in a fixed interval
${[-\lambda_\textsc{m},\lambda_\textsc{m}]}$
in every direction $\phi$, which implies 
\begin{equation}
\label{AB-QC}
\vec{\boldsymbol{\varepsilon}}\cdot\vec{\boldsymbol{\varepsilon}}\, 
\leq\lambda_\textsc{m}^2
\end{equation}
for $\vec{\boldsymbol{\varepsilon}}=(\langle X_\gamma\rangle,
\langle Y_\gamma\rangle)$.
Hence, the allowed region $\mathcal{E}$ of the mean vectors is the closed disk bounded by QC \eqref{AB-QC}.
By the same reasoning: whenever $\lambda_\textsc{m}(\theta,\phi)$ 
is the same for all the angles $\theta$'s and $\phi$'s,
then $\mathcal{E}$ is enclosed by a sphere of radius $\lambda_\textsc{m}$ centered at the origin [for instance,
see \eqref{JxJyJz,QC}, \eqref{AB-QC}, \eqref{A1A2-QC}, and \eqref{QC-A (3/2)}].

\begin{table}[]
	\centering
	\caption{For ${\gamma=2,3,}$ and 4, the greatest eigenvalues $\lambda_\textsc{m}$ of ${\Lambda_\gamma(\phi)}$ [of \eqref{Lgamma}] corresponding to different $j$-values are listed in the three columns on the right-hand-side. 
	Since the spectrum of ${\Lambda_\gamma(\phi)}$ does not depend on $\phi$,
	$\lambda_\textsc{m}$ is also an extreme eigenvalue of $X_\gamma$ as well as of $Y_\gamma$.
	The associated eigenkets ${|\phi\rangle_k}$ are arranged in 
	Tables~\ref{tab:lmax-kets gamma=2} and \ref{tab:lmax-kets gamma=3,4}
	in Appendix~\ref{sec:Sup-material}.
	}
	\label{tab:lmax}
	\begin{tabular}{c@{\hspace{2mm}} | @{\hspace{2mm}}c@{\hspace{6mm}}c@{\hspace{6mm}} c }
		\hline\hline
		\rule{0pt}{2ex}  
		$j$ & for $\gamma=2$ &  for $\gamma=3$ &  for $\gamma=4$ \\
		\hline\rule{0pt}{3ex} 	
$\tfrac{1}{2}\;$ & 0               & 0  & 0   \\[1mm]
$1$              & $2$ 	           & 0  & 0   \\[1mm]
$\tfrac{3}{2}$   & $2\sqrt{3}$     & 6  & 0   \\[1mm]
$2$              & ${4\sqrt{3}}$   & 12 & 24  \\[1mm]		
$\tfrac{5}{2}$   & ${4\sqrt{7}}$                &  24   & $24\sqrt{5}$ 
\\[1mm]	
$3$              & ${2\,(3+2\sqrt{6})}$         & $12\sqrt{10}$  & $120$ 
\\[1mm]	
$\tfrac{7}{2}$   & ${2\sqrt{3\,(21+4\sqrt{21})}}$  & $6\sqrt{115}$ & $120\sqrt{3}$ 
\\[1mm]	
$4$              & ${8\sqrt{13}}$                  & $12\sqrt{70}$ & ${360}$ 
\\[1.2mm]	
		\hline\hline
	\end{tabular}
\end{table}

Since ${J_\pm^{\,\gamma}\,|m\rangle}$ is directly proportional to
${|{m\pm\gamma}\rangle}$,
we can divide basis \eqref{basis-Jz} into $\gamma$ number of disjoint subsets 
in such a way that the linear span of a subset delivers an invariant subspace
\begin{equation}
\label{Inv-subspace}
\mathscr{S}^\gamma_k:=\text{span}\big\{ |{j-k}\rangle\,, |{j-k-\gamma}\rangle\,, 
|{j-k-2\gamma}\rangle,
\cdots\big\}
\end{equation}
of $J_\pm^{\,\gamma}$ and thus of $\Lambda_\gamma(\phi)$.
So, one can search eigenvector(s) ${|\phi\rangle}$ associated with $\lambda_\textsc{m}$ of ${\Lambda_\gamma(\phi)}$ in
these subspaces rather than in the whole Hilbert space $\mathscr{H}_d=\oplus_{k=0}^{\gamma-1}\mathscr{S}^\gamma_k$, 
where $\oplus$ symbolizes the direct sum.
In Appendix~\ref{sec:Sup-material}, Tables~\ref{tab:lmax-kets gamma=2} and
\ref{tab:lmax-kets gamma=3,4} carry
these eigenkets for $\gamma=2,3,$ and 4.
When $\lambda_\textsc{m}$ is twofold degenerate then two orthonormal eigenkets ${|\phi\rangle_k}$ and ${|\phi\rangle_{k'}}$ are registered in the tables, 
where the subscript $k$ illustrates ${|\phi\rangle_k\in\mathscr{S}^\gamma_k}$.

In the case of double-degeneracy, a general eigenket will be a superposition
\begin{equation}
\label{general-phi-ket}
|\phi \rangle=
\cos\mu\,|\phi\rangle_k+\sin\mu\,e^{\text{i}\nu}|\phi\rangle_{k'}
\in\mathscr{E}(\phi)\,,
\end{equation}
where $\mathscr{E}(\phi)$ is the eigenspace,
${\mu\in[0,\tfrac{\pi}{2}]}$, and ${\nu\in[0,2\pi)}$.
When we compute ${\langle \phi |X_\gamma|\phi \rangle}$ using \eqref{general-phi-ket}, then we encounter cross-terms such as
${{}_k\langle \phi |X_\gamma|\phi \rangle_{k'}}$ that all will be zero here.
It is because ${X_\gamma|\phi \rangle_{k'}}$ and $|\phi \rangle_k$ lie in mutually orthogonal invariant subspaces $\mathscr{S}^\gamma_{k'}$ and $\mathscr{S}^\gamma_k$, respectively, of $X_\gamma$.
Therefore, every
${\langle \phi |\vec{E}\,|\phi \rangle}$ will fall on a line segment
connecting the two extreme points 
\begin{equation}
\label{line-smt}
{}_k\langle \phi |\vec{E}\,|\phi \rangle_k
 \quad\mbox{and}\quad
{}_{k'}\langle \phi |\vec{E}\,|\phi \rangle_{k'}\,;
\end{equation}
${\vec{E}=(X_\gamma,Y_\gamma)}$ in this section.
The line segment forms a face $\mathcal{F}(\phi)$ [as per \eqref{E-para}]
of the permitted region $\mathcal{E}$.
If the degree of degeneracy is three, then
$\mathcal{F}(\phi)$ will be the convex hull of
three extreme points such as \eqref{line-smt}, and so on.
Here, both points in \eqref{line-smt} turn out to be the same, hence
every eigenket associated with $\lambda_\textsc{m}(\phi)$ produces a single extreme point 
\begin{equation}
\label{E-para Jpm gamma}
\langle \phi |\vec{E}\,|\phi \rangle=\lambda_\textsc{m}\,(\cos\phi\,,\,\sin\phi)
=\lambda_\textsc{m}\,\widehat{\eta}(\phi)
\end{equation}
of $\mathcal{E}$.
Taking Tables~\ref{tab:lmax}, \ref{tab:lmax-kets gamma=2}, and \ref{tab:lmax-kets gamma=3,4}, one can verify
\eqref{E-para Jpm gamma}.
Indeed, points \eqref{E-para Jpm gamma} saturate QC \eqref{AB-QC} and create the boundary ${\partial\mathcal{E}}$.

\begin{figure}[ ]
	\centering
	\includegraphics[width=0.2\textwidth]{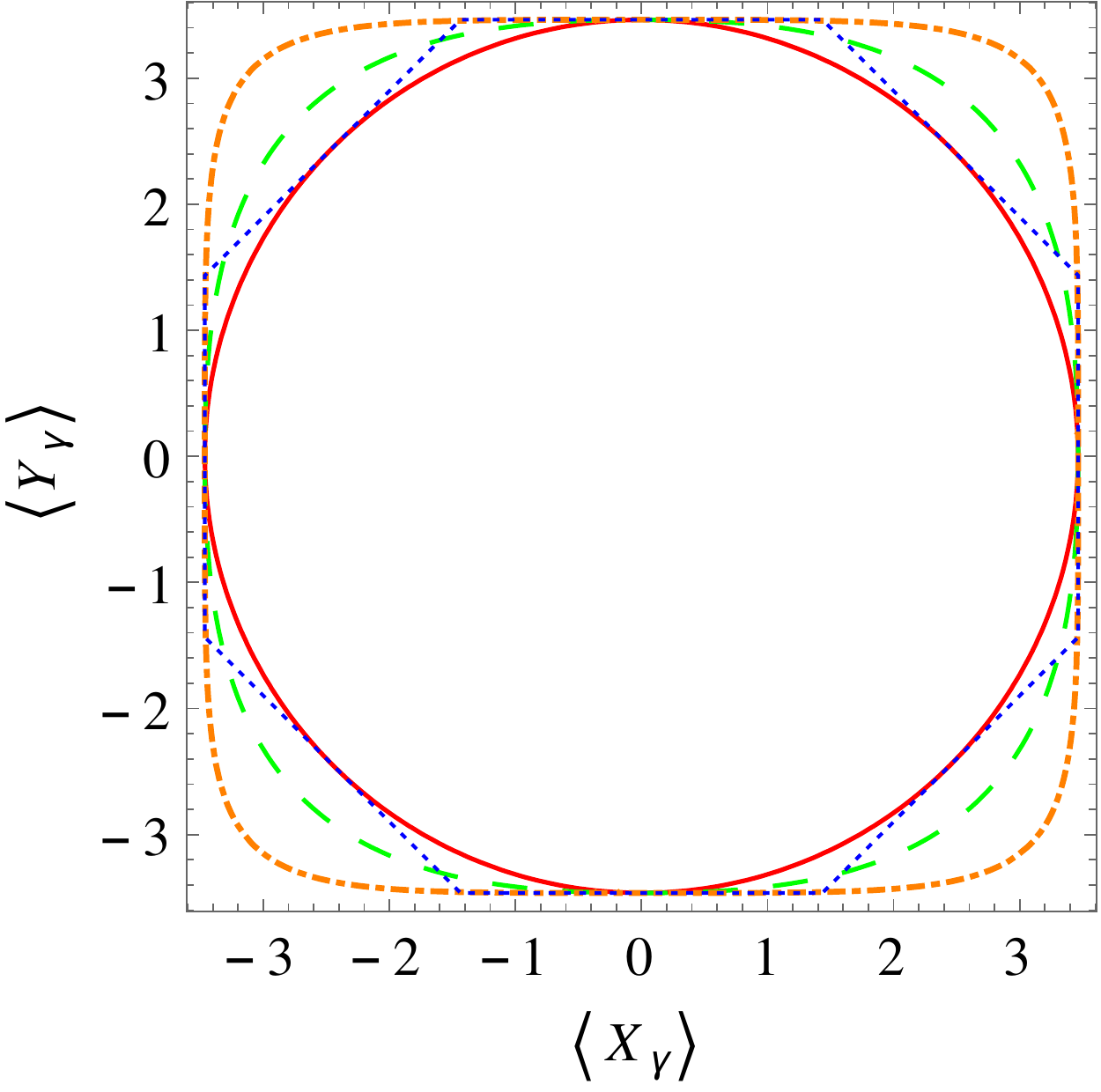} 
	\caption{(Color online)
		In the case ${j=\tfrac{3}{2}}$ and ${\gamma=2}$, ${\lambda_\textsc{m}=2\sqrt{3}}$, hyperrectangle~\eqref{hyperrectangle} is the square $[-\lambda_\textsc{m},\lambda_\textsc{m}]^{\times 2}$, and the regions $\mathcal{R}$'s are defined with respect to \eqref{R_u}. The boundaries of
		regions $\mathcal{R}_h,\mathcal{R}_{u_{\sfrac{1}{2}}},\mathcal{R}_{u_2},$ and $\mathcal{R}_{u_\text{max}}$ determined by \eqref{H-UR-ABgamma}--\eqref{umax-CR-ABgamma} are depicted through the 
		dashed, dot-dashed, solid, and dotted (green, orange, red, and blue) closed-curves. Here, ${\mathcal{R}_{u_2}=\mathcal{E}}$, which is clearly contained in all the other regions, and  ${\mathcal{R}_h\subset\mathcal{R}_{u_{\sfrac{1}{2}}}}$. 
	}
	\label{fig:regions}  
\end{figure}

Now, following the procedure laid out at the end of Sec.~\ref{sec:all-region},
we achieve tight relations
\begin{eqnarray}
\label{H-UR-ABgamma}
\ln 2&\leq&h(\vec{\boldsymbol{\varepsilon}}\,)=h(\langle X_\gamma\rangle)+h(\langle Y_\gamma\rangle)\,,\\
\label{u-UR-ABgamma}
1+\sqrt{2}&\leq&
u_{\sfrac{1}{2}}(\vec{\boldsymbol{\varepsilon}}\,)\,,\\
\label{u2-CR-ABgamma}
&& u_2(\vec{\boldsymbol{\varepsilon}}\,)\leq \tfrac{3}{2} \,,
\quad \mbox{and}\qquad\\
\label{umax-CR-ABgamma}
&& u_\text{max}(\vec{\boldsymbol{\varepsilon}}\,)\leq 
\tfrac{1+\sqrt{2}}{\sqrt{2}} \,,
\end{eqnarray}
which hold in every dimension ${d\geq 2}$ and for every power ${1\leq\gamma<d}$.
At boundary points \eqref{E-para Jpm gamma},
both ${h(\langle \phi |\vec{E}\,|\phi \rangle)}$ and 
${u_{\sfrac{1}{2}}(\langle \phi |\vec{E}\,|\phi \rangle)}$
hit their global minima given in \eqref{H-UR-ABgamma} and \eqref{u-UR-ABgamma}, respectively, at ${\phi=0,\tfrac{\pi}{2},\pi,\tfrac{3\pi}{2}}$.
Hence, for both the URs, minimum uncertainty states are 
the eigenstates of ${\{X_\gamma,Y_\gamma\}}$
corresponding to ${\pm\lambda_\textsc{m}}$.
Similarly, 
${u_\text{max}(\langle \phi |\vec{E}\,|\phi \rangle)}$ reaches its absolute maximum
at ${\phi=\tfrac{\pi}{4},\tfrac{3\pi}{4},\tfrac{5\pi}{4},\tfrac{7\pi}{4}}$,
thus CR \eqref{umax-CR-ABgamma} is saturated by the extreme-eigenvalue-states of ${\tfrac{1}{\sqrt{2}}(X_\gamma\pm Y_\gamma)}$.
Whereas CR \eqref{u2-CR-ABgamma} is the same as QC \eqref{AB-QC}, 
therefore it is \emph{optimal} in the sense \cite{Sehrawat17b} that it provides the smallest region 
$\mathcal{R}_{u_2}=\mathcal{E}$ as shown in Fig.~\ref{fig:regions}. 
Furthermore, \eqref{u2-CR-ABgamma} is saturated by ${|\phi\rangle\langle\phi|}$ for all ${\phi\in[0,2\pi)}$.

%===========================================
\section{Square of angular momentum operators}\label{sec:Jx2 and Jy2}

Squared angular momentum operators obey
\begin{equation}
\label{J^2}
J_x^2+J_y^2+J_z^2=j(j+1)I\,,
\end{equation}
hence we get only two independent real numbers out of the three
$\langle J_x^2\rangle$, $\langle J_y^2\rangle$, and $\langle J_z^2\rangle$.
So, in this section, we present the allowed region $\mathcal{E}$ for
${\vec{E}=(J_x^2,J_y^2)}$ and ${j=\tfrac{1}{2},\cdots,4,}$ and $\infty$ by taking
\begin{equation}
\label{Lphi Jxy2}
\Lambda(\phi)=
\cos\phi\, J_x^2 +\sin\phi\,J_y^2=\widehat{\eta}(\phi)\cdot\vec{E}
\end{equation}
according to \eqref{L}.
Here hyperrectangle~\eqref{hyperrectangle} is the square
\begin{align}
\label{rectangle J^2}
    [\tfrac{1}{4},j^2]^{\times 2}
     \quad &\mbox{or} \quad
    [0,j^2]^{\times 2}\,,\quad\mbox{and}\\
\label{Jdot J^2}
    \dot{J_i^{2}}=\frac{I-\tfrac{J_i^{2}}{j^2}}{1-\tfrac{1}{4j^2}}
    \quad &\mbox{or} \quad 
    \dot{J_i^{2}}=I-\tfrac{J_i^{2}}{j^2}\qquad (i=x,y)
\end{align}
for a half-integer or an integer $j$. 
The operator $\dot{J_i^{2}}$ is constructed with respect to \eqref{Xdot}.

\begin{figure}[ ]
	\centering
	\subfloat{\includegraphics[width=40mm]{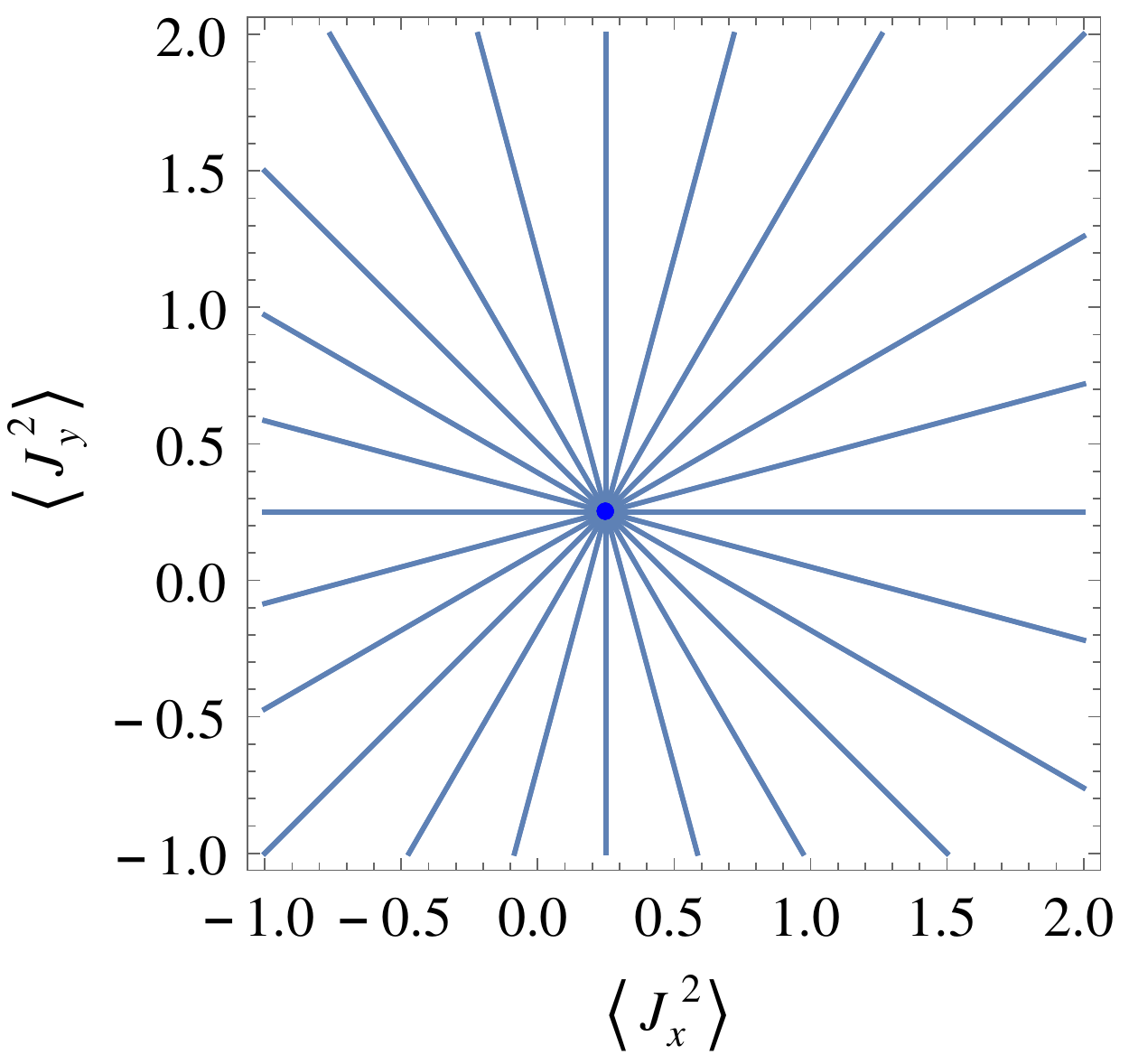}}\quad\
	\subfloat{\includegraphics[width=40mm]{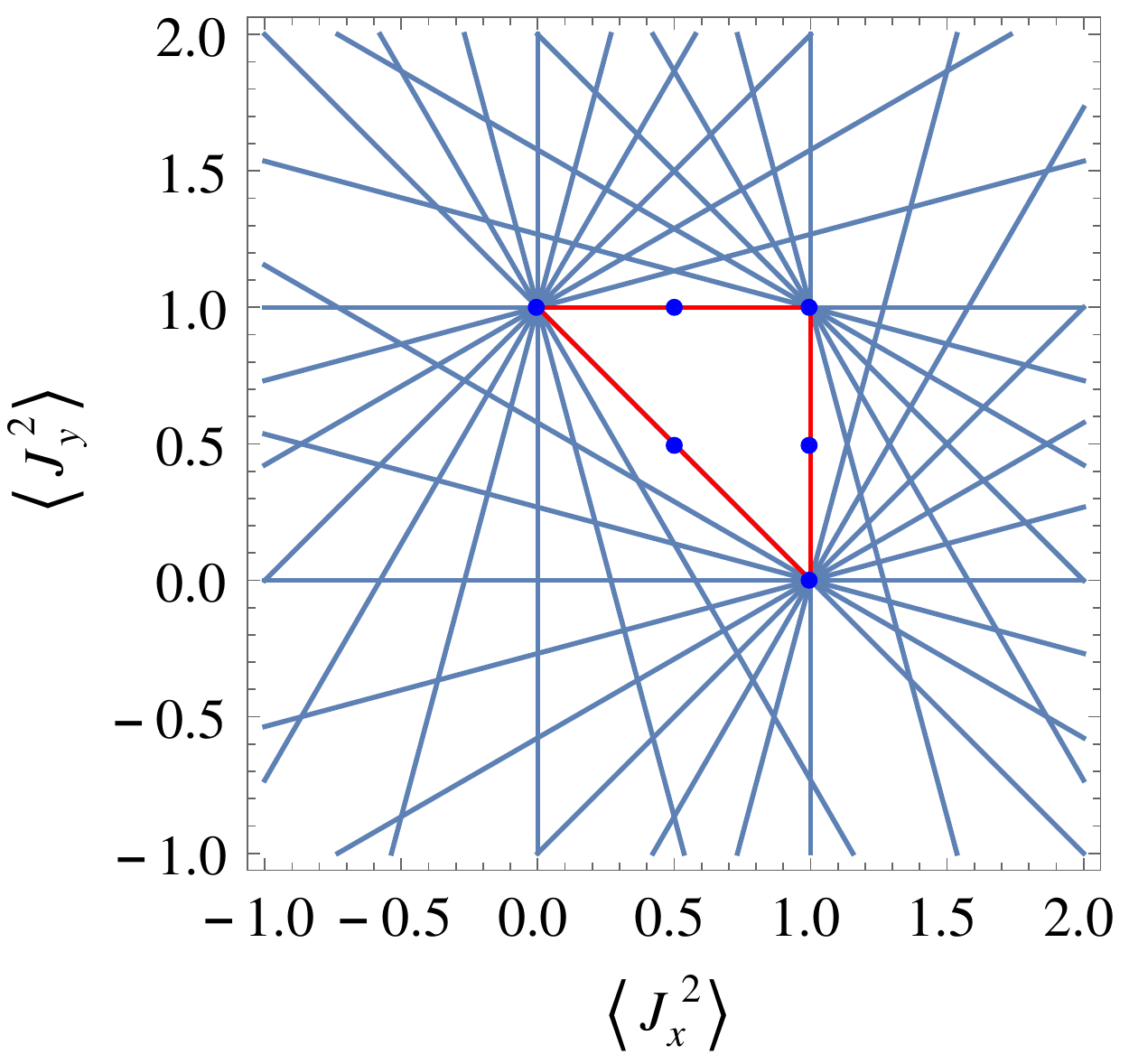}}
	\caption{ (Color online)
		For ${j=\tfrac{1}{2}}$ and 1, the permitted regions $\mathcal{E}$ of
		${(\langle J_x^2\rangle,\langle J_y^2\rangle)}$ are---exhibited in the left- and right-hand-side pictures---the single point ${(\tfrac{1}{4},\tfrac{1}{4})}$ and 
		the triangle whose boundary is shown in red color, respectively.
		The blue lines are the supporting hyperplanes \eqref{hyperplane}
		of $\mathcal{E}$ at $\phi=0,\tfrac{2\pi}{24},\cdots,23\tfrac{2\pi}{24}$.
	}
	\label{fig:E-J^2 for j=1/2and1}  
\end{figure}

In the case of ${j=\tfrac{1}{2}}$, ${J_x^2=J_y^2=\tfrac{1}{4}I}$, thus
$\mathcal{E}$ is just the single point ${(\tfrac{1}{4},\tfrac{1}{4})}$ exhibited in Fig.~\ref{fig:E-J^2 for j=1/2and1}.
For ${j>\tfrac{1}{2}}$, 
$J_x^2$, $J_y^2$, and their linear combinations \eqref{Lphi Jxy2}
have two mutually orthogonal invariant subspaces
$\mathscr{S}^2_0$ and $\mathscr{S}^2_1$ [defined in \eqref{Inv-subspace}] such that
$\mathscr{S}^2_0\oplus\mathscr{S}^2_1=\mathscr{H}_d$.
Like the previous section, we shall look for eigenkets of ${\Lambda(\phi)}$ in the
subspaces for a given ${j=\tfrac{d-1}{2}}$.

For ${j=1}$, $J_x^2$ and $J_y^2$ are rank-2 projectors that commute with each other, and
their common eigenbasis is the union of
\begin{equation}
|\textsf{e}_{1,3}\rangle=\tfrac{1}{\sqrt{2}}\big(\,|{+1}\rangle\pm|{-1}\rangle\,\big)\in\mathscr{S}^2_0
\quad \text{and} \quad
|\textsf{e}_2\rangle=|0\rangle\in\mathscr{S}^2_1\,.
\end{equation}
In this case, $\mathcal{E}$ is the convex hull of three points
${(1,0)}$, ${(1,1)}$, and ${(0,1)}$ as per \eqref{conv-hull}.
It means that $\mathcal{E}$
is a triangle displayed in Fig.~\ref{fig:E-J^2 for j=1/2and1}. 
The maximum eigenvalue
\begin{equation}
\label{lmax-J^2 (1)}
\lambda_\textsc{m}(\phi)
=
\max\,\{\cos\phi\,,\, \sin\phi \,,\, \cos\phi+\sin\phi\}
\end{equation}
of ${\Lambda(\phi)}$ becomes twofold degenerate at ${\phi=0,\tfrac{\pi}{2},}$
and $\tfrac{5\pi}{4}$.
Sides of the triangle are in fact three faces
$\mathcal{F}(\phi)$ of $\mathcal{E}$ at these angles, that is, they
are produced by the maximum-eigenvalue-states of $J_x^2$, $J_y^2$, and $J_z^2$.
One can realize that 
\begin{equation}
\label{lambda-5pi/4}
\Lambda(\tfrac{5\pi}{4})=\tfrac{1}{\sqrt{2}}[J_z^2-j(j+1)I]=
-\Lambda(\tfrac{\pi}{4})
\end{equation}
through \eqref{J^2} and \eqref{Lphi Jxy2}. 
As $J_x^2$ and $J_y^2$ commute in the case of ${j=\tfrac{1}{2},1}$, a UR or CR for them will be a trivial inequality.

\begin{figure}[ ]
	\centering
	\subfloat{\includegraphics[width=40mm]{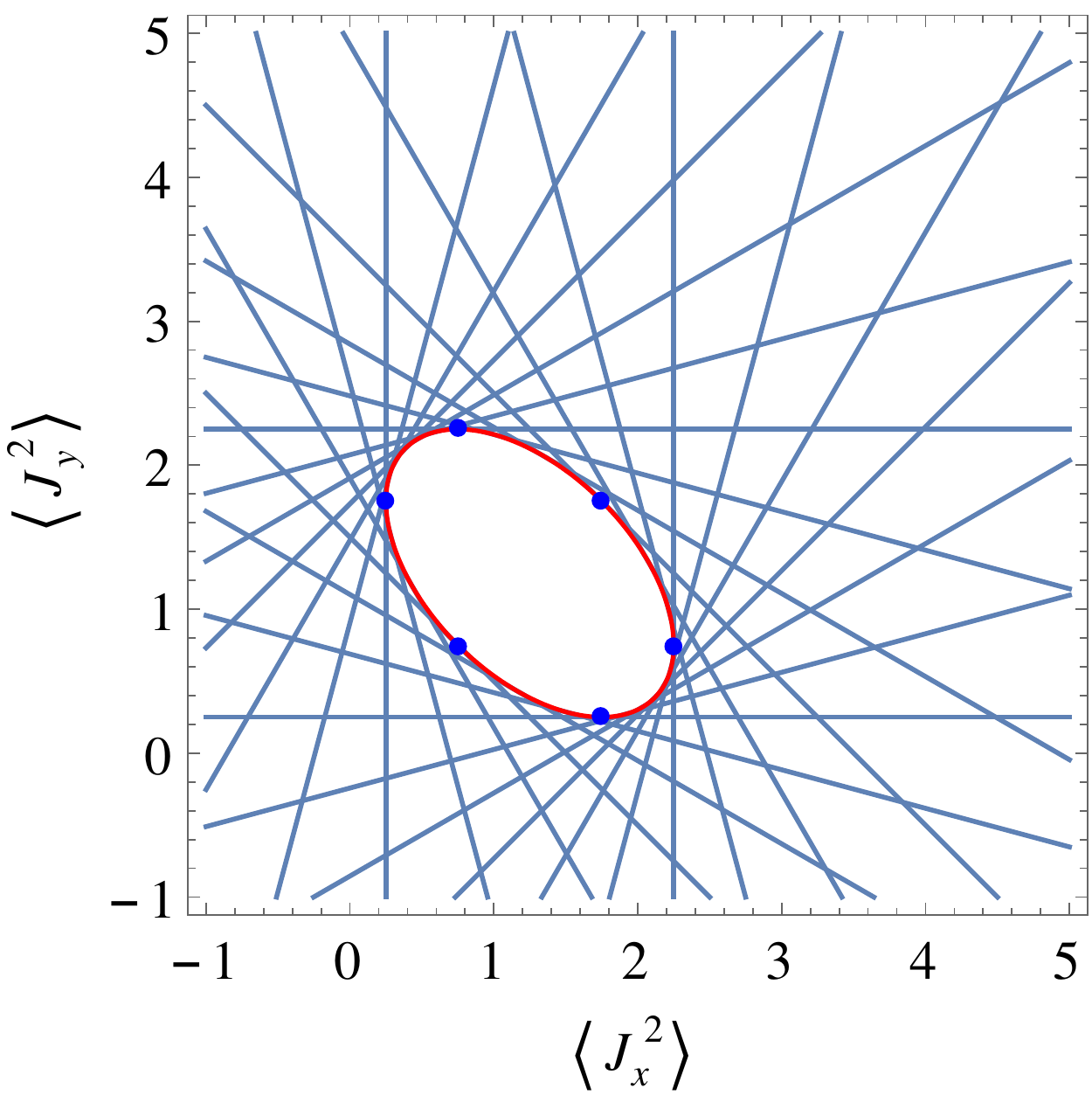}}\quad\
	\subfloat{\includegraphics[width=40mm]{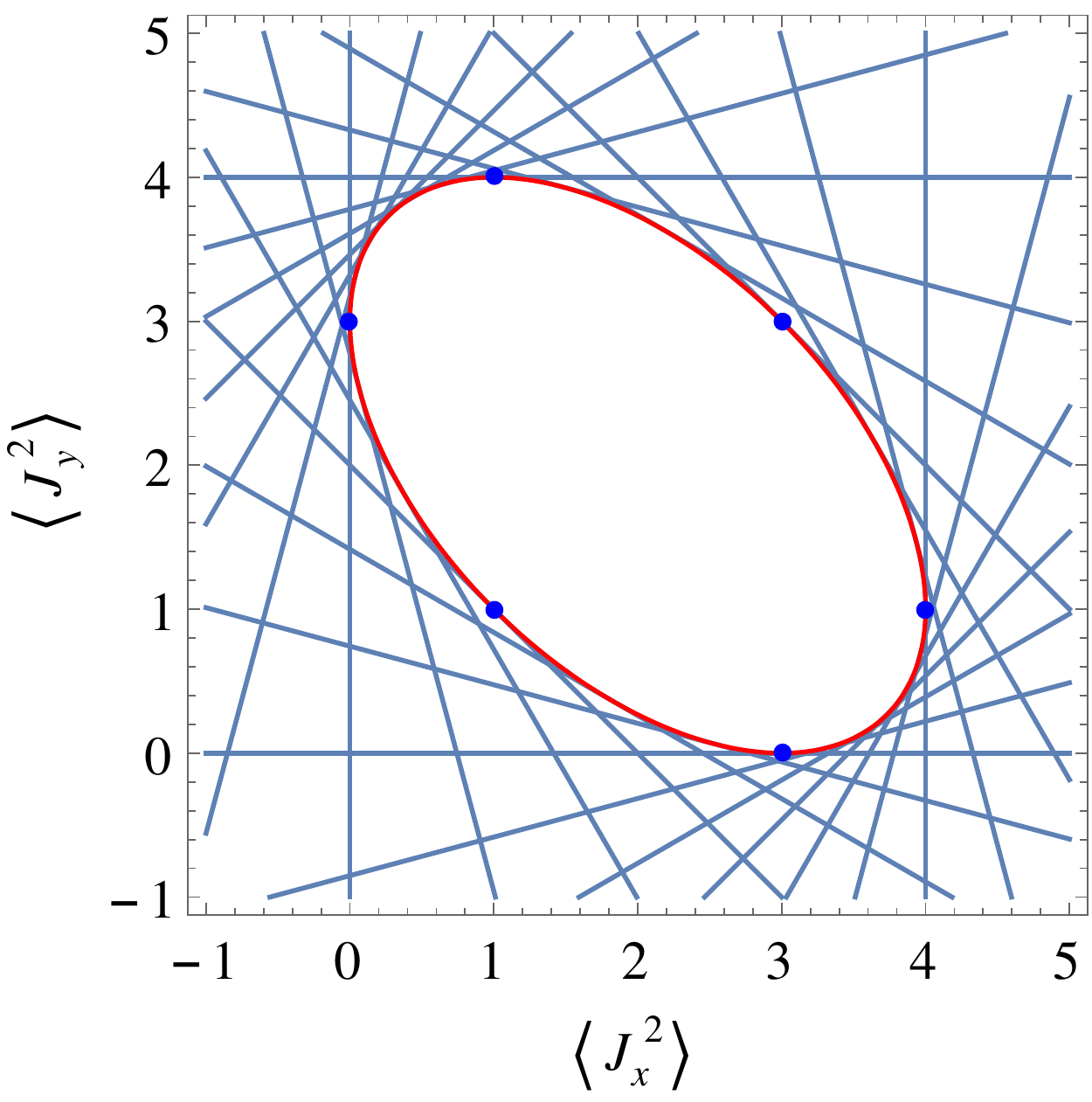}}
	\caption{(Color online)
	The left and right red ellipses with their interiors represent the allowed region $\mathcal{E}$ of ${\vec{\boldsymbol{\varepsilon}}=(\langle J_x^2\rangle,\langle J_y^2\rangle)}$
	in the case of ${j=\tfrac{3}{2}}$ and ${j=2}$, respectively.
	The points ${(1,1)}$, ${(4,1)}$, and ${(1,4)}$ on the right-hand-side ellipse come from the common eigenkets ${|\textsf{e}_1\rangle}$, ${|\textsf{e}_2\rangle}$, and ${|\textsf{e}_3\rangle}$, respectively, given in \eqref{e1 (2)}.
	}
	\label{fig:E-J^2 for j=3/2and2}  
\end{figure}

\begin{figure}[ ]
	\centering
	\subfloat{\includegraphics[width=40mm]{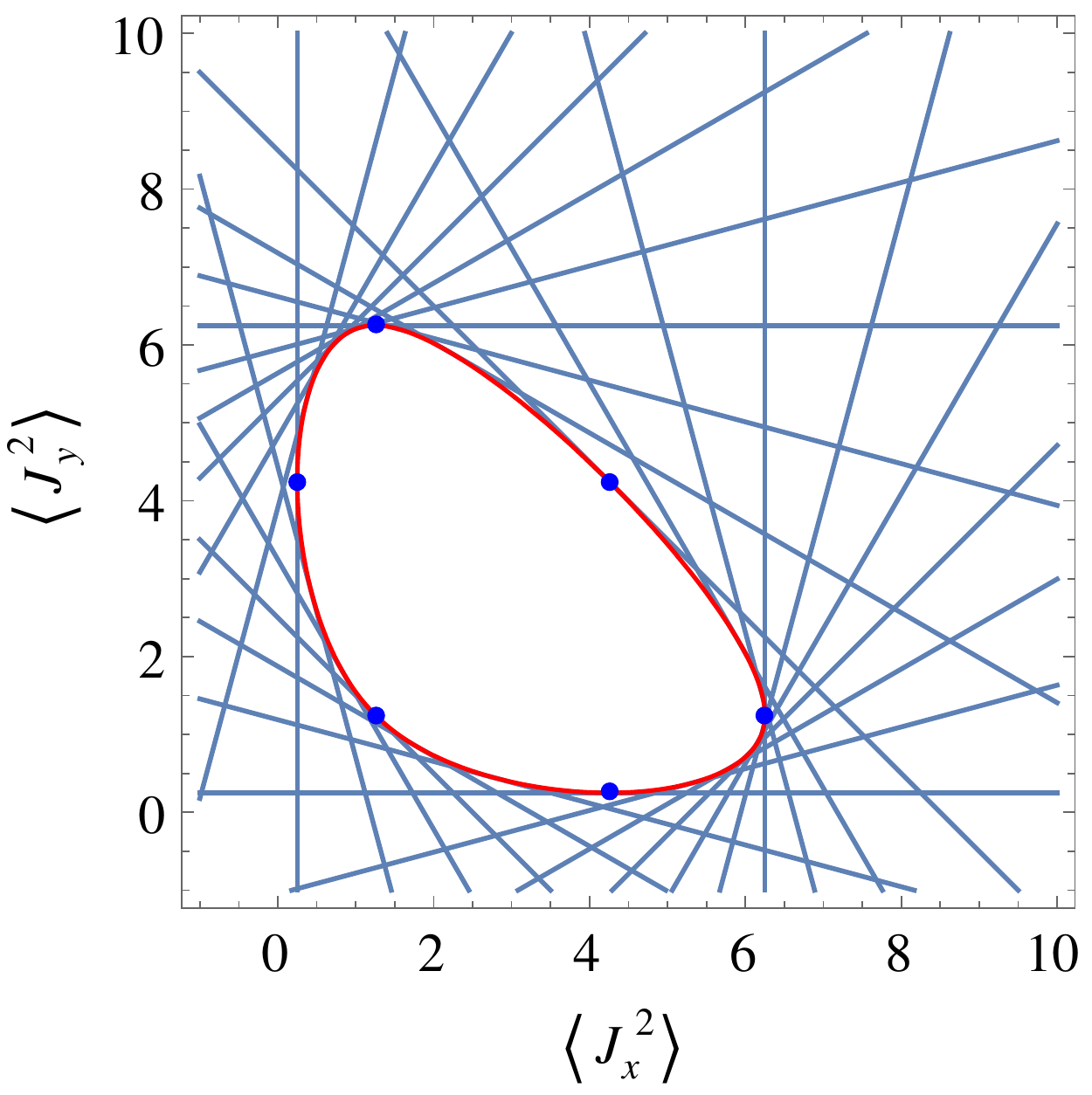}}\quad\
	\subfloat{\includegraphics[width=40mm]{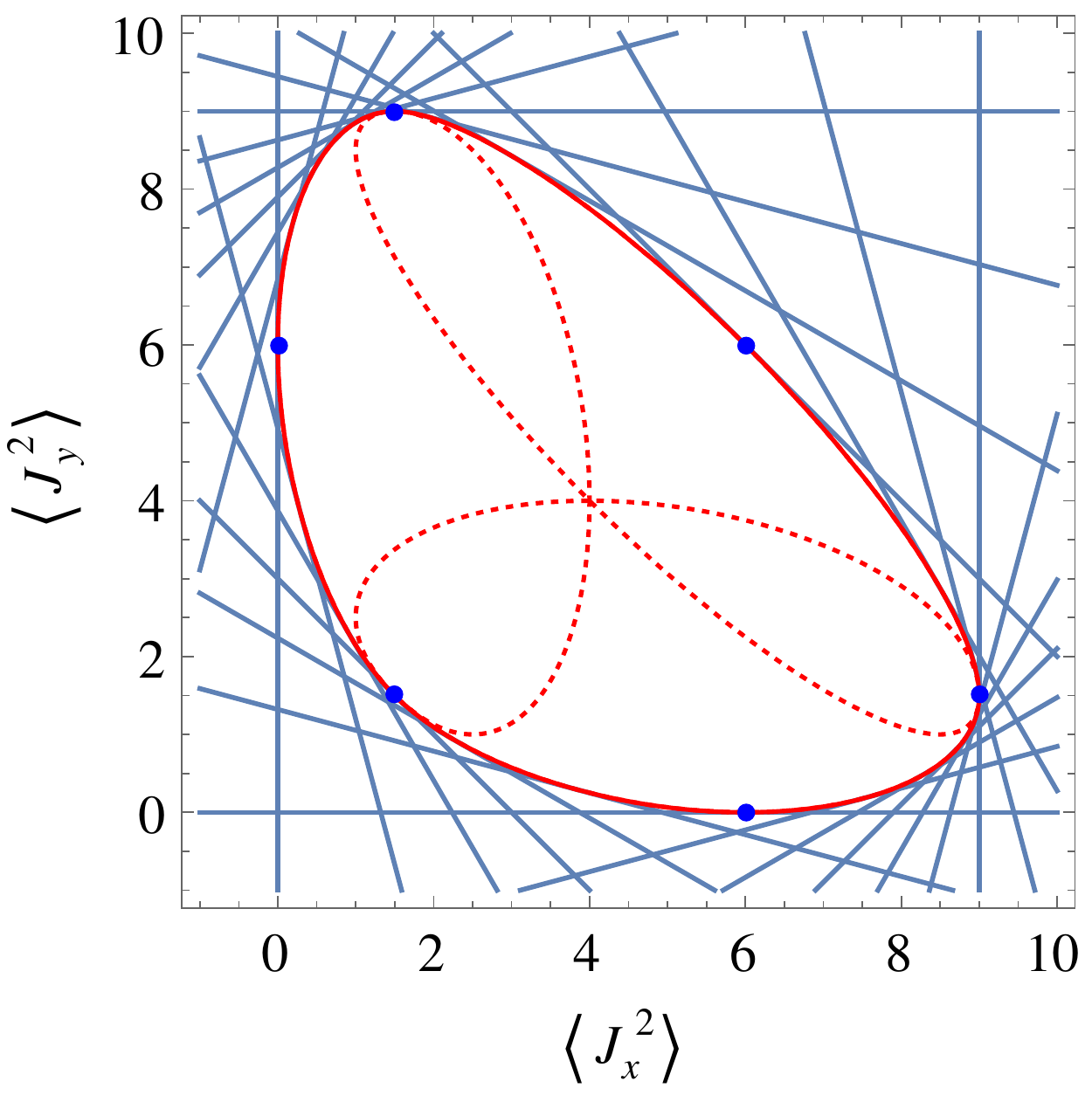}}
	\caption{ (Color online)
		For ${j=\tfrac{5}{2}}$ and ${j=3}$, the boundary 
		${\partial\mathcal{E}}$ of the joint numerical range of
		${\vec{E}=(J_x^2,J_y^2)}$ is exhibited by
		red solid-curves in
		the left- and right-hand-side panels.
		One can observe that $\mathcal{E}$ for ${j=3}$
		is the convex hull of three ellipses. A part of an ellipse that belongs to the boundary is shown by a solid-curve, and the remaining peace that falls inside $\mathcal{E}$ is illustrated by a dotted-curve.
		The ellipses are centered at ${(5,5),(2,5),}$ and ${(5,2)}$ [see their Eqs.~\eqref{ellipses-J^2 (3)}].
	}
	\label{fig:E-J^2 for j=5/2and3}  
\end{figure}

\begin{figure}[ ]
	\centering
	\subfloat{\includegraphics[width=40mm]{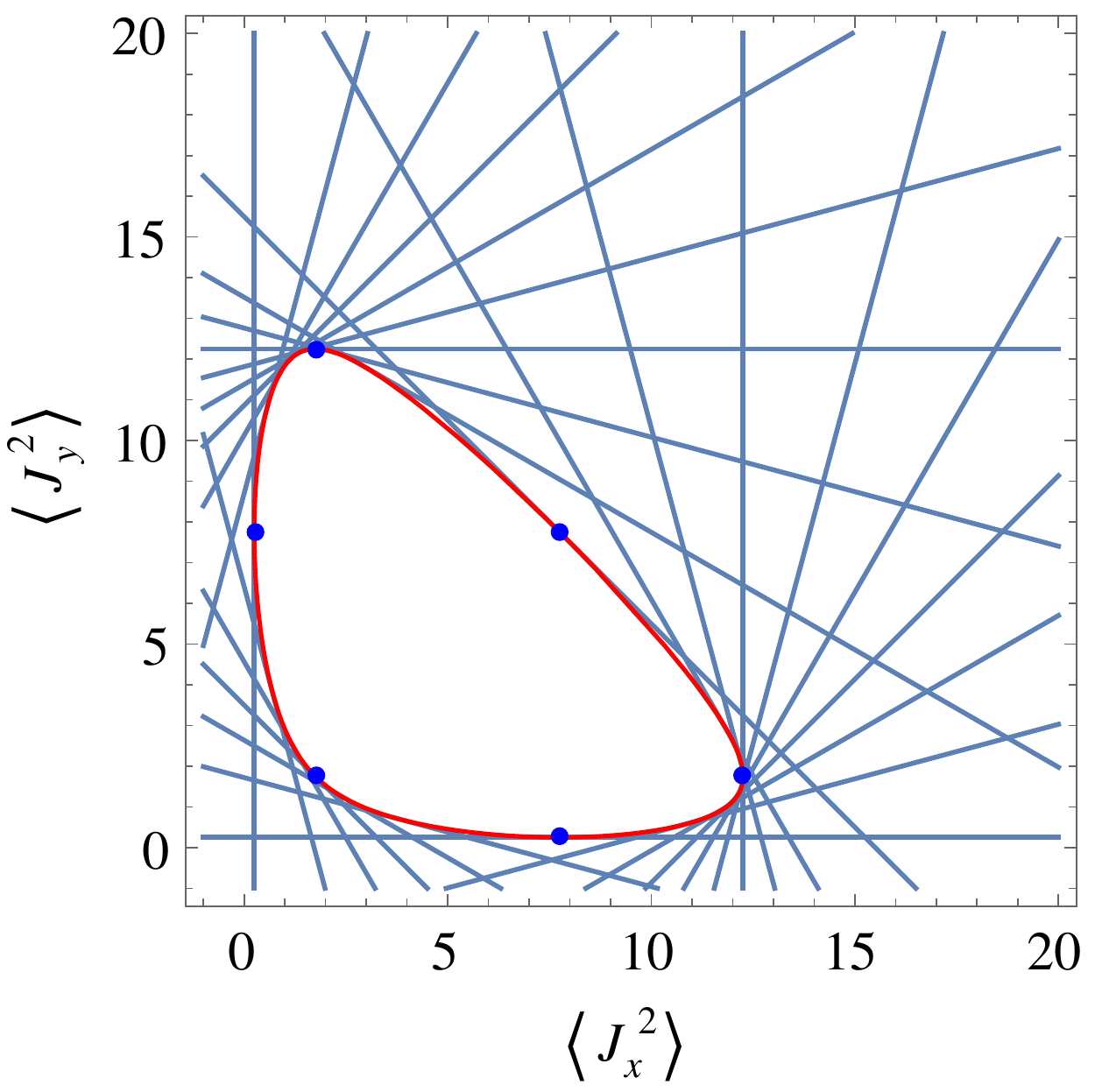}}\quad\
	\subfloat{\includegraphics[width=40mm]{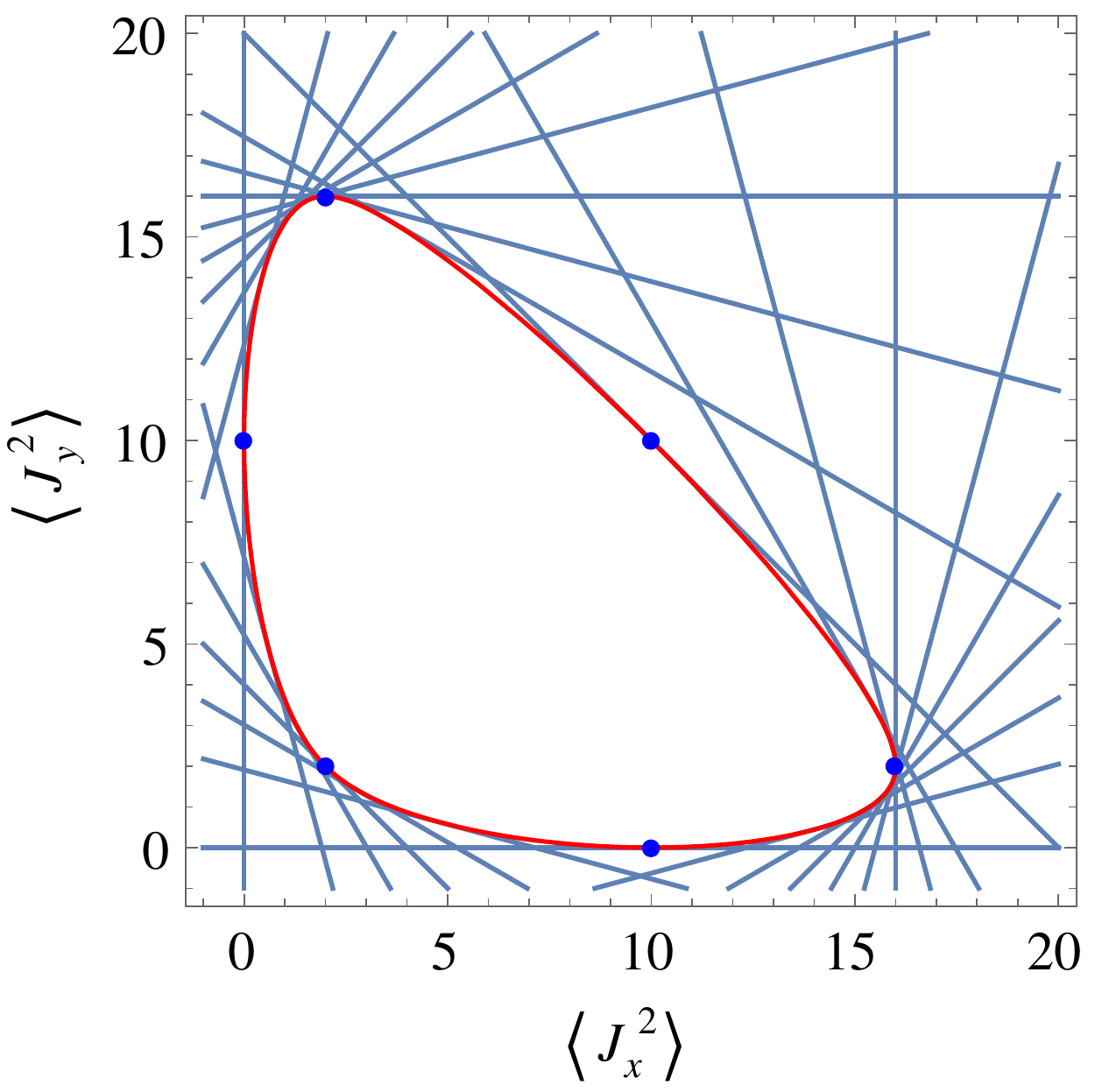}}
	\caption{(Color online)
	The left and right red closed-curves bound the allowed regions of the mean vectors ${\vec{\boldsymbol{\varepsilon}}=(\langle J_x^2\rangle,\langle J_y^2\rangle)}$ for ${j=\tfrac{7}{2}}$ and ${j=4}$, respectively. 
	There exists no quantum state for a point outside the closed-curves.
	}
	\label{fig:E-J^2 for j=7/2and4}  
\end{figure}

For ${j>1}$, the maximum-eigenvalue-states 
of $J_x^2$, $J_y^2$, and $J_z^2$ 
yield three separate extreme points
\begin{equation}
\label{max-ev-points}
\big(j^2,\tfrac{j}{2}\big)\,,\quad
\big(\tfrac{j}{2},j^2\big)\,,
\quad \text{and} \quad
\big(\tfrac{j}{2},\tfrac{j}{2}\big)
\end{equation}
of $\mathcal{E}$, respectively.
Whereas eigenstates related to the minimum characteristic value $\tfrac{1}{4}$
or $0$ of $J_x^2$, $J_y^2$, and $J_z^2$ 
provide the extreme points
\begin{equation}
\label{min-ev-points-half-int}
\big(\tfrac{1}{4},\tfrac{j(j+1)}{2}-\tfrac{1}{8}\big),
\big(\tfrac{j(j+1)}{2}-\tfrac{1}{8},\tfrac{1}{4}\big),
\ \text{and} \
\big(\tfrac{j(j+1)}{2}-\tfrac{1}{8},\tfrac{j(j+1)}{2}-\tfrac{1}{8}\big)
\end{equation}
for a half-integer $j$ and
\begin{equation}
\label{min-ev-points-int}
\big(0,\tfrac{j(j+1)}{2}\big)\,,\quad
\big(\tfrac{j(j+1)}{2},0\big)\,,
\quad \text{and} \quad
\big(\tfrac{j(j+1)}{2},\tfrac{j(j+1)}{2}\big)
\end{equation}
for an integer $j$, respectively.
These points are illustrated by blue dots in 
Figs.~\ref{fig:E-J^2 for j=1/2and1}--\ref{fig:E-J^2 for j=7/2and4}.

In the following, we present the eigenvalue $\lambda_\textsc{m}(\phi)$ of $\Lambda(\phi)$ [given in \eqref{Lphi Jxy2}], the corresponding eigenkets $|\phi\rangle$, and the expectation values ${\langle\phi|J_x^{2}|\phi\rangle}$ and ${\langle\phi|J_y^{2}|\phi\rangle}$ in that order for quantum numbers
${j=\tfrac{3}{2},\cdots,4}$.
$\lambda_\textsc{m}(\phi)$ is expressed in terms of
\begin{equation}
\label{fg}
f(\phi)\,,\,g(\phi)=\cos\phi\pm\sin\phi
\quad\text{for all}\quad \phi\in[0,2\pi)\,.
\end{equation}
In the case of half-integer ${j=\tfrac{3}{2},\tfrac{5}{2},\tfrac{7}{2}\,}$,
$\lambda_\textsc{m}$ is twofold degenerate at every $\phi$, hence two eigenkets 
${|\phi\rangle_{0,1}\in\mathscr{S}_{0,1}^2}$ are provided.
Both $|\phi\rangle_{0,1}$ produce a single extreme point
of $\mathcal{E}$.
As the two kets belong to mutually orthogonal invariant subspaces of $J_x^2$ and $J_y^2$, all their normalized linear combinations [see \eqref{general-phi-ket} and the text around it] will also deliver the same point ${\langle\phi|\vec{E}\,|\phi\rangle}$.
In the case of ${j=2,3,4}$,
$\lambda_\textsc{m}$ is twofold degenerate only at $\phi=0,\tfrac{\pi}{2},$ and $\tfrac{5\pi}{4}$, and the corresponding extreme points are already registered in
\eqref{max-ev-points}. So, for an integer $j$, only one eigenket is presented below.

For every ${j=\tfrac{3}{2},\cdots,4}$, the maximum-eigenvalue-kets $|\phi\rangle$ are stated in terms of coefficients $\alpha,\beta,\varsigma$ that are real functions of the parameter $\phi$.
As all these coefficients become zero or indeterminate at ${\phi=\tfrac{5\pi}{4}}$,
none of the formulas supplied below for $|\phi\rangle$, and thus for ${\langle\phi|\vec{E}\,|\phi\rangle}$ are applicable at that angle.
With \eqref{lambda-5pi/4}, we can directly realize 
${|{\phi=\tfrac{5\pi}{4}}\rangle=|{\pm j}\rangle}$, and 
${\langle {\pm j}|\vec{E}\,|{\pm j}\rangle=\big(\tfrac{j}{2},\tfrac{j}{2}\big)}$ 
is already listed in \eqref{max-ev-points}.

For ${j=\tfrac{3}{2},\cdots,4}$,
the boundary $\partial\mathcal{E}$ is represented graphically in Figs.~\ref{fig:E-J^2 for j=3/2and2}--\ref{fig:E-J^2 for j=7/2and4} by the (red) closed-curves and parametrically by ${\langle\phi|\vec{E}\,|\phi\rangle}$.
In Figs.~\ref{fig:E-J^2 for j=1/2and1}--\ref{fig:E-J^2 for j=7/2and4},
the supporting hyperplanes $\mathbb{H}(\phi)$ [defined in \eqref{hyperplane}] at equally spaced angles $\phi=k\tfrac{2\pi}{24}$, $k=0,\cdots,23$, are depicted by
the (blue) lines.
One can observe that each hyperplane touches $\mathcal{E}$ only at one point when $j>1$, which
implies that each face $\mathcal{F}(\phi)$ [defined in \eqref{face}] is then made of a single extreme point, and $\text{ext}(\mathcal{E})=\partial\mathcal{E}$.

Now let us start with ${j=\tfrac{3}{2}}$, where the maximum eigenvalue of $\Lambda(\phi)$ is
\begin{equation}
\label{lmax-J^2 (3/2)}
\lambda_\textsc{m}(\phi)
=
\tfrac{1}{4}\big(5f+2\sqrt{f^2+3g^2}\big)
\quad\text{for all}\quad \phi\in[0,2\pi)
\end{equation}
[for $f,g$ see \eqref{fg}].
Two mutually orthogonal eigenkets associated with $\lambda_\textsc{m}$
are
\begin{align}
\label{lmax-ket1-J^2 (3/2)}
|\phi\rangle_0&=
\tfrac{1}{\sqrt{\alpha^2+\beta^2}}
\left(
\alpha\,\big|{+\tfrac{3}{2}}\big\rangle +
\beta\,\big|{-\tfrac{1}{2}}\big\rangle
\right)\in\mathscr{S}_0^2
\ \ \text{and}  \nonumber\\
|\phi\rangle_1&=
\tfrac{1}{\sqrt{\alpha^2+\beta^2}}
\left(
\beta\, \big|{+\tfrac{1}{2}}\big\rangle +
\alpha\,\big|{-\tfrac{3}{2}}\big\rangle
\right)\in\mathscr{S}_1^2\,,
\ \ \text{where}  \\
\alpha(\phi)&=2\sqrt{3}\,g\quad\mbox{and}\quad
\beta(\phi)=4\,\lambda_\textsc{m}-3f\,.
\nonumber
\end{align}
One can show that both $|\phi\rangle_{0,1}$ and all their superpositions contribute a single extreme point
\begin{align}
\label{para-J^2 (3/2)}
\langle\phi|J_x^2|\phi\rangle&=
\tfrac{3\,\alpha^2+4\sqrt{3}\,\alpha\,\beta+7\beta^2}{4\,(\alpha^2+\beta^2)}\,,
\nonumber\\
\langle\phi|J_y^2|\phi\rangle&=
\tfrac{3\,\alpha^2-4\sqrt{3}\,\alpha\,\beta+7\beta^2}{4\,(\alpha^2+\beta^2)}
\end{align}
of $\mathcal{E}$.
The boundary ${\partial\mathcal{E}}$---characterized by 
its parametric form \eqref{para-J^2 (3/2)}---is an ellipse
[see Fig.~\ref{fig:E-J^2 for j=3/2and2}].
The QC
\begin{equation}
\label{QC-J^2 (3/2)}
\left(\langle J_x^2\rangle+\langle J_y^2\rangle-\tfrac{5}{2}\right)^2+
\frac{\left(\langle J_x^2\rangle-\langle J_y^2\rangle\right)^2}{3}\leq 1
\end{equation}
completely identifies the permitted region here, and the inequality is saturated by the maximum-eigenvalue-states of 
${\Lambda(\phi)}$ for all $\phi$'s.
The equality in \eqref{QC-J^2 (3/2)} describes the ellipse.

Next we pick ${j=2}$, where the largest characteristic value of ${\Lambda(\phi)}$ is
\begin{equation}
\label{lmax-J^2 (2)}
\lambda_\textsc{m}(\phi)=2f+\sqrt{f^2+3g^2}
\quad \text{for all}\quad 
0\leq\phi<2\pi\,.
\end{equation}
For an angle $\phi$ other than ${0,\tfrac{\pi}{2},}$ and $\tfrac{5\pi}{4}$,
$\lambda_\textsc{m}$ is non-degenerate, and thus we have only one largest-eigenvalue-ket
\begin{align}
\label{lmax-ket-J^2 (2)}
|\phi\rangle_0&=
\tfrac{1}{\sqrt{2\alpha^2+\beta^2}}
\left(\alpha\,|{+2}\rangle +\beta\,|0\rangle +\alpha\,|{-2}\rangle\right)
\in\mathscr{S}_0^2\,,\\
\alpha(\phi)&=\sqrt{6}\,g\quad\mbox{and}\quad
\beta(\phi)=2\,(\lambda_\textsc{m}-f)\,.
\nonumber
\end{align}
These kets generate the borderline  
\begin{equation}
\label{para-J^2 (2)}
\langle\phi|\vec{E}\,|\phi\rangle=
\big(\tfrac{2\,\alpha^2+2\sqrt{6}\,\alpha\,\beta+3\,\beta^2}{2\,\alpha^2+\beta^2}\,,
\tfrac{2\,\alpha^2-2\sqrt{6}\,\alpha\,\beta+3\,\beta^2}{2\,\alpha^2+\beta^2}\big)
\end{equation}
of $\mathcal{E}$, which is the other ellipse displayed in Fig.~\ref{fig:E-J^2 for j=3/2and2}.
All the pure states $|\phi\rangle_0\langle \phi|$ made of \eqref{lmax-ket-J^2 (2)} saturate QC
\begin{equation}
\label{QC-J^2 (2)}
\frac{\left(\langle J_x^2\rangle+\langle J_y^2\rangle-4\right)^2}{4}+
\frac{\left(\langle J_x^2\rangle-\langle J_y^2\rangle\right)^2}{12}
	\leq 1\,.
\end{equation}
That is to say that points \eqref{para-J^2 (2)} follow the equality in \eqref{QC-J^2 (2)}.
And, the same is true for the three extreme points in \eqref{max-ev-points}.

Before formulating a set of tight URs and CRs for ${j=2}$ by following the procedure described in the last part of Sec.~\ref{sec:all-region}, let us first note that operators ${J_x^2,J_y^2}$ and thus all their linear sums \eqref{Lphi Jxy2} possess three common orthogonal eigenkets
\begin{eqnarray}
\label{e1 (2)}
|\textsf{e}_1\rangle&=&\tfrac{1}{\sqrt{2}}
\left(\,|{+2}\rangle -|{-2}\rangle\right)
\in\mathscr{S}_0^2\subset\mathscr{H}_5\,,\nonumber\\
|\textsf{e}_2\rangle&=&\tfrac{1}{\sqrt{2}}
\left(\,|{+1}\rangle +|{-1}\rangle\right)
\in\mathscr{S}_1^2\subset\mathscr{H}_5\,,\quad\mbox{and}\quad\\
|\textsf{e}_3\rangle&=&\tfrac{1}{\sqrt{2}}
\left(\,|{+1}\rangle -|{-1}\rangle\right)
\in\mathscr{S}_1^2\subset\mathscr{H}_5\,.\nonumber
\end{eqnarray}
Now, taking \eqref{h(X)}--\eqref{u-half-sum} with \eqref{Jdot J^2}, we combine uncertainty or certainty measures
for ${\vec{\boldsymbol{\varepsilon}}=(\langle J_x^2\rangle,\langle J_y^2\rangle)}$
and obtain the tight relations
\begin{align}
\label{H-UR-J^2 (2)}
0<4\ln 2-\sqrt{3}\ln(2+\sqrt{3})&\leq h(\vec{\boldsymbol{\varepsilon}}\,)\,,
\\
\label{u-UR-J^2 (2)}
2<\tfrac{1}{2}(3+\sqrt{3})&\leq
u_{\sfrac{1}{2}}(\vec{\boldsymbol{\varepsilon}}\,)\,, 
\\
\label{u2-CR-J^2 (2)}
u_2(\vec{\boldsymbol{\varepsilon}}\,)
&\leq \tfrac{7}{4}<2 \,,\quad \mbox{and} 
\\
\label{umax-CR-J^2 (2)}
u_\text{max}(\vec{\boldsymbol{\varepsilon}}\,)&\leq 
1+\tfrac{\sqrt{3}}{2}<2 
\end{align}
by finding the absolute minima of concave and maxima of convex functions
on $\partial\mathcal{E}$ in \eqref{para-J^2 (2)}.
Relations \eqref{H-UR-J^2 (2)}, \eqref{u2-CR-J^2 (2)}, and \eqref{umax-CR-J^2 (2)}
are saturated by $\phi$-kets \eqref{lmax-ket-J^2 (2)} at
${\phi=\tfrac{3\pi}{4},\tfrac{7\pi}{4}}$.
Whereas, the extreme-eigenvalue-states of $J_x^2$ and $J_y^2$
are the minimum uncertainty states for UR \eqref{u-UR-J^2 (2)}.
By the way, we get the same relations \eqref{H-UR-J^2 (2)}--\eqref{umax-CR-J^2 (2)}
for ${j=\tfrac{3}{2}}$, and the states which saturate them will 
then be from \eqref{lmax-ket1-J^2 (3/2)}.

In \eqref{H-UR-J^2 (2)}--\eqref{umax-CR-J^2 (2)}, the strict inequality signs ($<$) represent the trivial lower or upper bounds.
Clearly, we achieve \emph{nontrivial} tight URs and CRs
despite the non-commuting operators $J_x^2$ and $J_y^2$ share three eigenvectors
\eqref{e1 (2)}. It is because
the situation described in the last two paragraphs of Sec.~\ref{sec:all-region} does not occur here.
On the other hand, a UR based on
the standard deviations of $J_x^2$ and $J_y^2$
will be a trivial one as ${\Delta J_x^2=0=\Delta J_y^2}$ when the system is in their common eigenstate.

As $J_x^2$ is a degenerate operator, there exist infinitely-many distinct projective measurements for it, and similarly for $J_y^2$.
If
${|\textsf{e}_l\rangle\langle\textsf{e}_l|}$ is a part of both the measurements for $J_x^2$ and $J_y^2$, then we also get 
a tight and trivial UR ${0\leq H(\vec{p}\,)+H(\vec{q}\,)}$ based on the Shannon entropy, where 
$\vec{p}$ and $\vec{q}$ are the probability vectors associated with the two measurements.
However, with a different choice of measurements, one can get a nontrivial entropic UR.
Since all such measurements provide the same mean vector 
${\vec{\boldsymbol{\varepsilon}}=(\langle J_x^2\rangle,\langle J_y^2\rangle)}$, 
the relation in \eqref{H-UR-J^2 (2)}--\eqref{umax-CR-J^2 (2)} are independent of
the measurement-settings used for $J_x^2$ and $J_y^2$.

Now we move to ${j=\tfrac{5}{2}}$, where the biggest eigenvalue of operator \eqref{Lphi Jxy2} is
\begin{align}
\label{lmax-J^2 (5/2)}
\lambda_\textsc{m}(\phi)&=
\tfrac{1}{4}
\left[\tfrac{35\,f}{3}+2\sqrt{\tfrac{p}{3}}
\cos\left(\tfrac{1}{3}\arccos
\left(\tfrac{-q}{2}\sqrt{\tfrac{27}{p^3}}\right)
\right)
\right],\nonumber\\
p(\phi)&=\tfrac{112}{3}(f^2+3g^2)\,,\
q(\phi)=\tfrac{1280}{27}(f^3-9fg^2)\,,
\end{align}
for all ${0\leq\phi<2\pi}$.
The two associated orthonormal eigenvectors are 
\begin{align}
\label{lmax-ket1-J^2 (5/2)}
&|\phi\rangle_0=
\tfrac{1}{\sqrt{\alpha^2+1+\beta^2}}
\left(
\alpha\,\big|{+\tfrac{5}{2}}\big\rangle +
\big|{+\tfrac{1}{2}}\big\rangle+
\beta\,\big|{-\tfrac{3}{2}}\big\rangle
\right),
\nonumber\\
&|\phi\rangle_1=
\tfrac{1}{\sqrt{\alpha^2+1+\beta^2}}
\left(
\beta\, \big|{+\tfrac{3}{2}}\big\rangle +
\big|{-\tfrac{1}{2}}\big\rangle+
\alpha\,\big|{-\tfrac{5}{2}}\big\rangle
\right),
\\
&\mbox{where}\quad
\alpha(\phi)=\tfrac{2\sqrt{10}\,g}{4\,\lambda_\textsc{m}-5f}
\quad\mbox{and}\quad
\beta(\phi)=\tfrac{6\sqrt{2}\,g}{4\,\lambda_\textsc{m}-13f}\,.
\nonumber
\end{align}
Both ${|\phi\rangle_{0,1}\in\mathscr{S}_{0,1}^2}$, all their superpositions, and all possible mixtures of ${|\phi\rangle_0\langle\phi|}$ and ${|\phi\rangle_1\langle\phi|}$ give a single point 
\begin{align}
\label{para-J^2 (5/2)}
\langle\phi|J_x^2|\phi\rangle&=
\tfrac{17+4\sqrt{10}\,\alpha+5\,\alpha^2+12\sqrt{2}\,\beta+13\,\beta^2}%
{4(\alpha^2+1+\beta^2)}\,,
\nonumber\\
\langle\phi|J_y^2|\phi\rangle&=
\tfrac{17-4\sqrt{10}\,\alpha+5\,\alpha^2-12\sqrt{2}\,\beta+13\,\beta^2}%
{4(\alpha^2+1+\beta^2)}
\end{align}
at the boundary $\partial\mathcal{E}$ displayed in Fig.~\ref{fig:E-J^2 for j=5/2and3}.

Next we take ${j=3}$, and the maximum eigenvalue is 
\begin{align}
\label{lmax-J^2 (3)}
\lambda_\textsc{m}(\phi)&=
\begin{cases} 
\lambda_1(\phi)  & \text{for  }\ 0\leq\phi\leq\tfrac{\pi}{2} \\
\lambda_2(\phi)  & \text{for  }\  \tfrac{\pi}{2}\leq\phi\leq\tfrac{5\pi}{4} \\
\lambda_3(\phi)  & \text{for  }\ \tfrac{5\pi}{4}\leq\phi\leq 2\pi
\end{cases},
\quad \mbox{where}  \nonumber\\
\lambda_1&=5f+\sqrt{f^2+15g^2}\,,\\
\lambda_2&=\tfrac{1}{2}
\left(
7f-3g+\sqrt{8}\sqrt{2f^2-3fg+3g^2}
\right),\  \mbox{and}\nonumber\\
\lambda_3&=\tfrac{1}{2}
\left(
7f+3g+\sqrt{8}\sqrt{2f^2+3fg+3g^2}
\right)\nonumber
\end{align}
are distinct eigenvalues of $\Lambda(\phi)$.
The corresponding eigenkets are
\begin{align}
\label{l1-ket-J^2 (3)}
&|\lambda_1(\phi)\rangle
\ \propto\ 
\alpha_1\, \left(\big|{+2}\big\rangle +\big|{-2}\big\rangle\right)+
\beta_1\,\big|0\big\rangle
\in\mathscr{S}_1^2,
\nonumber\\
&|\lambda_2(\phi)\rangle
\ \propto\
\alpha_2\, \left(\big|{+3}\big\rangle -\big|{-3}\big\rangle\right)+
\beta_2\,\left(\big|{+1}\big\rangle-\big|{-1}\big\rangle\right),
\nonumber\\
&|\lambda_3(\phi)\rangle
\ \propto\
\alpha_3\, \left(\big|{+3}\big\rangle +\big|{-3}\big\rangle\right)+
\beta_3\,\left(\big|{+1}\big\rangle+\big|{-1}\big\rangle\right),
\\
&\mbox{where}\quad\alpha_1=\sqrt{30}\,g\,,\quad \beta_1=2\lambda_1-8f\,,\nonumber\\
&\qquad\quad\ \ \alpha_2=\sqrt{15}\,g\,,\quad \beta_2=2\lambda_2-3f\,,\nonumber\\
&\qquad\quad\ \ \alpha_3=\sqrt{15}\,g\,,\quad \beta_3=2\lambda_3-3f\,,\nonumber
\end{align}
and ${|\lambda_{2,3}\rangle\in\mathscr{S}_0^2}$.
Since the three vectors on the right-hand-side in \eqref{l1-ket-J^2 (3)}
are not normalized, we put the proportionality sign $\propto$ there.
In \eqref{lmax-J^2 (3)}, one can observe that the degree of degeneracy of $\lambda_\textsc{m}$ is two only at ${\phi=0,\tfrac{\pi}{2},}$ and $\tfrac{5\pi}{4}$.

With the unit step function
\begin{equation}
\label{unit-step}
\zeta(\phi):=
\begin{cases} 
0 & \text{if  }\  \phi< 0 \\
1 & \text{if  }\ 0\leq\phi 
\end{cases}\,,
\end{equation}
we can compactly express $\lambda_\textsc{m}(\phi)$ and the corresponding eigenket
\begin{align}
\label{lmax-ket-J^2 (3)}
|\phi\rangle=\ &|\lambda_1(\phi)\rangle
\left(1-\zeta(\phi-\tfrac{\pi}{2})\right)+
\nonumber\\
&|\lambda_2(\phi)\,\rangle\,\zeta(\phi-\tfrac{\pi}{2})
\left(1-\zeta(\phi-\tfrac{5\pi}{4})\right)+
\nonumber\\
&|\lambda_3(\phi)\rangle\,\zeta(\phi-\tfrac{5\pi}{4})\in\mathscr{H}_7\,,
\end{align}
which provides $\text{ext}(\mathcal{E})=\partial\mathcal{E}$.
We present $\partial\mathcal{E}$ graphically by the (red) solid-curve in Fig.~\ref{fig:E-J^2 for j=5/2and3} and parametrically by joining
\begin{align}
\label{para-J^2 (3)}
\langle\lambda_1|\vec{E}\,|\lambda_1\rangle&=
\big(\tfrac{8\,\alpha_1^2+2\sqrt{30}\,\alpha_1\beta_1+6\,\beta_1^2}%
{2\alpha_1^2+\beta_1^2},
\tfrac{8\,\alpha_1^2-2\sqrt{30}\,\alpha_1\beta_1+6\,\beta_1^2}%
{2\alpha_1^2+\beta_1^2}\big),
\nonumber\\
\langle\lambda_2|\vec{E}\,|\lambda_2\rangle&=
\big(
\tfrac{3\,\alpha_2^2+2\sqrt{15}\,\alpha_2\beta_2+5\,\beta_2^2}%
{2(\alpha_2^2+\beta_2^2)},
\tfrac{3\,\alpha_2^2-2\sqrt{15}\,\alpha_2\beta_2+17\,\beta_2^2}%
{2(\alpha_2^2+\beta_2^2)}\big),
\nonumber\\
\langle\lambda_3|\vec{E}\,|\lambda_3\rangle&=
\big(
\tfrac{3\,\alpha_3^2+2\sqrt{15}\,\alpha_3\beta_3+17\,\beta_3^2}%
{2(\alpha_3^2+\beta_3^2)},
\tfrac{3\,\alpha_3^2-2\sqrt{15}\,\alpha_3\beta_3+5\,\beta_3^2}%
{2(\alpha_3^2+\beta_3^2)}
\big)
\end{align}
for ${\phi\in[0,\tfrac{\pi}{2})}$, ${\phi\in[\tfrac{\pi}{2},\tfrac{5\pi}{4})}$,
and ${\phi\in(\tfrac{5\pi}{4},2\pi)}$, respectively.
In Fig.~\ref{fig:E-J^2 for j=5/2and3}, one can see that these parts of the boundary $\partial\mathcal{E}$
come from three intersecting ellipses.
One can check with \eqref{para-J^2 (3)} that $\langle\lambda_k|\vec{E}\,|\lambda_k\rangle:=(a_k,b_k)$ 
for $k=1,2,$ and 3
satisfy the equations
\begin{align}
\label{ellipses-J^2 (3)}
&\Big(\frac{a_1+b_1-10}{2}\Big)^2 +\Big(\frac{a_1-b_1}{2\sqrt{15}}\Big)^2 = 1\,,
\nonumber\\
&\Big(\frac{a_2+b_2-7}{4}\Big)^2 +\Big(\frac{-7a_2+b_2+9}{4\sqrt{15}}\Big)^2 = 1\,,\
\mbox{and}
\\
&\Big(\frac{a_3+b_3-7}{4}\Big)^2 +\Big(\frac{a_3-7b_3+9}{4\sqrt{15}}\Big)^2 = 1\,,\
\nonumber
\end{align}
respectively, of the ellipses.
Like \eqref{e1 (2)}, here also
both the operators in ${(J_x^2,J_y^2)=\vec{E}}$ share a common eigenvector
${|\textsf{e}\rangle=\tfrac{1}{\sqrt{2}}
(|{+2}\rangle -|{-2}\rangle)}$.
The eigenvector provides the point $(4,4)$ at which all the three ellipses intersect.

Now we proceed to ${j=\tfrac{7}{2}}$,
where the maximum eigenvalue $\lambda_\textsc{m}$ is twofold degenerate
at all $\phi$'s:
\begin{align}
\label{lmax-J^2 (7/2)}
&  \lambda_\textsc{m}(\phi)=\tfrac{1}{4}
  \left(21f+s+\tfrac{1}{2}\sqrt{-4s^2-2p-\frac{q}{s}}\right), \nonumber\\
&  s(\phi)=\sqrt{\tfrac{1}{6}\left(p+\sqrt{\Upsilon_0}
	\cos\left(\tfrac{1}{3}\arccos\left(\tfrac{\Upsilon_1}{2\sqrt{\Upsilon_0^3}}\right)
	\right)\right)}\,,\nonumber\\
&  p(\phi)=168\left(f^2+3g^2\right),\
  q(\phi)=512\left(f^3-9fg^2\right),
  \\
&  \Upsilon_0(\phi)=48384\left(f^2+3g^2\right)^2,\quad \mbox{and} \nonumber\\
&  \Upsilon_1(\phi)=5971968\big(3f^6-5f^4g^2+145f^2g^4+49g^6\big).\nonumber
\end{align}
The related eigenkets are
\begin{align}
\label{lmax-ket1-J^2 (7/2)}
&|\phi\rangle_0\ \propto\ 
\alpha\,\big|{+\tfrac{7}{2}}\big\rangle +
\big|{+\tfrac{3}{2}}\big\rangle+
\beta\,\big|{-\tfrac{1}{2}}\big\rangle+
\varsigma\,\big|{-\tfrac{5}{2}}\big\rangle\in\mathscr{S}_0^2,
\nonumber\\
&|\phi\rangle_1\ \propto\ 
\varsigma\,\big|{+\tfrac{5}{2}}\big\rangle+
\beta\, \big|{+\tfrac{1}{2}}\big\rangle +
\big|{-\tfrac{3}{2}}\big\rangle+
\alpha\,\big|{-\tfrac{7}{2}}\big\rangle
\in\mathscr{S}_1^2\,,
\\
&\mbox{where}\ 
\alpha(\phi)=\tfrac{2\sqrt{21}\,g}{4\,\lambda_\textsc{m}-7f}\,,
\ 
\beta(\phi)=\tfrac{2\sqrt{21}\,g\,\alpha+27\,f-4\,\lambda_\textsc{m}}%
{-4\sqrt{15}\,g},\nonumber\nonumber\\
&\mbox{and}\ 
\varsigma(\phi)=\tfrac{6\sqrt{5}\,g}{4\,\lambda_\textsc{m}-19f}\beta\,.\nonumber
\end{align} 
All the maximum-eigenvalue-states of ${\Lambda{(\phi)}}$ give a single boundary point 
\begin{align}
\label{para-J^2 (7/2)}
\langle\phi|J_x^2|\phi\rangle&=
\tfrac{27+4\sqrt{21}\,\alpha+7\,\alpha^2+31\,\beta^2+19\,\varsigma^2+
	4\sqrt{5}\beta(2\sqrt{3}+3\varsigma)}%
{4(\alpha^2+\beta^2+\varsigma^2+1)}\,,
\nonumber\\
\langle\phi|J_y^2|\phi\rangle&=
\tfrac{27-4\sqrt{21}\,\alpha+7\,\alpha^2+31\,\beta^2+19\,\varsigma^2-
	4\sqrt{5}\beta(2\sqrt{3}+3\varsigma)}%
{4(\alpha^2+\beta^2+\varsigma^2+1)}\,,
\end{align}
and the boundary $\partial\mathcal{E}$ is showcased in
Fig.~\ref{fig:E-J^2 for j=7/2and4}.
Note that \eqref{lmax-ket1-J^2 (7/2)} and \eqref{para-J^2 (7/2)} are not applicable at
${\phi=\tfrac{\pi}{4},\tfrac{5\pi}{4}}$.
At these two angles, $|\phi\rangle$ is the minimum- and maximum-eigenvalue-states of $J_z^2$, and the corresponding boundary points are given in \eqref{min-ev-points-half-int} and \eqref{max-ev-points}.

Next, in the case of ${j=4}$, the largest eigenvalue of operator \eqref{Lphi Jxy2} is
\begin{align}
\label{lmax-J^2 (4)}
\lambda_\textsc{m}(\phi)&=
\tfrac{1}{2}
\left[\tfrac{40\,f}{3}+2\sqrt{\tfrac{p}{3}}
\cos\left(\tfrac{1}{3}\arccos
\left(\tfrac{-q}{2}\sqrt{\tfrac{27}{p^3}}\right)
\right)
\right],\nonumber\\
p(\phi)&=\tfrac{208}{3}(f^2+3g^2)\,,\
q(\phi)=\tfrac{4480}{27}(f^3-9fg^2)\,.
\end{align}
The associated eigenvector
\begin{equation}
\label{lmax-ket-J^2 (4)}
|\phi\rangle_0
\ \propto\ 
\alpha \left(\big|{+4}\big\rangle +\big|{-4}\big\rangle\right)+
\beta\left(\big|{+2}\big\rangle +\big|{-2}\big\rangle\right)+
\big|0\big\rangle
\end{equation}
provides an extreme point
\begin{align}
\label{para-J^2 (4)}
\langle\phi|J_x^2|\phi\rangle&=
\tfrac{2
	\left(5+2\,\alpha^2+3\sqrt{10}\,\beta+2\sqrt{7}\,\alpha\,\beta+8\,\beta^2
\right)}{2\,\alpha^2+2\,\beta^2+1}\,,
\nonumber\\
\langle\phi|J_y^2|\phi\rangle&=
\tfrac{2
	\left(5+2\,\alpha^2-3\sqrt{10}\,\beta-2\sqrt{7}\,\alpha\,\beta+8\,\beta^2
	\right)}{2\,\alpha^2+2\,\beta^2+1}
\end{align}
of the numerical range $\mathcal{E}$,
where
\begin{equation}
\label{alpha-beta j=4}
\beta(\phi)=\tfrac{-3\sqrt{10}\,g}%
{\tfrac{(2\sqrt{7}\,g)^2}{2\,\lambda_\textsc{m}-4f}+16\,f-2\,\lambda_\textsc{m}}\,,
\quad
\alpha(\phi)=\tfrac{2\sqrt{7}\,g}{2\,\lambda_\textsc{m}-4f}\beta
\end{equation}
for all except ${\phi=\tfrac{5\pi}{4}}$.
All the extreme points lie on the (red) closed-curve in Fig.~\ref{fig:E-J^2 for j=7/2and4}.

\begin{table}[]
	\centering
	\caption{From left, the angular momentum quantum numbers are listed in the first column. 
		The second and third columns contain the absolute minima of $h(\vec{\boldsymbol{\varepsilon}}\,)$ and $u_{\sfrac{1}{2}}(\vec{\boldsymbol{\varepsilon}}\,)$, respectively.
		Whereas, the third and forth columns carry the global maxima of $u_2(\vec{\boldsymbol{\varepsilon}}\,)$ and $u_\text{max}(\vec{\boldsymbol{\varepsilon}}\,)$ [for ${j>4}$, see Appendix~\ref{sec:Sup-material}].
		Each extremum is appended with the $\phi$-values at which it occurs on the boundary of $\mathcal{E}$. 
		Note that the numerical values given here are rounded to a few decimal places. 		
	}
	\label{tab:URs}
	\begin{tabular}{c@{\hspace{1mm}} | @{\hspace{1mm}}c@{\hspace{6mm}}
			c@{\hspace{6mm}} c @{\hspace{6mm}} c }
		\hline\hline
		\rule{0pt}{2ex}  
		$j$ & $\mathfrak{h}$ &  $\mathfrak{u_{\sfrac{1}{2}}}$ 
		&  $\mathfrak{u_2}$ &  $\mathfrak{u_{max}}$ \\
		\hline\rule{0pt}{3ex} 			
		\multirow{2}{*}{$\tfrac{5}{2}$}   
		& ${0.419}$ at    &  ${2.321}$ at           & $1.781$ at    & $1.882$ at \\
		& ${1.965,5.89}$ &  ${0,\tfrac{\pi}{2}}$   & ${2.29,5.57}$ & ${2.36,5.5}$ \\[2mm]
		\multirow{2}{*}{$3$}              
		& ${0.427}$ at   &  ${2.321}$ at           & $1.774$ at      & $1.878$ at \\
		& ${1.934,5.92}$ &  ${0,\tfrac{\pi}{2}}$   & ${2.281,5.573}$ & ${2.356,5.498}$ \\[2mm]
		\multirow{2}{*}{$\tfrac{7}{2}$}   
		& ${0.351}$ at    &  ${2.288}$ at          & $1.8225$ at     & $1.91$ at \\
		& ${1.981,5.873}$ &  ${0,\tfrac{\pi}{2}}$  & ${2.29,5.564}$ & ${2.356,5.498}$ \\[2mm]
		\multirow{2}{*}{$4$}              
		& ${0.356}$ at    &  ${2.288}$ at          & $1.8164$ at       & $1.9014$ at \\
		& ${1.951,5.903}$ &  ${0,\tfrac{\pi}{2}}$  & ${2.282,5.572}$ & ${2.356,5.498}$ \\[2mm]
		\hline\hline
	\end{tabular}
\end{table}

Like \eqref{H-UR-J^2 (2)}--\eqref{umax-CR-J^2 (2)}, we consider the combined uncertainty and certainty measures based on 
$\{h,u_{\sfrac{1}{2}}\}$ and
$\{u_2,u_\text{max}\}$ and achieve their
global minima $\{\mathfrak{h},\mathfrak{u_{\sfrac{1}{2}}}\}$
and maxima $\{\mathfrak{u_2},\mathfrak{u_{max}}\}$, respectively, by employing the parametric form
${\langle\phi|\vec{E}\,|\phi\rangle}$ of
the boundary ${\partial\mathcal{E}}$.
Table~\ref{tab:URs} carries all the extrema and the values of parameter 
$\phi$ at which they occur for ${j=\tfrac{5}{2},\cdots,4}$.
With the minima and maxima one can have tight URs and CRs such as \eqref{H-UR-J^2 (2)}--\eqref{umax-CR-J^2 (2)}, and with the $\phi$-values one can have the minimum-uncertainty or maximum-certainty states $|\phi\rangle\langle\phi|$.
Roughly, the extrema occur in those parts of the boundary which are near to the corners of hyperrectangle \eqref{rectangle J^2}.
As we increase $j$ these parts get closer to the corners, and as ${j\rightarrow\infty}$ the allowed region becomes a triangle that shares three corners with the hyperrectangle $\mathcal{H}$.
Moreover, all our URs and CRs become (more precisely, tends to) trivial in the limit $j\rightarrow\infty$ [for a justification, see the last two paragraphs of Sec.~\ref{sec:all-region}].

\begin{figure}
	\centering
	\subfloat{\includegraphics[width=40mm]{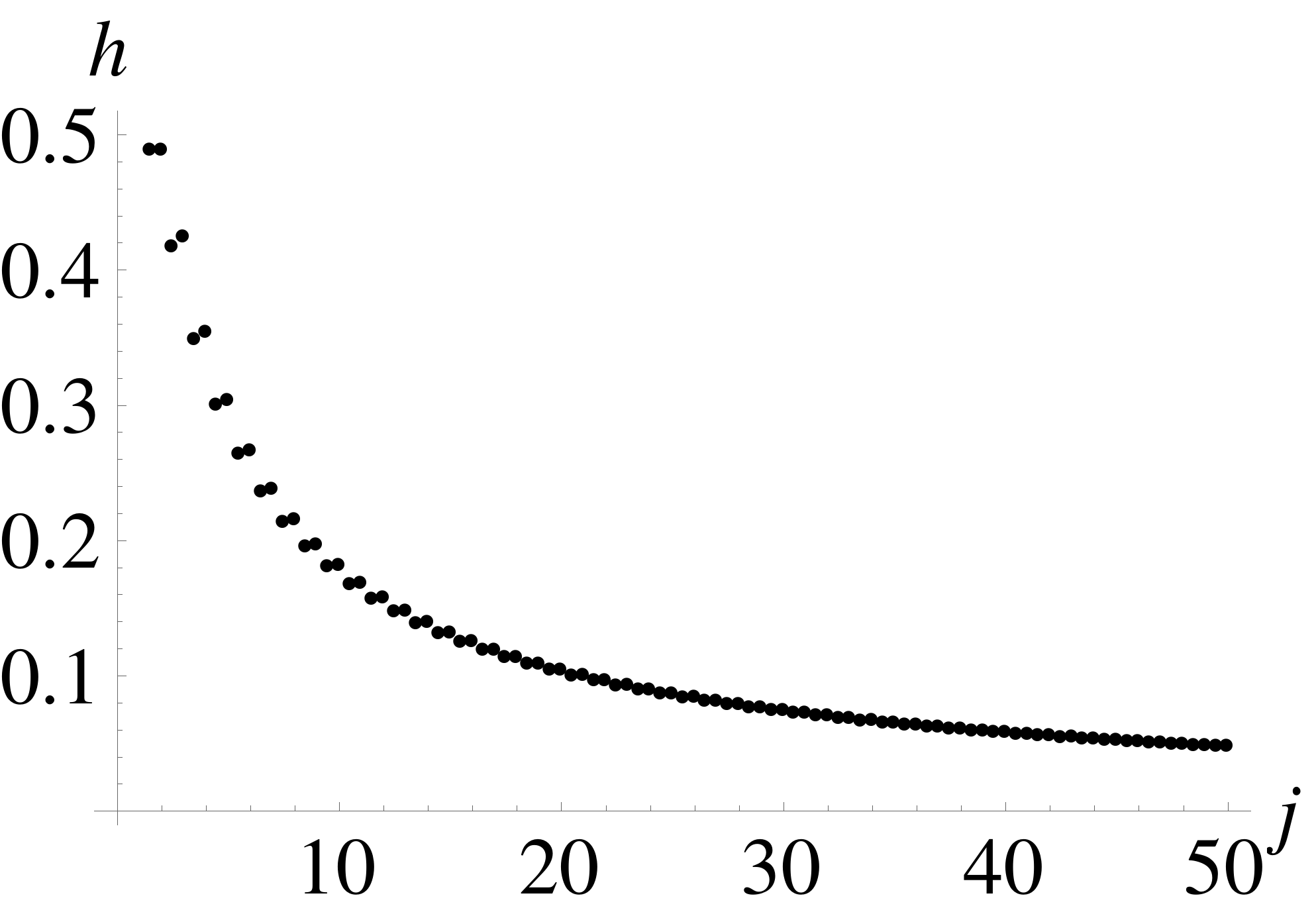}}\quad\
	\subfloat{\includegraphics[width=40mm]{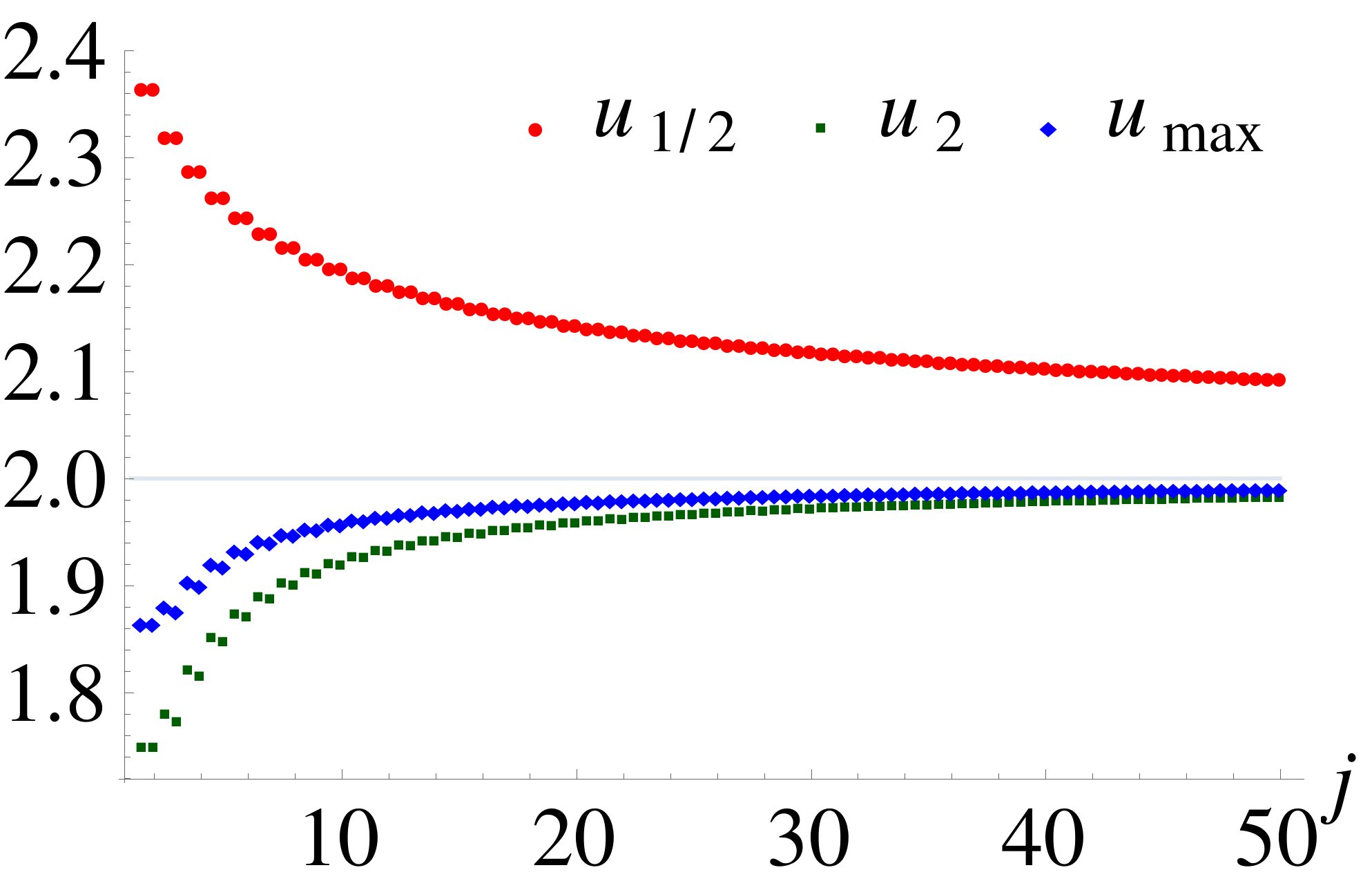}}
	\caption{	
		The left and right plots exhibit the values of $\mathfrak{h}$ and of $\{\mathfrak{u_{\sfrac{1}{2}}},\mathfrak{u_2},\mathfrak{u_{max}}\}$, which are obtained numerically for the quantum numbers $j=\tfrac{3}{2},2,\tfrac{5}{2}\cdots,50$. The values are documented in Appendix~\ref{sec:Sup-material}.
		These plots suggest slow convergence in $\mathfrak{h}\rightarrow 0$ and $\mathfrak{u_{\sfrac{1}{2}}}\rightarrow 2$ in comparison to
		$\mathfrak{u_2}\rightarrow 2$ and $\mathfrak{u_{max}}\rightarrow 2$
		as $j\rightarrow \infty$.
	}
	\label{fig:S2 UR bounds} 
\end{figure}

One can observe
that $\mathcal{E}$ takes the triangular shape as we move from  
Fig.~\ref{fig:E-J^2 for j=3/2and2} to Fig.~\ref{fig:E-J^2 for j=7/2and4}.
In Fig.~\ref{fig:S2 UR bounds} and Appendix~\ref{sec:Sup-material}, we present 
$\{\mathfrak{h},\mathfrak{u_{\sfrac{1}{2}}},\mathfrak{u_2},\mathfrak{u_{max}}\}$
as functions of $j=\tfrac{3}{2},2,\cdots,50$.
There one can see that the tight lower bounds 
$\mathfrak{h}$ and $\mathfrak{u_{\sfrac{1}{2}}}$ 
of $ h(\vec{\boldsymbol{\varepsilon}}\,)$ and $ u_{\sfrac{1}{2}}(\vec{\boldsymbol{\varepsilon}}\,)$ decrease toward 
their trivial lower bounds 0 and 2, respectively, as $j$ increases.
While the upper bounds 
$\mathfrak{u_2}$ and
$\mathfrak{u_{max}}$
of $u_2(\vec{\boldsymbol{\varepsilon}}\,)$ and $u_\text{max}(\vec{\boldsymbol{\varepsilon}}\,)$ increase
in the direction of their trivial upper bound 2.

To visualize $\mathcal{E}$ for a very large quantum number $j$, we first apply the uniform scaling
\begin{equation}
\label{uni-scaling-J}
(J_x,J_y,J_z)\mapsto\left(\frac{J_x}{j},\frac{J_y}{j},\frac{J_z}{j}\right)
\end{equation}
and then the limit
${j\rightarrow\infty}$.
After which hyperrectangle \eqref{rectangle J^2} becomes the square ${[0,1]^{\times 2}}$, and the extreme points in \eqref{max-ev-points} approach to
\begin{equation}
\label{max-ev-points, j->inf}
\big(1,0\big)\,,\quad
\big(0,1\big)\,,
\quad \text{and} \quad
\big(0,0\big)\,,
\end{equation}
respectively. In the limit ${j\rightarrow\infty}$,
the convex hull of points \eqref{max-ev-points, j->inf} is the allowed region of 
$\tfrac{1}{j^2}(\langle J_y^2\rangle,\langle J_y^2\rangle)$, which is indeed a triangle.
It is again justified in Sec.~\ref{sec:N-qubit} by applying the quantum de Finetti theorem \cite{Stormer69,Hudson76}. 
We want to stress that the boundary-plots in 
Figs.~\ref{fig:E-J^2 for j=1/2and1}--\ref{fig:E-J^2 for j=7/2and4} are obtained in \cite{Chen17} by taking a $N$-qubit system that we consider in Sec.~\ref{sec:N-qubit}.
By the way, the points in \eqref{min-ev-points-half-int} and \eqref{min-ev-points-int} fall on sides of the triangle after scaling \eqref{uni-scaling-J} and then the limit.

%===========================================
\section{Anticommutators of angular momentum operators}\label{sec:anticomm}

In this section, we consider the 
anticommutators
\begin{align}
\label{anti-commutators}
A_1&:=J_xJ_z+J_zJ_x\,,\nonumber\\
A_2&:=J_yJ_z+J_zJ_y\,,\quad \mbox{and}\\
A_3&:=J_xJ_y+J_yJ_x=-\tfrac{1}{2}\,Y_2\nonumber
\end{align}
of the angular momentum operators; $Y_\gamma$ is defined in \eqref{ABgamma}. 
In Sec.~\ref{sec:J+^2 and J-^2}, we learned that the maximum and minimum eigenvalues of
$Y_2$ are the same in magnitude but opposite in sign.
Hence, the extreme eigenvalues $\pm a_\textsc{m}$ of $A_3$
are ${\pm\tfrac{\lambda_\textsc{m}}{2}}$, where $\lambda_\textsc{m}$ 
is---an extreme eigenvalue of $Y_2$---listed in Table~\ref{tab:lmax}.
Since the anticommutators in \eqref{anti-commutators} are unitarily equivalent, they all share a common spectrum.
Consequently, in the case of ${\vec{E}=(A_1,A_2,A_3)}$, hyperrectangle~\eqref{hyperrectangle}
is the cube ${[-a_\textsc{m},a_\textsc{m}]^{\times 3}}$.

First, let us only take $A_1$ and $A_2$.
Like in Sec.~\ref{sec:J+^2 and J-^2},
one can show that the trace of different powers of 
\begin{equation}
\label{Lambda two anti-comm}
\Lambda(\phi)=\cos\phi\, A_1+\sin\phi\, A_2
\end{equation}
and thus all its eigenvalues are independent of $\phi$.
As a result, the permitted region of $(\langle A_1\rangle,\langle A_2\rangle)$ is completely identified by the QC
\begin{equation}
\label{A1A2-QC}
\langle A_1\rangle^2 +
\langle A_2\rangle^2 
\leq a_\textsc{m}^2\,.
\end{equation}
Since we can cyclically transform one anticommutator into other by a unitary conjugation, the above statements are true for any pair of the three anticommutators.
One can check that the orthogonal projection of $\mathcal{E}$---given below for 
${\vec{E}=(A_1,A_2,A_3)}$---onto the $\langle A_t\rangle\langle A_{t'}\rangle$-plane, ${t\neq t'}$,
is the circular disk specified by a QC such as \eqref{A1A2-QC}.
Moreover, URs and CRs for $A_t$ and $A_{t'}$ will be same as  \eqref{H-UR-ABgamma}--\eqref{umax-CR-ABgamma} for $X_\gamma$ and $Y_\gamma$.

\begin{figure}
	\centering
	\subfloat{\includegraphics[width=40mm]{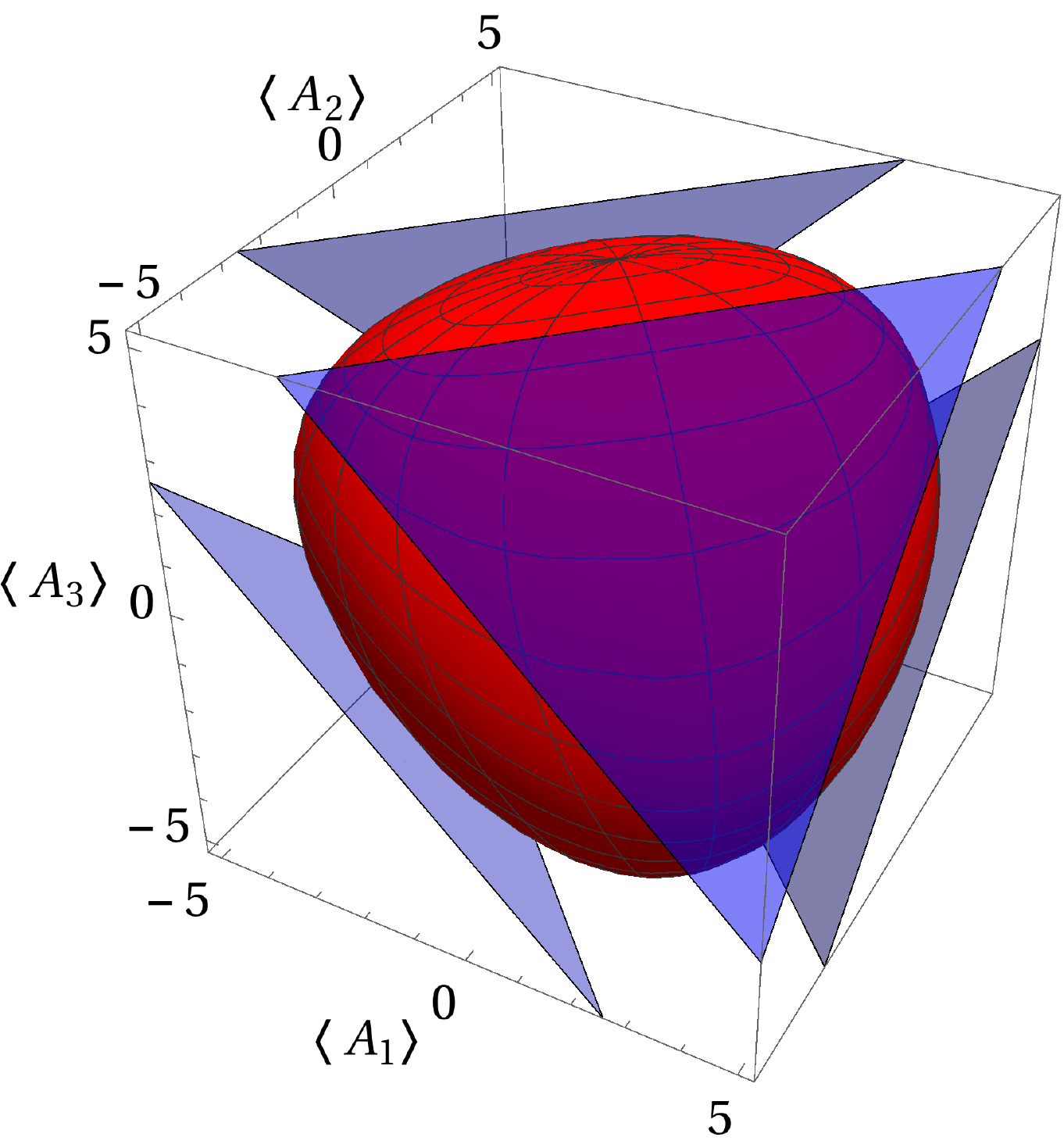}}\quad\
	\subfloat{\includegraphics[width=40mm]{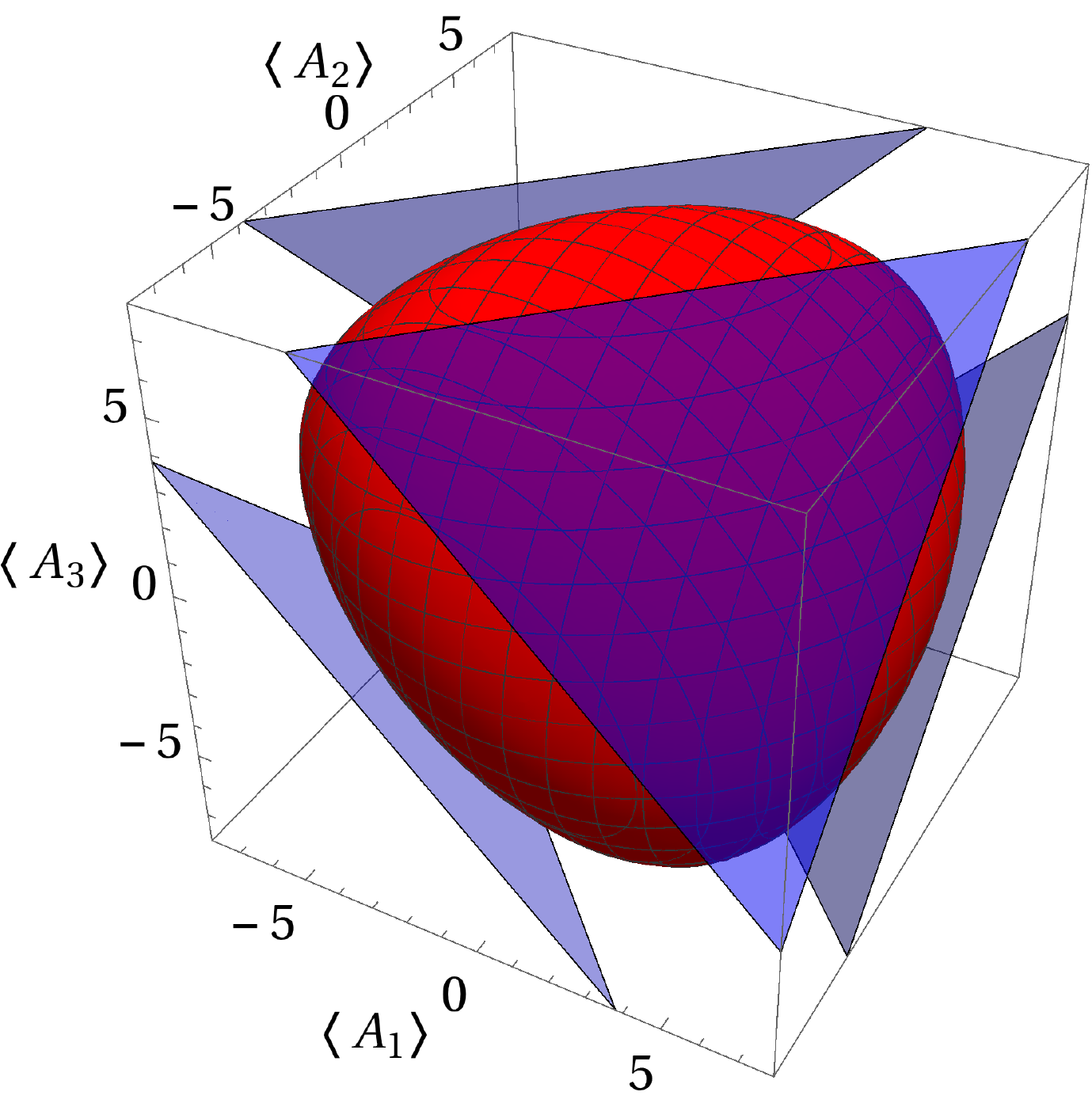}}
	\hspace{0mm}
	\subfloat{\includegraphics[width=40mm]{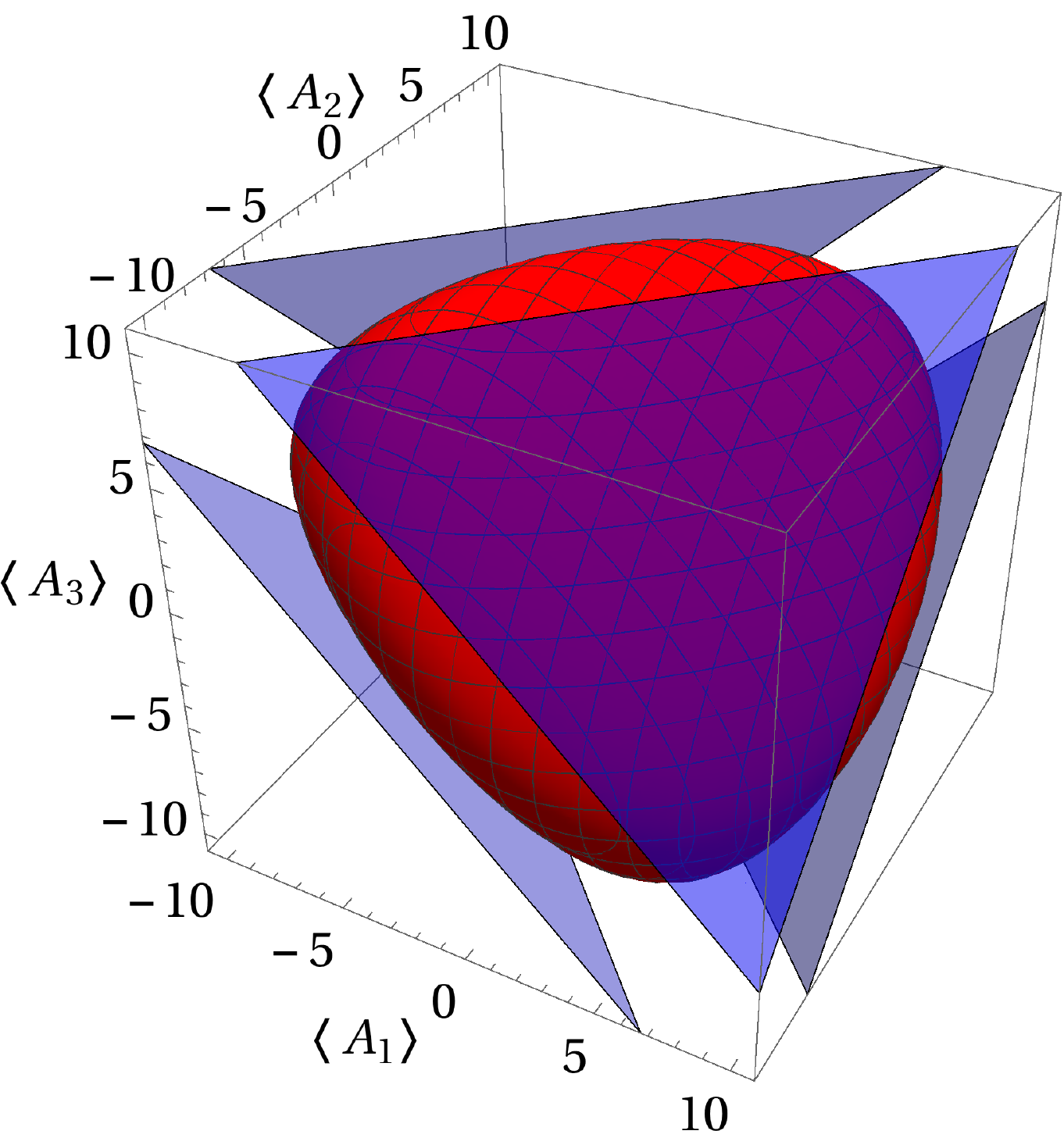}}\quad\
	\subfloat{\includegraphics[width=40mm]{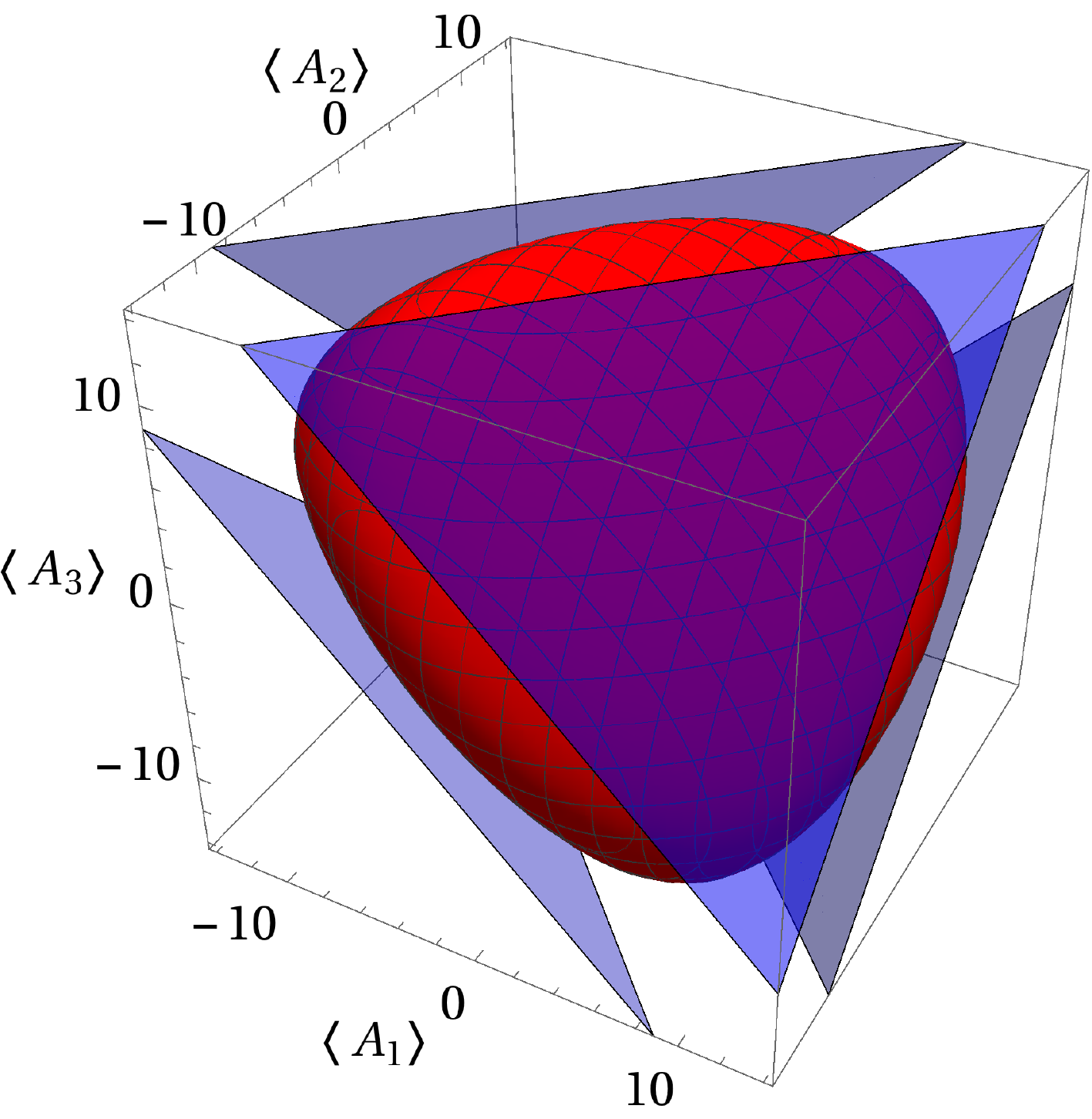}}
	\caption{(Color online) From top-left to bottom-right, moving horizontally, 
		the allowed regions $\mathcal{E}$ for the three anticommutators in~\eqref{anti-commutators} are illustrated by the red convex bodies for $j=\tfrac{5}{2},3,\tfrac{7}{2},$ and 4.
		The four supporting hyperplanes shown in blue color are normal to the four unit-vectors $-\widehat{\eta}_l$ registered in Table~\ref{tab:theta-phi-for(-a)}.
	}
	\label{fig:E-A for j=5/2---4} 
\end{figure}

Now let us take
\begin{equation}
\label{Lambda three anti-comm}
\Lambda(\theta,\phi)=
\sin\theta\left(\cos\phi\, A_1+\sin\phi\, A_2\right)+
\cos\theta\,A_3
\end{equation}
as per \eqref{L} and present the allowed region $\mathcal{E}$ of the mean vector 
${\vec{\boldsymbol{\varepsilon}}=
	(\langle A_1\rangle,\langle A_2\rangle,\langle A_3\rangle)}$.
In the case of ${j=\tfrac{1}{2}}$, all the anticommutators are the null operator, thus
$\mathcal{E}$ only carries the origin ${(0,0,0)}$.
For ${j=1,\cdots,\tfrac{5}{2}}$, in the following, the maximum eigenvalue ${\lambda_\textsc{m}(\theta,\phi)}$ and the associated eigenkets ${|\theta,\phi\rangle}$ of ${\Lambda(\theta,\phi)}$ are provided as functions of ${\theta\in[0,\pi]}$ and ${\phi\in[0,2\pi)}$.
Then it becomes cumbersome to deliver analytic expressions for them.
Therefore,
in the case of ${j=3,\tfrac{7}{2},4,25,}$ and ${50}$, we divide
both the intervals ${[0,\pi]}$ and ${[0,2\pi]}$
into, say, \textsc{k} and $\textsc{k}'$ numbers of equal parts.
And, only the angles
${\theta_k=k\tfrac{\pi}{\textsc{k}}}$ and 
${\phi_{k'}=k'\tfrac{2\pi}{\textsc{k}'}}$ are chosen, where 
${k=0,1,\cdots,\textsc{k}}$ and 
${k'=0,1,\cdots,\textsc{k}'}$.
For every
${(\theta_k,\phi_{k'})}$, we compute $\lambda_\textsc{m}$,  ${|\theta,\phi\rangle}$, face \eqref{E-para} of $\mathcal{E}$, and then by joining these faces
we draw boundary \eqref{boundary} of $\mathcal{E}$
in Figs.~\ref{fig:E-A for j=5/2---4} and \ref{fig:E-A for j=25,50}.
In the next section, almost all plots are generated in this way.

In an even dimension ${d={2j+1}}$, ${\lambda_\textsc{m}(\theta,\phi)}$ is twofold degenerate at every $\theta$ and $\phi$.
Hence, two orthogonal eigenkets ${|\theta,\phi\rangle_\mathit{1,2}}$ are provided for ${j=\tfrac{3}{2},\tfrac{5}{2}}$.
In this section, the subscript of a ket, say, $|\theta,\phi\rangle_\mathit{1}$ is just an index, and it does not represent that the ket belongs to an invariant subspace  $\mathscr{S}_1$ like before.
In odd dimensions ${d=3,5,7,}$ and 9, ${\lambda_\textsc{m}(\theta,\phi)}$ is twice degenerate only at ${\{(\theta_l,\phi_l)\}_{l=1}^4}$, which are recorded in Table~\ref{tab:theta-phi}.

\begin{table}[]
	\centering
	\caption{
	Four $(\theta_l,\phi_l)$ and the corresponding unit vectors \eqref{n-vector}
	are listed in the right and left columns.
	By replacing $\theta$ and $\phi$ with ${\pi-\theta}$ and ${\pi+\phi}$, respectively, one can have the angles for the antipodal vectors ${\{-\widehat{\eta}_l\}_{l=1}^{4}}$.  	
		}
	\label{tab:theta-phi}
\begin{tabular}{l@{\hspace{2mm}} | @{\hspace{2mm}}l@{\hspace{3mm}} l }
		\hline\hline\rule{0pt}{2.5ex}  
$\quad\ \widehat{\eta}(\theta,\phi)$ & $\quad\ \theta$ & $\quad\ \phi$  
        \\
		\hline\rule{0pt}{2.5ex} 
	
${\widehat{\eta}_1=\tfrac{1}{\sqrt{3}}(1,1,1)}\ $ & 
$\theta_1=\arccos(\tfrac{1}{\sqrt{3}})$ & $\phi_1=\tfrac{\pi}{4}$
		\\[2mm]
${\widehat{\eta}_2=\tfrac{1}{\sqrt{3}}(-1,1,-1)}$ & 
$\theta_2=\pi-\theta_1$ & $\phi_2=\tfrac{3\pi}{4}$ 
		\\[2mm]
${\widehat{\eta}_3=\tfrac{1}{\sqrt{3}}(-1,-1,1)}$ & 
$\theta_3=\theta_1$ & $\phi_3=\tfrac{5\pi}{4}$
		\\[2mm]
${\widehat{\eta}_4=\tfrac{1}{\sqrt{3}}(1,-1,-1)}$ & 
$\theta_4=\theta_2$ & $\phi_4=\tfrac{7\pi}{4}$   \\[1.5mm]			
		\hline\hline
	\end{tabular}
	\label{tab:theta-phi-for(-a)}
\end{table}

\begin{figure}[ ]
	\centering
	\includegraphics[width=0.25\textwidth]{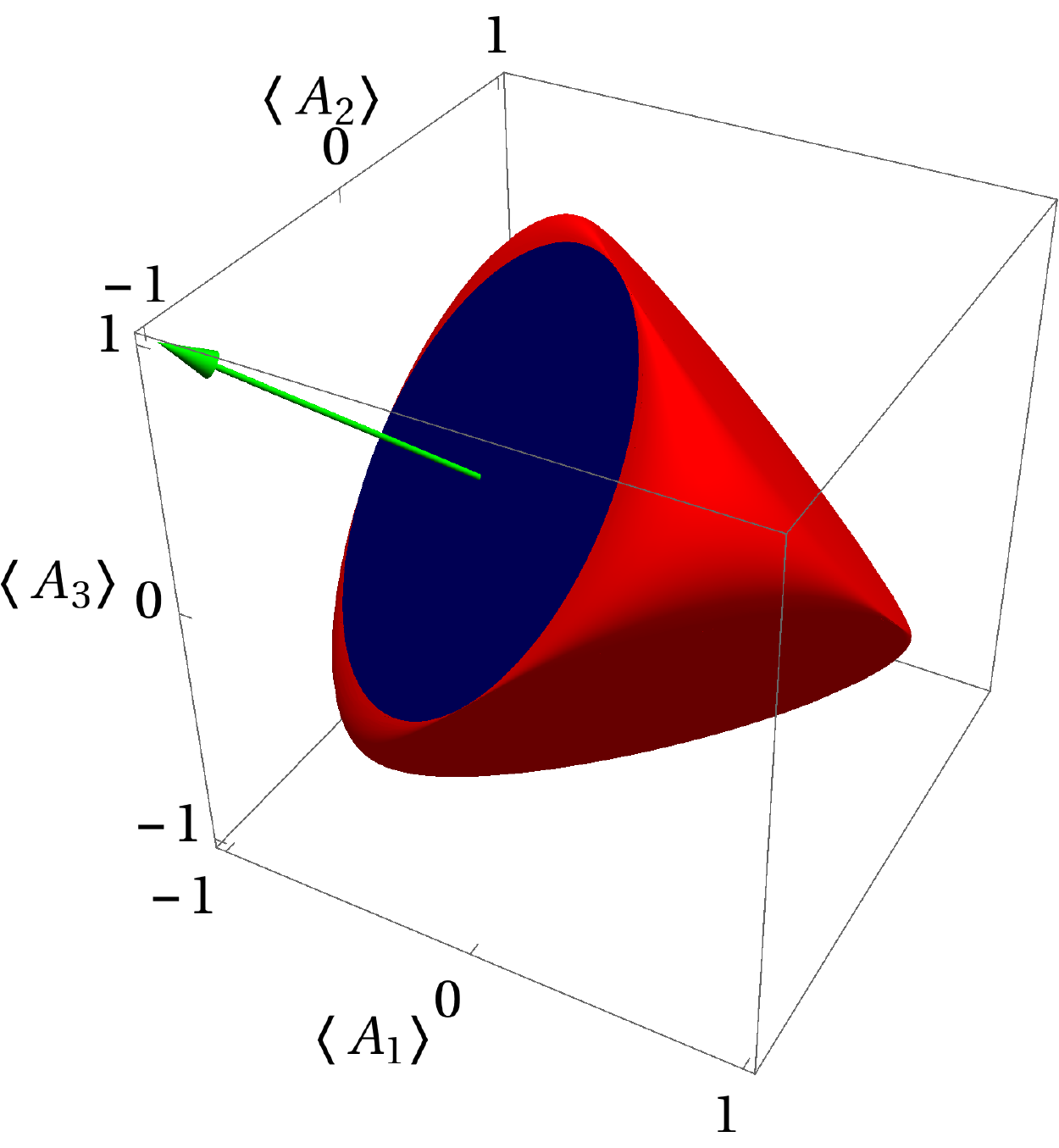} 
	\caption{ (Color online) In the case of ${j=1}$, ${a_\textsc{m}=1}$, and the numerical range $\mathcal{E}$ of ${\vec{E}=(A_1,A_2,A_3)}$ is the red strawberry-shaped region with the blue circular disk.
	The disk is the face $\mathcal{F}(\theta_3,\phi_3)$ of $\mathcal{E}$, and the arrow is in the direction of its normal $\widehat{\eta}_3$.
	There are total four such disks---normal to each unit vector $\widehat{\eta}_l$ in Table~\ref{tab:theta-phi-for(-a)}---that are parts of the boundary ${\partial\mathcal{E}}$.
	}
	\label{fig:E-A for j=3}  
\end{figure}

Now we begin with ${j=1}$, where 
\begin{equation}
\label{lmax-A (1)}
\lambda_\textsc{m}(\theta,\phi)=
\tfrac{2}{\sqrt{3}}
\cos\big(\tfrac{1}{3}\arccos
\big(\tfrac{-\sqrt{27}}{2}\cos\theta(\sin\theta)^2\sin(2\phi)\big)
\big).
\end{equation}
At all the four degeneracy points ${\lambda_\textsc{m}=\tfrac{1}{\sqrt{3}}}$, and the pair ${|\theta_l,\phi_l\rangle_\mathit{1,2}}$ of eigenkets are 
\begin{align}
\label{lmax-ket(1)-A (1)}
|\theta_1,\phi_1\rangle_\mathit{1}&=
\tfrac{1}{\sqrt{2}}\left(|{+1}\rangle +\text{i}\,|{-1}\rangle\right)
=|\theta_3,\phi_3\rangle_\mathit{1}\,,
\nonumber\\
|\theta_2,\phi_2\rangle_\mathit{1}&=
\tfrac{1}{\sqrt{2}}\left(|{+1}\rangle -\text{i}\,|{-1}\rangle\right)
=|\theta_4,\phi_4\rangle_\mathit{1}\,,
\nonumber\\
|\theta_1,\phi_1\rangle_\mathit{2}&=
\tfrac{1}{\sqrt{6}}
\left(e^{-\text{i}\frac{\pi}{4}}|{+1}\rangle +
2\,|0\rangle-
e^{\text{i}\frac{\pi}{4}}|{-1}\rangle\right),
\\
|\theta_2,\phi_2\rangle_\mathit{2}&=
\tfrac{1}{\sqrt{6}}
\left(-e^{\text{i}\frac{\pi}{4}}|{+1}\rangle +
2\,|0\rangle+
e^{-\text{i}\frac{\pi}{4}}|{-1}\rangle\right),
\nonumber\\
|\theta_3,\phi_3\rangle_\mathit{2}&=
\tfrac{1}{\sqrt{6}}
\left(-e^{-\text{i}\frac{\pi}{4}}|{+1}\rangle +
2\,|0\rangle+
e^{\text{i}\frac{\pi}{4}}|{-1}\rangle\right),
\quad\mbox{and}
\nonumber\\
|\theta_4,\phi_4\rangle_\mathit{2}&=
\tfrac{1}{\sqrt{6}}
\left(e^{\text{i}\frac{\pi}{4}}|{+1}\rangle +
2|0\rangle-
e^{-\text{i}\frac{\pi}{4}}|{-1}\rangle\right).
\nonumber
\end{align}
At each $(\theta_l,\phi_l)$, by varying ${\mu\in[0,\tfrac{\pi}{2}]}$ and ${\nu\in[0,2\pi)}$ in a general eigenket
\begin{equation}
\label{theta-phi-l-ket}
|\theta_l,\phi_l\rangle=
\cos\mu\,|\theta_l,\phi_l\rangle_\mathit{1}+
\sin\mu\,e^{\text{i}\nu}|\theta_l,\phi_l\rangle_\mathit{2}\,,
\end{equation}
one can generate the face
$\mathcal{F}(\theta_l,\phi_l)$ of allowed region $\mathcal{E}$
according to \eqref{E-para}.
These four faces are circular disks,
$\mathcal{F}(\theta_3,\phi_3)$ on $\mathcal{E}$ is shown in Fig.~\ref{fig:E-A for j=3} with its normal vector $\widehat{\eta}_3$.
Moreover, the four circles (boundary of the disks) are intersections of $\partial\mathcal{E}$ and the unit sphere centered at the origin.

The maximum-eigenvalue-ket
\begin{equation}
\label{lmax-ket(2)-A (1)}
|\theta,\phi\rangle=
\begin{cases}
|\theta_1,\phi_1\rangle_\mathit{1}
& \text{for  }  \theta\in[0,\theta_1)
\  \mbox{and}\  
\phi\in\{\phi_1,\phi_3\} \\
|\theta_2,\phi_2\rangle_\mathit{1}
& \text{for  }  \theta\in(\theta_2,\pi]
\  \mbox{and}\  
\phi\in\{\phi_2,\phi_4\} \\
\end{cases},
\end{equation}
and for the rest of angles, the eigenkets are
\begin{align}
\label{lmax-ket(3)-A (1)}
&|\theta,\phi\rangle\ \propto\ 
\alpha\,|{+1}\rangle +
\beta\,|0\rangle+
\varsigma\,|{-1}\rangle\,,
\quad\mbox{where}
\nonumber\\
&\alpha=(e^{-\text{i}\phi}\lambda_\textsc{m}+\text{i}\,e^{\text{i}\phi}\cos\theta)
\sin\theta\,,\\
&\beta=\sqrt{2}\,(\lambda_\textsc{m}^2-(\cos\theta)^2)\,,
\qquad\qquad\quad\mbox{and}
\nonumber\\
&\varsigma=(-e^{\text{i}\phi}\lambda_\textsc{m}+\text{i}\,e^{-\text{i}\phi}\cos\theta)
\sin\theta\,.
\nonumber
\end{align}
The eigenkets in \eqref{lmax-ket(3)-A (1)} provide
the parametric form
\begin{align}
\label{para-A (1)}
\langle\theta,\phi|A_1|\theta,\phi\rangle&=
\tfrac{\beta\big((\alpha+\overline{\alpha})-(\varsigma+\overline{\varsigma})\big)}
{\sqrt{2}\big(|\alpha|^2+\beta^2+|\varsigma|^2\big)}\,,
\nonumber\\
\langle\theta,\phi|A_2|\theta,\phi\rangle&=
\tfrac{\text{i}\,\beta\big((\alpha-\overline{\alpha})+(\varsigma-\overline{\varsigma})\big)}
{\sqrt{2}\big(|\alpha|^2+\beta^2+|\varsigma|^2\big)}\,,\\
\langle\theta,\phi|A_3|\theta,\phi\rangle&=
\tfrac{\text{i}\,(\alpha\overline{\varsigma}-\overline{\alpha}\varsigma)}
{|\alpha|^2+\beta^2+|\varsigma|^2}\nonumber
\end{align}
of the part---other than the four circular disks---of the boundary ${\partial\mathcal{E}}$. The overline in $\overline{\alpha}$ denotes the complex conjugation of $\alpha$.
With \eqref{para-A (1)}, one can realize that $\mathcal{E}$ is the convex hull of Steiner's Roman surface described by
\begin{equation}
\label{Roman j=1}
(\texttt{a}_1\texttt{a}_2)^2+
(\texttt{a}_2\texttt{a}_3)^2+
(\texttt{a}_3\texttt{a}_1)^2=
-2\,\texttt{a}_1\texttt{a}_2\texttt{a}_3
\end{equation}
in \cite{Henrion11} (see also \cite{Weis11,Szymanski18}),
here ${-1\leq \texttt{a}_t=\tfrac{\langle A_t\rangle}{a_\textsc{m}} \leq 1}$ 
and $t=1,2,3$.

We combine (un)certainty measures for the three anticommutators with respect to \eqref{Xdot}--\eqref{u-half-sum}.
Then, to reach their absolute extrema
on the boundary, we vary the two angles in \eqref{para-A (1)} and thus
obtain the tight URs and CRs
\begin{eqnarray}
\label{H-UR-A (1)}
\tfrac{1}{2}\big(6\ln 6-5\ln 5\big)
&\leq&\sum_{t=1}^3 h(\langle A_t\rangle)=h(\vec{\boldsymbol{\varepsilon}}\,)\,,\\
\label{u-UR-A (1)}
1+2\sqrt{2}&\leq& \sum_{t=1}^3 
u_{\sfrac{1}{2}}(\langle A_t\rangle)\,,\\
\label{u2-UR-A (1)}
&& \sum_{t=1}^3 u_2(\langle A_t\rangle)\leq \tfrac{13}{6} \,,
\quad \mbox{and}\qquad\\
\label{umax-UR-A (1)}
&& \sum_{t=1}^3 u_\text{max}(\langle A_t\rangle)\leq 
\tfrac{5}{2}\,.
\end{eqnarray}
Inequalities~\eqref{H-UR-A (1)}, \eqref{u2-UR-A (1)}, and \eqref{umax-UR-A (1)}
are saturated by
the maximum-eigenvalue-states of ${-\widehat{\eta}_l \cdot \vec{E}}$ for ${l=1,\cdots,4}$ [for $\widehat{\eta}_l$, see Table~\ref{tab:theta-phi-for(-a)}].
Whereas \eqref{u-UR-A (1)} is saturated by the six extreme-eigenvalue-states of $A_1,A_2,$ and $A_3$.

Next we pick ${j=\tfrac{3}{2}}$, where ${\lambda_\textsc{m}(\theta,\phi)=\sqrt{3}=a_\textsc{m}}$ is the same for all angles. One can check that both the orthonormal eigenkets
\begin{align}
\label{lmax-ket-A (3/2)}
|\theta,\phi\rangle_\mathit{1}&=\tfrac{1}{\sqrt{2}}
\left(
e^{-\text{i}\phi}\sin\theta|{+\tfrac{3}{2}}\rangle + 
|{+\tfrac{1}{2}}\rangle 
+\text{i}\cos\theta|{-\tfrac{3}{2}}\rangle
\right),
\nonumber\\
|\theta,\phi\rangle_\mathit{2}&=\tfrac{1}{\sqrt{2}}
\left(-\text{i}\cos\theta|{+\tfrac{3}{2}}\rangle + 
|{-\tfrac{1}{2}}\rangle 
-e^{\text{i}\phi}\sin\theta|{-\tfrac{3}{2}}\rangle
\right)
\end{align}
and all there normalized linear combinations such as \eqref{theta-phi-l-ket} deliver a single point 
on the sphere of radius $\sqrt{3}$ centered at the origin.
In a nutshell, here the numerical range is bounded by the QC
\begin{equation}
\label{QC-A (3/2)}
\langle A_1\rangle^2+\langle A_2\rangle^2+\langle A_3\rangle^2\leq a_\textsc{m}^2\,.
\end{equation}

Now we move to ${j=2}$, where ${\lambda_\textsc{m}(\theta,\phi)=\sqrt{12}=a_\textsc{m}}$
is also independent of both the angles, hence $\mathcal{E}$ is completely determined by
QC~\eqref{QC-A (3/2)}.
Except for ${\theta\in\{\theta_1,\theta_2\}}$, here the maximum-eigenvalue-ket is given by
\begin{align}
\label{lmax-ket-A (2)}
&|\theta,\phi\rangle\ \propto\
\alpha|{+2}\rangle+\beta|{+1}\rangle+
|0\rangle+
\varsigma|{-1}\rangle+\xi|{-2}\rangle,
\ \mbox{where}\nonumber\\
&\alpha=\tfrac{3e^{-\text{i}\phi}(\sin\theta)\,\beta-\text{i}\sqrt{6}\cos\theta}
{\sqrt{12}},\nonumber\\
&\xi=\tfrac{-3e^{\text{i}\phi}(\sin\theta)\,\varsigma+\text{i}\sqrt{6}\cos\theta}
{\sqrt{12}},\\
&\beta=\tfrac{
\big(\left(-\text{i}3e^{\text{i}\phi}\cos\theta+\sqrt{3}e^{-\text{i}\phi}\right)
c-\text{i}3\cos\theta
\left(-\text{i}3e^{-\text{i}\phi}\cos\theta-\sqrt{3}e^{\text{i}\phi}\right)\big)
\sin\theta}
{\sqrt{2}(c^2-9(\cos\theta)^2)},\nonumber\\
&\varsigma=\tfrac{
\big(\left(-\text{i}3e^{-\text{i}\phi}\cos\theta-\sqrt{3}e^{\text{i}\phi}\right)
c+\text{i}3\cos\theta
\left(-\text{i}3e^{\text{i}\phi}\cos\theta+\sqrt{3}e^{-\text{i}\phi}\right)\big)
\sin\theta}
{\sqrt{2}(c^2-9(\cos\theta)^2)},\nonumber\\
&c=\sqrt{12}-\tfrac{9 (\sin\theta)^2}{\sqrt{12}}.\nonumber
\end{align}
At ${\theta=\theta_1}$ and ${\theta=\theta_2}$, the denominators of $\beta$ and $\varsigma$ become zero, then \eqref{lmax-ket-A (2)} is no more applicable.
In these cases, the eigenkets are
\begin{align}
\label{lmax-ket(2)-A (2)}
&|\theta_1,\phi\rangle=\tfrac{1}{\sqrt{6}}
\big(
e^{-\text{i}\phi}|{+2}\rangle+\sqrt{2}|{+1}\rangle+
\text{i}\sqrt{2}|{-1}\rangle-\text{i}e^{\text{i}\phi}|{-2}\rangle
\big),\nonumber\\
&|\theta_2,\phi\rangle=\tfrac{1}{\sqrt{6}}
\big(
e^{-\text{i}\phi}|{+2}\rangle+\sqrt{2}|{+1}\rangle-
\text{i}\sqrt{2}|{-1}\rangle+\text{i}e^{\text{i}\phi}|{-2}\rangle
\big).
\end{align}
As $\lambda_\textsc{m}$ is twofold degenerate in the four directions.
For each ${(\theta_l,\phi_l)}$ in Table~\ref{tab:theta-phi-for(-a)}, one eigenket can be retrieved from \eqref{lmax-ket(2)-A (2)} and the other orthonormal eigenkets are 
\scriptsize
\begin{align}
\label{lmax-ket(3)-A (2)}
|\theta_1,\phi_1\rangle_\mathit{2}&=\tfrac{1}{3}
\left(
\sqrt{2}|{+2}\rangle+
e^{\text{i}\frac{\pi}{4}}|{+1}\rangle+
\text{i}\sqrt{3}|{0}\rangle+
e^{-\text{i}\frac{\pi}{4}}|{-1}\rangle
-\sqrt{2}|{-2}\rangle
\right),\nonumber\\
|\theta_2,\phi_2\rangle_\mathit{2}&=\tfrac{1}{3}
	\left(
	-\sqrt{2}|{+2}\rangle+
	e^{-\text{i}\frac{\pi}{4}}|{+1}\rangle
	+\text{i}\sqrt{3}|{0}\rangle+
	e^{\text{i}\frac{\pi}{4}}|{-1}\rangle
	+\sqrt{2}|{-2}\rangle
	\right),
\nonumber\\
|\theta_3,\phi_3\rangle_\mathit{2}&=\tfrac{1}{3}
	\left(
	-\sqrt{2}|{+2}\rangle+
	e^{\text{i}\frac{\pi}{4}}|{+1}\rangle
	-\text{i}\sqrt{3}|{0}\rangle+
	e^{-\text{i}\frac{\pi}{4}}|{-1}\rangle
	+\sqrt{2}|{-2}\rangle
	\right),
\nonumber\\
|\theta_4,\phi_4\rangle_\mathit{2}&=\tfrac{1}{3}
	\left(
	-\sqrt{2}|{+2}\rangle
	-e^{-\text{i}\frac{\pi}{4}}|{+1}\rangle
   +\text{i}\sqrt{3}|{0}\rangle
	-e^{\text{i}\frac{\pi}{4}}|{-1}\rangle
	+\sqrt{2}|{-2}\rangle
	\right).
\end{align}
\normalsize
One can verify that the eigenkets in \eqref{lmax-ket-A (2)}--\eqref{lmax-ket(3)-A (2)} yield different points
\begin{equation}
\label{para-A (2)}
\langle \theta,\phi |\vec{E}\,|\theta,\phi \rangle= a_\textsc{m}\,\widehat{\eta}(\theta,\phi)
\end{equation}
of the boundary ${\partial\mathcal{E}}$, which is a sphere.

Like~\eqref{H-UR-A (1)}--\eqref{umax-UR-A (1)},
here we achieve
\begin{eqnarray}
\label{H-UR-A (2)}
2\ln 2=\mathfrak{h}&\leq&h(\vec{\boldsymbol{\varepsilon}}\,)\\
\label{u-UR-A (2)}
1+2\sqrt{2}=\mathfrak{u_{\sfrac{1}{2}}}&\leq& 
u_{\sfrac{1}{2}}(\vec{\boldsymbol{\varepsilon}}\,)\,,\\
\label{u2-UR-A (2)}
&&u_2(\vec{\boldsymbol{\varepsilon}}\,)\leq\mathfrak{u_2}= 2 \,,
\quad \mbox{and}\qquad\\
\label{umax-UR-A (2)}
&&u_\text{max}(\vec{\boldsymbol{\varepsilon}}\,)\leq 
\mathfrak{u_{max}}=
\tfrac{1}{2}(3+\sqrt{3}) \qquad\
\end{eqnarray}
by using parametric form \eqref{para-A (2)} for both ${j=\tfrac{3}{2}}$ as well as ${j=2}$.
Since ${\partial\mathcal{E}}$ specified by \eqref{JxJyJz-para} is also a sphere, the same URs and CRs \eqref{H-UR-A (2)}--\eqref{umax-UR-A (2)} are reported in \cite{Sehrawat17b} for the three angular momentum operators.
QC \eqref{QC-A (3/2)} and CR \eqref{u2-UR-A (2)} are equal, therefore every 
${|\theta,\phi \rangle\langle \theta,\phi|}$ saturates \eqref{u2-UR-A (2)}.
For URs~\eqref{H-UR-A (2)} and \eqref{u-UR-A (2)},
the minimum uncertainty states are 
the eigenstates of $A_1,A_2,A_3$ related to their extreme-eigenvalues $\pm a_\textsc{m}$. 
The maximum- and minimum-eigenvalue-states of ${\Lambda(\theta_l,\phi_l)}$
for every $l=1,\cdots,4$ saturate CR \eqref{umax-UR-A (2)}.

Next we take ${j=\tfrac{5}{2}}$, where $\lambda_\textsc{m}(\theta,\phi)$ is
\begin{equation}
\label{lmax-A (5/2)}
4
\sqrt{\tfrac{7}{3}}
\cos\Big(\tfrac{1}{3}\arccos
\Big(5\cos\theta(\sin\theta)^2\sin(2\phi)\sqrt{\tfrac{3^3}{7^3}}\,
\Big)
\Big)
\end{equation}
for all $\theta$ and $\phi$.
The two orthonormal eigenkets associated with $\lambda_\textsc{m}$ are
 \begin{align}
 \label{lmax-ket-A (5/2)}
 |\theta,\phi\rangle_\mathit{1}&\ \propto\ 
 \alpha_\mathit{1}|{+\tfrac{5}{2}}\rangle+\beta_\mathit{1}|{+\tfrac{3}{2}}\rangle+
 |{+\tfrac{1}{2}}\rangle+\varsigma_\mathit{1}|{-\tfrac{3}{2}}\rangle+
\xi_\mathit{1}|{-\tfrac{5}{2}}\rangle\,,
 \nonumber\\
 |\theta,\phi\rangle_\mathit{2}&\ \propto\ 
 \alpha_\mathit{2}|{+\tfrac{5}{2}}\rangle+\beta_\mathit{2}|{+\tfrac{3}{2}}\rangle+
 |{-\tfrac{1}{2}}\rangle+\varsigma_\mathit{2}|{-\tfrac{3}{2}}\rangle+
\xi_\mathit{2}|{-\tfrac{5}{2}}\rangle\,,
 \end{align}
 where
 \begin{align}
 \label{lmax-ket(2)-A (5/2)}
&\alpha_\mathit{1}=\tfrac{2\sqrt{5}\,e^{-\text{i}\phi}(\sin\theta)\,\beta_\mathit{1}
	-\text{i}\sqrt{10}\cos\theta} 
        {\lambda_\textsc{m}}\,,\nonumber\\
&\xi_\mathit{1}=\tfrac{-2\sqrt{5}e^{\text{i}\phi}(\sin\theta)\,\varsigma_\mathit{1}}
          {\lambda_\textsc{m}}\,,\nonumber\\
&\beta_\mathit{1}=\tfrac{2\left(-\text{i}\sqrt{50}\,e^{\text{i}\phi}\cos\theta+
	\sqrt{2}\,e^{-\text{i}\phi}\lambda_\textsc{m}\right)\sin\theta}
          {\lambda_\textsc{m}^2-20(\sin\theta)^2}\,,\nonumber\\
&\varsigma_\mathit{1}=\tfrac{\text{i}\,3\sqrt{2}\cos\theta\lambda_\textsc{m}}
          {\lambda_\textsc{m}^2-20(\sin\theta)^2}\,,\\
&\alpha_\mathit{2}=\tfrac{2\sqrt{5}e^{-\text{i}\phi}(\sin\theta)\,\beta_\mathit{2}}
          {\lambda_\textsc{m}}\,,\nonumber\\
&\xi_\mathit{2}=\tfrac{-2\sqrt{5}\,e^{\text{i}\phi}(\sin\theta)\,\varsigma_\mathit{2}
	+\text{i}\sqrt{10}\cos\theta} 
        {\lambda_\textsc{m}}\,,\nonumber\\
&\beta_\mathit{2}=\tfrac{-\text{i}\,3\sqrt{2}\cos\theta\lambda_\textsc{m}}
         {\lambda_\textsc{m}^2-20(\sin\theta)^2}\,,
\qquad\mbox{and}
\nonumber\\
&\varsigma_\mathit{2}=\tfrac{2\left(-\text{i}\sqrt{50}\,e^{-\text{i}\phi}\cos\theta-
	\sqrt{2}\,e^{\text{i}\phi}\lambda_\textsc{m}\right)\sin\theta}
          {\lambda_\textsc{m}^2-20(\sin\theta)^2}\,.\nonumber
 \end{align}
Both $|\theta,\phi\rangle_\mathit{1,2}$ and all their superpositions provide a single boundary-point ${\langle\theta,\phi|\vec{E}|\theta,\phi\rangle}$ of $\mathcal{E}$.
The permitted regions for ${\vec{E}=(A_1,A_2,A_3)}$---with their four supporting hyperplanes~\eqref{hyperplane} at $\{(\pi-\theta_l,\pi+\phi_l)\}_{l=1}^4$---are displayed in Fig.~\ref{fig:E-A for j=5/2---4} for ${j=\tfrac{5}{2},3,\tfrac{7}{2},}$
and ${4}$.
To draw ${\partial\mathcal{E}}$ for ${j=\tfrac{5}{2}}$,
\eqref{lmax-A (5/2)}--\eqref{lmax-ket(2)-A (5/2)} are used.
While, for $j=3,\tfrac{7}{2},$ and ${4}$, we obtain ${\partial\mathcal{E}}$ numerically as described at the beginning of this section.

Like \eqref{uni-scaling-J}, now we apply the uniform scaling
\begin{equation}
\label{uni-scaling-A}
(A_1,A_2,A_3)\mapsto\tfrac{1}{a_\textsc{m}}(A_1,A_2,A_3)=
\tfrac{1}{a_\textsc{m}}\vec{E}\,,
\end{equation}
and then the allowed region $\mathcal{E}^\textsc{us}$ of ${\tfrac{1}{a_\textsc{m}}\vec{\boldsymbol{\varepsilon}}}$
will be in the hyperrectangle ${\mathcal{H}^\textsc{us}=[-1,1]^{\times 3}}$
provided ${j\geq 1}$.
Up to ${j=50}$ we have checked that 
${\langle {\pm a_\textsc{m}^{\scriptscriptstyle(t)}}|A_{t'}|
	{\pm a_\textsc{m}^{\scriptscriptstyle(t)}}\rangle
=\pm a_\textsc{m}\delta_{t\,t'}}$ holds true, where
${|{\pm a_\textsc{m}^{\scriptscriptstyle(t)}}\rangle}$
are the eigenkets of $A_t$
corresponding to the extreme eigenvalues $\pm a_\textsc{m}$,
and $\delta_{t\,t'}$ is the Kronecker delta function.
The above statement seems to hold for all ${j\geq 1}$, hence
\begin{equation}
\label{pm 1 0 0}
(\pm 1,0,0),\ (0,\pm 1,0),\ \mbox{and}\ (0,0,\pm 1)
\end{equation}
will be the extreme points of $\mathcal{E}^\textsc{us}$.

\begin{figure}
	\centering
	\subfloat{\includegraphics[width=40mm]{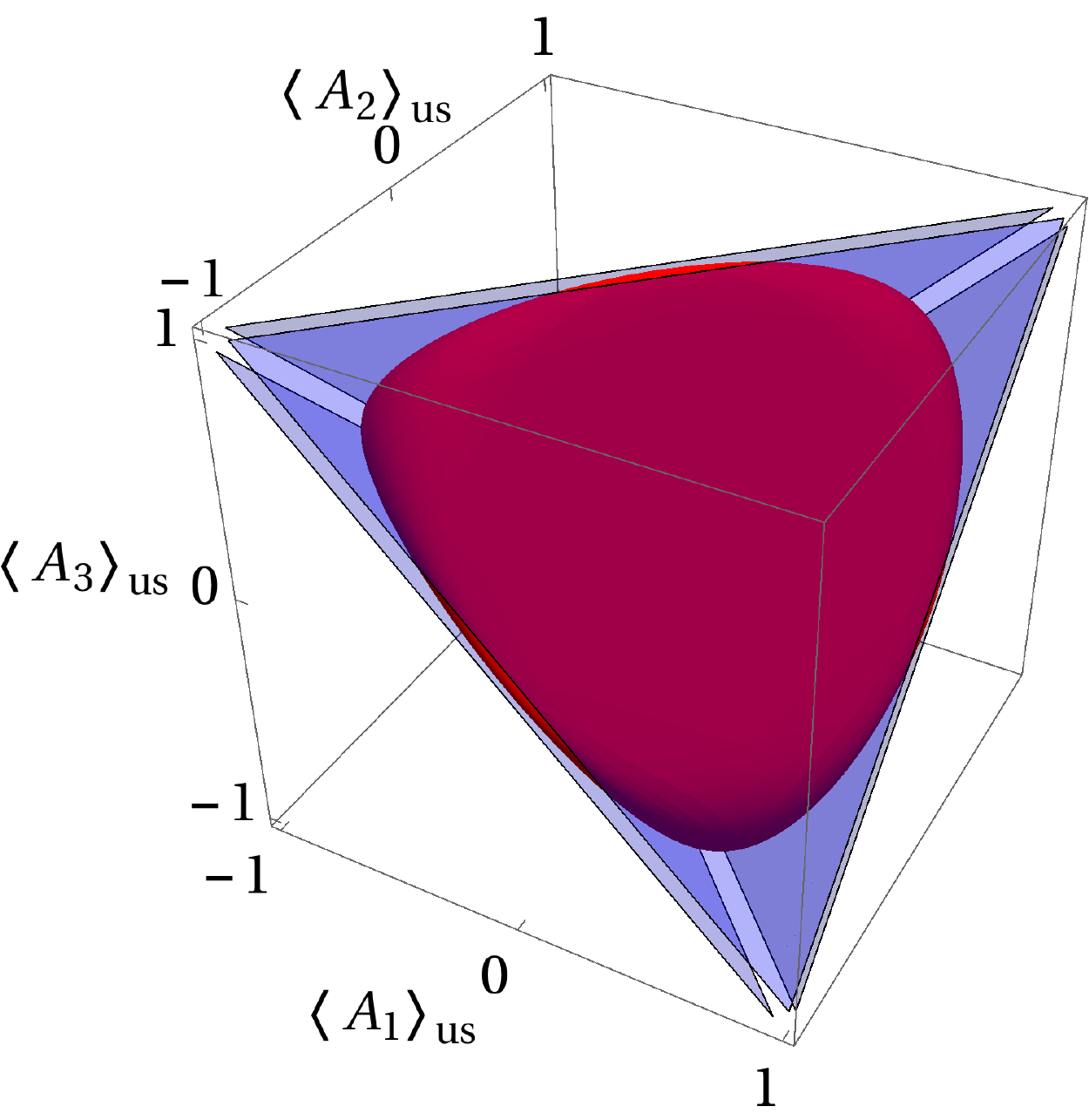}}\quad\
	\subfloat{\includegraphics[width=40mm]{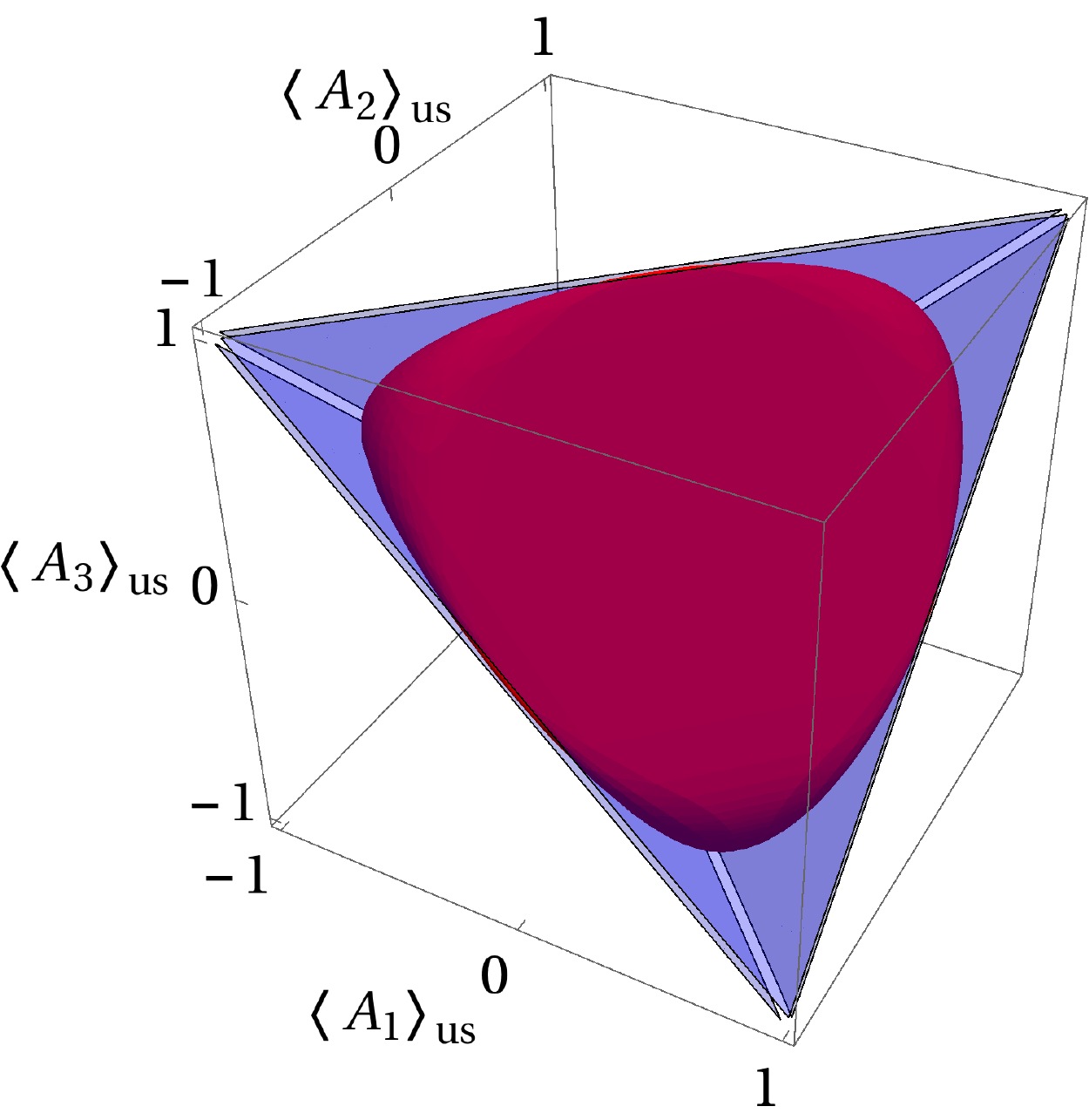}}
	\caption{(Color online) The red convex bodies on left- and right-hand-side
		represent regions $\mathcal{E}^\textsc{us}$ of the average vectors ${\tfrac{1}{a_\textsc{m}}\vec{\boldsymbol{\varepsilon}}}$ for the large quantum numbers
		${j=25}$ and ${j=50}$, respectively, for which ${a_\textsc{m}=614.689}$ and ${a_\textsc{m}=2479.333}$.
		Here $\langle A\rangle_\textsc{us}$ denotes $\tfrac{\langle A\rangle}{a_\textsc{m}}$.
		Like in Fig.~\ref{fig:E-A for j=5/2---4}, the four tangent hyperplane
		are painted in blue color.}
	\label{fig:E-A for j=25,50} 
\end{figure}

According to \eqref{QC-A (3/2)},
$\mathcal{E}^\textsc{us}$ is the unit ball for $j=\tfrac{3}{2},2$,
and as we increase $j$ one can observe in Figs.~\ref{fig:E-A for j=5/2---4} and 
\ref{fig:E-A for j=25,50} that $\mathcal{E}^\textsc{us}$ monotonically contracts in
the four directions $\{-\widehat{\eta}_l\}_{l=1}^4$
and expands in their antipodal directions $\{\widehat{\eta}_l\}_{l=1}^4$.
Using the quantum de Finetti theorem \cite{Stormer69,Hudson76}
in the limit ${j\rightarrow \infty}$, we show in Sec.~\ref{sec:N-qubit} that
$\mathcal{E}^\textsc{us}$ is the convex hull of 
Steiner's Roman surface characterized by
\begin{equation}
\label{Roman j=inf}
(\texttt{a}_1\texttt{a}_2)^2+
(\texttt{a}_2\texttt{a}_3)^2+
(\texttt{a}_3\texttt{a}_1)^2=
2\,\texttt{a}_1\texttt{a}_2\texttt{a}_3\,,
\end{equation}
recall that ${-1\leq \texttt{a}_t=\tfrac{\langle A_t\rangle}{a_\textsc{m}} \leq 1}$ for ${t=1,2,}$ and 3.
Even at a finite $j$ one can clearly recognize the Roman surface, \eqref{Roman j=inf}, 
in Fig.~\ref{fig:E-A for j=25,50} that displays 
$\mathcal{E}^\textsc{us}$ for ${j=25,50}$.

\begin{figure}
	\centering
	\subfloat{\includegraphics[width=40mm]{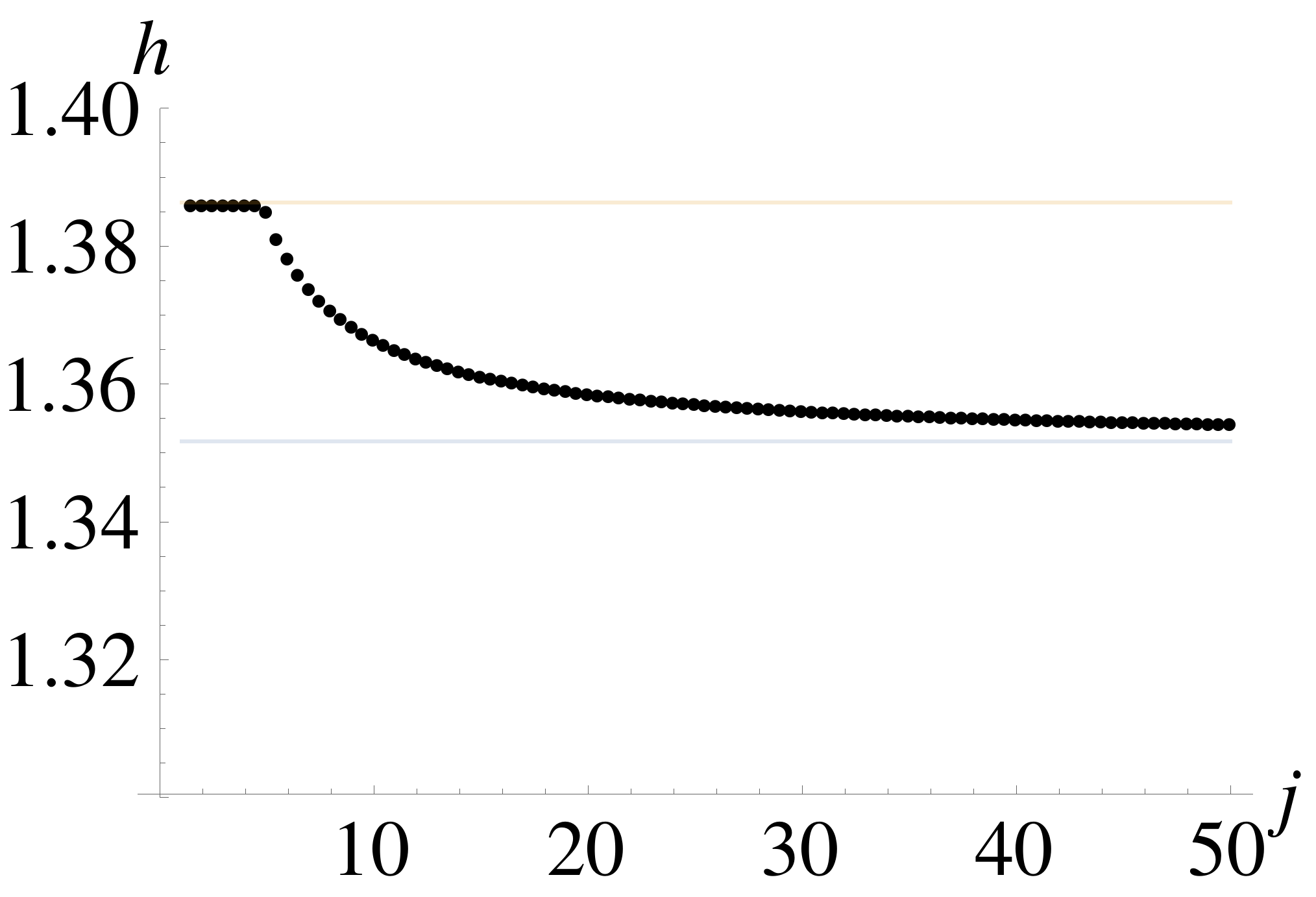}}\quad\
	\subfloat{\includegraphics[width=40mm]{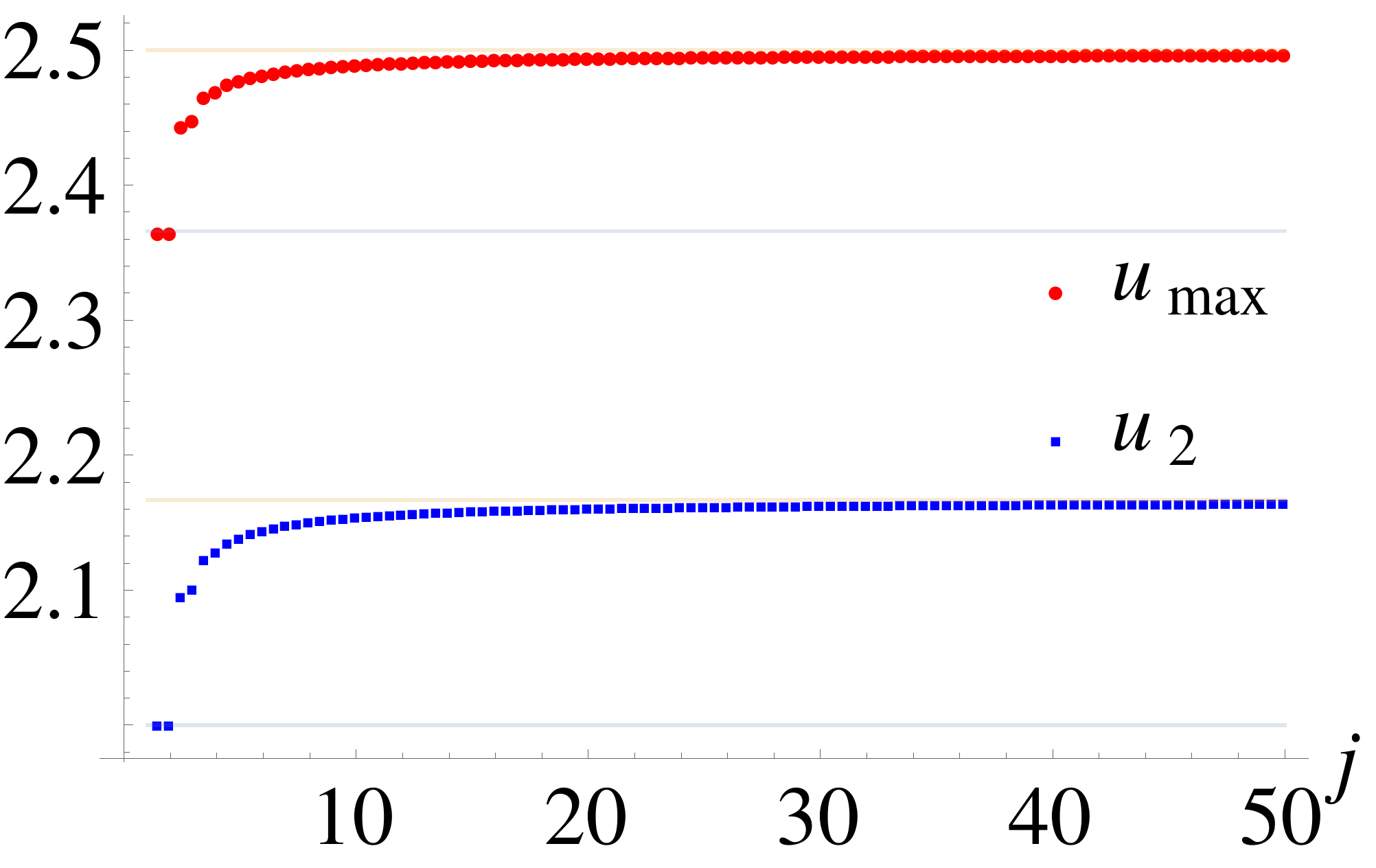}}
	\caption{
		We present list-plots of the tight lower bound $\mathfrak{h}$	
		and the tight upper bounds $\{\mathfrak{u_2},\mathfrak{u_{max}}\}$
		in the left and right panels
		for ${j=\tfrac{3}{2},2,\tfrac{5}{2}\cdots,50}$.
		These bounds are printed in Appendix~\ref{sec:Sup-material}, with which
		one can enjoy the tight UR
		${\mathfrak{h}\leq h(\vec{\boldsymbol{\varepsilon}}\,)}$
		and the CRs
		${\mathfrak{u_2}\leq u_2(\vec{\boldsymbol{\varepsilon}}\,)}$
		and
		${\mathfrak{u_{max}}\leq u_{\text{max}}(\vec{\boldsymbol{\varepsilon}}\,)}$
		for $\{A_1,A_2,A_3\}$.
	}
	\label{fig:h,u2,umax for A} 
\end{figure}

An uncertainty measure such as $h(\vec{\boldsymbol{\varepsilon}}\,)$
on $\mathcal{E}$ is basically 
$h(\tfrac{\vec{\boldsymbol{\varepsilon}}}{a_\textsc{m}})$
on $\mathcal{E}^\textsc{us}$.
For $j=\tfrac{3}{2},2,\cdots,\tfrac{9}{2}$,
$h(\tfrac{\vec{\boldsymbol{\varepsilon}}}{a_\textsc{m}})$
reaches its absolute minimum ${\mathfrak{h}=2\ln 2}$
at the six points in \eqref{pm 1 0 0}, and for all other $j\geq5$
the minimum $\mathfrak{h}$ occurs at the four points
\begin{align}
\label{point at t,f A}
\tfrac{\langle\theta_l,\phi_l |\vec{E}|\theta_l,\phi_l \rangle}{a_\textsc{m}}
=
\tfrac{\langle\theta_1,\phi_1 |A_1|\theta_1,\phi_1 \rangle}{a_\textsc{m}}
\sqrt{3}\,\widehat{\eta}_l
\end{align}
in the directions $\{\widehat{\eta}_l\}_{l=1}^4$.
Although the maximum eigenvalue of $\widehat{\eta}_l\cdot\vec{E}$ is double degenerate 
but all its associated eigenkets ${|\theta_l,\phi_l \rangle}$ give a single point \eqref{point at t,f A} as long as $j>1$.
The mean value
\begin{align}
\label{mean value t,f A}
\tfrac{\langle\theta_1,\phi_1 |A_1|\theta_1,\phi_1 \rangle}{a_\textsc{m}}
\in
\big[\tfrac{1}{\sqrt{3}},\tfrac{2}{3}\big]
\quad\mbox{for all}\quad j> 1\,,
\end{align}
and it is given in Appendix~\ref{sec:Sup-material} for 
$j=\tfrac{3}{2},\cdots,50.$
Taking those numerical values,
we compute $h(\vec{\boldsymbol{\varepsilon}}\,)$ at an extreme point \eqref{point at t,f A} in order to have
the tight lower bound $\mathfrak{h}$ for $j\geq5$.
The values of $\mathfrak{h}$ for ${j=\tfrac{3}{2},\cdots,50}$ are then plotted in Fig.~\ref{fig:h,u2,umax for A}
and recorded in Appendix~\ref{sec:Sup-material}.

The global minimum
$\mathfrak{u_{\sfrac{1}{2}}}=1+2\sqrt{2}$
of
combined uncertainty measure
$u_{\sfrac{1}{2}}(\vec{\boldsymbol{\varepsilon}}\,) \equiv u_{\sfrac{1}{2}}(\tfrac{\vec{\boldsymbol{\varepsilon}}}{a_\textsc{m}})$
occurs at points \eqref{pm 1 0 0} for every ${j\geq 1}$,
hence tight UR~\eqref{u-UR-A (2)} [see also \eqref{u-UR-A (1)}]
holds for all ${j\geq 1}$.
The extreme-eigenvalue-kets 
${|{\pm a_\textsc{m}^{\scriptscriptstyle(t)}}\rangle}$
of the anticommutators are the minimum uncertainty kets for this UR.
Next, both 
certainty measures
${u_2(\vec{\boldsymbol{\varepsilon}}\,)}$
and
${u_{\text{max}}(\vec{\boldsymbol{\varepsilon}}\,)}$
achieve their absolute maxima at
the four points in \eqref{point at t,f A}
for ${j>1}$, and thus
the CRs
${\mathfrak{u_2}\leq u_2(\vec{\boldsymbol{\varepsilon}}\,)}$
and
${\mathfrak{u_{max}}\leq u_{\text{max}}(\vec{\boldsymbol{\varepsilon}}\,)}$
are saturated by
the maximum-eigenvalue-states of ${\widehat{\eta}_l\cdotp\vec{E}}$,
$l=1,\cdots,4$.
Like $h(\vec{\boldsymbol{\varepsilon}}\,)$, taking 
mean value \eqref{mean value t,f A} from Appendix~\ref{sec:Sup-material}, we compute
the upper bounds $\mathfrak{u_2}$ and $\mathfrak{u_{max}}$ for $j=\tfrac{3}{2},\cdots,50$,
record them in Appendix~\ref{sec:Sup-material} and plot them in 
Fig.~\ref{fig:h,u2,umax for A}.

In a nutshell, 
the lower- and upper-bounds of combined uncertainty- and certainty-measures
obey:
\begin{align}
\label{H,u,u2,umax-UR-A (3)}
&\mathfrak{h}\in\big[\tfrac{1}{2}\big(6\ln 6-5\ln 5\big),2\ln 2\big]
\,,\quad
\mathfrak{u_{\sfrac{1}{2}}}=1+2\sqrt{2}\,,
\qquad\nonumber\\
&\mathfrak{u_2}\in \big[2,\tfrac{13}{6}\big],
\quad \mbox{and}\quad
\mathfrak{u_{max}}\in
\big[\tfrac{1}{2}(3+\sqrt{3}),\tfrac{5}{2}\big]
\end{align}
for all ${j=1,\tfrac{3}{2},\cdots,\infty}$.
In Fig.~\ref{fig:h,u2,umax for A} [see also Appendix~\ref{sec:Sup-material}], one can perceive that as $j$ increases
$\mathfrak{h}$ monotonically decreases whereas both $\mathfrak{u_2}$ and $\mathfrak{u_{max}}$ increases, and in the limit 
${j\rightarrow\infty}$ we have
\begin{align}
\label{point at nl-A}
\tfrac{\langle\theta_1,\phi_1 |A_1|\theta_1,\phi_1 \rangle}{a_\textsc{m}}
&\rightarrow\tfrac{2}{3}\,,
\nonumber\\
\mathfrak{h}
&\rightarrow\tfrac{1}{2}\big(6\ln 6-5\ln 5\big)\,,
\\
\mathfrak{u_2}
&\rightarrow\tfrac{13}{6},
\quad \mbox{and}\quad
\nonumber\\
\mathfrak{u_{max}}
&\rightarrow\tfrac{5}{2}\,.
\nonumber
\end{align}
Although relations \eqref{H-UR-A (1)}--\eqref{umax-UR-A (1)} are not always tight but they
are nontrivial and hold for all ${j=1,\tfrac{3}{2},\cdots,\infty}$.

%===========================================
\section{Powers of angular momentum operators}\label{sec:S-power}

\begin{figure}
	\centering
	\subfloat{\includegraphics[width=40mm]{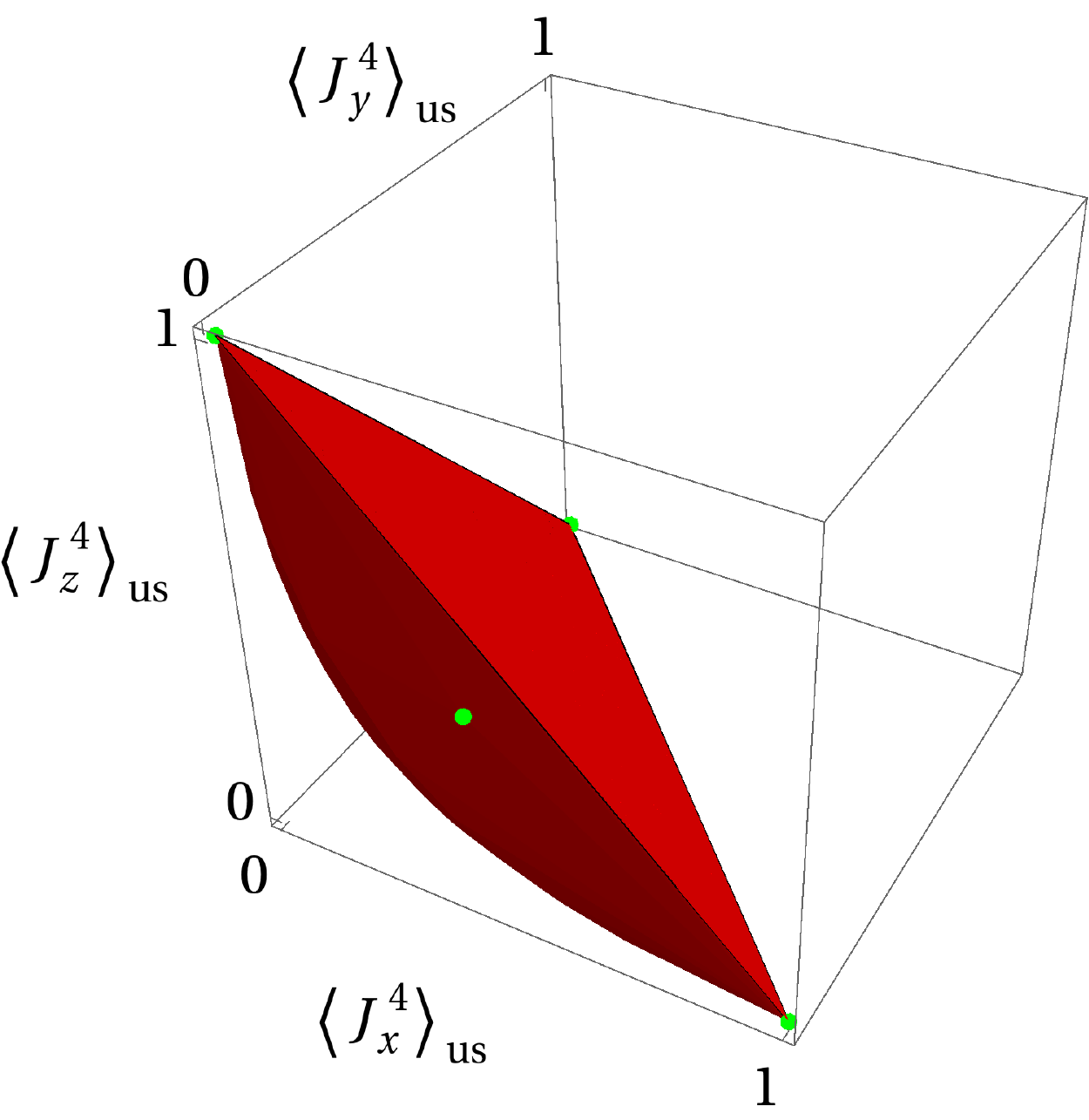}}\quad\
	\subfloat{\includegraphics[width=40mm]{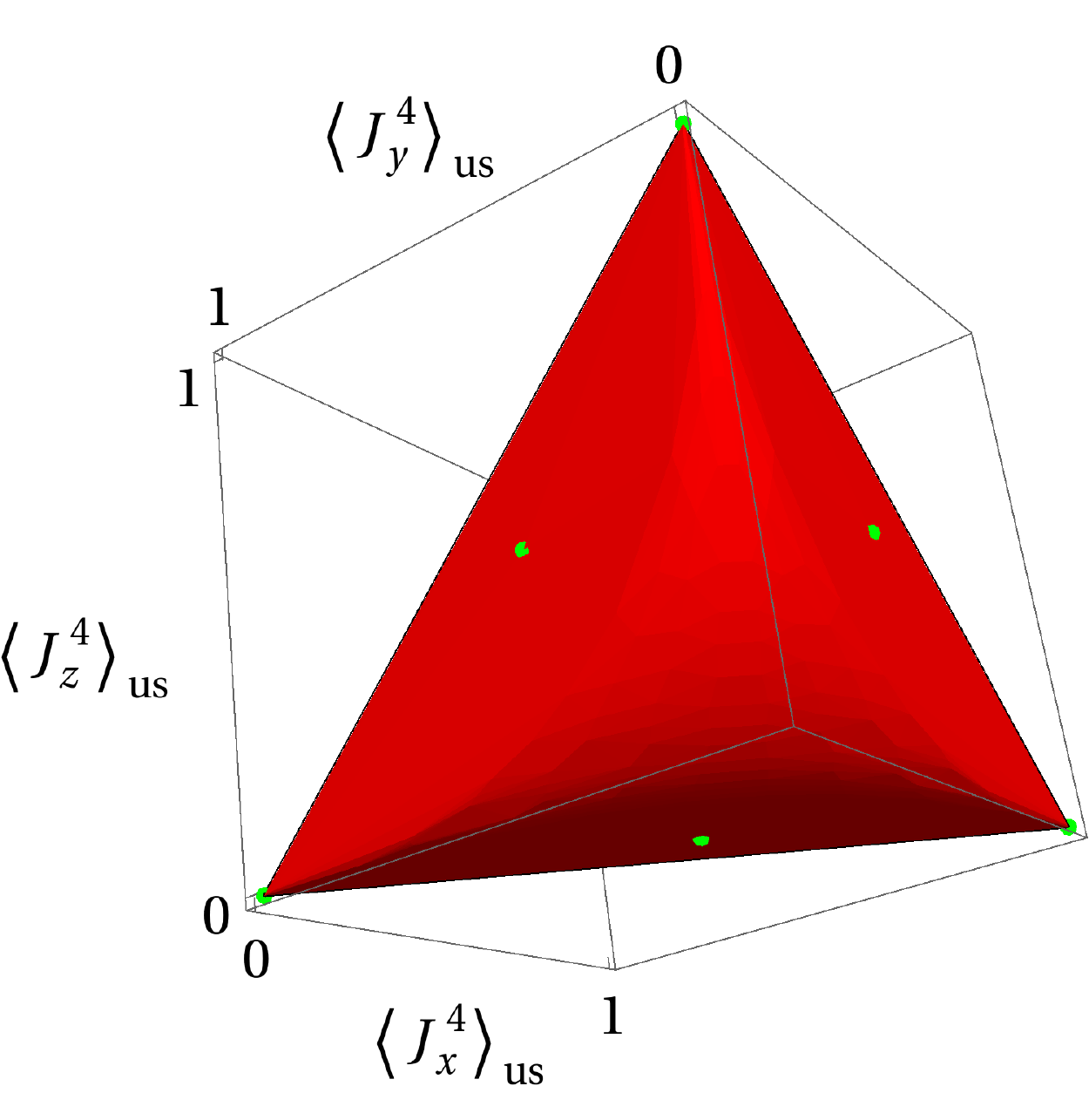}}
	\caption{(color online) For ${j=25}$,
		the numerical range $\mathcal{E}_4^\textsc{us}$ of $\tfrac{1}{j^4}\vec{E}_4$ is shown through---the red convex body---two different directions.
		Here we present the mean values
		${\langle J_x^4\rangle_\textsc{us}:=\big\langle(\tfrac{J_x}{j})^4\big\rangle}$, $\langle J_y^4\rangle_\textsc{us}$, and $\langle J_z^4\rangle_\textsc{us}$ after
		the uniform scaling~\eqref{uni-scaling-J}.
		As per \eqref{max-min-ev-points Jxyz^4}, here two of the green points are 
		${(0.001184, 0.001184, 1)}$ and ${(0.405184, 0.405184, 0)}$.
	}
	\label{fig:E-J4 for j=25} 
\end{figure}

The first and second powers of angular momentum operators are studied in 
Secs.~\ref{sec:spin-ops} and \ref{sec:Jx2 and Jy2}, respectively, here we consider
\begin{equation}
\label{J-gamma}
\vec{E}_\gamma=\big((J_x)^\gamma,(J_y)^\gamma,(J_z)^\gamma\big)
\end{equation}
with higher but finite integer powers ${\gamma=3,4,\cdots}$.
Since analytic expressions of  
${\lambda_\textsc{m}(\theta,\phi)}$ and 
${|\theta,\phi\rangle}$
are cumbersome, 
we compute here the boundary points
${\langle\theta,\phi|\vec{E}_\gamma|\theta,\phi\rangle}$
numerically by taking only a finite set of ${(\theta_k,\phi_{k'})}$.
Then, by connecting these points as per \eqref{boundary} and \eqref{E-para}, we draw boundaries of the allowed regions
in Figs.~\ref{fig:E-J3 for j=3/2,2,5/2,10}, \ref{fig:E-J4 for j=2,5/2}, 
\ref{fig:E-J4 for j=3,7/2}, and
\ref{fig:E-J4 for j=25}.

Let us first take odd powers ${\gamma=3,5,\cdots}$, where
hyperrectangle~\eqref{hyperrectangle} for $\vec{E}_\gamma$ will be
the cube ${[-j^\gamma,j^\gamma]^{\times 3}}$.
Since
$J_x^\gamma$ is directly proportional to $J_x$ when ${j=\tfrac{1}{2},1}$,
the allowed region $\mathcal{E}_\gamma$ is bounded by the
sphere---centered at the origin $(0,0,0)$---of radius $j^\gamma$ for the two quantum numbers.
By picking ${\gamma=3}$, we display $\mathcal{E}_3$ for 
$j=\tfrac{3}{2},2,\tfrac{5}{2},$ and ${10}$
in Fig.~\ref{fig:E-J3 for j=3/2,2,5/2,10}.
One can observe that as we increase $j$ 
the (boundary of) allowed region changes its shape from 
a sphere to an octahedron, in fact, it is true for all finite odd powers ${\gamma>1}$.

Now, let us apply uniform scaling ~\eqref{uni-scaling-J}, where 
the permissible region for ${\tfrac{1}{j^\gamma}\vec{E}_\gamma}$ is denoted
by $\mathcal{E}_\gamma^\textsc{us}$.
For every $j$ and all positive odd powers,
\begin{equation}
\label{octahedron-vertices}
(\pm 1,0,0),\ (0,\pm 1,0),\ \mbox{and}\ (0,0,\pm 1)
\end{equation}
lie on the boundary of $\mathcal{E}_\gamma^\textsc{us}$.
These six extreme points [see Fig.~\ref{fig:E-J3 for j=3/2,2,5/2,10}] come from the
eigenkets of $J_x,J_y,$ and $J_z$ corresponding to the eigenvalues ${\pm j}$.
In the case of an odd ${\gamma>1}$, as we increase $j$ from $\tfrac{1}{2}$
to $\infty$, 
$\mathcal{E}_\gamma^\textsc{us}$
monotonically shrinks from the unit sphere centered at the origin to the octahedron with vertices~\eqref{octahedron-vertices}.
With the help of quantum de Finetti theorem \cite{Stormer69,Hudson76},
it is shown in Sec.~\ref{sec:N-qubit} that indeed
$\mathcal{E}_\gamma^\textsc{us}$ becomes the octahedron
in the limit ${j\rightarrow\infty}$.
The eight unit vectors ${\{\pm\widehat{\eta}_l\}^4_{l=1}}$ in Table~\ref{tab:theta-phi-for(-a)}
are normals to eight faces of the octahedron.
With these vectors, we define another family of regions
\begin{equation}
\label{R_p}
\mathcal{G}_\vartheta:=
\Big\{\vec{r}\in[-1,1]^{\times 3}
\ \Big|\ 
\sum_{l=1}^4(\sqrt{3}\,\widehat{\eta}_l\cdot\vec{r}\,)^{2\vartheta}\leq 4
\Big\}
\end{equation} 
such that as we increase ${\vartheta\in\{1,2,3,\cdots, \infty\}}$
$\mathcal{G}_\vartheta$ also contracts from the unit sphere to the octahedron, and
points \eqref{octahedron-vertices} lie on the boundary of every $\mathcal{G}_\vartheta$
like $\mathcal{E}_\gamma^\textsc{us}$.
Moreover, 
there always exit two ${\vartheta\leq \vartheta'}$ such that
${\mathcal{G}_\vartheta\supseteq\mathcal{E}_\gamma^\textsc{us}
	\supseteq\mathcal{G}_{\vartheta'}}$,
hence $\mathcal{G}_\vartheta$ and $\mathcal{G}_{\vartheta'}$ are outer and inner approximations of 
the numerical range $\mathcal{E}_\gamma^\textsc{us}$.
In particular, the boundary $\partial\mathcal{E}_\gamma^\textsc{us}$ lies between 
the octahedron
$\partial\mathcal{G}_\infty$ and the unit sphere $\partial\mathcal{G}_1$.

\begin{figure}
	\centering
	\subfloat{\includegraphics[width=40mm]{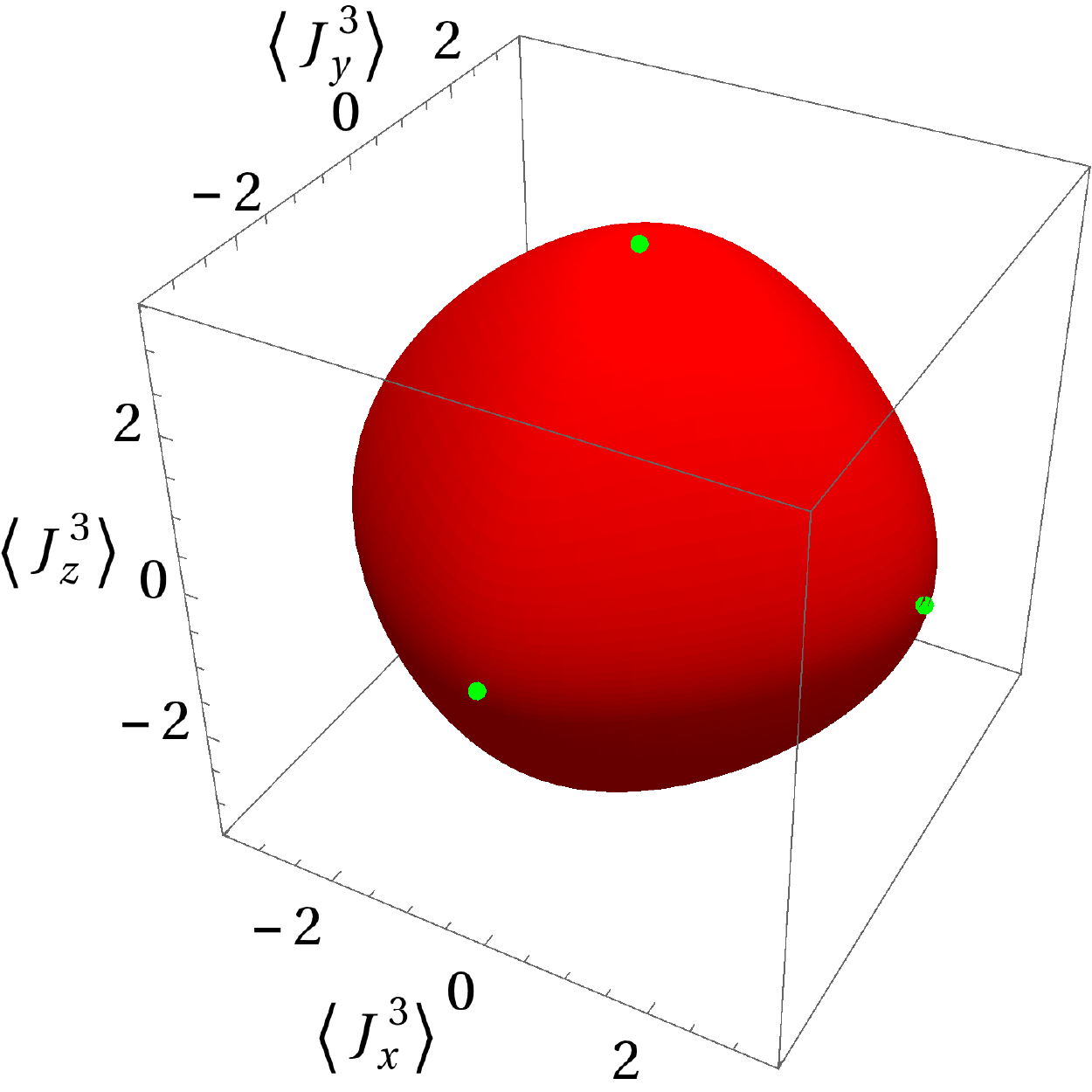}}\quad\
	\subfloat{\includegraphics[width=40mm]{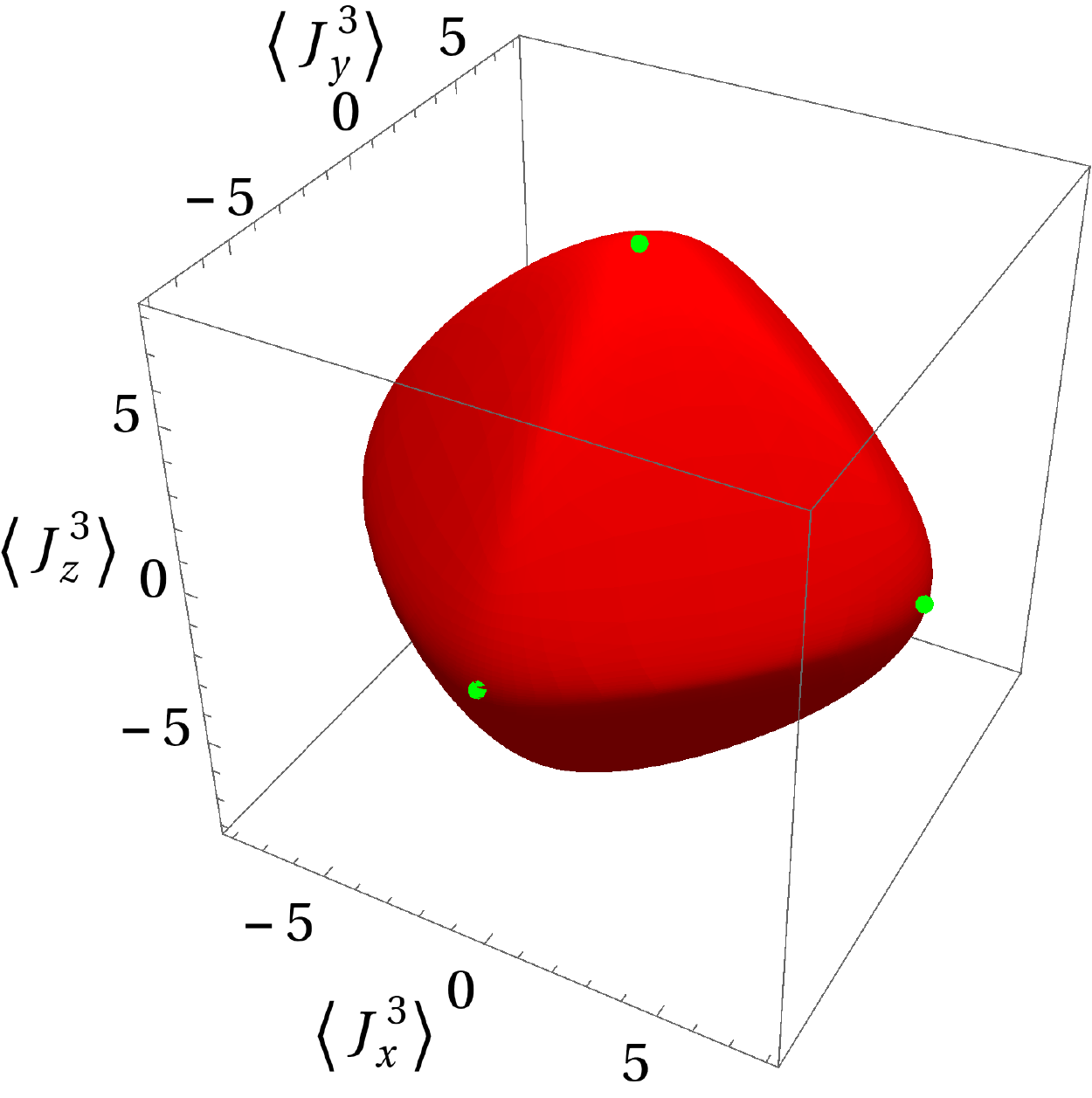}}
	\\%\vspace{0mm}
	\subfloat{\includegraphics[width=40mm]{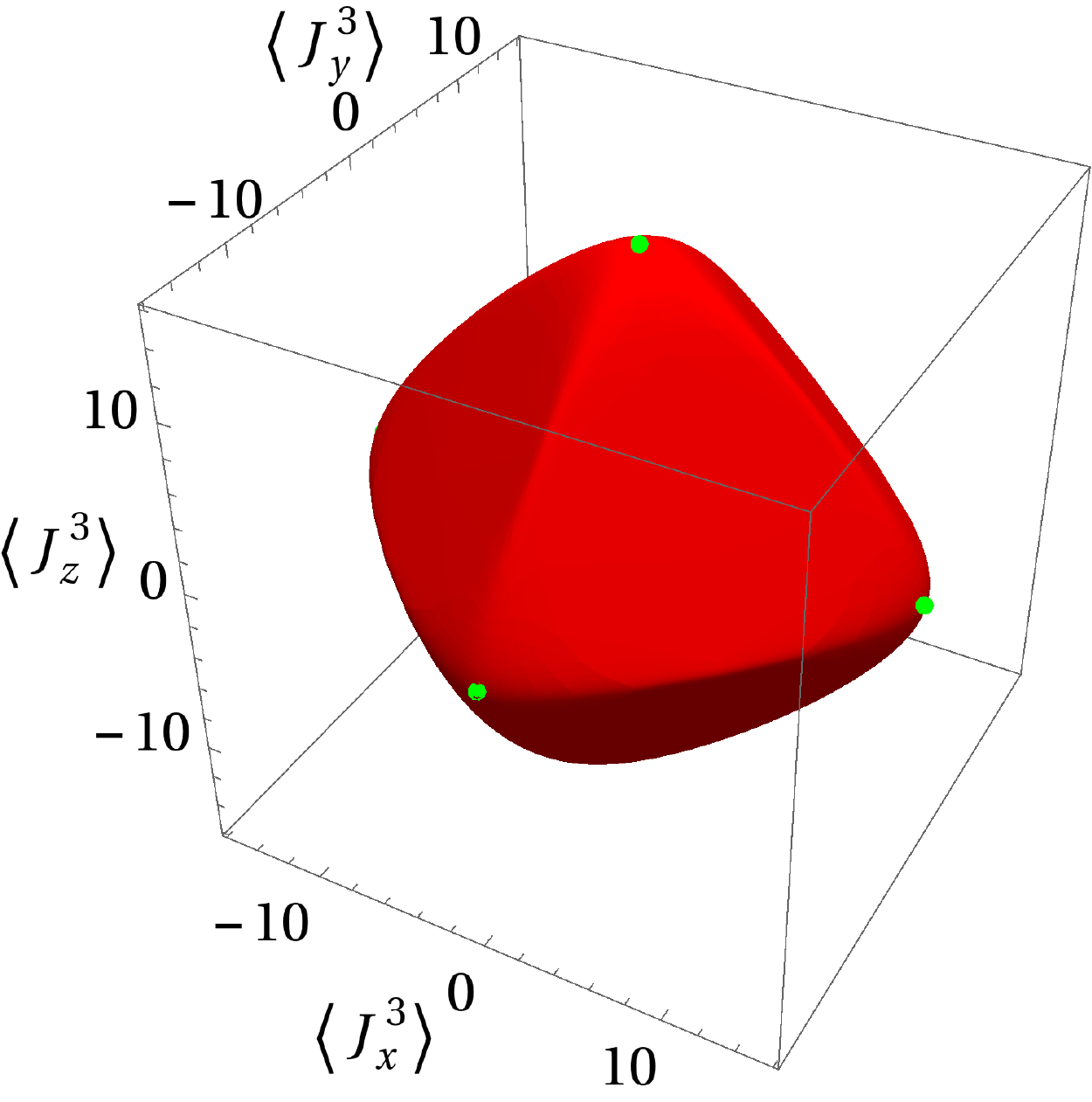}}\quad\
	\subfloat{\includegraphics[width=40mm]{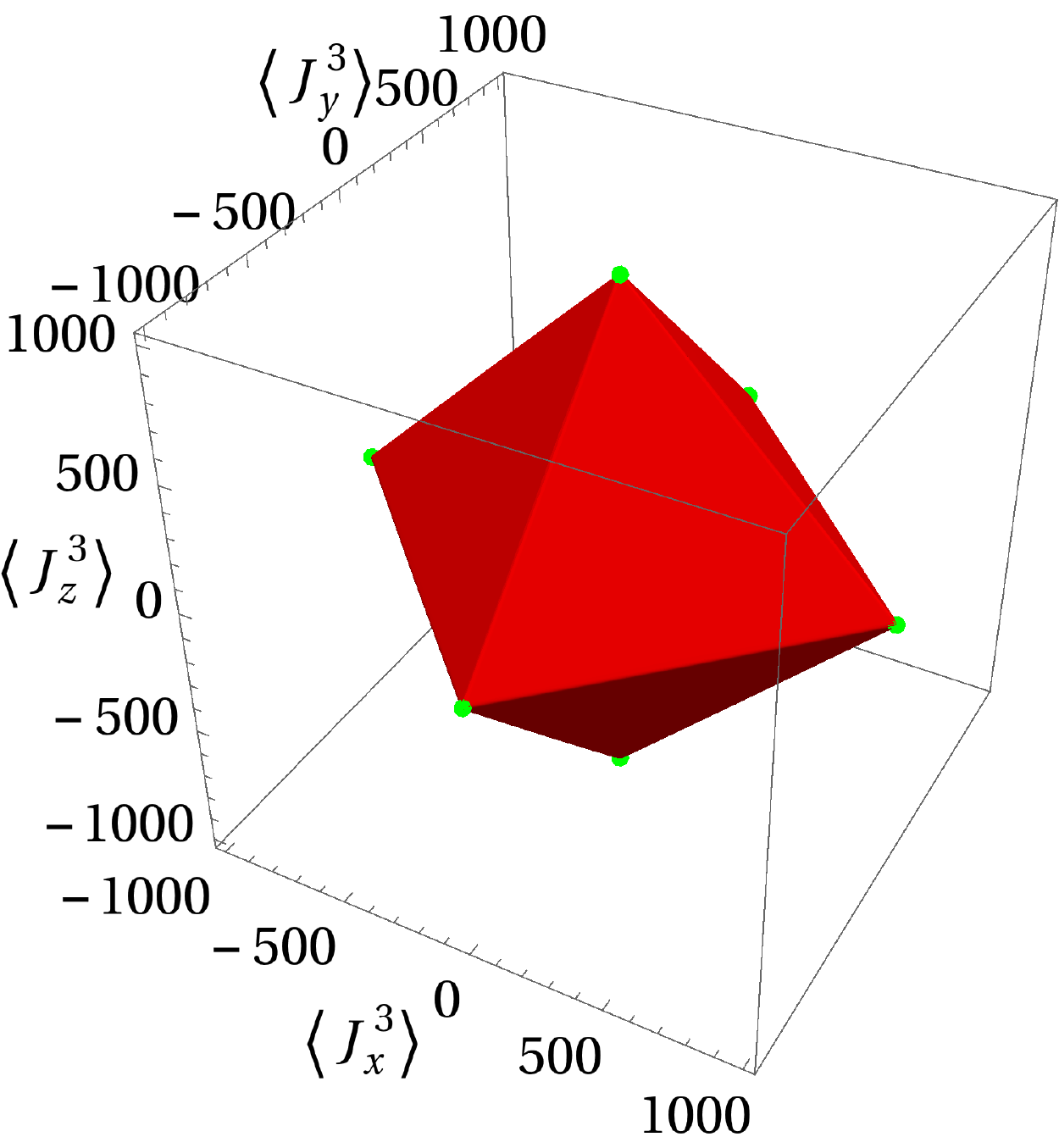}}
	\caption{(Color online) From top-left to bottom-right, moving row by row, 
	the numerical ranges of $\vec{E}_{\gamma=3}$ of \eqref{J-gamma} for 
	${j=\tfrac{3}{2}, 2,\tfrac{5}{2},}$ and ${10}$ are the red convex bodies.
	For each $j$, the green extreme points ${(\pm j^3,0,0)}$,
	${(0,\pm j^3,0)}$, and ${(0,0,\pm j^3)}$ are provided by the
	extreme-eigenvalue-kets of $J_x,J_y,$ and $J_z$.
	}
	\label{fig:E-J3 for j=3/2,2,5/2,10} 
\end{figure}

Recall that
the two extreme eigenvalues
of an operator in ${\tfrac{1}{j^\gamma}\vec{E}_\gamma}$ [for $\vec{E}_\gamma$, see \eqref{J-gamma}] are
${\pm 1}$ for an odd number $\gamma$.
Using \eqref{Xdot}--\eqref{u-half-sum}, we establish uncertainty and certainty functions of the mean vector ${\vec{\boldsymbol{\varepsilon}}=
\tfrac{1}{j^\gamma}\langle\vec{E}_\gamma\rangle}$
and achieve
\begin{eqnarray}
\label{H-UR-j^gamma-inf}
2\ln 2&\leq&h(\vec{\boldsymbol{\varepsilon}}\,)\,,\\
\label{u-UR-j^gamma-inf}
1+2\sqrt{2}&\leq& 
u_{\sfrac{1}{2}}(\vec{\boldsymbol{\varepsilon}}\,)\,,\\
\label{u2-UR-j^gamma-inf}
&&u_2(\vec{\boldsymbol{\varepsilon}}\,)\leq 2 \,,
\quad \mbox{and}\qquad\\
\label{umax-UR-j^gamma-inf}
&&u_\text{max}(\vec{\boldsymbol{\varepsilon}}\,)\leq 
\mathfrak{u_{max}}\in
\big[2,\tfrac{3+\sqrt{3}}{2}\,\big]
\qquad\
\end{eqnarray}
on ${\mathcal{E}_\gamma^\textsc{us}\subseteq\mathcal{G}_1}$.
The functions $h,u_{\sfrac{1}{2}},$ and $u_2$ reach their lower and upper bounds 
at the six extreme points in \eqref{octahedron-vertices}.
Hence, inequalities~\eqref{H-UR-j^gamma-inf}--\eqref{u2-UR-j^gamma-inf}
are saturated by the extreme-eigenvalue-states of $J_x$, $J_y$, and $J_z$.
These three relations
hold for all $j$ and positive odd powers $\gamma$.

The CR \eqref{umax-UR-j^gamma-inf}
is saturated by the extreme-eigenvalue-states
of the operators, $\pm{\widehat{\eta}_l\cdot\vec{E}_\gamma}$, in the eight directions ${\{\pm\widehat{\eta}_l\}^4_{l=1}}$.
Since the numerical range $\mathcal{E}_\gamma^\textsc{us}$ shrinks
in these directions, the upper bound $\mathfrak{u_{max}}$
decreases monotonically from $\tfrac{3+\sqrt{3}}{2}$ to $2$
as $j$ goes from $\tfrac{1}{2}$ to $\infty$.
In Fig.~\ref{fig:lmax,Umax} and Appendix~\ref{sec:Sup-material}, taking ${\gamma=3}$, we present the values of $\mathfrak{u_{max}}$ for ${j=1,\tfrac{3}{2},2,\cdots,50}$.
One can clearly observe predictions of the quantum de Finetti theorem
even at ${j=10}$:
${\mathcal{E}_3^\textsc{us}}$ in Fig.~\ref{fig:E-J3 for j=3/2,2,5/2,10} becomes almost the octahedron,
and 
${\mathfrak{u_{max}}\approx 2.00759}$ [see Appendix~\ref{sec:Sup-material} and Fig.~\ref{fig:lmax,Umax}] approaches 2.
If one wants a CR based on $u_\text{max}$ that holds for all the quantum numbers $j$ then
${u_\text{max}(\vec{\boldsymbol{\varepsilon}}\,)\leq \tfrac{3+\sqrt{3}}{2}}$ 
can be adopted, it is not tight for ${j>1}$ but it is non-trivial.
Besides, one can safely employ
${u_\text{max}(\vec{\boldsymbol{\varepsilon}}\,)\leq 2.00759}$ as a legitimate CR for all ${j\geq10}$ and $\gamma=3$.

\begin{figure}
	\centering
	\subfloat{\includegraphics[width=40mm]{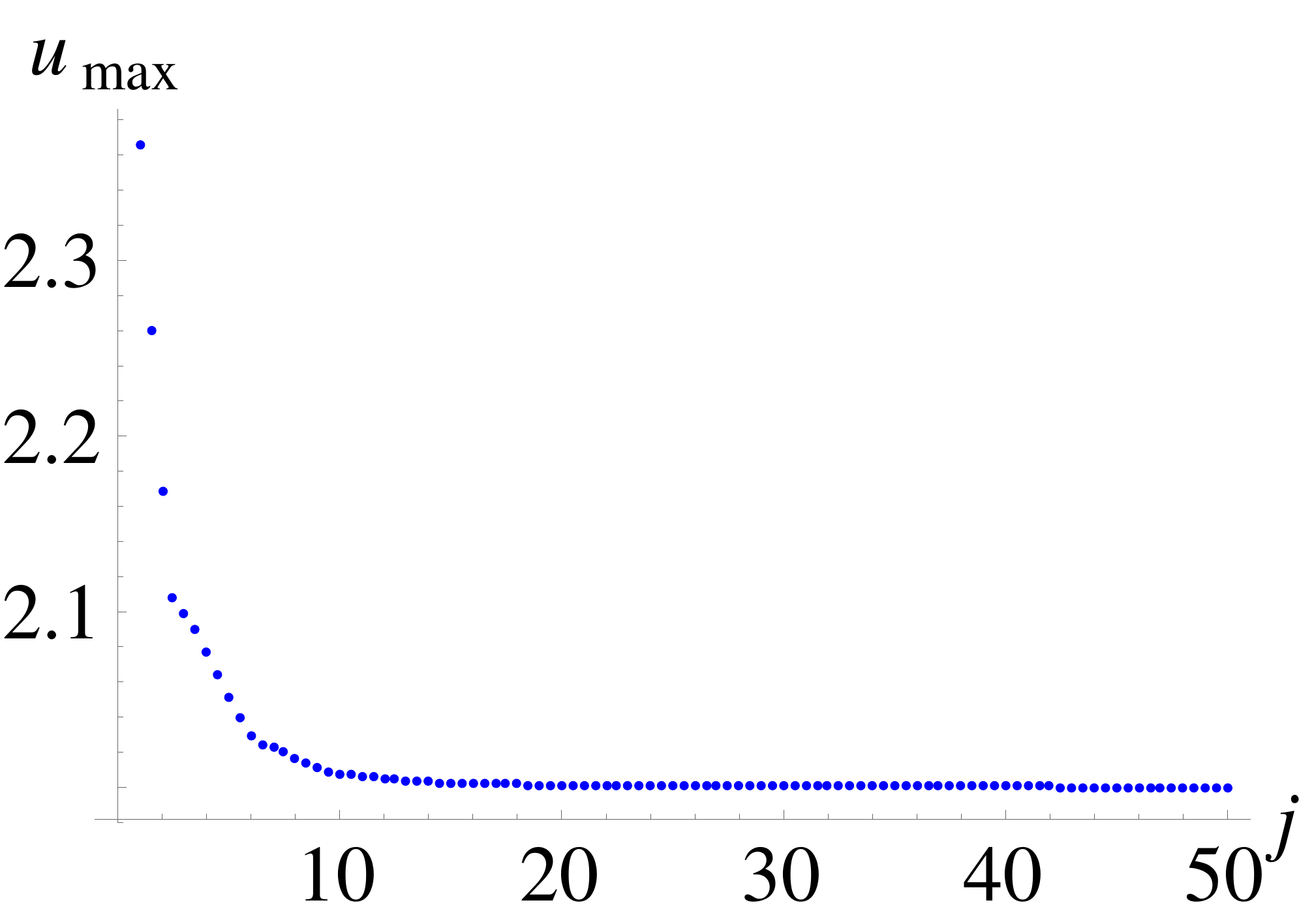}}\quad\
	\subfloat{\includegraphics[width=40mm]{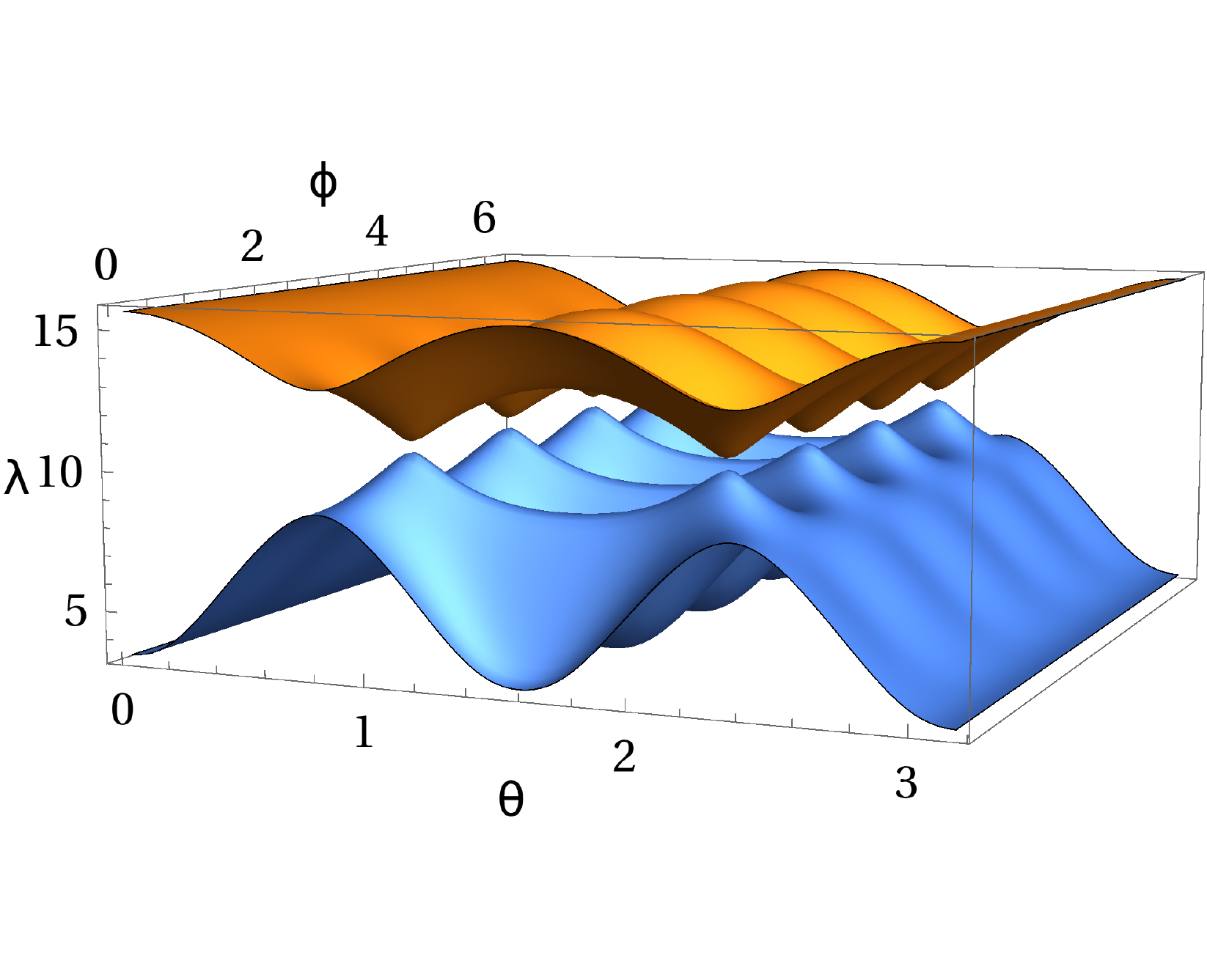}}
	\caption{ 
		On the left-hand-side, taking ${\gamma=3}$, we plot the values of upper-bound $\mathfrak{u_{max}}$---appears in tight CR \eqref{umax-UR-j^gamma-inf}---for ${j=1,\tfrac{3}{2},2,\cdots,50}$. The values are filed in Appendix~\ref{sec:Sup-material}. 
		On the right-hand-side, choosing ${\gamma=3}$ and ${j=\tfrac{5}{2}}$, we plot
		the largest and second-largest eigenvalues
		$\lambda(\theta,\phi)$
		of $\widehat{\eta}(\theta,\phi)\cdot\vec{E}_\gamma$
		as functions of $\theta$ and $\phi$ [for $\vec{E}_\gamma$, see
		\eqref{J-gamma}].
		One can recognize that the gap between eigenvalues is the smallest at 
		${\{(\theta_l,\phi_l),(\pi-\theta_l,\pi+\phi_l)\}_{l=1}^4}$ [for $(\theta_l,\phi_l)$, see Table~\ref{tab:theta-phi-for(-a)}].
	}
	\label{fig:lmax,Umax} 
\end{figure}

In Fig.~\ref{fig:lmax,Umax}, for $\gamma=3$ and $j=\tfrac{5}{2}$,
we also plot the largest and second-largest eigenvalues
of $\Lambda(\theta,\phi)=\widehat{\eta}(\theta,\phi)\cdot\vec{E}_\gamma$.
There one can perceive that the gap between the eigenvalues is 
least at eight different $(\theta,\phi)$ that correspond to ${\{\pm\widehat{\eta}_i\}_{i=1}^4}$.
The gap at these eight places reduces as $j$ grows, and it seems to disappear at ${j\rightarrow\infty}$.
By the way, such a situation appears in the study of quantum phase transitions.
The gap is already very small for ${j=10}$, and 
the two eigenvalues are,
${586.116}$ and ${585.098}$, almost equal at ${(\theta_l,\phi_l)}$.
In other words, the maximum eigenvalue $\lambda_\textsc{m}(\theta_l,\phi_l)$
turns almost degenerate for a large quantum number $j$, and the
degeneracy is a necessary requirement for the allowed region to have a flat face (that has more than one distinct points).
Therefore, eight flat faces appear on the boundary
${\partial\mathcal{E}_\gamma^\textsc{us}}$ for 
an odd power ${\gamma\geq 3}$.

Now we take even powers $\gamma=4,6\cdots$ for $\vec{E}_\gamma$ of \eqref{J-gamma}, then hyperrectangle \eqref{hyperrectangle} becomes
\begin{equation}
\label{H gamna-even}
\mathcal{H}_\gamma=
\begin{cases}
\big[(\tfrac{1}{2})^\gamma,j^\gamma\big]^{\times 3}
& \text{for a half-integer } j \vspace{1mm}\\
[0,j^\gamma]^{\times 3}
& \text{for an integer } j
\end{cases}.
\end{equation}
In the case of 
${j=\tfrac{1}{2},1,}$
and $\tfrac{3}{2}$,
one can compute the mean value ${\langle (J_x)^\gamma\rangle}$
from ${\langle J_x^2\rangle}$ by using the Cayley-Hamilton theorem for $J_x$, and similarly for the other two operators in $\vec{E}_\gamma$.
As a result, the numerical range $\mathcal{E}_\gamma$ of $\vec{E}_\gamma$---directly obtained from the numerical range of ${(J_x^2,J_y^2)}$ in Sec.~\ref{sec:Jx2 and Jy2}---will be a single point, a triangle, and an elliptical disk for ${j=\tfrac{1}{2},1,}$
and $\tfrac{3}{2}$, respectively.

Recall from Sec.~\ref{sec:Jx2 and Jy2} that
$\mathscr{S}^2_0$ and $\mathscr{S}^2_1$ [stated in \eqref{Inv-subspace}] 
are two mutually orthogonal invariant subspaces of $J_x^2$ and $J_y^2$.
Hence, they will also be invariant subspaces of 
${\Lambda(\theta,\phi)=\widehat{\eta}(\theta,\phi)\cdot\vec{E}_\gamma}$
for all even ${\gamma\geq 2}$.
Moreover, one can find eigenvectors of $\Lambda$ in these subspaces
like Sec.~\ref{sec:Jx2 and Jy2}.
Here also, for every half-integer $j$,
the maximum eigenvalue ${\lambda_\textsc{m}(\theta,\phi)}$
of ${\Lambda(\theta,\phi)}$ is at least twofold degenerate in all the directions
${(\theta,\phi)}$.

Now we only focus on ${\gamma=4}$.
The maximum-eigenvalue-kets ${|{\pm j}\rangle}$
and the minimum-eigenvalue-kets ${|{\pm \tfrac{1}{2}}\rangle}$
or ${|0\rangle}$ of $J_z^4$ provide
\begin{align}
\label{max-min-ev-points Jxyz^4}
\langle{\pm j}|(J_{x,y})^4|{\pm j}\rangle&=
\tfrac{j(3j-1)}{4}\quad\mbox{and}\nonumber\\
\big\langle{\pm \tfrac{1}{2}}\big|(J_{x,y})^4\big|{\pm \tfrac{1}{2}}\big\rangle&=
\tfrac{j(j+1)(6 j (j+1)-7) +\frac{23}{8}}{16}\quad\mbox{or} \\
\langle 0|(J_{x,y})^4|0\rangle&=
\tfrac{j(j+1)(3j(j+1)-2)}{8}\,. \nonumber
\end{align}
One will have similar relations with the extreme-eigenvalue-kets
of $J_x^4$ and $J_y^4$.
In this way, analogues to \eqref{max-ev-points}--\eqref{min-ev-points-int}, we have six extreme points of the allowed region $\mathcal{E}_4$.
These points are illustrated in Figs.~\ref{fig:E-J4 for j=2,5/2}, 
\ref{fig:E-J4 for j=3,7/2}, and 
\ref{fig:E-J4 for j=25}, where we present $\mathcal{E}_4$ for 
$j=2,\tfrac{5}{2},3,\tfrac{7}{2},$ and ${25}$.
Let us analyze these one by one.

In the case of ${j=2}$, the operator ${\Lambda(\theta_1,\phi_1)=\widehat{\eta}_1\cdot\vec{E}_4}$ 
has only two distinct eigenvalues $8\sqrt{3}$ and $6\sqrt{3}$ [for $\widehat{\eta}_1$, see Table.~\ref{tab:theta-phi-for(-a)}].
With the eigenvalues, one can have two parallel supporting hyperplanes whose outward-pointing normals are ${\pm\widehat{\eta}_1}$
as per \eqref{L-exp} and \eqref{hyperplane}.
Here the numerical range $\mathcal{E}_4$ is the convex hull of an ellipse
in one of the planes and the three extreme points
\begin{equation}
\label{3-pts, j=2, J4}
(16,1,1)\,,\quad(1,16,1)\,,\quad \mbox{and}\quad (1,1,16)
\end{equation}
lie in the other plane.
The ellipse is specified by the equality in 
\begin{align}
\label{ell plane, j=2, J4}
&\bigg(\frac{\langle J_x^4\rangle+\langle J_y^4\rangle-16}{8}\bigg)^2+
\bigg(\frac{\langle J_x^4\rangle-\langle J_y^4\rangle}{8\sqrt{3}}\bigg)^2
\leq 1\,,\ \mbox{and}\quad\nonumber\\
&\sqrt{3}\,\langle \widehat{\eta}_1\cdot\vec{E}_4\rangle=\langle J_x^4\rangle+\langle J_y^4\rangle+\langle J_z^4\rangle=24
\end{align}
describes its plane.
According to \eqref{E-para}, the elliptical face ${\mathcal{F}(\theta_1,\phi_1)}$ identified by \eqref{ell plane, j=2, J4}
and the triangular face ${\mathcal{F}(\pi-\theta_1,\pi+\phi_1)}$
made of the points in \eqref{3-pts, j=2, J4} are the images of 
the two sets of eigenstates that are
associated with the maximum and minimum eigenvalues 
of ${\widehat{\eta}_1\cdot\vec{E}_4}$.
In other words, the conditions in~\eqref{ell plane, j=2, J4} are only met by the
maximum-eigenvalue-states of  ${\widehat{\eta}_1\cdot\vec{E}_4}$.
One can see in Fig.~\ref{fig:E-J4 for j=2,5/2} that there are three more triangles on the boundary ${\partial\mathcal{E}_4}$ whose outward normals are 
$\vec{\eta}=-(1,1,7)$ and the two obtained by permuting the entries in $\vec{\eta}$.
The lowest eigenvalue ${24}$ of ${J_x^4+J_y^4+7J_z^4}$ is threefold degenerate, 
hence ${\mathcal{E}_4}$ gets a triangular face in the direction of $\vec{\eta}$.

\begin{figure}
	\centering
	\subfloat{\includegraphics[width=40mm]{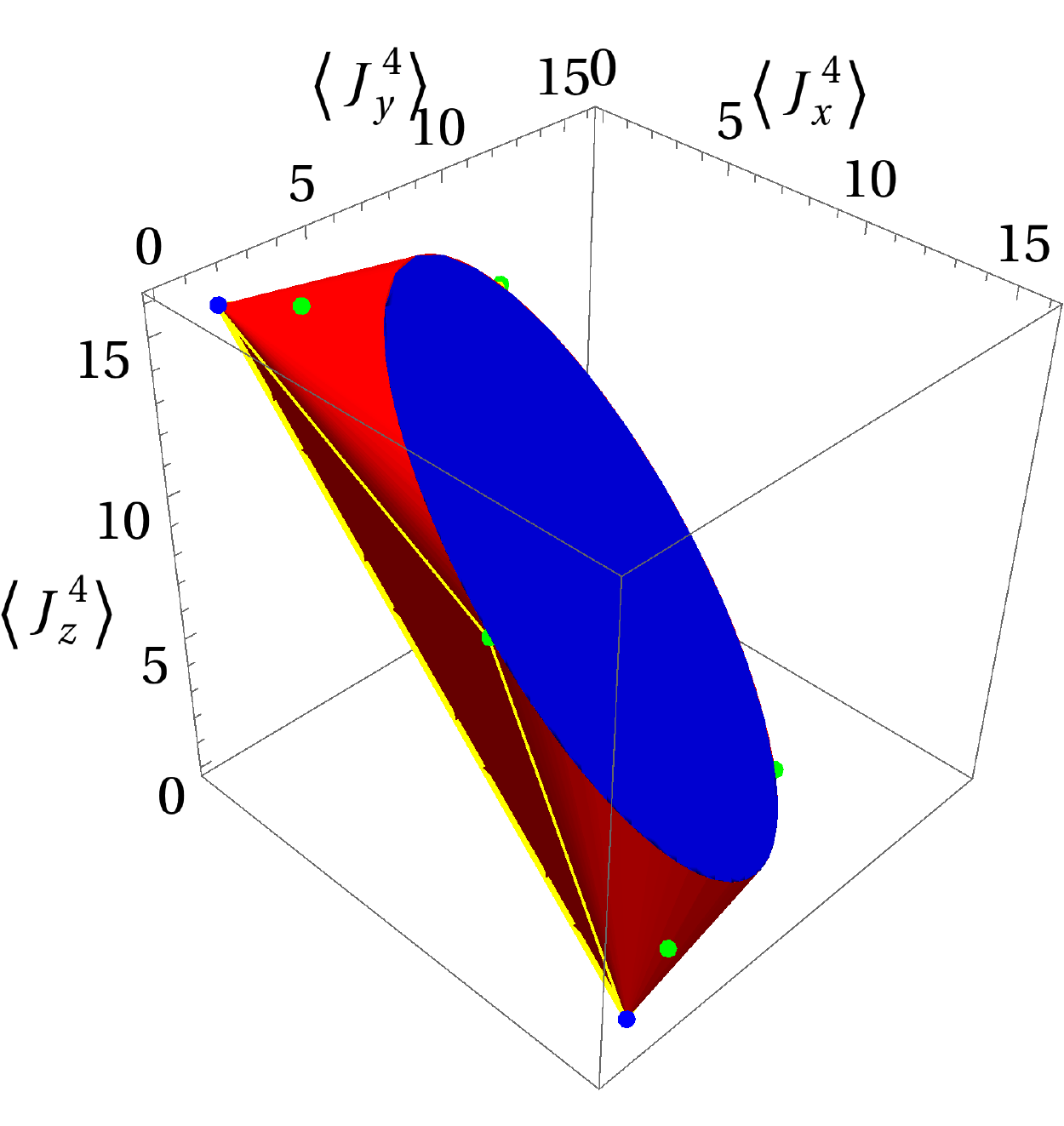}}\quad\
	\subfloat{\includegraphics[width=40mm]{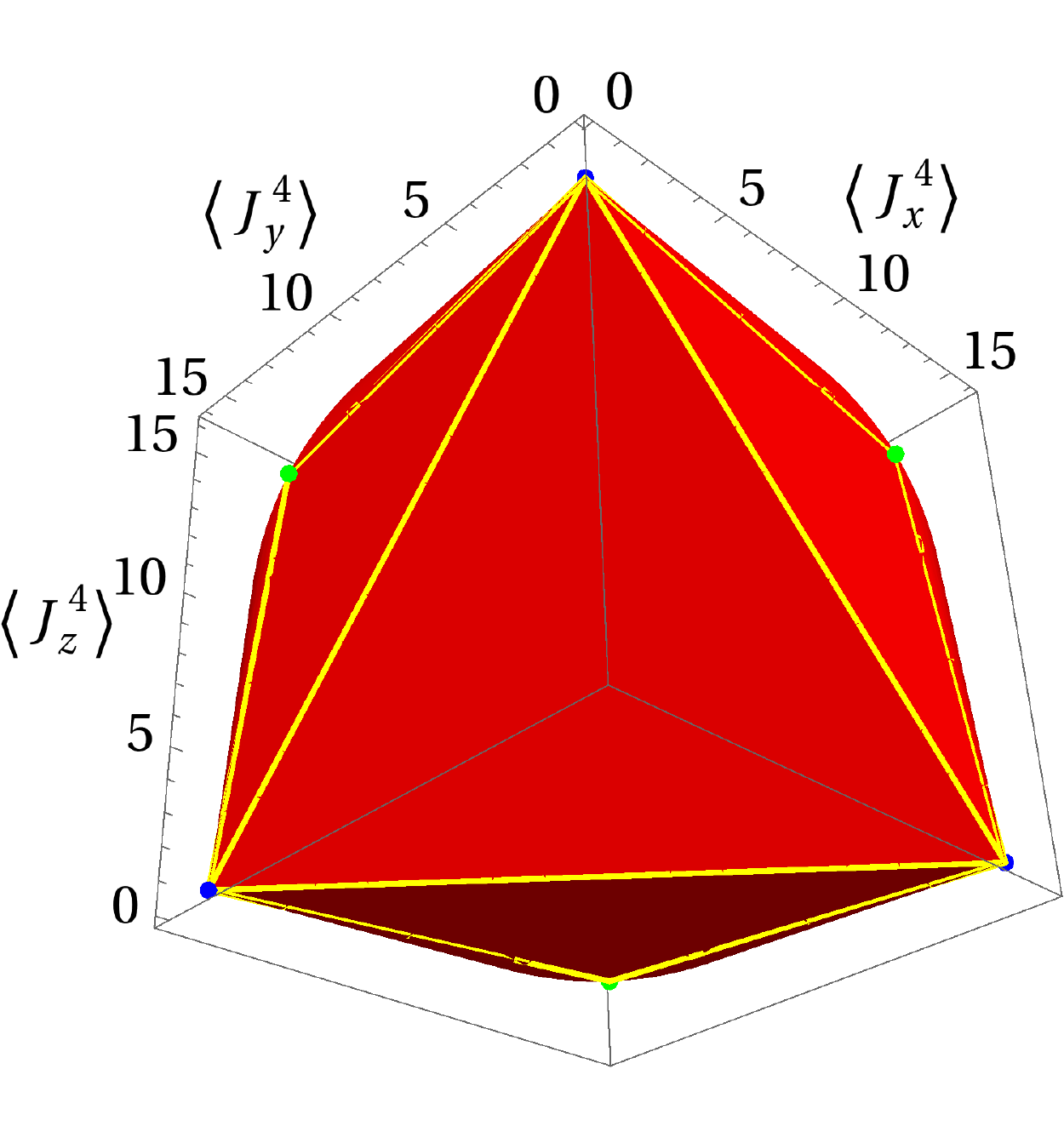}}
	\\
	\subfloat{\includegraphics[width=40mm]{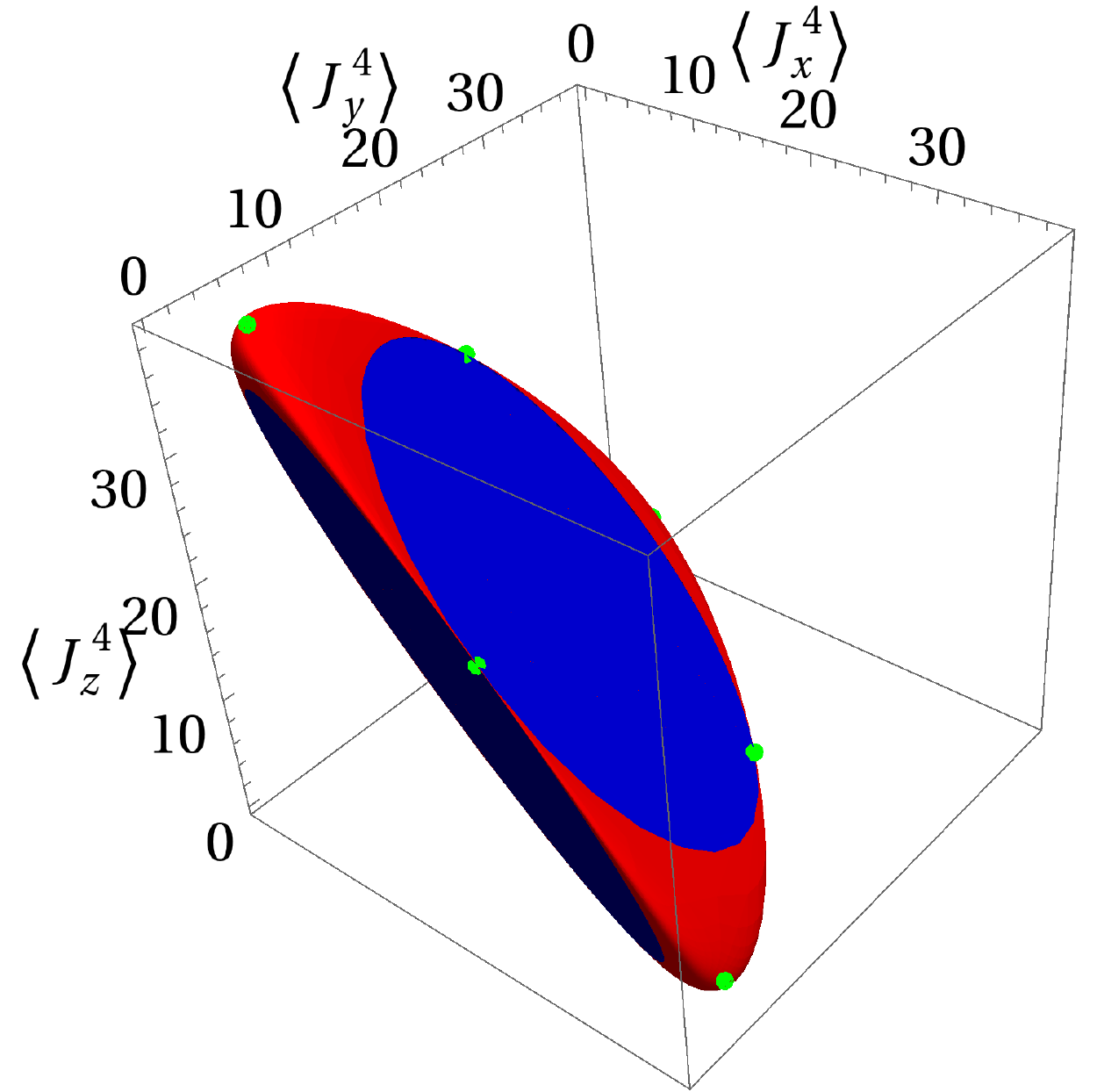}}\quad\
	\subfloat{\includegraphics[width=40mm]{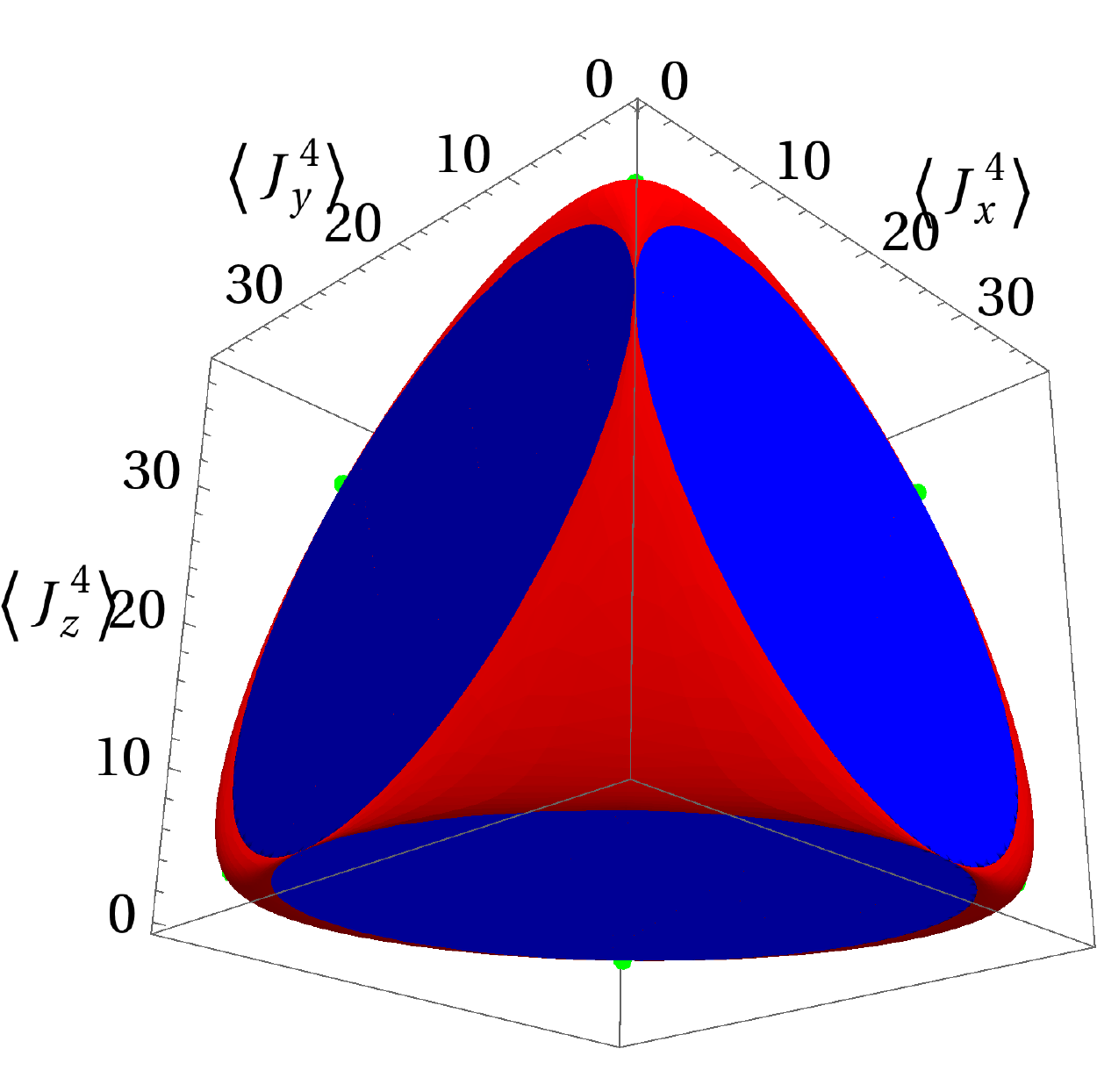}}
	\caption{(Color online) The red convex bodies in the top and bottom rows 
	are the allowed regions $\mathcal{E}_4$ for $\vec{E}_4$ [in \eqref{J-gamma}] when ${j=2}$ and ${j=\tfrac{5}{2}}$, respectively.
	In each row, the same region is shown from different viewpoints.
	In all the pictures, the green points come from the extreme-eigenvalue-states of 
	$J_x^4$, $J_y^4$, and $J_z^4$ as per \eqref{max-min-ev-points Jxyz^4}.
	In the case of ${j=2}$, the top row, the three blue points and
	the blue elliptical disk represent \eqref{3-pts, j=2, J4} and \eqref{ell plane, j=2, J4}, respectively.
	The blue points and boundary of the disk form the set of all extreme points of $\mathcal{E}_4$. 
	Furthermore, there are four triangles---illustrated by the yellow lines---on the boundary $\partial\mathcal{E}_4$, and their 
	vertices are the blue points and the three green points $(0,12,12)$, $(12,0,12)$, and $(12,12,0)$.
	In the case of ${j=\tfrac{5}{2}}$, the bottom row, the boundary $\partial\mathcal{E}_4$
	has four elliptical disks painted in blue color.
	}
	\label{fig:E-J4 for j=2,5/2} 
\end{figure}

\begin{figure}
	\centering
	\subfloat{\includegraphics[width=40mm]{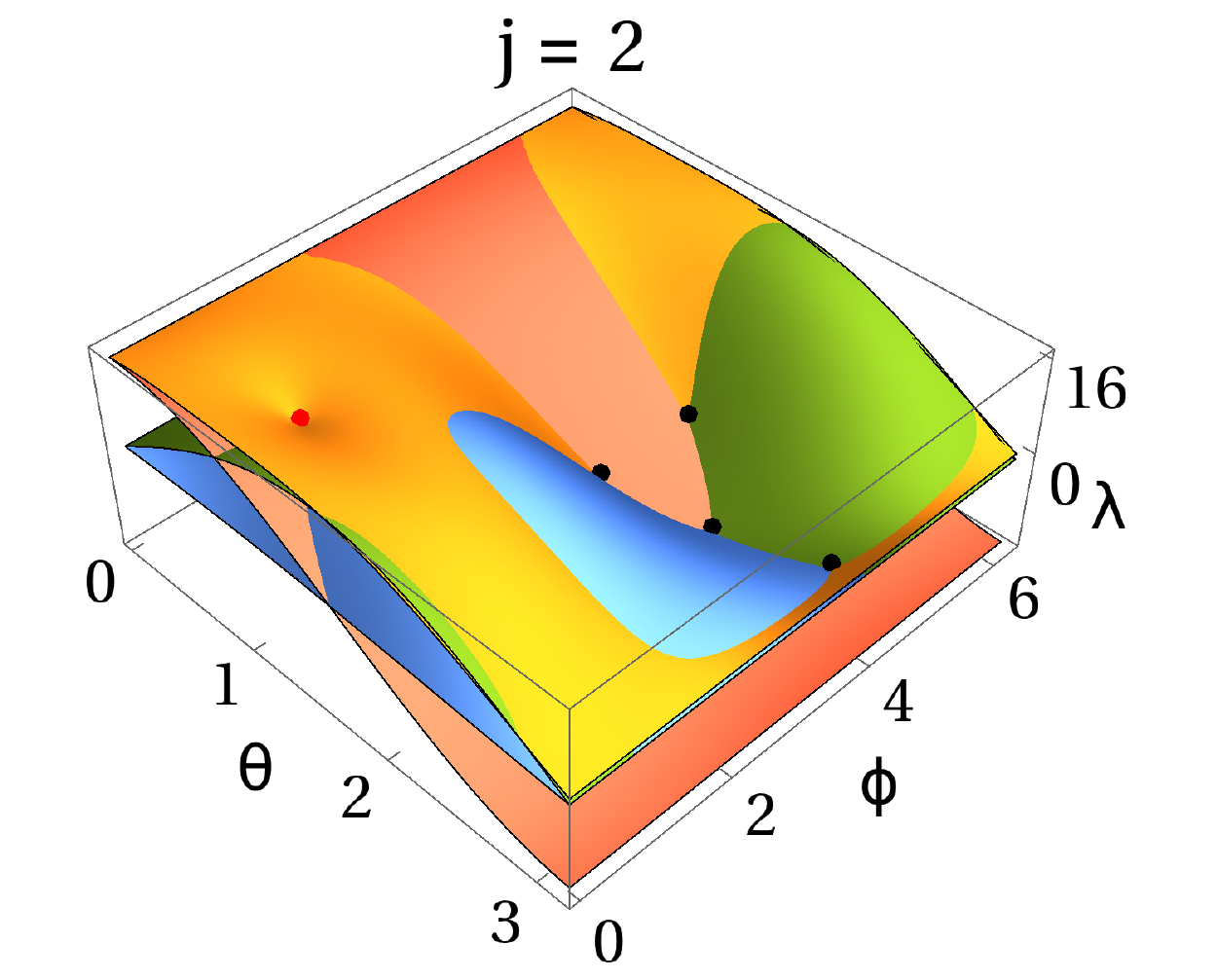}}\quad\
	\subfloat{\includegraphics[width=40mm]{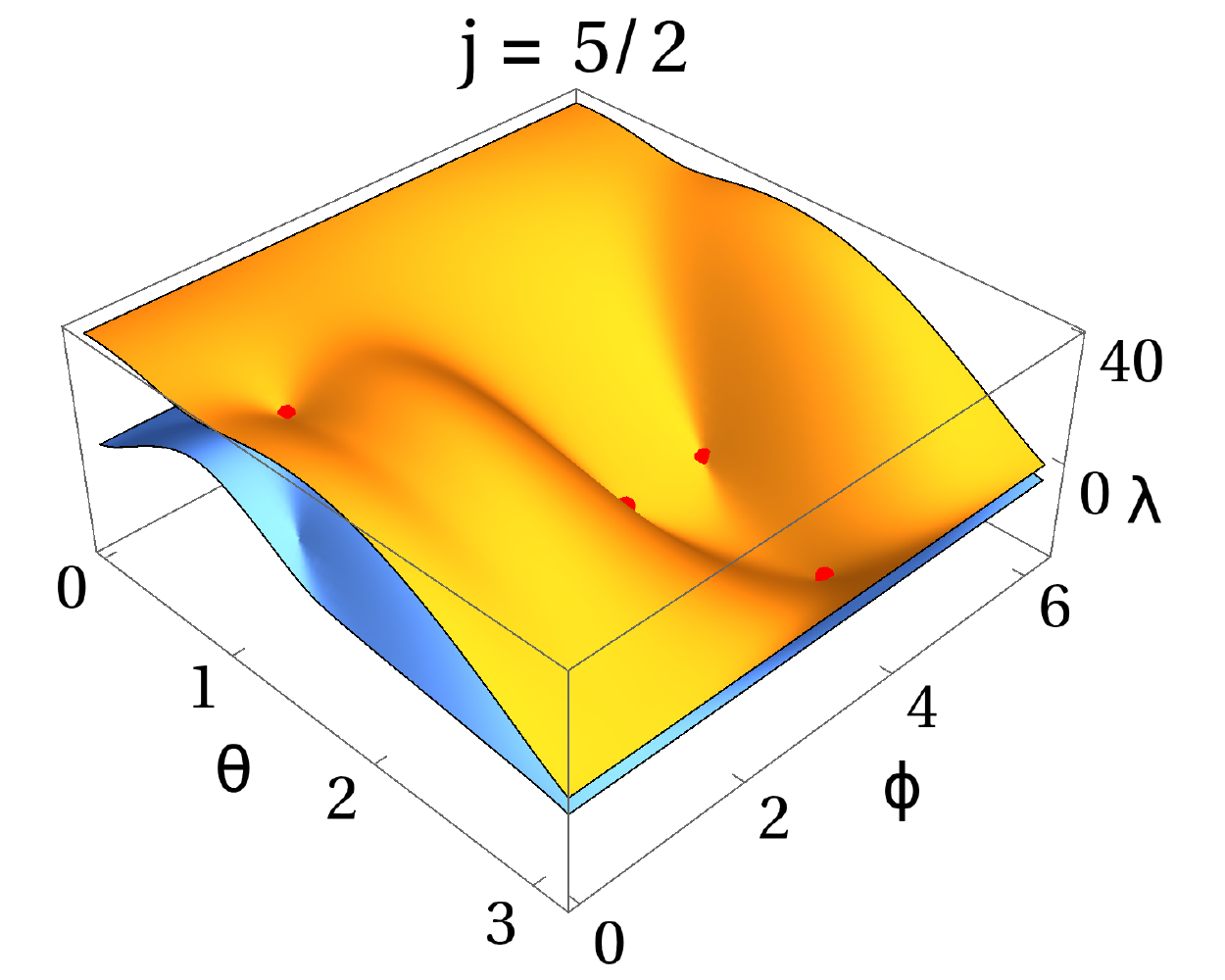}}
	\\
	\subfloat{\includegraphics[width=40mm]{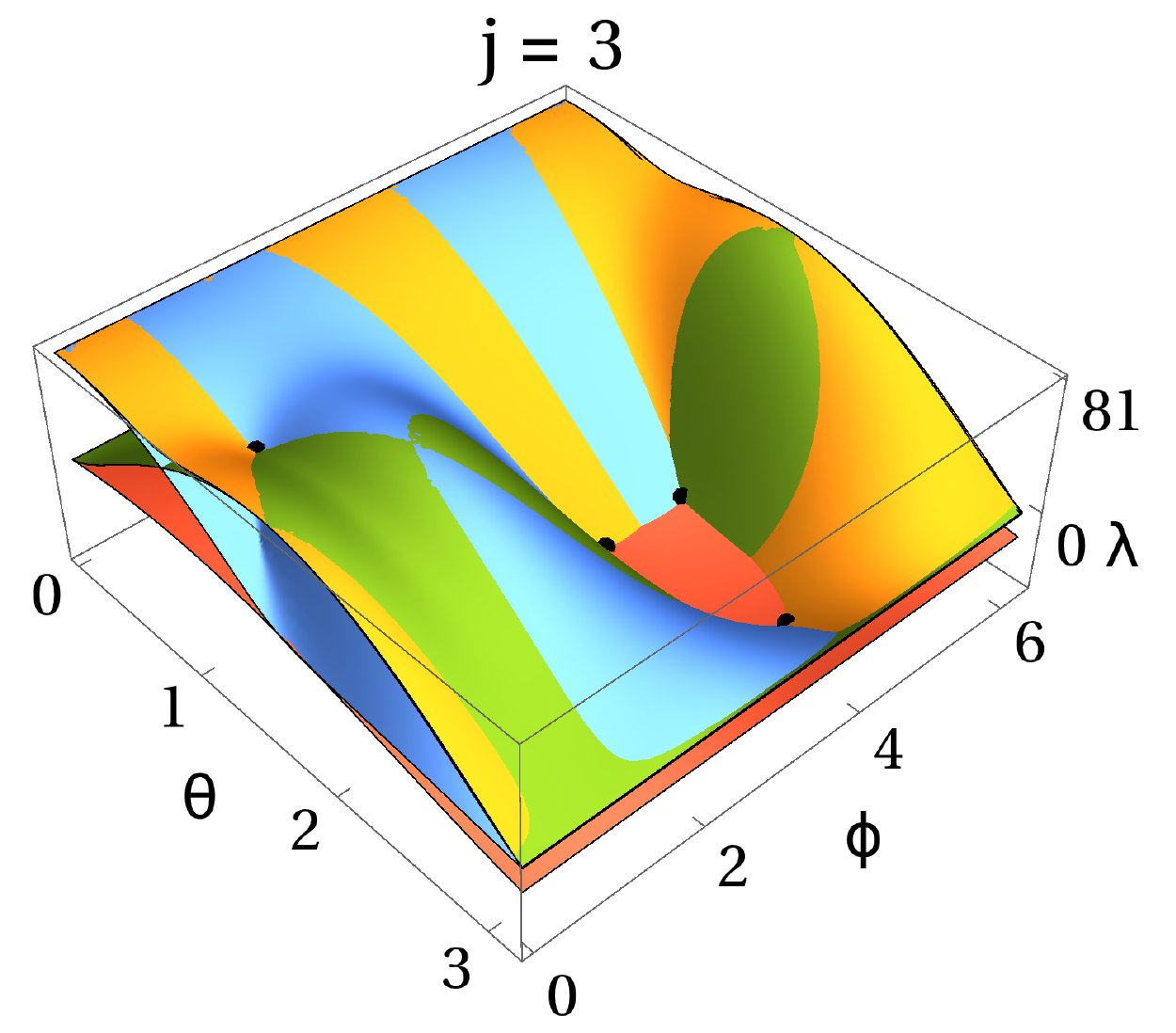}}\quad\
	\subfloat{\includegraphics[width=40mm]{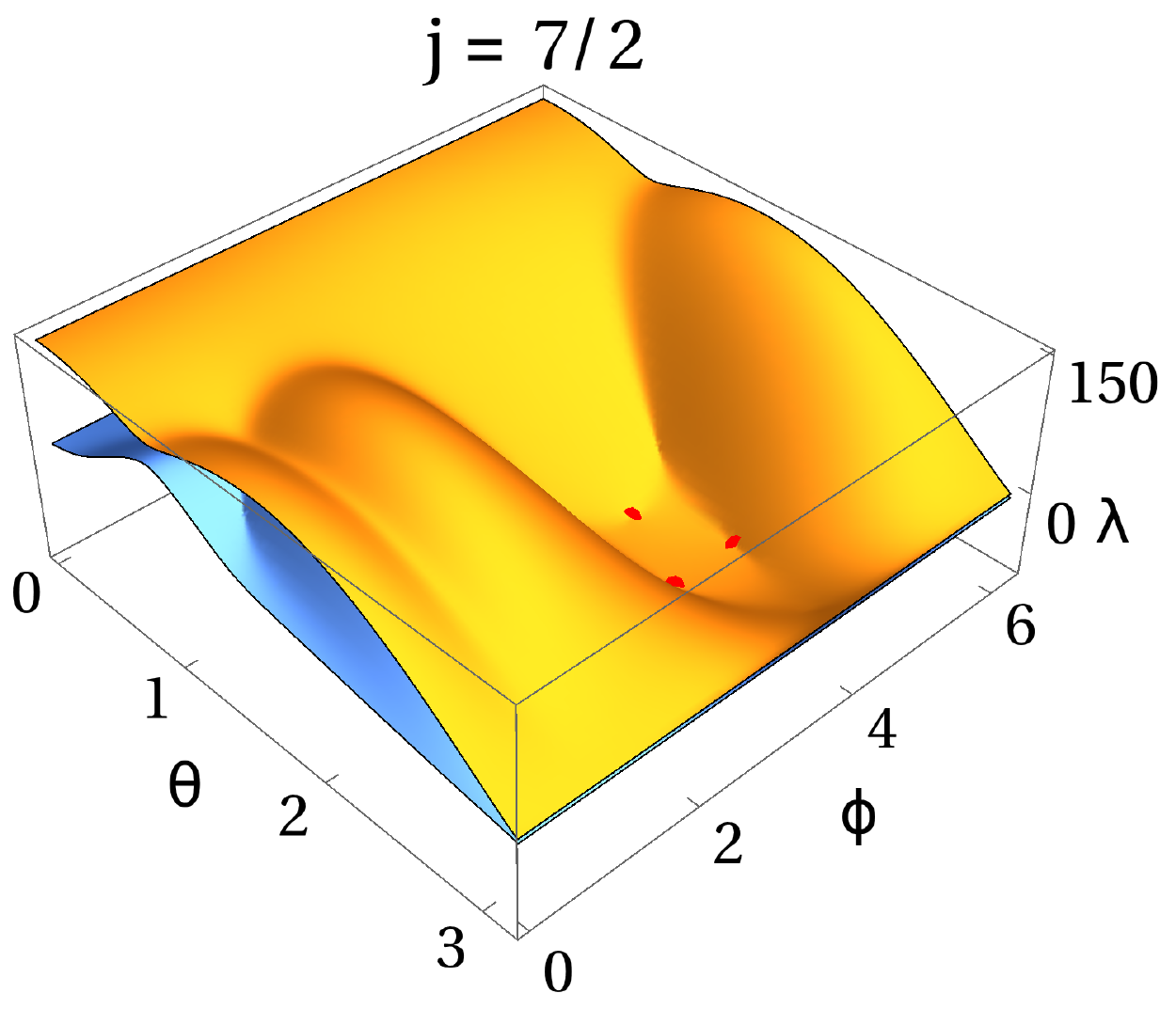}}
	\caption{(Color online) 
		For ${j=3,\cdots,\tfrac{7}{2}}$, we plot different eigenvalues of
		${\Lambda(\theta,\phi)=\widehat{\eta}(\theta,\phi)\cdot\vec{E}_4}$ 
		in separate colors on the parameter space ${[0,\pi]\times[0,2\pi)}$.
		$\vec{E}_{\gamma=4}$ is given in \eqref{J-gamma}.
		One can recognize the maximum eigenvalue $\lambda_\textsc{m}(\theta,\phi)$
		from the top view.
		In the case of integer $j$-values, one out of four eigenvalues of $\Lambda$
		dominates in a region of the parameter space, and one can observe the level crossings in the two left-hand-side pictures.
		There one can easily spot points where---three colors join---$\lambda_\textsc{m}$
		becomes threefold degenerate, some of them are indicated by black dots.
		The black dots correspond to the triangular faces of the allowed regions
		$\mathcal{E}_4$ in Figs.~\ref{fig:E-J4 for j=2,5/2} and \ref{fig:E-J4 for j=3,7/2}.
		Whereas, in the case of ${j=\tfrac{5}{2}}$ and ${j=\tfrac{7}{2}}$, the largest and the second-largest eigenvalues meet only at four and six individual points, respectively.
		These four and three out of the six points are marked in red color.
		The red dots in all the above pictures have link with the elliptical faces of
		${\mathcal{E}_4}$ in Figs.~\ref{fig:E-J4 for j=2,5/2} and \ref{fig:E-J4 for j=3,7/2}.
	}
	\label{fig:lmax J4} 
\end{figure}

For ${j=2}$, there is not a single eigenvalue of $\Lambda(\theta,\phi)$
that stays the largest throughout the parameter space
${[0,\pi]\times[0,2\pi)}$
of $\theta$ and $\phi$.
Four distinct eigenvalues ${\lambda(\theta,\phi)}$ compete and cross each other, known as the \emph{level crossing}, in different parts of the parameter space as shown in Fig.~\ref{fig:lmax J4}.
There we highlight points one in red and four in black color---that correspond to the ellipse and triangles---where 
the maximum eigenvalue ${\lambda_\textsc{m}(\theta,\phi)}$ becomes double and triple degenerate, respectively.

Next, in the case of ${j=\tfrac{5}{2}}$, $\Lambda(\theta,\phi)$ has three distinct eigenvalues, and each one is double degenerate for all $\theta$ and $\phi$. 
Although one of the eigenvalues stays the largest in the whole parameter space ${[0,\pi]\times[0,2\pi)}$, it meets the second-largest eigenvalue at four different ${(\theta,\phi)}$.
In other words, the maximum eigenvalue $\lambda_\textsc{m}(\theta,\phi)$ becomes fourfold degenerate only at these 4 points featured in red color in Fig.~\ref{fig:lmax J4}. 
Consequently, we observe four flat faces [see Fig.~\ref{fig:E-J4 for j=2,5/2} for  ${j=\tfrac{5}{2}}$] on the boundary
of the numerical range $\mathcal{E}_4$.
All these faces are elliptical in shape.
One of these is described by
\begin{align}
\label{ell plane 1, j=5/2, J4}
&\bigg(\frac{\langle J_x^4\rangle+\langle J_y^4\rangle-\tfrac{803}{24}}{\tfrac{50}{3}}\bigg)^2+
\bigg(\frac{\langle J_x^4\rangle-\langle J_y^4\rangle}{\tfrac{50}{\sqrt{3}}}\bigg)^2
\leq 1\ \mbox{and}\quad\nonumber\\
&\sqrt{3}\,\langle \widehat{\eta}_1\cdot\vec{E}_4\rangle=\langle J_x^4\rangle+\langle J_y^4\rangle+\langle J_z^4\rangle=\tfrac{803}{16}\,,
\end{align}
and another is by
\begin{align}
\label{ell plane 2, j=5/2, J4}
&\bigg(\frac{\langle J_x^4\rangle+\langle J_y^4\rangle-39.7056}{10.4194}\bigg)^2+
\bigg(\frac{\langle J_x^4\rangle-\langle J_y^4\rangle}{30.7721}\bigg)^2
\leq 1\ \mbox{and}\quad\nonumber\\
&\quad \langle J_x^4\rangle+\langle J_y^4\rangle+4.076923\,\langle J_z^4\rangle=50.37981\,.
\end{align}

Like \eqref{ell plane, j=2, J4},
the second equations in \eqref{ell plane 1, j=5/2, J4} and
\eqref{ell plane 2, j=5/2, J4}
identify the tangent hyperplanes of $\mathcal{E}_4$ in which the associated ellipses reside.
One can check that only the eigenstates attached to the biggest eigenvalue $\tfrac{803}{16\sqrt{3}}$ of 
${\widehat{\eta}_1\cdot\vec{E}_4}$
and ${-50.37981}$ of ${-(J_x^4+J_y^4+4.076923\,J_z^4)}$
satisfy \eqref{ell plane 1, j=5/2, J4} and \eqref{ell plane 2, j=5/2, J4}, respectively.
Since the operator ${J_x^4+4.076923\,J_y^4+J_z^4}$ is unitarily equivalent to ${J_x^4+J_y^4+4.076923\,J_z^4}$, by the cyclic permutations of $\{x,y,z\}$ in \eqref{ell plane 2, j=5/2, J4}
one can obtain relations that characterize the two remaining elliptical faces of the allowed region. 
Note that numbers such as ${4.076923}$ are rounded to a few decimal places.

Next, we take ${j=3}$, where the allowed region is the convex hull of three intersecting ellipses and the point ${(16, 16, 16)}$ that is marked in blue color in 
Fig.~\ref{fig:E-J4 for j=3,7/2}.
There one can notice blue curves, which are the parts---of the ellipses---that lie on the boundary
$\partial\mathcal{E}_4$.
The remaining (unseen) parts of ellipses fall inside the numerical range $\mathcal{E}_4$.
One of the ellipses can be characterized by the equality in
\begin{align}
\label{ell plane, j=3, J4}
&\bigg(\frac{\langle J_x^4\rangle+\langle J_y^4\rangle-82}{20}\bigg)^2+
\bigg(\frac{\langle J_x^4\rangle-\langle J_y^4\rangle}{20\sqrt{15}}\bigg)^2
\leq 1\ \mbox{and}\quad\nonumber\\
&\quad \langle\, 2 J_x^4 + 2 J_y^4 + 5 J_z^4\,\rangle=204\,,
\end{align}
and the other two by permuting $\{x,y,z\}$ in these relations. 
The second equation in \eqref{ell plane, j=3, J4} describes the plane where the ellipse
stays.
Note that this plane is not a supporting plane of $\mathcal{E}_4$ but it passes through  
the allowed region.
Moreover the eigenvalue ${204}$ of ${2 J_x^4 + 2 J_y^4 + 5 J_z^4}$ is not an extreme eigenvalue.

Like ${j=2}$, the level crossings occur also in the case of ${j=3}$.
In Fig.~\ref{fig:lmax J4}, we mark four points in black color where
$\lambda_\textsc{m}(\theta,\phi)$ turns triple degenerate.
In these four directions---$\widehat{\eta}_1$, ${\vec{\eta}=-(1,1,3.55739)}$, and the two obtained by permuting the entries in
$\vec{\eta}$---$\mathcal{E}_4$ has four triangular faces [see Fig.~\ref{fig:E-J4 for j=3,7/2}].
One can check that
the biggest eigenvalues ${34 \sqrt{3}}$ and ${-88.9183}$ of 
the operators ${\widehat{\eta}_1\cdot\vec{E}_4}$ [for $\widehat{\eta}_1$, see Table~\ref{tab:theta-phi-for(-a)}] and
${\vec{\eta}\cdot\vec{E}_4}$ are threefold degenerate, and the associated eigenstates generate the triangular faces.

\begin{figure}
	\centering
	\subfloat{\includegraphics[width=40mm]{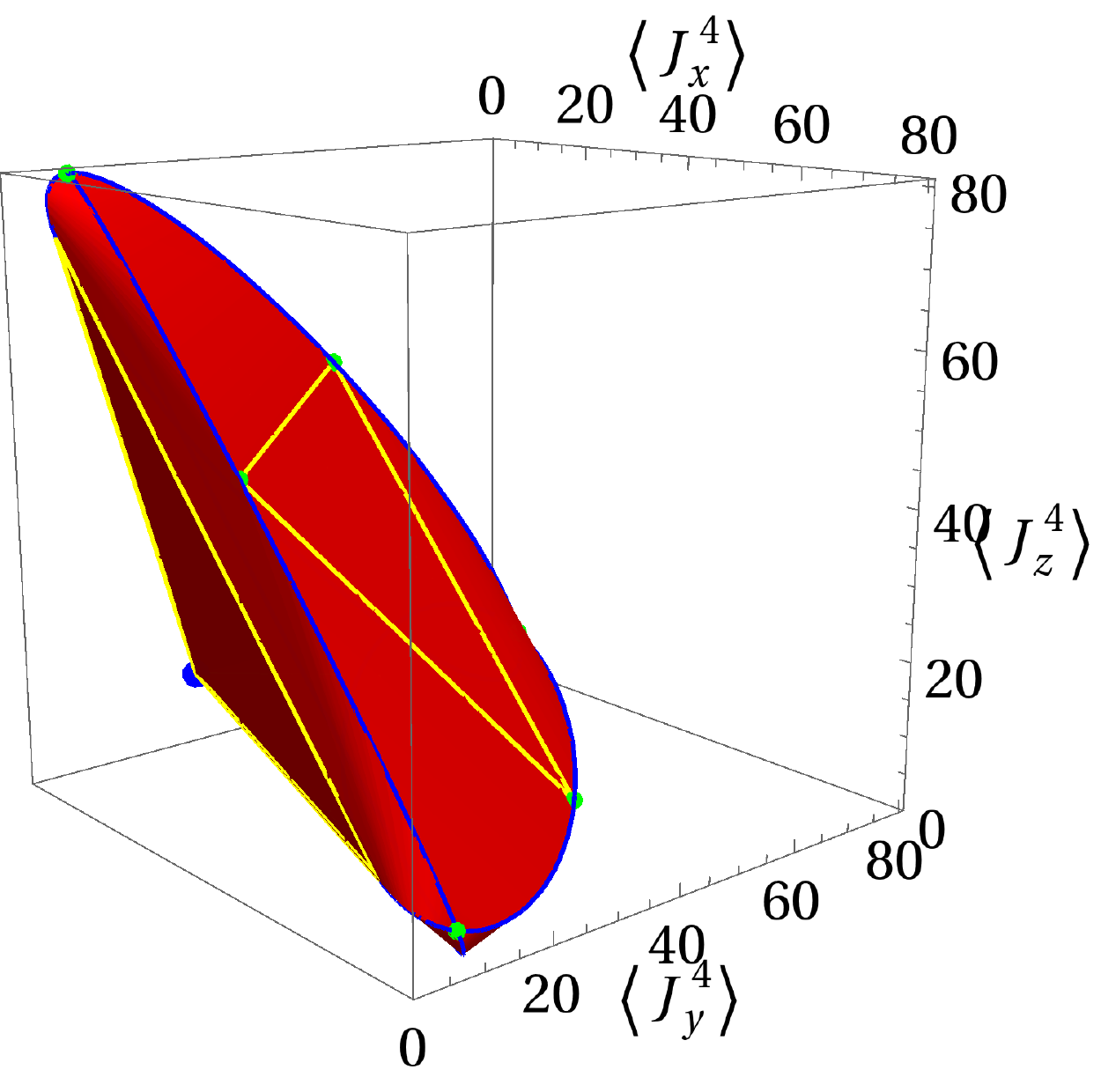}}\quad\
	\subfloat{\includegraphics[width=40mm]{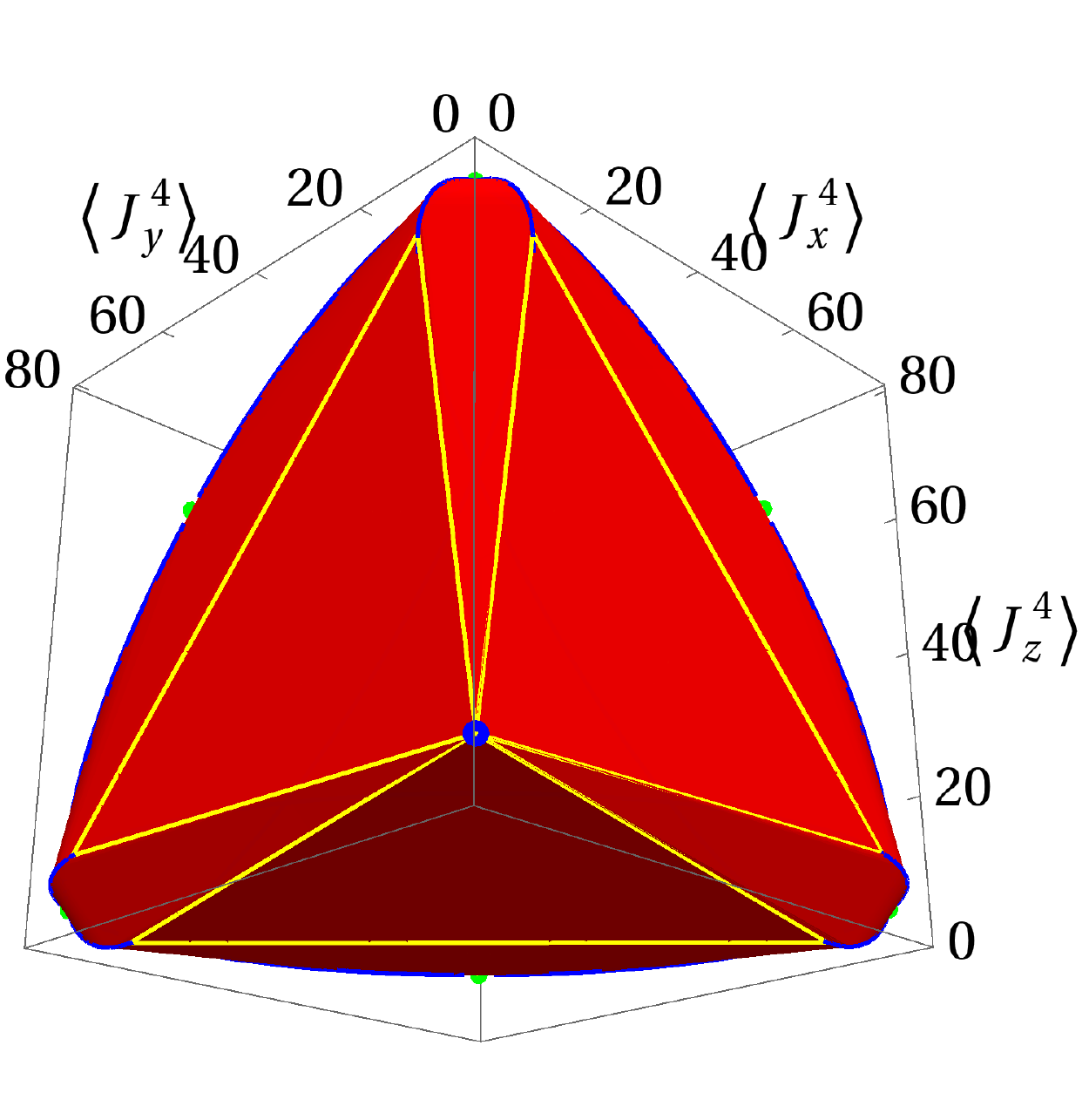}}\quad\
	\\
	\subfloat{\includegraphics[width=40mm]{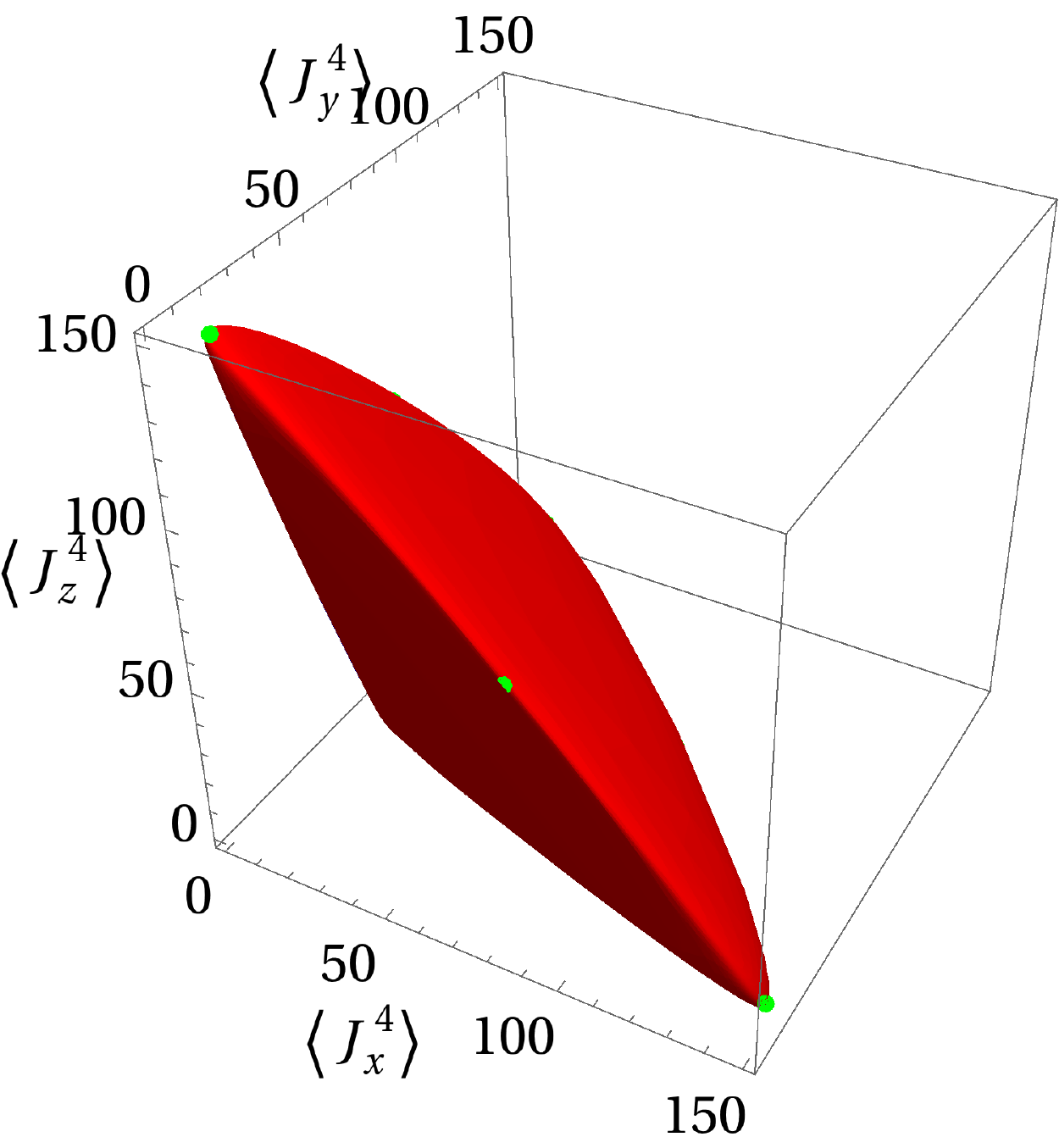}}
	\subfloat{\includegraphics[width=40mm]{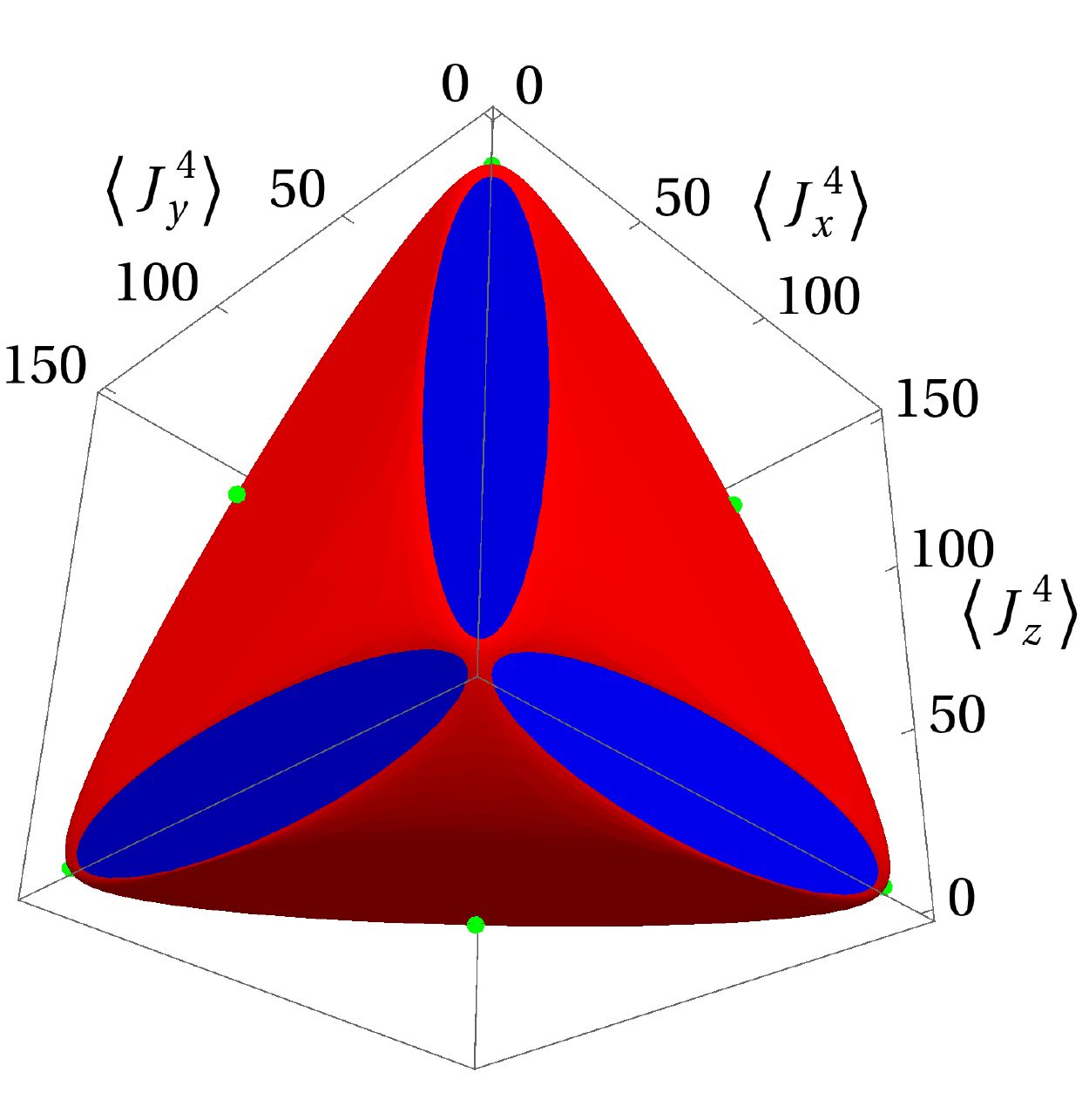}}
	\caption{(Color online) The red convex bodies in the top and bottom rows
	depict the numerical range $\mathcal{E}_4$ of $\vec{E}_{\gamma=4}$ in the case of quantum numbers ${j=3}$ and ${j=\tfrac{7}{2}}$, respectively.
	Like Fig.~\ref{fig:E-J4 for j=2,5/2}, each row carries the same $\mathcal{E}_4$.
	In the top-pictures, the blue curves are parts of three ellipses, the blue point is ${16(1,1,1)}$, and the yellow lines constitute four different triangles on the boundary
	$\partial\mathcal{E}_4$.
	The ellipses cross each other at the three green points, such as ${({81,6,6})}$, 
	that come from the maximum-eigenvalue states of $J_x^4$, $J_y^4$, and $J_z^4$
	[see \eqref{max-min-ev-points Jxyz^4}].
	The blue point is a vertex of the triangles, and the other 
	vertices---fall on the ellipses---are ${(73.6649, 1.24132, 10.8375)}$,
	the green point
	${({51, 51,0})}$, and the rest obtained by the permutations of their entries.	
    In the bottom-pictures, for ${j=\tfrac{7}{2}}$, one can see three elliptical
    disks in blue color on the boundary of
    $\mathcal{E}_4$.
    The vector ${-(1,1, 0.3890792)}$ is an outward normal to the supporting hyperplane in which one of the disks resides.
	}
	\label{fig:E-J4 for j=3,7/2} 
\end{figure}

Now we move to ${j=\tfrac{7}{2}}$,
where each eigenvalue of 
$\Lambda(\theta,\phi)$ is at least double degenerate, and  
there are at most four distinct eigenvalues
over the whole parameter space.
Its maximum eigenvalue $\lambda_\textsc{m}(\theta,\phi)$ becomes fourfold degenerate
at six different $(\theta,\phi)$, and three of them are indicated by the red-dots in Fig.~\ref{fig:lmax J4}.
The allowed region $\mathcal{E}_4$ has three elliptical faces in the directions 
$(\theta,\phi)$ associated with the red-dots.
One of the faces is characterized by
\begin{align}
\label{ell plane, j=7/2, J4}
&\bigg(
\frac{\langle J_x^4\rangle+\langle J_y^4\rangle-36.96675}{21.71172}\bigg)^2+
\bigg(\frac{\langle J_x^4\rangle-\langle J_y^4\rangle}{20.731196}\bigg)^2
\leq 1\ \mbox{and}\quad
\nonumber\\
&\quad\langle J_x^4 + J_y^4 + 0.3890792\, J_z^4\,\rangle=71.8851\,,
\end{align}
and the remaining two by the permutations of $\{x,y,z\}$ 
in these relations.
Only the eigenstates associated with the smallest eigenvalue ${71.8851}$
of $J_x^4 + J_y^4 + {0.3890792}\, J_z^4$ meet all the conditions in 
\eqref{ell plane, j=7/2, J4}.

Finally, after the uniform scaling~\eqref{uni-scaling-J}, we present the allowed $\mathcal{E}_4^\textsc{us}$
in Fig.~\ref{fig:E-J4 for j=25} 
in the case of a large quantum number ${j=25}$.
One can notice a curved surface at the boundary $\partial\mathcal{E}_4^\textsc{us}$, which is predicted by the quantum de Finetti theorem in the limit ${j\rightarrow\infty}$ [see the next section].
If we apply this limit to the mean values in \eqref{max-min-ev-points Jxyz^4} then we realize
\begin{align}
\label{max-min-ev-points Jxyz^4, j-inf}
\lim\limits_{j\rightarrow\infty}\,
\tfrac{1}{j^4}
\langle{\pm j}|\vec{E}_4|{\pm j}\rangle&=(0,0,1)
\quad\mbox{and}\\
\lim\limits_{j\rightarrow\infty}\,
\tfrac{1}{j^4}
\big\langle{\pm \tfrac{1}{2}}\big|\vec{E}_4\big|{\pm \tfrac{1}{2}}\big\rangle&=
(\tfrac{3}{8},\tfrac{3}{8},0)
=
\lim\limits_{j\rightarrow\infty}\,
\tfrac{1}{j^4}
\langle 0|\vec{E}_4|0\rangle\,.
\nonumber
\end{align}
One can recognize that two of the green points in Fig.~\ref{fig:E-J4 for j=25} are indeed close to ${(\tfrac{3}{8},\tfrac{3}{8},0)}$
and a corner $(0,0,1)$ of the hypercube ${
[0,1]^{\times 3}=
\lim\limits_{j\rightarrow\infty}\tfrac{1}{j^4}\mathcal{H}_4}$
[for $\mathcal{H}_{\gamma=4}$, see \eqref{H gamna-even}].

Like before, according to \eqref{Xdot}--\eqref{u-half-sum}, here we build our combined uncertainty $\{h,u_{\sfrac{1}{2}}\}$ and certainty $\{u_2,u_\text{max}\}$ functions of  
the mean vector ${\vec{\boldsymbol{\varepsilon}}=\langle\vec{E}_4\rangle}$.
Then, for every $j=1,\tfrac{3}{2},\cdots,50$, we compute values of these functions on
a finite set of boundary points of $\mathcal{E}_4$ and consider the minimum values 
of $h,u_{\sfrac{1}{2}}$ as their lower bounds $\mathfrak{h},\mathfrak{u_{\sfrac{1}{2}}}$
and the maximum values 
of $u_2,u_\text{max}$ as their upper bounds $\mathfrak{u_2},\mathfrak{u_\text{max}}$.
With a bound one has a UR or CR.
We plot these bounds in Fig.~\ref{fig:S4 UR bounds} and record them in Appendix~\ref{sec:Sup-material}.
These bounds are accurate up to a first few decimal places, and one can improve them by taking more boundary points or adopting a procedure such as reported in \cite{Schwonnek17}.

The uncertainty and certainty measures reach their extreme
values 
$\{\mathfrak{h},\mathfrak{u_{\sfrac{1}{2}}},
\mathfrak{u_2},\mathfrak{u_\text{max}}\}$ 
in those parts of the boundary which are close to the corners of hyperrectangle~\eqref{H gamna-even}.
As these parts move toward the corners 
[see Figs.~\ref{fig:E-J4 for j=2,5/2}, \ref{fig:E-J4 for j=3,7/2}, and \ref{fig:E-J4 for j=25}] as $j$ grows, the lower bounds decrease and the upper bounds increase
[see Fig.~\ref{fig:S4 UR bounds}].
In the limit ${j\rightarrow\infty}$, the allowed region will share three
vertices of the hyperrectangle 
[see \eqref{max-min-ev-points Jxyz^4, j-inf}] and 
$\mathfrak{h},\mathfrak{u_{\sfrac{1}{2}}},\mathfrak{u_2},$
and 
$\mathfrak{u_\text{max}}$
will hit their trivial values 0, 3, 3, and 3, respectively.

\begin{figure}
	\centering
	\subfloat{\includegraphics[width=40mm]{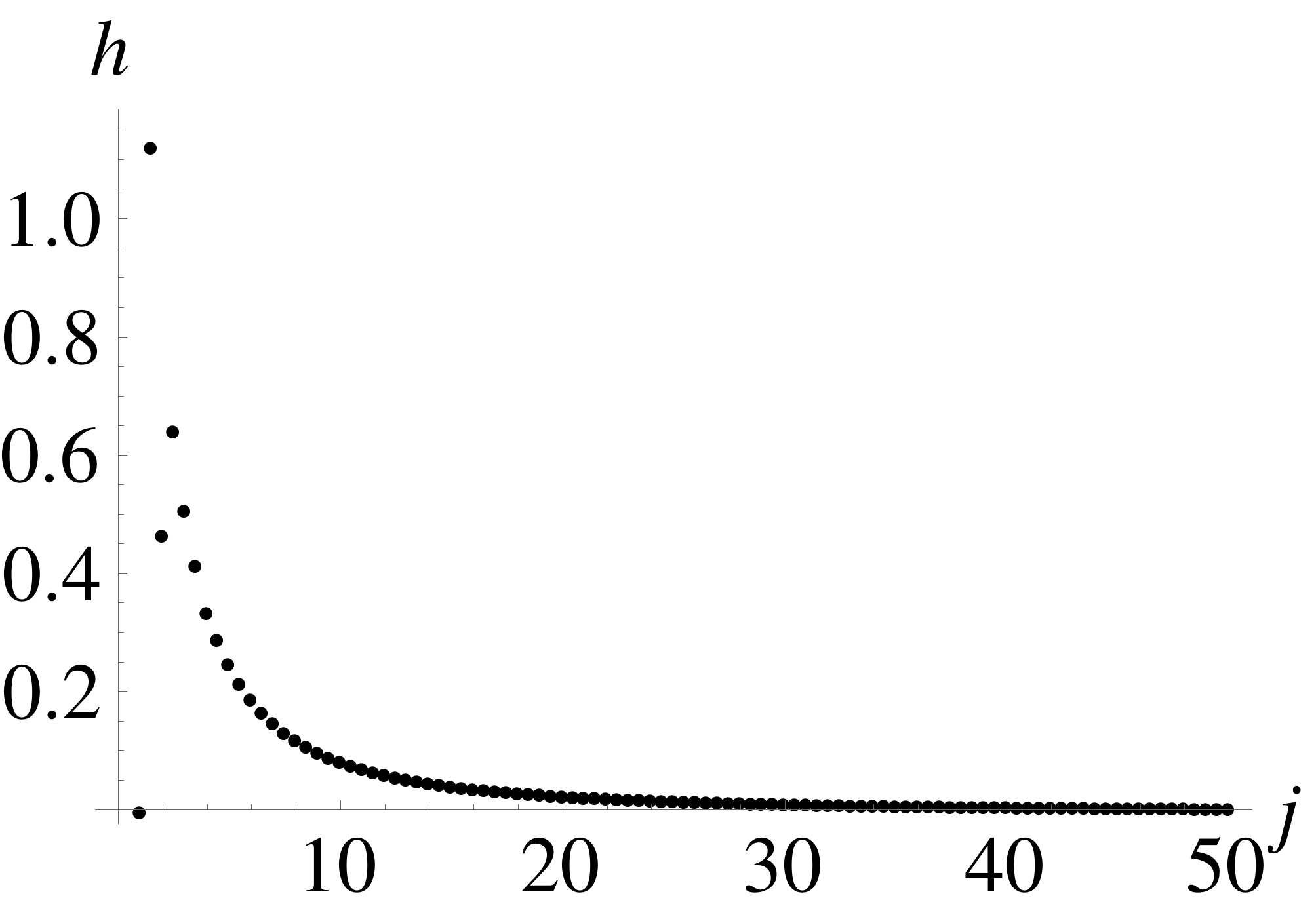}}\quad\
	\subfloat{\includegraphics[width=40mm]{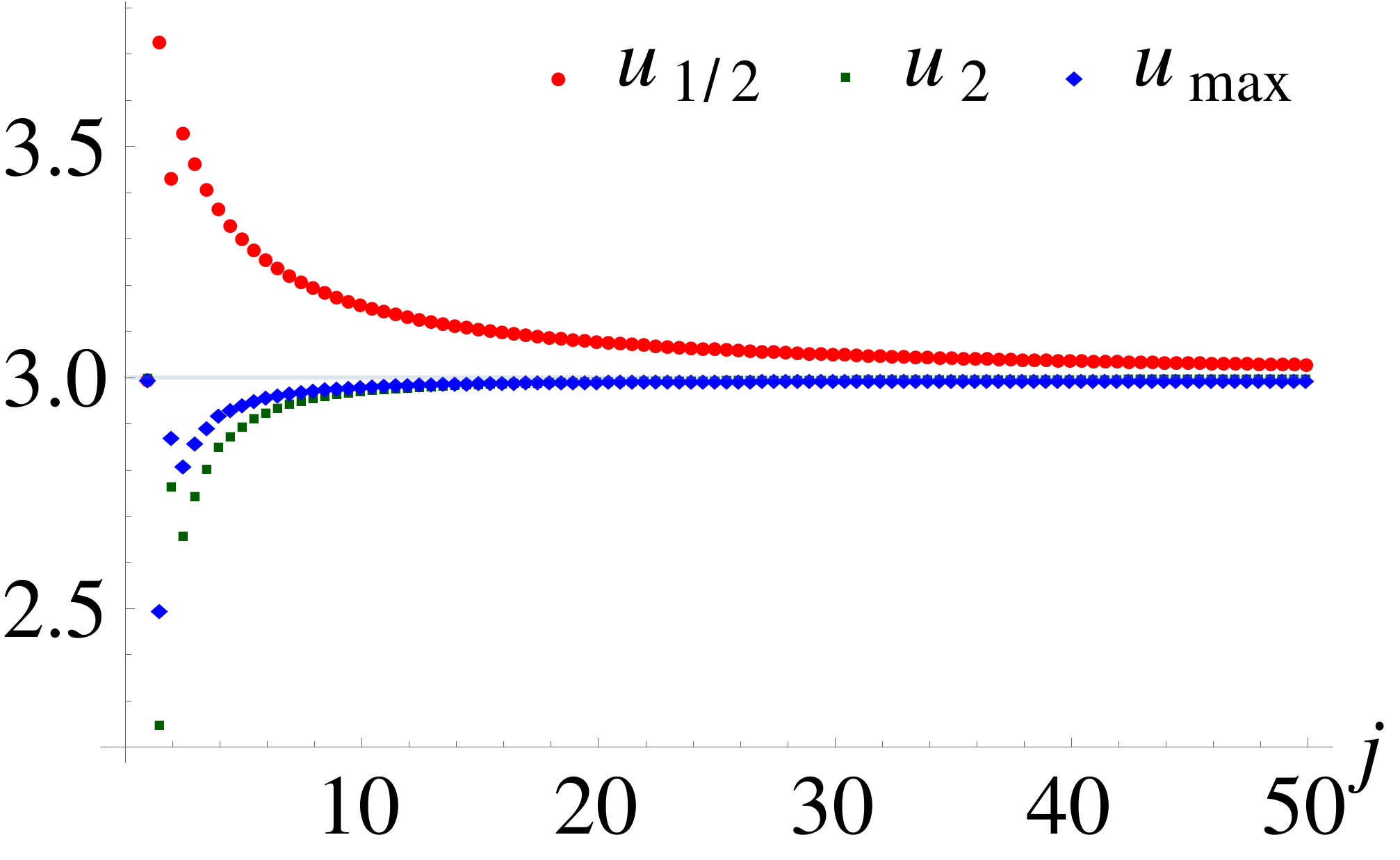}}
	\caption{Like Fig.~\ref{fig:S2 UR bounds}, 
	the plots exhibit the lower bounds of 
	$h(\vec{\boldsymbol{\varepsilon}})$ and $u_{\sfrac{1}{2}}(\vec{\boldsymbol{\varepsilon}})$
	and the upper bounds of
	$u_2(\vec{\boldsymbol{\varepsilon}})$ and 
	$u_\text{max}(\vec{\boldsymbol{\varepsilon}})$
	on $\mathcal{E}_4$
	as functions of the quantum number 
	$j=1,\tfrac{3}{2},\cdots,50$.
	}
	\label{fig:S4 UR bounds} 
\end{figure}

%===========================================
\section{\textit{N}-qubit system}\label{sec:N-qubit}

In this section, one by one, we consider \eqref{J-gamma} and 
\begin{align}
\label{A-gamma}
\vec{E}_\gamma&=(A_{1,\gamma}\,,\,A_{2,\gamma}\,,\,A_{3,\gamma})\;,
\qquad\quad\mbox{where}\nonumber\\
A_{1,\gamma}&:=(J_x)^\gamma (J_z)^\gamma+(J_z)^\gamma (J_x)^\gamma\,,
\nonumber\\
A_{2,\gamma}&:=(J_y)^\gamma (J_z)^\gamma+(J_z)^\gamma (J_y)^\gamma\,,
\quad\mbox{and}
\\
A_{3,\gamma}&:=(J_x)^\gamma (J_y)^\gamma+(J_y)^\gamma (J_x)^\gamma\,.
\nonumber
\end{align}
As before $\gamma$ is a finite positive integer.
Here, in both the cases,
the main task is to achieve the allowed region for $\vec{E}_\gamma$ in the limit ${j\rightarrow\infty}$.
To complete the task, we take a system of $N$ spin-$\tfrac{1}{2}$ particles (qubits),
and applying the famous quantum de Finetti theorem \cite{Stormer69,Hudson76}
in the limit ${N\rightarrow\infty}$.

Let us begin with the vector Pauli operator ${\vec{\sigma}_i=(\textsf{X}_i,\textsf{Y}_i,\textsf{Z}_i)}$ that acts on $i$th qubit's Hilbert space, and 
\begin{equation}
\vec{\textbf{J}}=(\textbf{J}_x,\textbf{J}_y,\textbf{J}_z):=
\tfrac{1}{2}\sum_{i=1}^{N}\vec{\sigma}_i
\end{equation}
is the total angular momentum vector operator on 
$N$-qubit Hilbert space ${{\mathscr{H}_2}^{\otimes N}}$.
The operator ${\textbf{J}^2=\vec{\textbf{J}}\cdot\vec{\textbf{J}}}$
has eigenvalues $j(j+1)$ where 
$j=\tfrac{N}{2},\tfrac{N}{2}-1,\cdots,0\,\mbox{or}\, \tfrac{1}{2}$ when
$N$ is an even or odd number.
In an eigenbasis of (degenerate operator)
$\textbf{J}^2$, the components of $\vec{\textbf{J}}$ 
reveal their block-diagonal forms, for instance,
\begin{equation}
\label{Sz}
\textbf{J}_z=\tfrac{1}{2}\sum_{i=1}^{N}\textsf{Z}_i
=J_z^{({\scriptscriptstyle\frac{N}{2}})}\boldsymbol\oplus 
J_z^{({\scriptscriptstyle\frac{N}{2}}-1)}
\boldsymbol\oplus\cdots\boldsymbol\oplus
 J_z^{(0\,\mbox{or}\, \frac{1}{2})}\,,
\end{equation}
where ${\boldsymbol\oplus J_z^{({\scriptscriptstyle\frac{N}{2}}-1)}}$ denotes the direct sum of multiple copies of 
the angular momentum operator $J_z$ corresponding to the quantum number ${j=\frac{N}{2}-1}$, and so on.
There is only one copy of $J_z^{({\scriptscriptstyle\frac{N}{2}})}$ in the direct sum. 
For more details, we point the reader to \cite{Dicke54,Arecchi72}.
Polynomials of $\textbf{J}_x,\textbf{J}_y,$ and $\textbf{J}_z$ such as
\begin{align}
\label{J-N}
&\qquad (\textbf{J}_x)^\gamma\ ,\; (\textbf{J}_y)^\gamma\ ,\; (\textbf{J}_z)^\gamma\ ,\\
\label{A-N}
\textbf{A}_{1,\gamma}&:=
(\textbf{J}_x)^\gamma(\textbf{J}_z)^\gamma+(\textbf{J}_z)^\gamma(\textbf{J}_x)^\gamma\,,
\nonumber\\
\textbf{A}_{2,\gamma}&:=
(\textbf{J}_y)^\gamma(\textbf{J}_z)^\gamma+(\textbf{J}_z)^\gamma(\textbf{J}_y)^\gamma\,,
\quad\mbox{and}
\\
\textbf{A}_{3,\gamma}&:=
(\textbf{J}_x)^\gamma(\textbf{J}_y)^\gamma+(\textbf{J}_y)^\gamma(\textbf{J}_x)^\gamma
\nonumber
\end{align}
carry the same block-diagonal structure.

\begin{figure}
	\centering
	\subfloat{\includegraphics[width=40mm]{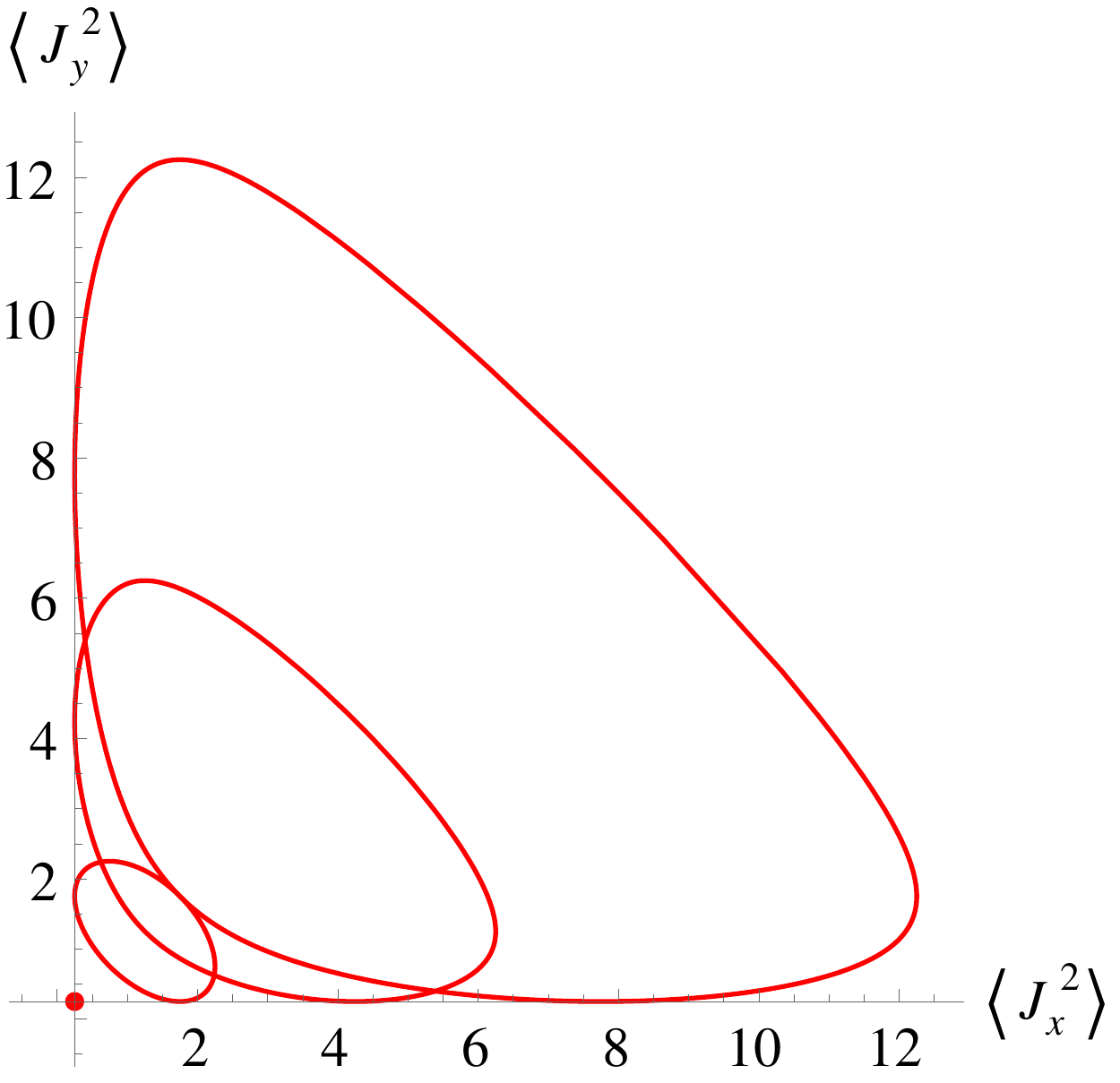}}\quad\
	\subfloat{\includegraphics[width=40mm]{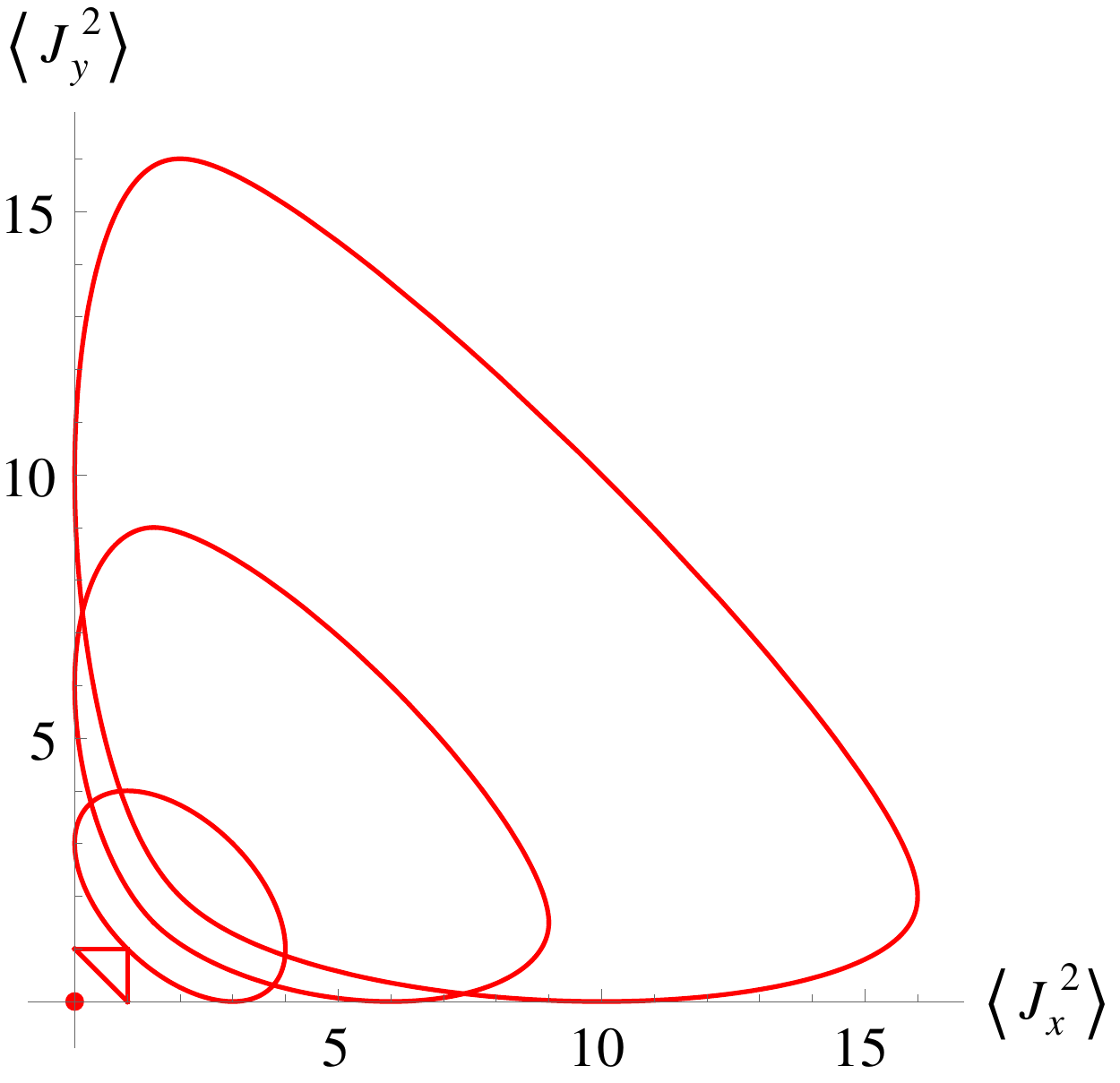}}
	\caption{The left and right portrayals are for ${N=7}$ and ${N=8}$
		number of qudits. The red closed curves and the red points
		are the boundaries of 
		$\mathcal{E}\big((J_x^{\scriptscriptstyle(j)})^2,
		(J_y^{\scriptscriptstyle(j)})^2\big)$
		for different ${j=\tfrac{N}{2},\tfrac{N}{2}-1,\cdots,0\, \mbox{or}\, \tfrac{1}{2}}$.
		The curves are taken from Figs.~\ref{fig:E-J^2 for j=1/2and1}--\ref{fig:E-J^2 for j=7/2and4}.
	}
	\label{fig:J2-diff-j} 
\end{figure}

\textit{A side remark:} The numerical range of a direct sum of operators, say,
${(A\oplus A',B\oplus B')}$
is the convex hull of the numerical ranges of
${(A,B)}$ and ${(A',B')}$ provided $A$ and $B$ act on the same Hilbert space, and similarly for $A'$ and $B'$ \cite{Binding91,Chen17b}.
Hence, we can express
the numerical range of ${(\textbf{J}_x^2,\textbf{J}_y^2)}$ as
\begin{equation}
\mathcal{E}(\textbf{J}_x^2,\textbf{J}_y^2)=
\text{Conv}\Bigg\{
\bigcup_{j=0\, \mbox{or}\, \frac{1}{2}}^{\frac{N}{2}}
\mathcal{E}\big((J_x^{\scriptscriptstyle(j)})^2,(J_y^{\scriptscriptstyle(j)})^2\big)
\Bigg\}\,.
\end{equation}
By looking at Fig.~\ref{fig:J2-diff-j} one can tell, at least up to ${N=8}$, that
$\mathcal{E}(\textbf{J}_x^2,\textbf{J}_y^2)$
is the convex hull of 
$\mathcal{E}\big((J_x^{\scriptscriptstyle(N/2)})^2,
(J_y^{\scriptscriptstyle(N/2)})^2\big)$
and the point ${(0,0)}$ or ${(\frac{1}{4},\frac{1}{4})}$
when $N$ is an even or odd number.
The point ${(0,0)}$ 
is the allowed region of 
$(J_x^2,J_y^2)$ when the quantum number ${j=0}$.
Since ${(0,0)}$ or ${(\frac{1}{4},\frac{1}{4})}$
is a vertex of the hyperrectangle~\eqref{hyperrectangle}
in the case of $\textbf{J}_x^2$ and
$\textbf{J}_y^2$,
we do not get a nontrivial UR and CR for them
based on the procedure given in Sec.~\ref{sec:all-region}.
Whereas, in the case of anticommutators \eqref{A-gamma} and \eqref{A-N}, 
we observe a nested sequence of regions
and thus 
${\mathcal{E}(\textbf{A}_{1,1},\textbf{A}_{2,1},\textbf{A}_{3,1})=
	\mathcal{E}(
	A_{1,1}^{\scriptscriptstyle(N/2)},
	A_{2,1}^{\scriptscriptstyle(N/2)}
	A_{3,1}^{\scriptscriptstyle(N/2)})}$, 
which we have checked up to ${N=50}$.

Now coming back to our main task, we only need the eigenspace of $\textbf{J}^2$ that
corresponds to the largest quantum number ${j=\tfrac{N}{2}}$.
Since $\textbf{J}_z$ commutes with $\textbf{J}^2$, we choose
their common eigenkets, known as the Dicke kets \cite{Dicke54,Arecchi72},
\begin{align}
\label{Dicke kets}
&|j,m\rangle=\tfrac{1}{(j-m)!}\tfrac{1}{\sqrt{{2j}\choose{j-m}}}
(\textbf{J}_-)^{j-m}|{+\tfrac{1}{2}}\rangle^{\otimes N}\,,\ \mbox{where}
\\
&\textbf{J}_\pm=\textbf{J}_x\pm\text{i}\textbf{J}_y\;,\
m\in\{j,j-1,\cdots,-j\}\,,\,j=\tfrac{N}{2}\,,
\end{align}
${{2j}\choose{j-m}}$ is the binomial coefficient, and 
${|{+\tfrac{1}{2}}\rangle=\textsf{Z}|{+\tfrac{1}{2}}\rangle}$.
Dicke kets satisfy the eigenvalue equations
\begin{align}
\label{Dicke kets eigen-eqn J2}
\textbf{J}^2\,|j,m\rangle&=j(j+1)\,|j,m\rangle\,,\\
\label{Dicke kets eigen-eqn Jz}
\textbf{J}_z\,|j,m\rangle&=m\,|j,m\rangle\,,\quad\mbox{and}\\
\label{Dicke kets eigen-eqn Up}
U_\textsc{p}\,|j,m\rangle&=|j,m\rangle \quad\mbox{for all }\textsc{p}\,,
\end{align}
where $U_\textsc{p}$ is the unitary operator associated with a permutation \textsc{p} of qubits' indices ${\{1,\cdots,N\}}$.
Equation \eqref{Dicke kets eigen-eqn Jz} is same as \eqref{Jz}.
Due to Eq.~\eqref{Dicke kets eigen-eqn Up}, all Dicke kets \eqref{Dicke kets}
have the Bose-Einstein symmetry \cite{Hudson76}, and they
form an orthonormal basis of the symmetric subspace
\begin{equation}
\label{sym-subspace}
\text{Sym}(\mathscr{H}_2^{\otimes N}):=
\text{span}\big\{\big|\tfrac{N}{2},m\big\rangle
\,\big|\,m=\tfrac{N}{2},\cdots,-\tfrac{N}{2} \big\}
\end{equation}
of the $N$-qubit Hilbert space.
In short, $\text{Sym}(\mathscr{H}_2^{\otimes N})$ is the eigenspace 
$\textbf{J}^2$ and $U_\textsc{p}$ for all \textsc{p}'s.

The $N$-qubit operators $(\textbf{J}_x,\textbf{J}_y,\textbf{J}_z)$
restricted to this subspace behave as $(J_x,J_y,J_z)$ [see \eqref{Jx,y,z}--\eqref{basis-Jz}], that is,
\begin{equation}
\label{restriction-J}
\vec{\textbf{J}}\Big|_{\text{Sym}(\mathscr{H}_2^{\otimes N})}
\equiv\vec{J}=(J_x,J_y,J_z)\quad \mbox{on}\quad \mathscr{H}_d\,,
\end{equation}
where the dimension ${d=2j+1=N+1}$.
We put the same restriction, \eqref{restriction-J},
on the $N$-qubit operators in \eqref{J-N} and \eqref{A-N}.
Now let us form a set
\begin{equation}
\label{state-space-BE}
\boldsymbol{\Omega}_\textsc{be}= \text{Conv}\Big\{ |\psi\rangle\langle\psi|\; \Big|\; 
|\psi\rangle\in\text{Sym}(\mathscr{H}_2^{\otimes N})\Big\}
\end{equation}
of all those $N$-qubit density operators whose supports lie in symmetric subspace \eqref{sym-subspace}.
$\boldsymbol{\Omega}_\textsc{be}$ is analogous to the state space---mentioned in \eqref{set-of-expt} for a $d$-level system---but here every state 
also has the Bose-Einstein symmetry $U_\textsc{p}\boldsymbol{\rho}=\boldsymbol{\rho}\in\boldsymbol{\Omega}_\textsc{be}$ as described in \cite{Hudson76}.

Here we quote a special case of the celebrated
quantum de Finetti theorem \cite{Stormer69,Hudson76}:
\begin{equation}
\label{de Finetti theorem}
\parbox{0.8\columnwidth}
{
In the limit ${N\rightarrow\infty}$,
$\boldsymbol{\Omega}_\textsc{be}$ becomes a Choquat simplex, extreme points of which are 
pure product states.
}
\end{equation} 
In a simplex, every point has a unique decomposition in terms of its extreme points, and 
the word ``pure'' in \eqref{de Finetti theorem} is attributed to 
the Bose-Einstein symmetry.
Every pure qubit's state
\begin{align}
\label{qubit-state}
\rho&=
\tfrac{1}{2}(\textsf{I}+\vec{\textsf{r}}\cdot\vec{\sigma})\,, \qquad \mbox{where}
\nonumber\\
\vec{\textsf{r}}&=\langle\vec{\sigma}\rangle=(\textsf{x},\textsf{y},\textsf{z})=
(\sin\mu\cos\nu,\sin\mu\sin\nu,\cos\mu),\qquad\
\nonumber\\
\mu&\in[0,\pi]\,,\quad\mbox{and}\quad 
\nu\in[0,2\pi)\,,
\end{align}
provides a symmetric product state ${\rho^{\otimes N}}$ that is an extreme point of $\boldsymbol{\Omega}_\textsc{be}$ for all $N=1,2,\cdots$, particularly,
in the limit ${N\rightarrow\infty}$ thanks to \eqref{de Finetti theorem}.
Now we can present our allowed regions.

\begin{figure*}
	\centering
	\subfloat{\includegraphics[width=40mm]{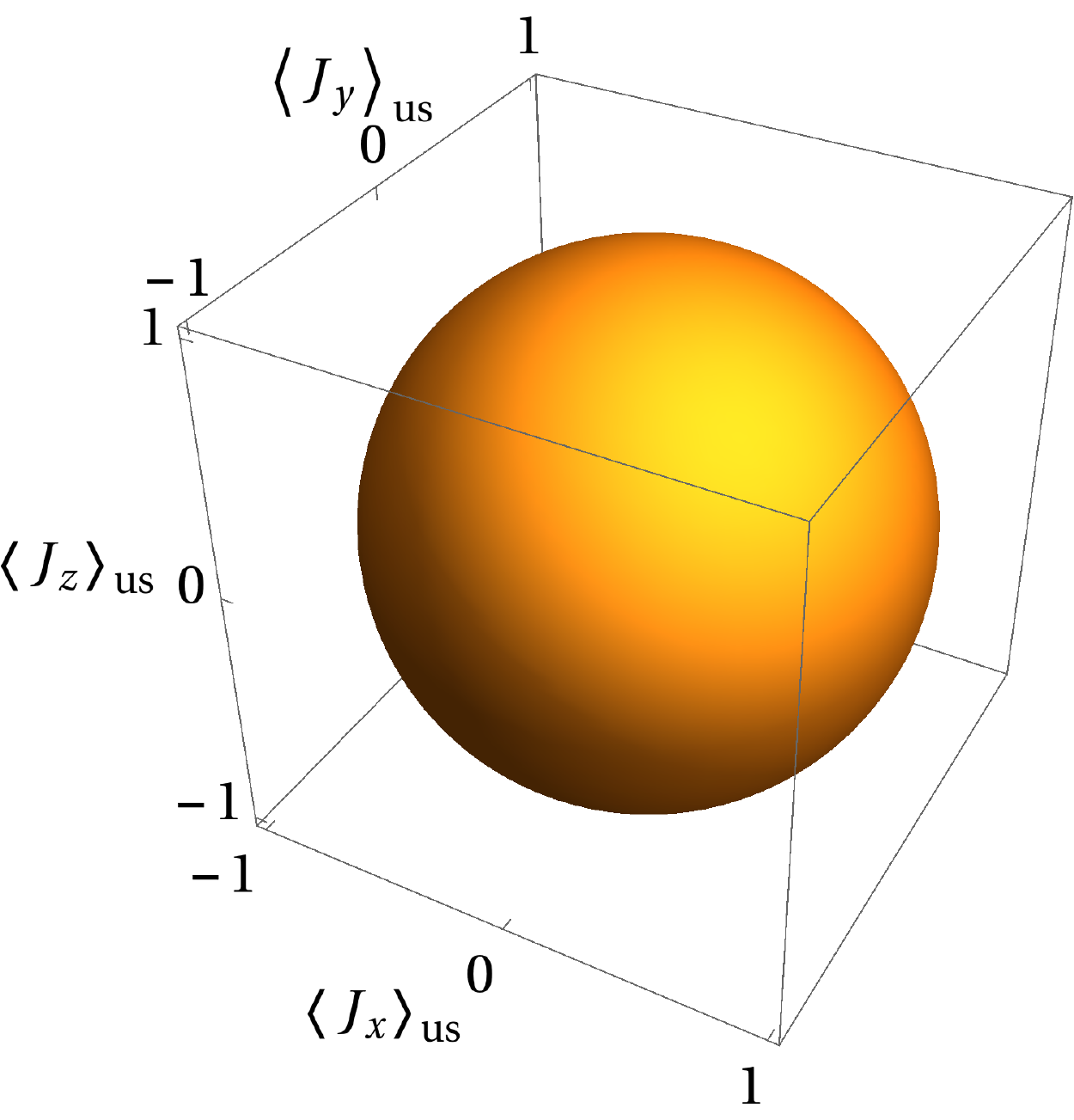}}\quad\
	\subfloat{\includegraphics[width=40mm]{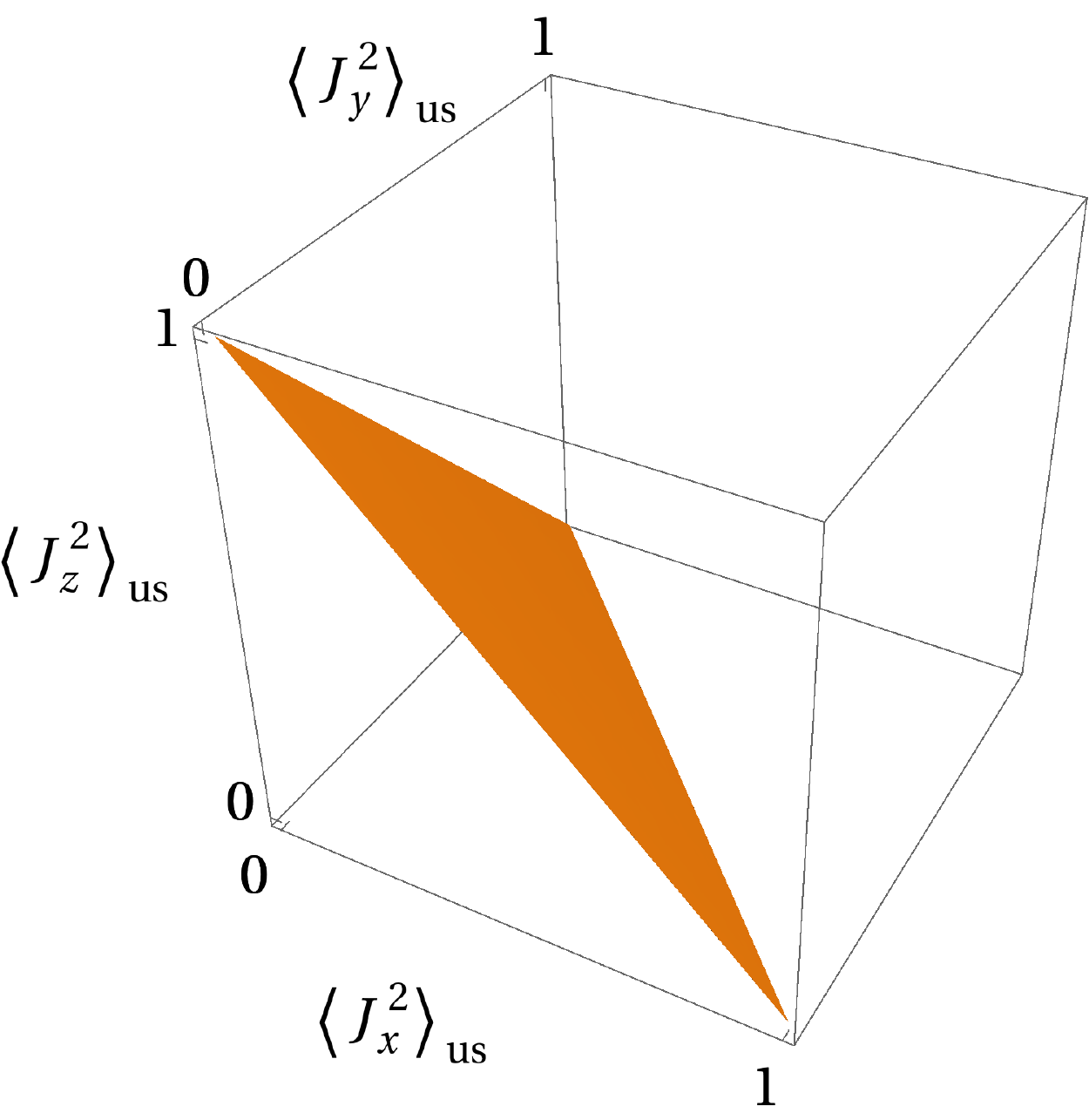}}\quad\
	\subfloat{\includegraphics[width=40mm]{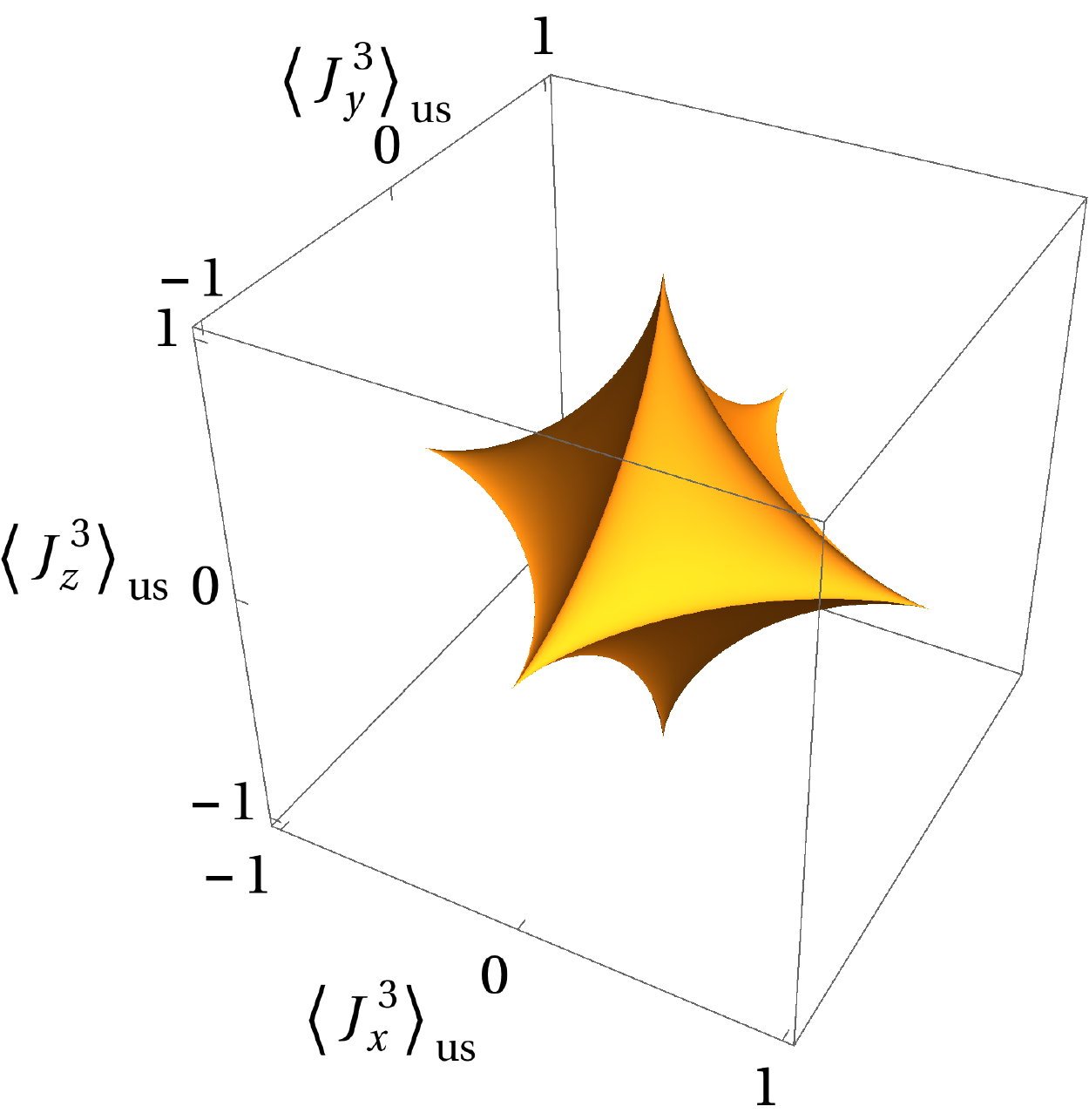}}\quad\
	\subfloat{\includegraphics[width=40mm]{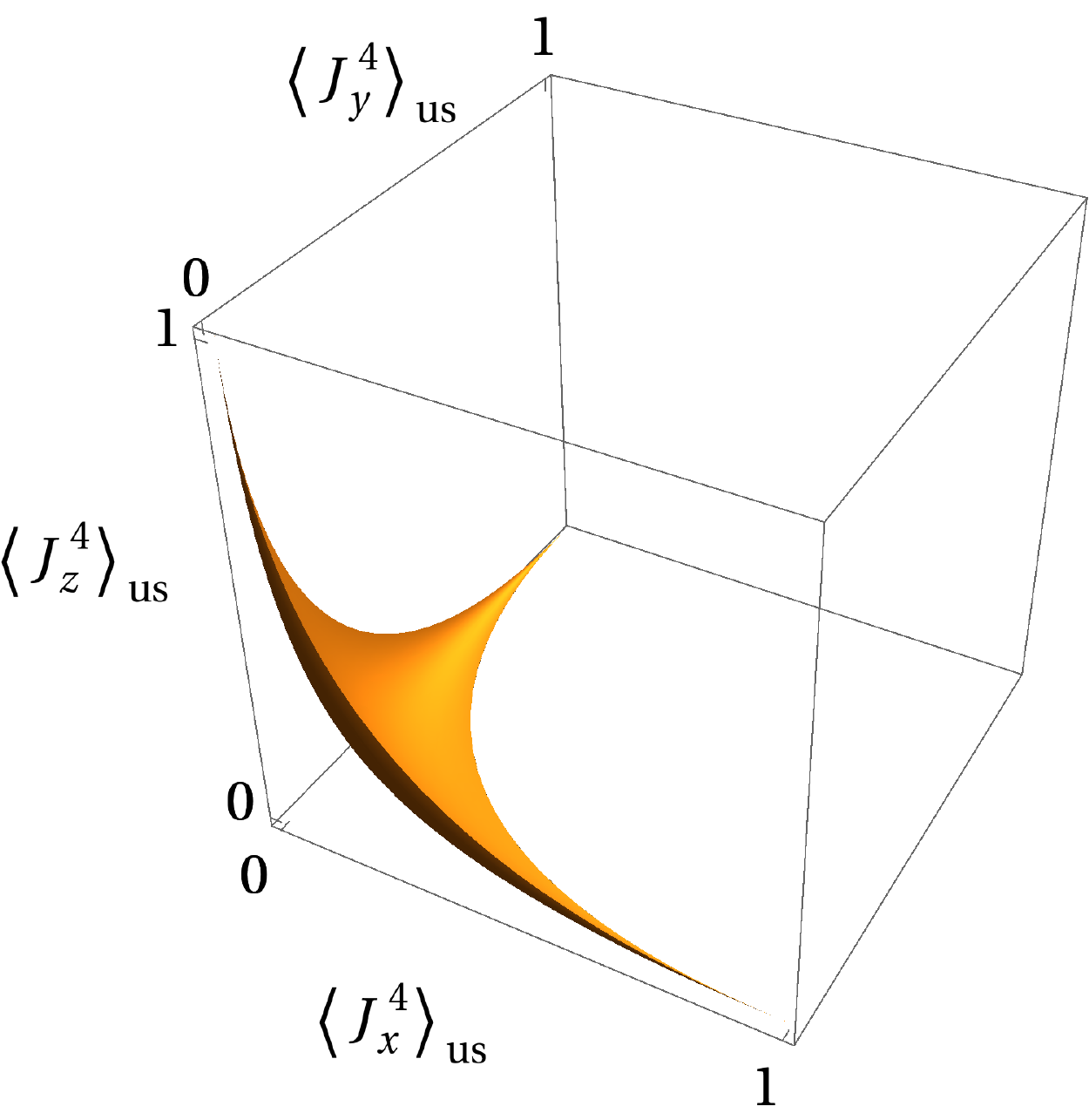}}
	\\\vspace{1mm}
	\subfloat{\includegraphics[width=40mm]{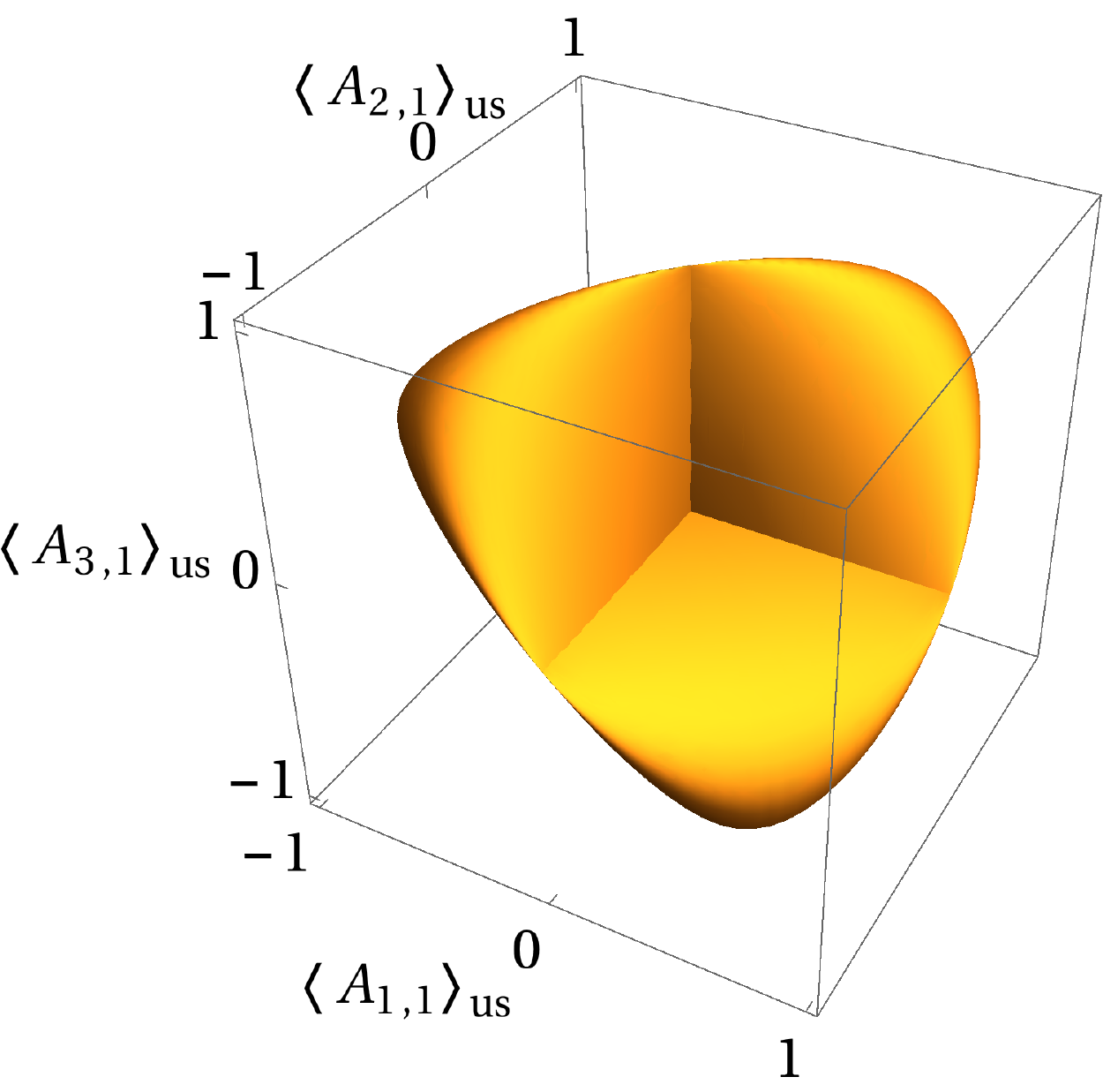}}\quad\
	\subfloat{\includegraphics[width=40mm]{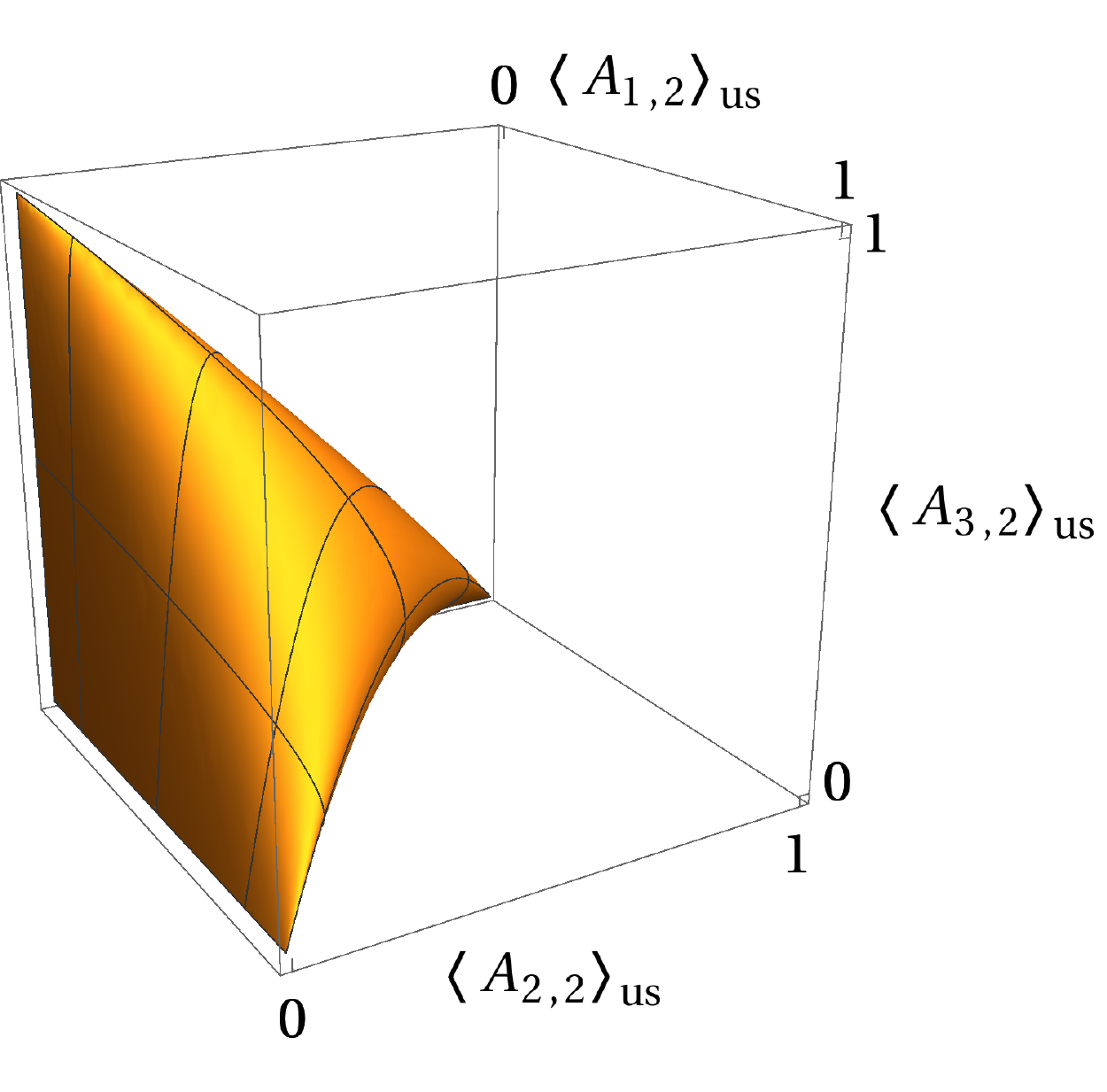}}\quad\
	\subfloat{\includegraphics[width=40mm]{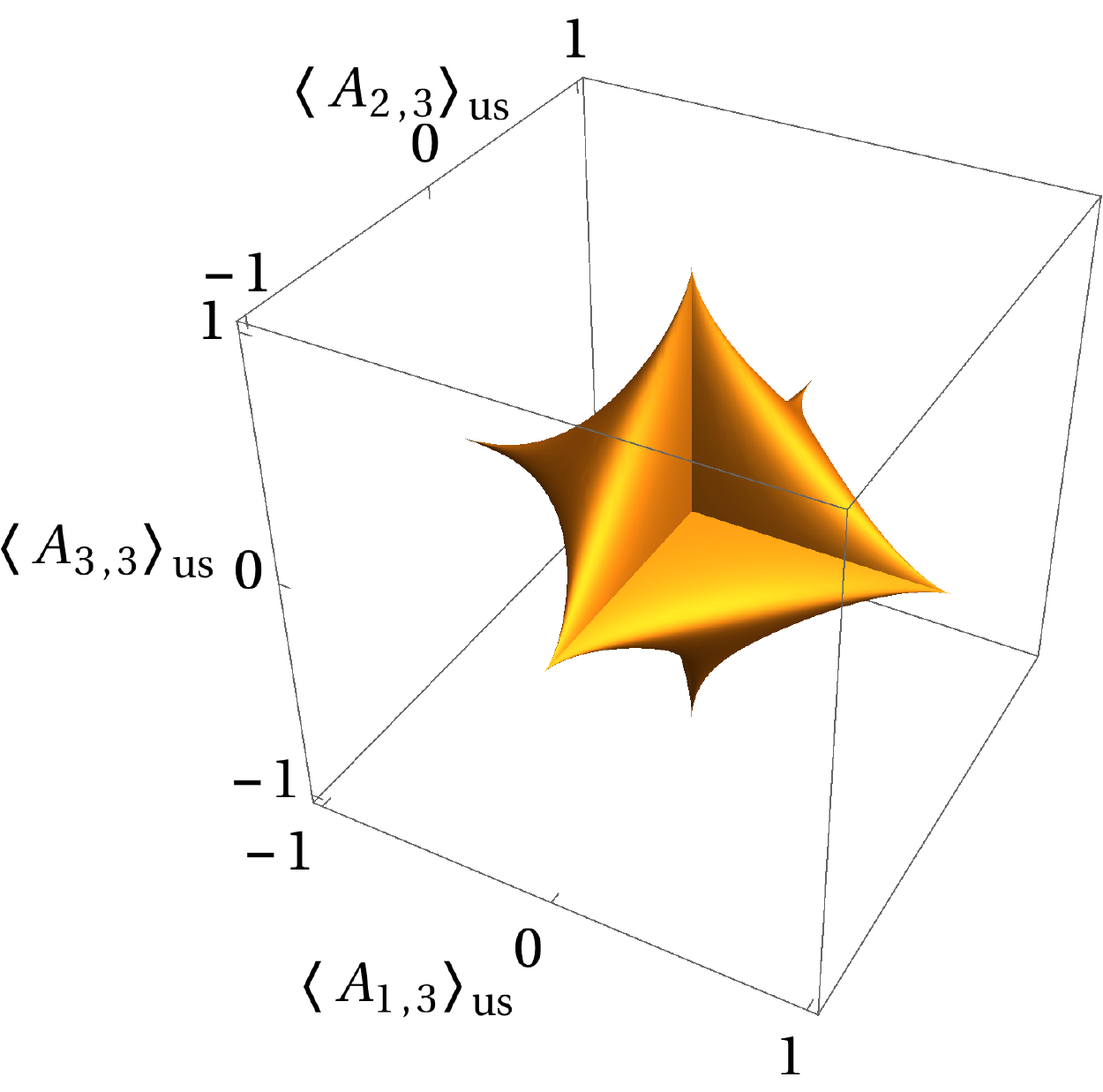}}\quad\
	\subfloat{\includegraphics[width=40mm]{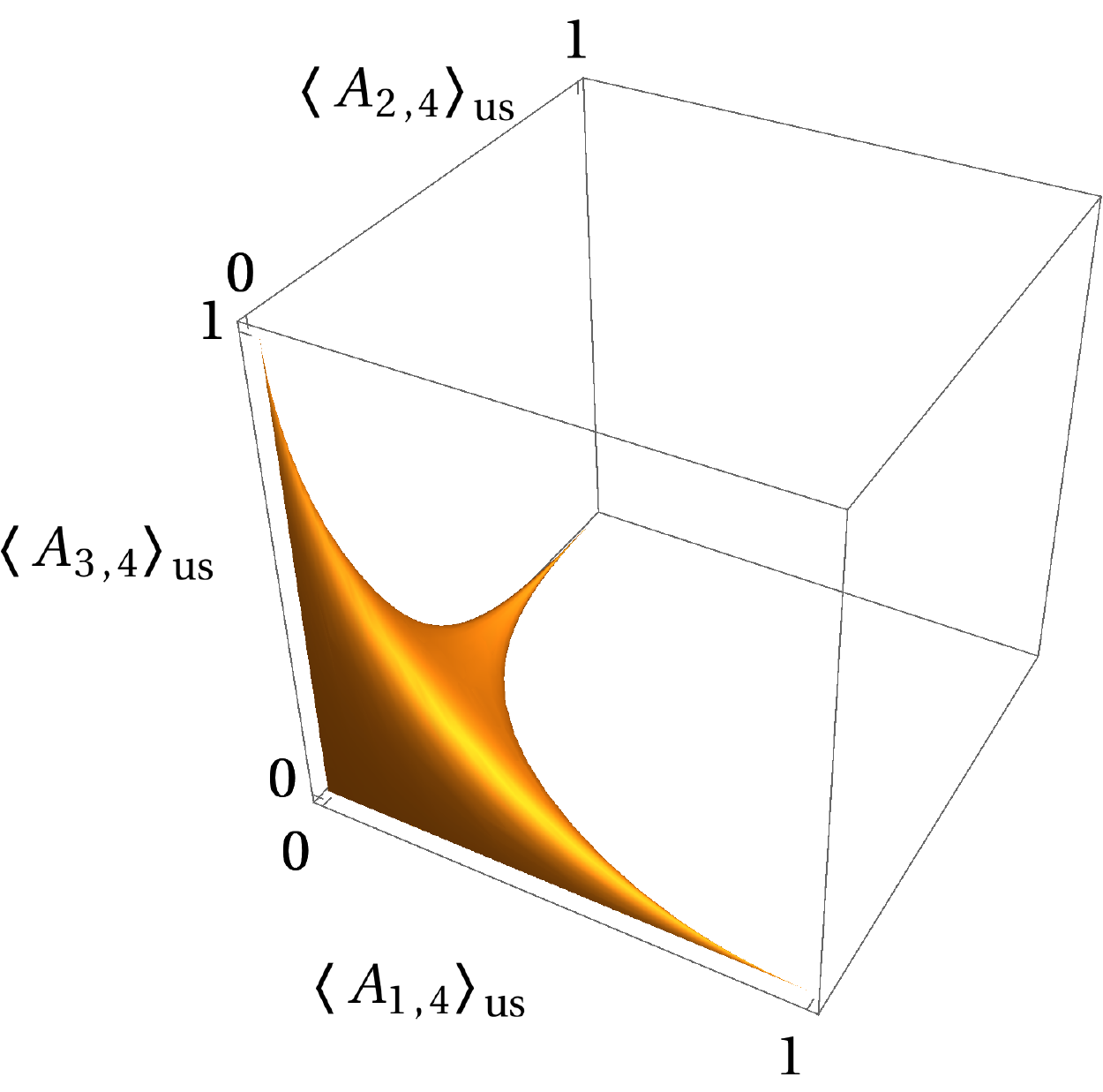}}
	\caption{In top and bottom rows,
		the parametric plots of
		$\mathcal{J}_\gamma$ of \eqref{J-gamma-para}
		and 
		$\mathcal{A}_\gamma$ of \eqref{A-gamma-para}
		are displayed sequentially for ${\gamma=1,\cdots,4}$.
		The convex hull of $\mathcal{J}_\gamma$ and $\mathcal{A}_\gamma$
		are the numerical range of 
		${\big(
			\big(\tfrac{J_x}{j}\big)^\gamma,
			\big(\tfrac{J_y}{j}\big)^\gamma,
			\big(\tfrac{J_z}{j}\big)^\gamma
			\big)}$
		and
		${\big(
			\tfrac{A_{1,\gamma}}{a_{\textsc{m},\gamma}},
			\tfrac{A_{2,\gamma}}{a_{\textsc{m},\gamma}},
			\tfrac{A_{3,\gamma}}{a_{\textsc{m},\gamma}}
			\big)}$,
		respectively, in the limit $j\rightarrow\infty$.
		Here ${\langle J_x^\gamma\rangle_\textsc{us}}$
		and ${\langle A_{1,\gamma}\rangle_\textsc{us}}$
		represent
		${\langle \big(\tfrac{J_x}{j}\big)^\gamma\rangle}$
		and ${\langle \tfrac{A_{1,\gamma}}{a_{\textsc{m},\gamma}}\rangle}$,
		correspondingly.
	}
	\label{fig:Efor j=inf} 
\end{figure*}

For a finite $N$, one can realize
\begin{align}
\label{Jx-N}
&(2\,\textbf{J}_x)^\gamma=
\sum_{i_1\neq i_2\neq\cdots\neq i_\gamma}
\textsf{X}_{i_1}\textsf{X}_{i_2}\cdots\textsf{X}_{i_\gamma}
\ +\ \text{Rest}\,,
\nonumber\\
&\left\langle\big(\tfrac{\textbf{J}_x}{\frac{N}{2}}\big)^\gamma\right\rangle_{\rho^{\otimes N}}
=
\tfrac{1}{N^\gamma}\tfrac{N!}{(N-\gamma)!}\,
\textsf{x}^\gamma
\ +\ \big\langle\tfrac{\text{Rest}}{N^\gamma}\big\rangle_{\rho^{\otimes N}}\,,
\end{align}
where $i_1,\cdots,i_\gamma$ are qubits' indices and
${\textsf{x}=\langle\textsf{X}\rangle_\rho}$ [see \eqref{qubit-state}].
Then we take the limit and obtain
\begin{align}
\label{J-N-Inf}
&
\lim\limits_{N\rightarrow\infty}
\left\langle\big(\tfrac{\textbf{J}_x}{\frac{N}{2}}\big)^\gamma\right\rangle_{\rho^{\otimes N}}
=
\textsf{x}^\gamma\,,\qquad\mbox{likewise}\qquad\nonumber\\
&\lim\limits_{N\rightarrow\infty}
\left\langle\big(\tfrac{\textbf{J}_y}{\frac{N}{2}}\big)^\gamma\right\rangle_{\rho^{\otimes N}}
=
\textsf{y}^\gamma\qquad\mbox{and}\qquad\\
&\lim\limits_{N\rightarrow\infty}
\left\langle\big(\tfrac{\textbf{J}_z}{\frac{N}{2}}\big)^\gamma\right\rangle_{\rho^{\otimes N}}
=
\textsf{z}^\gamma\,.\nonumber
\end{align}
In Fig.~\ref{fig:Efor j=inf}, we plot 
\begin{align}
\label{J-gamma-para}
\mathcal{J}_\gamma&=
\{\,(\textsf{x}^\gamma,\textsf{y}^\gamma,\textsf{z}^\gamma)\;| \
0\leq\mu\leq\pi
\ \mbox{and}\
0\leq\nu<2\pi
\}
\nonumber\\
&=\{\,(\texttt{x},\texttt{y},\texttt{z})\in\mathcal{H}^\textsc{us}\;| \
\texttt{x}^{\frac{2}{\gamma}}+\texttt{y}^{\frac{2}{\gamma}}+\texttt{z}^{\frac{2}{\gamma}}=1
\}
\end{align}
for ${\gamma=1,\cdots,4}$.
The hyperrectangle $\mathcal{H}^\textsc{us}$ in \eqref{J-gamma-para}
is ${[0,1]^{\times 3}}$ and ${[-1,1]^{\times 3}}$ for an even and odd $\gamma$, and 
${\texttt{x}=\textsf{x}^\gamma}$ and so on.
Basically, $\mathcal{J}_\gamma$ is an image of the Bloch sphere, which is
identified with ${\textsf{x}^2+\textsf{y}^2+\textsf{z}^2=1}$.
In the case of ${\gamma>4}$, the shape of $\mathcal{J}_\gamma$ is same as $\mathcal{J}_3$ and $\mathcal{J}_4$
for an odd and even $\gamma$, respectively.

\begin{figure}
	\centering
	\subfloat{\includegraphics[width=40mm]{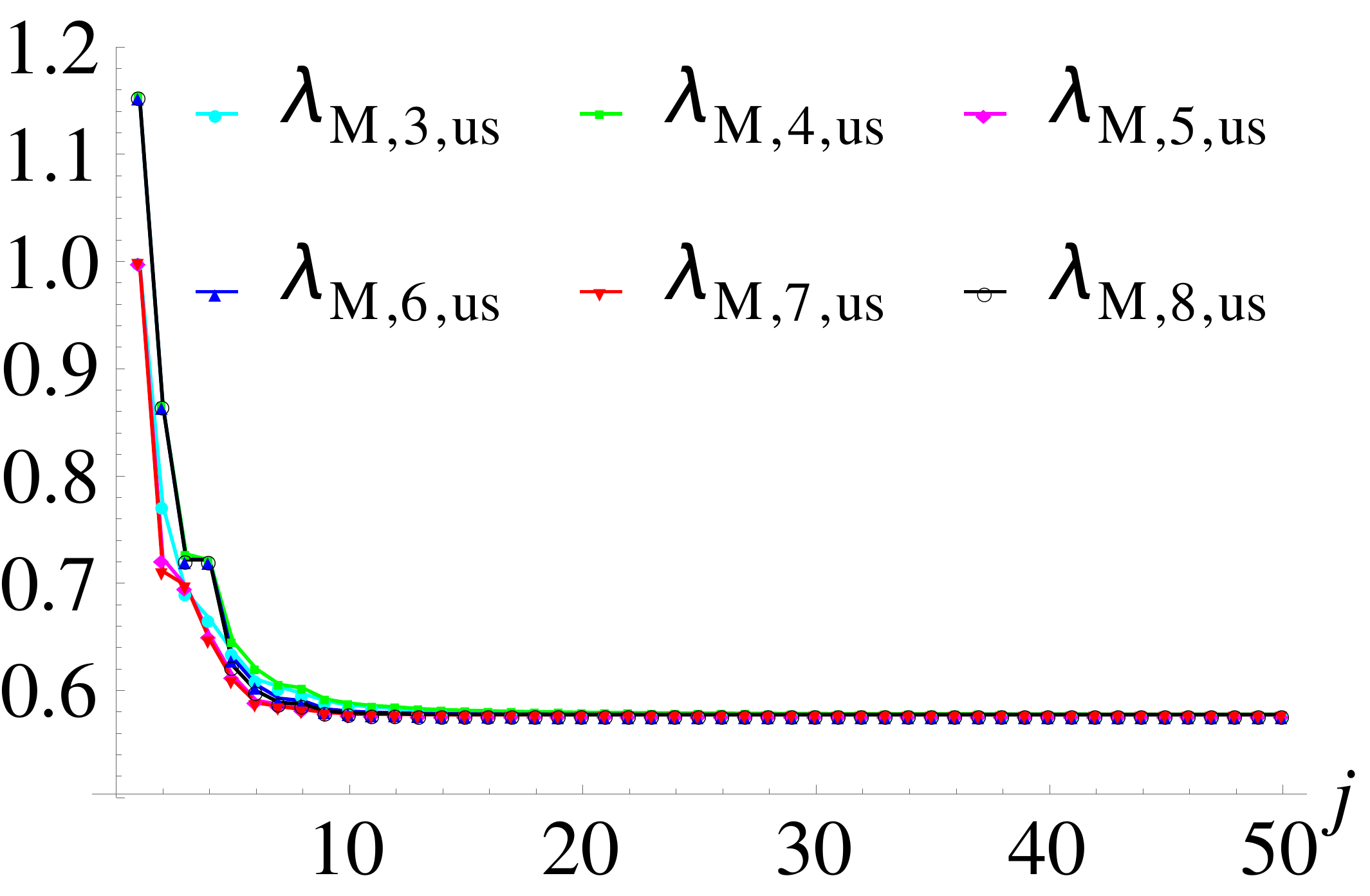}}\quad\
	\subfloat{\includegraphics[width=40mm]{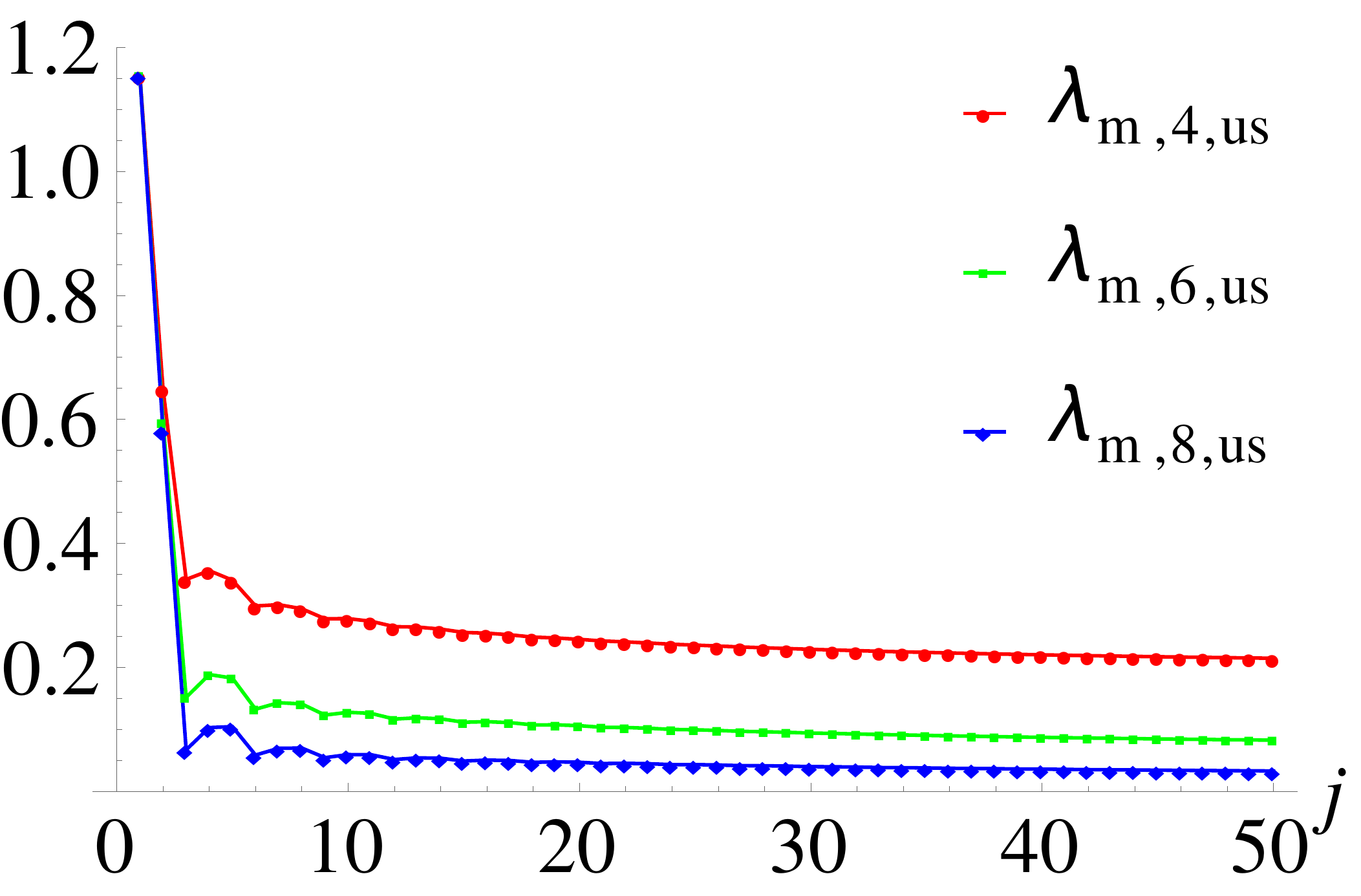}}
	\caption{
	On the left-hand-side, we present 	
	list-line-plots of $\{\lambda_{\textsc{m},\gamma,\textsc{us}}\}_{j=1}^{50}$
	for ${\gamma=3,\cdots,8}$. $\lambda_{\textsc{m},\gamma,\textsc{us}}$ denotes
	$\tfrac{\lambda_{\textsc{m},\gamma}(\theta_1,\phi_1)}{j^{\gamma}}$, where
	$\lambda_{\textsc{m},\gamma}(\theta_1,\phi_1)$
	is the maximum eigenvalue of ${\widehat{\eta}_1\cdot\vec{E}_\gamma}$.
	The unit vector $\widehat{\eta}_1$ and the vector operator ${\vec{E}_\gamma}$ are given in Table~\ref{tab:theta-phi-for(-a)} and \eqref{J-gamma}, correspondingly.
	On the right-hand-side, we show 	
	list-line-plots of $\{\lambda_{\text{m},\gamma,\textsc{us}}\}_{j=1}^{50}$
	for ${\gamma=4,6,8}$, where $\lambda_{\text{m},\gamma,\textsc{us}}=\tfrac{\lambda_{\text{m},\gamma}(\theta_1,\phi_1)}{j^{\gamma}}$, and
	$\lambda_{\text{m},\gamma}(\theta_1,\phi_1)$
	is the minimum eigenvalue of ${\widehat{\eta}_1\cdot\vec{E}_\gamma}$.
	}
	\label{fig:EV J-gamma j=inf} 
\end{figure}

Since $\boldsymbol{\Omega}_\textsc{be}$ is the convex hull of
pure product states as ${\tfrac{N}{2}=j\rightarrow\infty}$ [see \eqref{state-space-BE} and \eqref{de Finetti theorem}],
the numerical range $\mathcal{E}^\textsc{us}_\gamma$ of 
$\tfrac{1}{j^\gamma}\vec{E}_\gamma$ [${\vec{E}_\gamma}$ is given in \eqref{J-gamma}]
is the convex hull of $\mathcal{J}_\gamma$ in that limit.
For ${\gamma=1}$ and ${\gamma=2}$, 
$\mathcal{E}$ is the unit ball and the triangle, respectively, as reported
in Secs.~\ref{sec:spin-ops} and \ref{sec:Jx2 and Jy2}
as well as in \cite{Sehrawat17b,Chen17}.
For an odd and even ${\gamma>1}$,
the allowed region is
\begin{align}
\label{E-j^gamma-inf}
&
\mathcal{E}^\textsc{us}_\gamma
=\text{Conv}\{(\pm 1,0,0),(0,\pm 1,0),(0,0,\pm 1)\}
\quad \mbox{and}\nonumber\\
&\mathcal{E}^\textsc{us}_\gamma
=\{(\texttt{x},\texttt{y},\texttt{z})\in[0,1]^{\times 3}\,|\,
1\leq 
\texttt{x}^{\frac{2}{\gamma}}+
\texttt{y}^{\frac{2}{\gamma}}+
\texttt{z}^{\frac{2}{\gamma}}
\ \mbox{and}\qquad \nonumber\\
&\qquad\qquad\qquad\qquad\qquad\qquad
\texttt{x}+\texttt{y}+\texttt{z}\leq 1
\}\,,
\end{align}
respectively.
It means that, in the case of every odd power ${\gamma>1}$,
$\mathcal{E}^\textsc{us}_\gamma$ is the octahedron, and tight URs and CRs are listed in \eqref{H-UR-j^gamma-inf}--\eqref{umax-UR-j^gamma-inf}.
In CR \eqref{umax-UR-j^gamma-inf}, recall that 
$\mathfrak{u_{max}}$ approaches $2$ as ${j\rightarrow\infty}$.
While, in the case of an even power $\gamma$,
three corners of the hyperrectangle ${\mathcal{H}^\textsc{us}=[0,1]^{\times 3}}$ 
lie in the numerical range $\mathcal{E}^\textsc{us}_\gamma$, therefore we do not get a non-trivial UR or CR
for $\tfrac{1}{j^\gamma}\vec{E}_\gamma$ [for a justification, see the last two paragraphs
in Sec.~\ref{sec:all-region}].

Taking the unit vector 
$\widehat{\eta}_1$ and ${\vec{E}_\gamma}$ from Table~\ref{tab:theta-phi-for(-a)}
and \eqref{J-gamma}, respectively, we find the extreme eigenvalues
of the operator ${\widehat{\eta}_1\cdot\vec{E}_\gamma}$ for
${j=1,\cdots,50}$ and $\gamma=3,\cdots,8$. 
With the largest $\lambda_{\textsc{m},\gamma}(\theta_1,\phi_1)$
and smallest $\lambda_{\text{m},\gamma}(\theta_1,\phi_1)$
eigenvalues 
one can draw two parallel supporting hyperplanes~\eqref{hyperplane} of $\mathcal{E}^\textsc{us}_\gamma$ whose outward normal vectors are $\pm\widehat{\eta}_1$.
We plot the eigenvalues in Fig.~\ref{fig:EV J-gamma j=inf}.
Since $\lambda_{\text{m},\gamma}=-\lambda_{\textsc{m},\gamma}$ for an odd power $\gamma$,
we present $\lambda_{\text{m},\gamma}$ for even $\gamma$'s.
The plots suggest
\begin{align}
\label{J^gamma lambda/j inf}
&\lim\limits_{j\rightarrow\infty}
\tfrac{\lambda_{\textsc{m},\gamma}(\theta_1,\phi_1)}{j^{\gamma}}
=\tfrac{1}{\sqrt{3}}\quad\mbox{for}\quad \gamma>1
\quad\mbox{and}
\nonumber\\
&\lim\limits_{j\rightarrow\infty}
\tfrac{\lambda_{\text{m},\gamma}(\theta_1,\phi_1)}{j^{\gamma}}
=\tfrac{1}{\sqrt{3}}\,\tfrac{1}{3^{\frac{\gamma}{2}-1}}
\quad\mbox{for an even}\ \gamma>1\,,
\end{align}
which agree with what is achieved by applying the quantum de Finetti theorem on a $N$-qubit system.

Now we move to our second example:
the anticommutators from \eqref{A-gamma} and
\eqref{A-N}.
Like \eqref{Jx-N} and \eqref{J-N-Inf}, first we acquire
\begin{align}
\label{A1-N}
2^{2\gamma}\textbf{A}_{1,\gamma}&=
\sum_{i_1\neq\cdots\neq k_\gamma}
\big(
\textsf{X}_{i_1}\cdots\textsf{X}_{i_\gamma}\textsf{Z}_{k_1}\cdots\textsf{Z}_{k_\gamma}+
\nonumber\\
&\hspace{5mm}
\textsf{Z}_{k_1}\cdots\textsf{Z}_{k_\gamma}\textsf{X}_{i_1}\cdots\textsf{X}_{i_\gamma}
\big)
\ +\ \text{Rest}\,,
\\
\Big\langle
\tfrac{\textbf{A}_{1,\gamma}}{\big(\frac{N}{2}\big)^{2\gamma}}
\Big\rangle_{\rho^{\otimes N}}
&=
2\tfrac{1}{N^{2\gamma}}\tfrac{N!}{(N-2\gamma)!}\,
\textsf{x}^\gamma\textsf{z}^\gamma
\ +\ \big\langle\tfrac{\text{Rest}}{N^{2\gamma}}\big\rangle_{\rho^{\otimes N}}\,,
\hspace{2cm}
\nonumber
\end{align}
and then we apply the limit
\begin{align}
\label{A-N-Inf}
&
\lim\limits_{N\rightarrow\infty}
\Big\langle
\tfrac{\textbf{A}_{1,\gamma}}{\big(\frac{N}{2}\big)^{2\gamma}}
\Big\rangle_{\rho^{\otimes N}}
=
2\,(\textsf{x}\,\textsf{z})^\gamma\,,\quad\mbox{similarly}\qquad\nonumber\\
&
\lim\limits_{N\rightarrow\infty}
\Big\langle
\tfrac{\textbf{A}_{2,\gamma}}{\big(\frac{N}{2}\big)^{2\gamma}}
\Big\rangle_{\rho^{\otimes N}}
=
2\,(\textsf{y}\,\textsf{z})^\gamma\quad\mbox{as well as}
\\
&
\lim\limits_{N\rightarrow\infty}
\Big\langle
\tfrac{\textbf{A}_{3,\gamma}}{\big(\frac{N}{2}\big)^{2\gamma}}
\Big\rangle_{\rho^{\otimes N}}
=
2\,(\textsf{x}\,\textsf{y})^\gamma\,.
\nonumber
\end{align}
The numerical range of
${\tfrac{1}{j^{2\gamma}}\vec{E}_\gamma}$, where $\vec{E}_\gamma$ is from \eqref{A-gamma}, 
is the convex hull of points 
${2((\textsf{x}\,\textsf{z})^\gamma,(\textsf{y}\,\textsf{z})^\gamma,(\textsf{x}\,\textsf{y})^\gamma)}$.
However, we want the range 
$\mathcal{E}_\gamma^\textsc{us}$ of
${\tfrac{1}{a_{\textsc{m},\gamma}}\vec{E}_\gamma}$, where
$a_{\textsc{m},\gamma}$ is the maximum eigenvalue 
of the (unitary equivalent) anticommutators $A_{t,\gamma}$ and $t=1,2,3.$

\begin{figure}
	\centering
	\subfloat{\includegraphics[width=40mm]{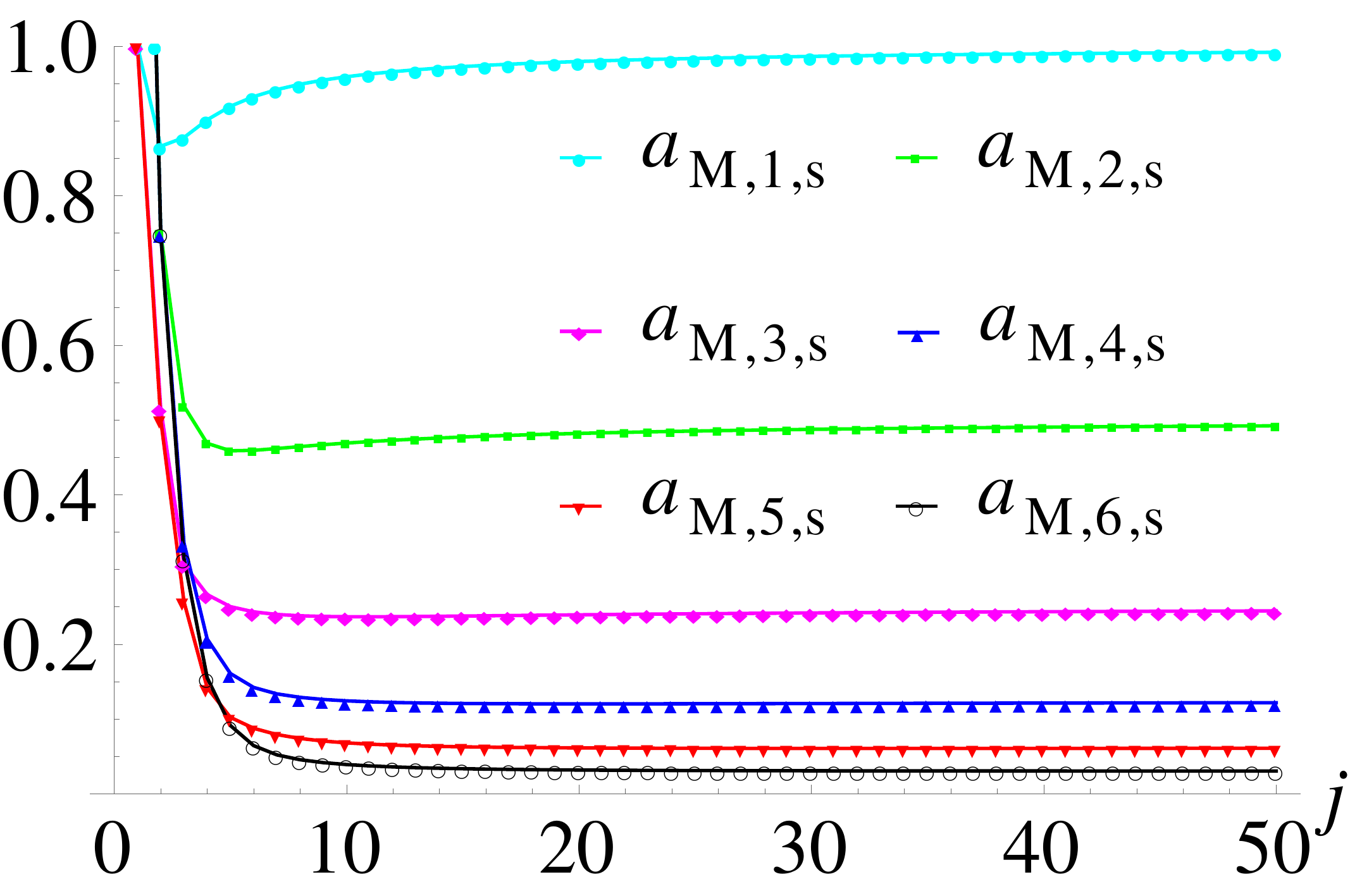}}\quad\
	\subfloat{\includegraphics[width=40mm]{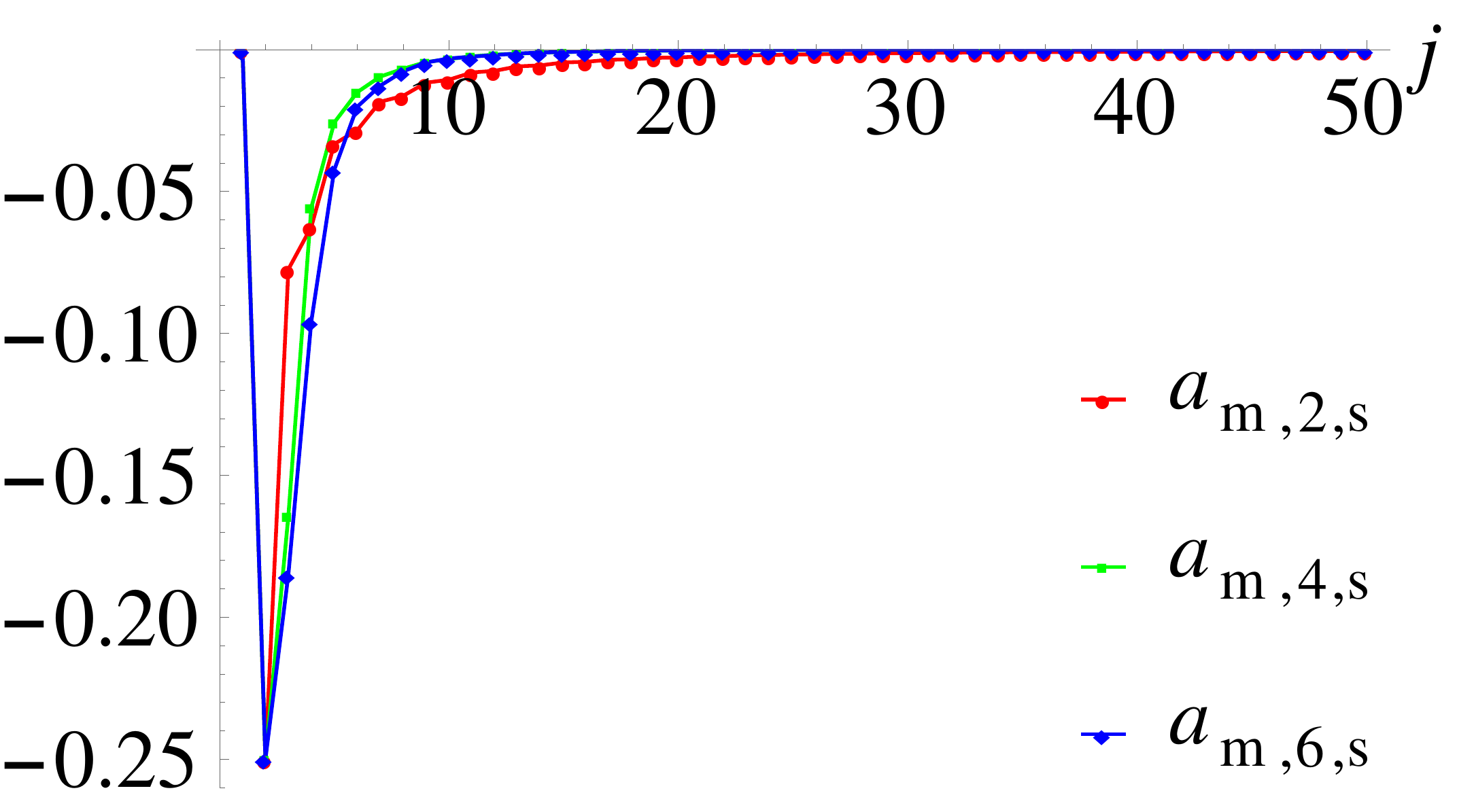}}
	\\\vspace{1mm}
	\subfloat{\includegraphics[width=40mm]{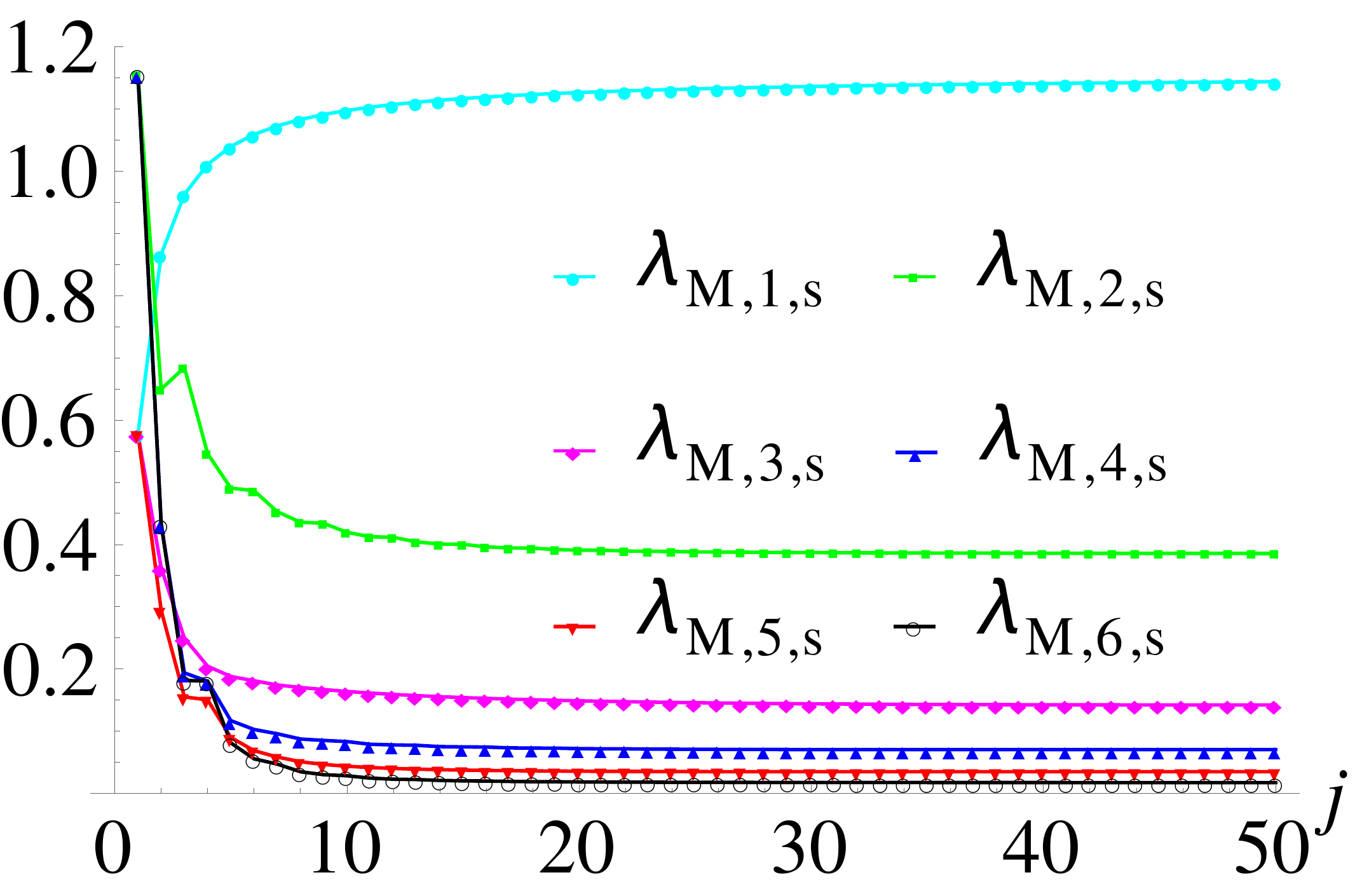}}\quad\
	\subfloat{\includegraphics[width=40mm]{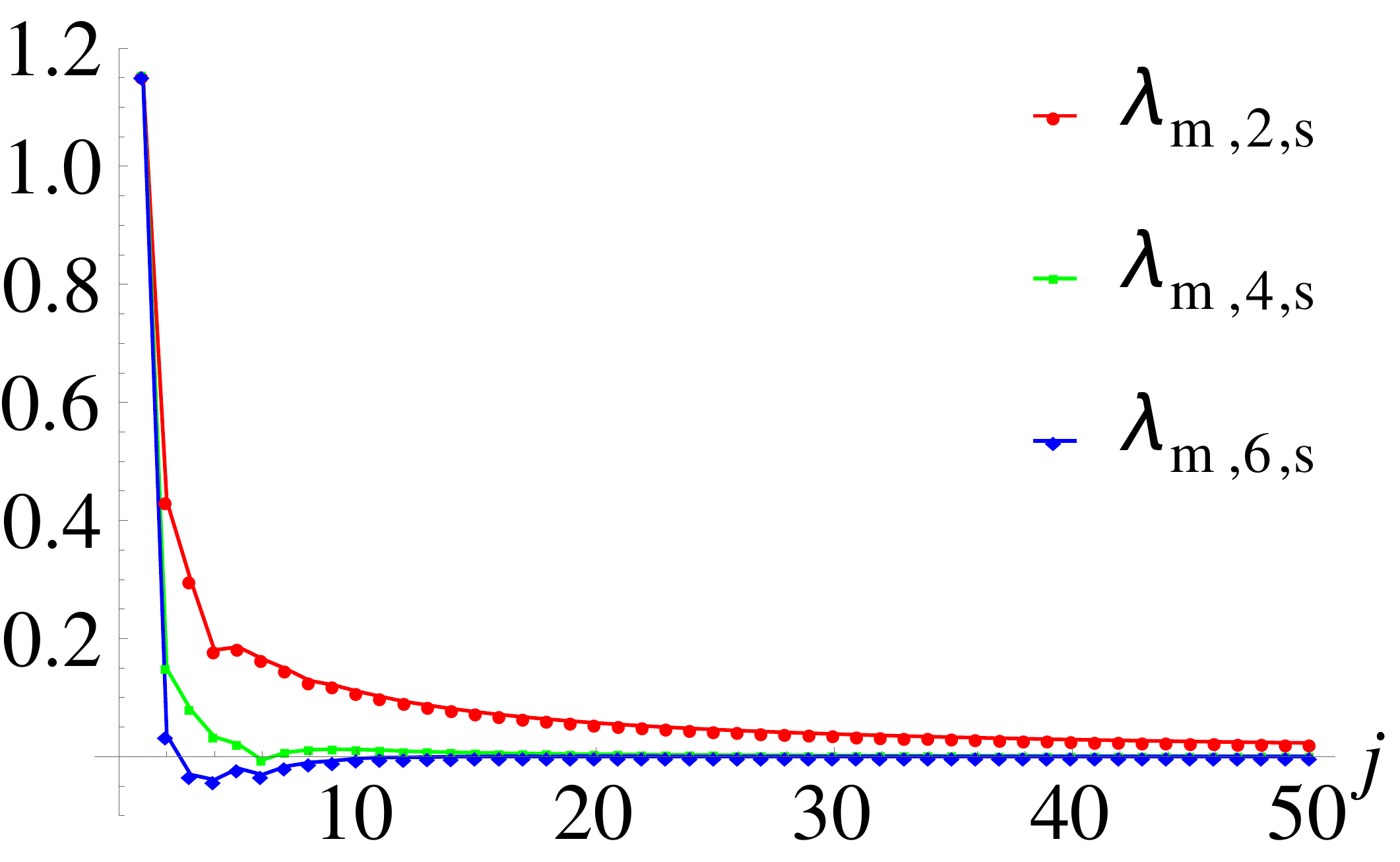}}
	\caption{
		We numerically compute 
		$a_{\textsc{m},\gamma,\text{s}}$,	
		$a_{\text{m},\gamma,\text{s}}$,
		$\lambda_{\textsc{m},\gamma,\text{s}}$, and
		$\lambda_{\text{m},\gamma,\text{s}}$
		for ${j=1,2,\cdots,50}$ and ${\gamma=1,\cdots,6}$, and 
		show their list-line-plots here.
		$a_{\textsc{m},\gamma,\text{s}}$ and	
		$a_{\text{m},\gamma,\text{s}}$ symbolize
		$\tfrac{a_{\textsc{m},\gamma}}{j^{2\gamma}}$ and $\tfrac{a_{\text{m},\gamma}}{j^{2\gamma}}$, respectively, where 
		$a_{\textsc{m},\gamma}$ and $a_{\text{m},\gamma}$ are the maximum and minimum characteristic values of the anticommutators in \eqref{A-gamma}.
		Likewise, 
		$\lambda_{\textsc{m},\gamma,\text{s}}$, and
		$\lambda_{\text{m},\gamma,\text{s}}$ express
		$\tfrac{\lambda_{\textsc{m},\gamma}(\theta_1,\phi_1)}{j^{2\gamma}}$ and
		$\tfrac{\lambda_{\text{m},\gamma}(\theta_1,\phi_1)}{j^{2\gamma}}$,
		respectively, where 
		${\lambda_{\textsc{m},\gamma}(\theta_1,\phi_1)}$ and ${\lambda_{\text{m},\gamma}(\theta_1,\phi_1)}$
		are the largest and smallest eigenvalues of the operator
		${\widehat{\eta}_1\cdot\vec{E}_\gamma}$.
		The unit vector ${\widehat{\eta}_1}$
		and the vector operator ${\vec{E}_\gamma}$ are registered in
		Table~\ref{tab:theta-phi-for(-a)} and \eqref{A-gamma}, correspondingly.
	}
	\label{fig:EV anti j=inf} 
\end{figure}

In Fig.~\ref{fig:EV anti j=inf}, we display
$\tfrac{a_{\textsc{m},\gamma}}{j^{2\gamma}}$
as a function of $j$ for ${\gamma=1,\cdots,6}$,
which indicates that
\begin{equation}
\label{aM/j inf}
\lim\limits_{j\rightarrow\infty}\tfrac{a_{\textsc{m},\gamma}}{j^{2\gamma}}
=\tfrac{1}{2^{\gamma-1}}\,.
\end{equation}
Hence, in the limit ${j\rightarrow\infty}$, the numerical range $\mathcal{E}_\gamma^\textsc{us}$ of
${\tfrac{1}{a_{\textsc{m},\gamma}}\vec{E}_\gamma}$
is the convex hull of
\begin{align}
\label{A-gamma-para}
\mathcal{A}_\gamma&=
\{
(
(2\,\textsf{x}\,\textsf{z})^\gamma,
(2\,\textsf{y}\,\textsf{z})^\gamma,
(2\,\textsf{x}\,\textsf{y})^\gamma)| 
0\leq\mu\leq\pi,
0\leq\nu<2\pi
\}\nonumber\\
&=
\{
(\texttt{a}_1,\texttt{a}_2,\texttt{a}_3)\in\mathcal{H}^\textsc{us}\,|\, 
(\texttt{a}_1\texttt{a}_2)^{\frac{2}{\gamma}}+
(\texttt{a}_2\texttt{a}_3)^{\frac{2}{\gamma}}+
(\texttt{a}_3\texttt{a}_1)^{\frac{2}{\gamma}}
\nonumber\\
&\hspace{3cm}
=
2\,(\texttt{a}_1\texttt{a}_2\texttt{a}_3)^{\frac{1}{\gamma}}
\}\,,
\end{align}
where the hyperrectangle $\mathcal{H}^\textsc{us}$ is ${[0,1]^{\times 3}}$ and ${[-1,1]^{\times 3}}$
for an even and odd $\gamma$.
The equation in second expression of $\mathcal{A}_\gamma$
comes from the normalization condition, ${\vec{\textsf{r}}\cdot\vec{\textsf{r}}=1}$, of the Bloch vector in \eqref{qubit-state}.
In Fig.~\ref{fig:Efor j=inf}, we exhibit $\mathcal{A}_\gamma$
for ${\gamma=1,\cdots,4}$.

One can perceive that, for $\gamma=1$,
$\mathcal{A}_\gamma$
is Steiner's Roman surface characterized by Eq.~\eqref{Roman j=inf}.
In this case, tight URs and CRs will be \eqref{H-UR-A (1)}--\eqref{umax-UR-A (1)}
[see \eqref{point at nl-A}].
Like the previous example, in the case of an even power $\gamma$,
we do not get a non-trivial UR or CR because 
four corners of the hyperrectangle ${\mathcal{H}^\textsc{us}=[0,1]^{\times 3}}$ 
fall in the range $\mathcal{E}^\textsc{us}_\gamma$.
Next one can check that, for $\gamma>4$, the shape of $\mathcal{A}_\gamma$
is similar to $\mathcal{A}_3$ and $\mathcal{A}_4$
in the case of an odd and even $\gamma$, respectively.
Hence, the permitted region is the octahedron and tetrahedron,
\begin{align}
\label{E-A^gamma-inf}
&\mathcal{E}^\textsc{us}_\gamma
=\text{Conv}\{(\pm 1,0,0),(0,\pm 1,0),(0,0,\pm 1)\}
\ \mbox{and}\qquad \nonumber\\
&
\mathcal{E}^\textsc{us}_\gamma
=\text{Conv}\{(0,0,0), (1,0,0),(0,1,0),(0,0,1)\}\,,
\end{align}
for an odd and even ${\gamma \geq 3}$, respectively.
In the case of the octahedron, one will have the same
tight URs and CRs \eqref{H-UR-j^gamma-inf}--\eqref{umax-UR-j^gamma-inf}
for the anticommutators.

Figure~\ref{fig:EV anti j=inf} also carries the plots of
$\tfrac{a_{\text{m},\gamma}}{j^{2\gamma}}$,
$\tfrac{\lambda_{\textsc{m},\gamma}(\theta_1,\phi_1)}{j^{2\gamma}}$,
and $\tfrac{\lambda_{\text{m},\gamma}(\theta_1,\phi_1)}{j^{2\gamma}}$
as functions of $j=1,2,\cdots,50$ for different $\gamma$'s.
$a_{\text{m},\gamma}$ represents the minimum eigenvalue of the anticommutators
$A_{t,\gamma}$, and ${a_{\text{m},\gamma}=-a_{\textsc{m},\gamma}}$ when $\gamma$ is a positive odd number.
Similarly, for an odd $\gamma$, the minimum eigenvalue ${\lambda_{\text{m},\gamma}(\theta_1,\phi_1)}$
is same in magnitude but opposite in sign of 
the maximum eigenvalue
$\lambda_{\textsc{m},\gamma}(\theta_1,\phi_1)$
of the operator ${\widehat{\eta}_1\cdot\vec{E}_\gamma}$
[$\vec{E}_\gamma$ is given in \eqref{A-gamma}].
The plots in Fig.~\ref{fig:EV anti j=inf}
indicate that
\begin{align}
\label{am/j inf}
&\lim\limits_{j\rightarrow\infty}\tfrac{a_{\text{m},\gamma}}{j^{2\gamma}}
=0 \quad\mbox{for an even}\ \gamma>1\,,
\nonumber\\
&\lim\limits_{j\rightarrow\infty}
\tfrac{\lambda_{\textsc{m},\gamma}(\theta_1,\phi_1)}{j^{2\gamma}}
=\begin{cases}
\tfrac{2}{\sqrt{3}} & \mbox{ for } \gamma=1 \\
\tfrac{2}{3\sqrt{3}} & \mbox{ for } \gamma=2 \\
\tfrac{1}{2^{\gamma-1}\sqrt{3}} & \mbox{ for } \gamma\geq 3 
\end{cases}\,,\quad\mbox{and}\\
&\lim\limits_{j\rightarrow\infty}
\tfrac{\lambda_{\text{m},\gamma}(\theta_1,\phi_1)}{j^{2\gamma}}
=0\quad\mbox{for an even}\ \gamma>1\,,
\nonumber
\end{align}
which supports the above results secured via the quantum de Finetti theorem.

%===========================================
\section{Summary and outlook}\label{sec:conclusion}

In this paper, we studied three kinds of Hermitian operators:
the combinations ${(X_\gamma,Y_\gamma)}$ of powers of the ladder operators,
powers of the angular momentum operators ${(J^\gamma_x,J^\gamma_y,J^\gamma_z)}$,
and their anticommutator
${(A_{1,\gamma},A_{2,\gamma},A_{3,\gamma})}$.
In each case, we presented the joint numerical range $\mathcal{E}$ and tight lower and upper bounds for URs and CRs, respectively.
Essentially, all the main results are displayed in 
Figs.~\ref{fig:regions}--\ref{fig:EV anti j=inf}.

Boundary $\partial\mathcal{E}$ of the allowed region is entirely generated by
the maximum-eigenvalue-kets of $\Lambda(\theta,\phi)$.
In simple cases,
where the angular momentum quantum number $j$ is small, we provided
analytical expressions of its maximum eigenvalue $\lambda_\textsc{m}(\theta,\phi)$,
the associated eigenkets ${|\theta,\phi\rangle}$,
and the boundary $\partial\mathcal{E}$.
For large quantum numbers, we obtained these numerically.
Up to ${j=50}$, the bounds for tight URs and CRs are also gained numerically
by exploiting a finite set of boundary points.
The limiting case ${j\rightarrow\infty}$ is handled by applying
the quantum de Finetti theorem on a $N$-qubit system, where 
${N\rightarrow\infty}$, and
the allowed regions as well as tight URs and CRs are achieved.

In the case of ${(J_x,J_y,J_z)}$, ${(X_\gamma,Y_\gamma)}$,
and for a pair of the anticommutators ${(A_t,A_{t'\neq t})}$, recall that $A_t=A_{t,\gamma=1}$, the allowed region is bounded by
a sphere (circle, in the case of two operators) centered at the origin for all the quantum numbers $j$.
There lower and upper bounds in the tight URs and CRs do not change with $j$.
In the case of ${(J^2_x,J^2_y)}$, the numerical range changes its shape from triangular to elliptical to triangular as $j$ goes from 1 to infinity.
In this case, the lower bounds decrease and upper bounds increase as $j$ grows, and they reach their trivial values in the limit ${j\rightarrow\infty}$.

In the case of anticommutators, after the uniform scaling, 
the allowed region $\mathcal{E}^\text{us}$ for ${\tfrac{1}{a_\textsc{m}}(A_1,A_2,A_3)}$
is the convex hull of a Roman surface for both $j=1$ and ${j\rightarrow\infty}$.
Whereas, for ${j=\tfrac{3}{2},2}$, $\mathcal{E}^\text{us}$ is a unit ball centered at $(0,0,0)$, then
it starts contracting in four directions and expanding in their antipodal directions as 
$j$ increases.
One of our tight URs for ${(A_1,A_2,A_3)}$ does not change with $j$, whereas the other grows weaker because its
lower bound decreases with $j$ but it never becomes trivial even in the limit
${j\rightarrow\infty}$.
Likewise, our tight CRs for the anticommutators grow weaker with the expansion of $\mathcal{E}^\text{us}$ but they stay nontrivial for all $j$s.

In the case of odd powers ${\gamma>1}$, the numerical range $\mathcal{E}^\text{us}$ of
${\tfrac{1}{j^\gamma}(J^\gamma_x,J^\gamma_y,J^\gamma_z)}$ shrinks from a ball
to an octahedron as the quantum number $j$ rises.
With the rise of $j$, some of our tight URs and CRs stay as they are, whereas one CR becomes stronger because of its upper bound decreases with the contraction of $\mathcal{E}^\text{us}$.

In the case of an even power ${\gamma>2}$, particularly for ${(J^4_x,J^4_y,J^4_z)}$, 
we discovered that distinct eigenvalues of ${\Lambda(\theta,\phi)}$ cross each other 
and dominate in different parts of the parameter space.
As a result of this level crossing and the disappearance of the gap between eigenvalues,
the largest eigenvalue $\lambda_\textsc{m}(\theta,\phi)$ of $\Lambda$ turns degenerate in different sections of the parameter space, and thus we observe flat faces on the boundary of the allowed region.
The degeneracy is a necessary but not sufficient requirement for $\mathcal{E}$ to has a flat face.
For small quantum numbers, it is difficult to predict the shape of $\mathcal{E}$ for ${(J^4_x,J^4_y,J^4_z)}$, but around ${j=10}$ the allowed region roughly takes the shape that is suggested by the quantum de Finetti theorem for ${j\rightarrow\infty}$.
For ${(J^4_x,J^4_y,J^4_z)}$,
our tight URs and CRs turn weaker as $j$ grows, and they become trivial in the limit ${j\rightarrow\infty}$.

With the de Finetti theorem, when ${\gamma>2}$ and ${j\rightarrow\infty}$, we realized that
the numerical range of
${\tfrac{1}{j^\gamma}(J^\gamma_x,J^\gamma_y,J^\gamma_z)}$
is an octahedron for an odd $\gamma$
and is bounded by
one curved and four plane surfaces for an even $\gamma$, and the numerical range of
${\tfrac{1}{a_{\textsc{m},\gamma}}
	(A_{1,\gamma},A_{2,\gamma},A_{3,\gamma})}$
is an octahedron and tetrahedron for an odd and even $\gamma$.

Recently, quantum phase transitions are explored through 
the joint numerical range of certain observables
in \cite{Chen15,Chen16,Zauner16,Chen17,Chen17b,Szymanski17}.
Results from this paper may become useful for such an investigation
as operators, for example, $\textbf{J}^\gamma_x$ and
$\textbf{A}_{1,\gamma}$ represent $\gamma$- and ${2\gamma}$-body interactions between qubits.

%===========================================
\begin{acknowledgments} 
	
I am very grateful to Aditi Sen(De), Ujjwal Sen, and Arun Kumar Pati for many helpful discussions.

\end{acknowledgments}

%===========================================
%===========================================
\appendix*

\section{Supplementary material}\label{sec:Sup-material}

\begin{table*}[]
	\centering
	\caption{For ${\gamma=2}$, the
		eigenkets ${|\phi\rangle_k\in\mathscr{S}^\gamma_k}$ 
		are presented below. They correspond to the maximum eigenvalue
		$\lambda_\textsc{m}$ [listed in Table~\ref{tab:lmax}] of ${\Lambda_2(\phi)}$.
		Note that, for a given $j$, an invariant space $\mathscr{S}^2$ of ${\Lambda_2(\phi)}$ belongs to the Hilbert space $\mathscr{H}_d$, where $d=2j+1$. Therefore, for example, $\mathscr{S}^2_0$ for distinct $j$-values are distinct subspaces.
	}
	\label{tab:lmax-kets gamma=2}
	\begin{tabular}{c@{\hspace{2mm}} | @{\hspace{2mm}}c}
		\hline\hline\rule{0pt}{2ex}  
		$j$ & Eigenkets  \\
		\hline\rule{0pt}{3ex} 
		%----------------------------------------------------------				
		$\tfrac{1}{2}\;$ 
		& All kets in the Hilbert space $\mathscr{H}_2$  \\[3mm]
		
		%----------------------------------------------------------	
		$1$              
		& ${|\phi\rangle_0=\tfrac{1}{\sqrt{2}}
			\left(e^{\text{i}\phi}|{+1}\rangle +\,|{-1}\rangle\right)
			\in\mathscr{S}^2_0\subset\mathscr{H}_3}$
		\\[3mm]
		
		%----------------------------------------------------------	
		
		\multirow{2}{*}{$\tfrac{3}{2}$}   
		& ${|\phi\rangle_0=\tfrac{1}{\sqrt{2}}
			\left(e^{\text{i}\phi}\big|{+\tfrac{3}{2}}\big\rangle +\,\big|{-\tfrac{1}{2}}\big\rangle\right)
			\in\mathscr{S}^2_0\subset\mathscr{H}_4}$ \\[1mm]
		& ${|\phi\rangle_1=\tfrac{1}{\sqrt{2}} 
			\left(e^{\text{i}\phi}\big|{+\tfrac{1}{2}}\big\rangle +\,\big|{-\tfrac{3}{2}}\big\rangle\right)
			\in\mathscr{S}^2_1\subset\mathscr{H}_4}$ \\[3mm]
		
		%----------------------------------------------------------	
		
		$2$              
		&  ${|\phi\rangle_0=\tfrac{1}{2}
			\left(e^{\text{i}2\phi}|{+2}\rangle +\sqrt{2}\,e^{\text{i}\phi}|0\rangle +|{-2}\rangle\right)\in\mathscr{S}^2_0\subset\mathscr{H}_5}$
		\\[3mm]	
		
		%----------------------------------------------------------	
		
		\multirow{2}{*}{$\tfrac{5}{2}$}   
		& ${|\phi\rangle_0=\tfrac{1}{\sqrt{28}}
			\left(\sqrt{5}\,e^{\text{i}2\phi}\big|{+\tfrac{5}{2}}\big\rangle +\sqrt{14}\,e^{\text{i}\phi}\big|{+\tfrac{1}{2}}\big\rangle +3\,\big|{-\tfrac{3}{2}}\big\rangle\right)
			\in\mathscr{S}^2_0\subset\mathscr{H}_6}$ \\[1mm]
		& ${|\phi\rangle_1=\tfrac{1}{\sqrt{28}}
			\left(3\,e^{\text{i}2\phi}\big|{+\tfrac{3}{2}}\big\rangle +\sqrt{14}\,e^{\text{i}\phi}\big|{-\tfrac{1}{2}}\big\rangle +\sqrt{5}\,\big|{-\tfrac{5}{2}}\big\rangle\right)
			\in\mathscr{S}^2_1\subset\mathscr{H}_6}$ \\[3mm]
		
		%----------------------------------------------------------	
		
		$3$              
		& ${|\phi\rangle_0=\tfrac{1}{\sqrt{8(4+\sqrt{6})}}
			\left(\sqrt{5}\,e^{\text{i}3\phi}|{+3}\rangle
			+(2\sqrt{2}+\sqrt{3})e^{\text{i}2\phi}|{+1}\rangle  
			+(2\sqrt{2}+\sqrt{3})e^{\text{i}\phi}|{-1}\rangle  +\sqrt{5}\,|{-3}\rangle\right)\in\mathscr{S}^2_0\subset\mathscr{H}_7}$\\[4mm]	
		
		%----------------------------------------------------------	
		
		\multirow{2}{*}{$\tfrac{7}{2}$}   
		& ${|\phi\rangle_0=\tfrac{
				(6-\sqrt{21})\sqrt{21+4\sqrt{21}}\,e^{\text{i}3\phi}\big|{+\tfrac{7}{2}}\big\rangle
				+3(\sqrt{3}+2\sqrt{7})\,e^{\text{i}2\phi}\big|{+\tfrac{3}{2}}\big\rangle
				+\sqrt{15}\sqrt{21+4\sqrt{21}}\,e^{\text{i}\phi}\big|{-\tfrac{1}{2}}\big\rangle +15\,\big|{-\tfrac{5}{2}}\big\rangle}{\sqrt{72(14+\sqrt{21})}}
			\in\mathscr{S}^2_0\subset\mathscr{H}_8}$ \\[1mm]
		&${\quad\ |\phi\rangle_1=\tfrac{
				(14-\sqrt{21})\sqrt{21+4\sqrt{21}}\,e^{\text{i}3\phi}\big|{+\tfrac{5}{2}}\big\rangle
				+7\sqrt{5}(2\sqrt{3}+\sqrt{7})\,e^{\text{i}2\phi}\big|{+\tfrac{1}{2}}\big\rangle
				+5\sqrt{7}\sqrt{21+4\sqrt{21}}\,e^{\text{i}\phi}\big|{-\tfrac{3}{2}}\big\rangle +35\,\big|{-\tfrac{7}{2}}\big\rangle}{\sqrt{1960(6+\sqrt{21})}}
			\in\mathscr{S}^2_1\subset\mathscr{H}_8}$
		\\[4mm]
		
		%----------------------------------------------------------	
		
		$4$   
		& ${|\phi\rangle_0=\tfrac{1}{\sqrt{208}}
			\left(\sqrt{7}\,e^{\text{i}4\phi}|{+4}\rangle
			+2\sqrt{13}\,e^{\text{i}3\phi}|{+2}\rangle  
			+3\sqrt{10}\,e^{\text{i}2\phi}|{0}\rangle 
			+2\sqrt{13}\,e^{\text{i}\phi}|{-2}\rangle  +\sqrt{7}\,|{-4}\rangle\right)\in\mathscr{S}^2_0
			\subset\mathscr{H}_9}$
		\\[2mm]
		
		\hline\hline
	\end{tabular}
\end{table*}

%============================================================

\begin{table*}[]
	\centering
	\caption{Like Table~\ref{tab:lmax-kets gamma=2}, the maximum-eigenvalue-kets ${|\phi\rangle_k\in\mathscr{S}^\gamma_k}$ of ${\Lambda_\gamma(\phi)}$ are registered here for ${\gamma=3}$ and ${\gamma=4}$.  
	}
	\label{tab:lmax-kets gamma=3,4}
	\begin{tabular}{c@{\hspace{2mm}} | @{\hspace{2mm}}c@{\hspace{2mm}} | @{\hspace{2mm}}c}
		%----------------------------------------------------------		
		\hline\hline\rule{0pt}{2ex}  
		$j$ & Eigenkets for ${\gamma=3}$ & Eigenkets for ${\gamma=4}$ \\
		\hline\rule{0pt}{3ex}
		%----------------------------------------------------------		
		$\tfrac{1}{2}\;$ 
		& All kets in the Hilbert space $\mathscr{H}_2$
		& All kets in the Hilbert space $\mathscr{H}_2$  \\[3mm]
		
		%----------------------------------------------------------
		
		$1$ 
		& All kets in the Hilbert space $\mathscr{H}_3$             
		& All kets in the Hilbert space $\mathscr{H}_3$ \\[3mm]
		
		%----------------------------------------------------------
		
		$\tfrac{3}{2}$  
		& ${|\phi\rangle_0=\tfrac{1}{\sqrt{2}}
			\left(e^{\text{i}\phi}\big|{+\tfrac{3}{2}}\big\rangle +\,\big|{-\tfrac{3}{2}}\big\rangle\right)\in\mathscr{S}^3_0
			\subset\mathscr{H}_4}$ 
		& All kets in the Hilbert space $\mathscr{H}_4$  \\[4mm]
		
		%----------------------------------------------------------
		
		$2$            
		&  
		\begin{tabular}{ l }
			${|\phi\rangle_0=\tfrac{1}{\sqrt{2}}
				\left(e^{\text{i}\phi}|{+2}\rangle +|{-1}\rangle\right)
				\in\mathscr{S}^3_0\subset\mathscr{H}_5}$
			\\[1mm]	 
			${|\phi\rangle_1=\tfrac{1}{\sqrt{2}}
				\left(e^{\text{i}\phi}|{+1}\rangle +|{-2}\rangle\right)
				\in\mathscr{S}^3_1\subset\mathscr{H}_5}$\\
		\end{tabular}
		&
		${|\phi\rangle_0=\tfrac{1}{\sqrt{2}}
			\left(e^{\text{i}\phi}|{+2}\rangle +|{-2}\rangle\right)
			\in\mathscr{S}^4_0\subset\mathscr{H}_5}$ \\[4mm]
		
		%----------------------------------------------------------
		
		$\tfrac{5}{2}$ &
		${|\phi\rangle_1=\tfrac{1}{\sqrt{2}}\left(
			e^{\text{i}\phi}\big|{+\tfrac{3}{2}}\big\rangle +\big|{-\tfrac{3}{2}}\big\rangle
			\right)
			\in\mathscr{S}^3_1\subset\mathscr{H}_6}$
		&
		\begin{tabular}{ l }
			${|\phi\rangle_0=\tfrac{1}{\sqrt{2}}
				\left(e^{\text{i}\phi}\big|{+\tfrac{5}{2}}\big\rangle +
				\big|{-\tfrac{3}{2}}\big\rangle\right)\in\mathscr{S}^4_0\subset\mathscr{H}_6}$
			\\[1mm]	 
			${|\phi\rangle_1=\tfrac{1}{\sqrt{2}}
				\left(e^{\text{i}\phi}\big|{+\tfrac{3}{2}}\big\rangle +
				\big|{-\tfrac{5}{2}}\big\rangle\right)\in\mathscr{S}^4_1\subset\mathscr{H}_6}$\\
		\end{tabular}
		\\[4mm]
		
		%----------------------------------------------------------
		
		$3$             
		& 
		\begin{tabular}{ l }
			${|\phi\rangle_0=\tfrac{1}{2}
				\left(e^{\text{i}2\phi}|{+3}\rangle+\sqrt{2}\,e^{\text{i}\phi}|0\rangle  +|{-3}\rangle\right)\in\mathscr{S}^3_0\subset\mathscr{H}_7}$\\[1mm]
			${|\phi\rangle_1=\tfrac{1}{\sqrt{2}}
				\left(e^{\text{i}\phi}|{+2}\rangle+|{-1}\rangle\right)
				\in\mathscr{S}^3_1\subset\mathscr{H}_7}$\\[1mm]
			${|\phi\rangle_2=\tfrac{1}{\sqrt{2}}
				\left(e^{\text{i}\phi}|{+1}\rangle+|{-2}\rangle \right)
				\in\mathscr{S}^3_2\subset\mathscr{H}_7}$\\
		\end{tabular}
		&
		${|\phi\rangle_1=\tfrac{1}{\sqrt{2}}
			\left(e^{\text{i}\phi}|{+2}\rangle+|{-2}\rangle\right)
			\in\mathscr{S}^4_1\subset\mathscr{H}_7}$
		\\[8mm]
		
		%----------------------------------------------------------

		$\tfrac{7}{2}$
		&
		\begin{tabular}{ l }
			${|\phi\rangle_0=\tfrac{
					\sqrt{7}\,e^{\text{i}2\phi}\big|{+\tfrac{7}{2}}\big\rangle
					+\sqrt{23}\,e^{\text{i}\phi}\big|{+\tfrac{1}{2}}\big\rangle
					+4\,\big|{-\tfrac{5}{2}}\big\rangle}{\sqrt{46}}
				\in\mathscr{S}^3_0\subset\mathscr{H}_8}$ 
			\\[1mm]
			${|\phi\rangle_1=\tfrac{
					4\,e^{\text{i}2\phi}\big|{+\tfrac{5}{2}}\big\rangle
					+\sqrt{23}\,e^{\text{i}\phi}\big|{-\tfrac{1}{2}}\big\rangle
					+\sqrt{7}\,\big|{-\tfrac{7}{2}}\big\rangle}{\sqrt{46}}
				\in\mathscr{S}^3_1\subset\mathscr{H}_8}$
			\\
		\end{tabular}
		& 
		\begin{tabular}{ l }
			${|\phi\rangle_1=\tfrac{1}{\sqrt{2}}
				\left(e^{\text{i}\phi}\big|{+\tfrac{5}{2}}\big\rangle
				+\big|{-\tfrac{3}{2}}\big\rangle\right)\in\mathscr{S}^4_1\subset\mathscr{H}_8}$ 
			\\[1mm]
			${|\phi\rangle_2=\tfrac{1}{\sqrt{2}}
				\left(e^{\text{i}\phi}\big|{+\tfrac{3}{2}}\big\rangle
				+\big|{-\tfrac{5}{2}}\big\rangle\right)\in\mathscr{S}^4_2\subset\mathscr{H}_8}$
			\\
		\end{tabular}
		\\[8mm]

		%----------------------------------------------------------
		$4$   
		& ${|\phi\rangle_1=\tfrac{1}{2}
			\left(e^{\text{i}2\phi}|{+3}\rangle+\sqrt{2}e^{\text{i}\phi}|0\rangle  +|{-3}\rangle\right)\in\mathscr{S}^3_1\subset\mathscr{H}_9}$
		&
		${|\phi\rangle_2=\tfrac{1}{\sqrt{2}}
			\left(e^{\text{i}\phi}|{+2}\rangle+|{-2}\rangle\right)
			\in\mathscr{S}^4_2\subset\mathscr{H}_9}$
		\\[2mm]
		
		%----------------------------------------------------------
		
		\hline\hline
		
	\end{tabular}
\end{table*}

For ${j=\tfrac{3}{2},2,\tfrac{5}{2},\cdots,50,}$ the values of 
$\mathfrak{h}$ displayed in Fig.~\ref{fig:S2 UR bounds}
are
\\
\texttt{0.491551, 0.491551, 0.41986, 0.427261, 0.351636, 0.356853, 0.302929, 
	0.30647, 0.266717, 0.269349, 0.238819, 0.240789, 0.216628, 0.218167, 
	0.198523, 0.199764, 0.183432, 0.184472, 0.170624, 0.171533, 0.15963, 
	0.160408, 0.15008, 0.150755, 0.141701, 0.142292, 0.134283, 0.134806, 
	0.127667, 0.128132, 0.121724, 0.122142, 0.116355, 0.116733, 0.111478, 
	0.111821, 0.107026, 0.107339, 0.102945, 0.103232, 0.0991885, 
	0.0994531, 0.0957188, 0.0959634, 0.0925031, 0.0927299, 0.0895138, 
	0.0897248, 0.0867271, 0.0869239, 0.0841192, 0.0843066, 0.0816738, 
	0.0818534, 0.0793778, 0.0795465, 0.0772175, 0.0773764, 0.0751809, 
	0.0753308, 0.0732574, 0.0733991, 0.0714375, 0.0715717, 0.0697129, 
	0.0698402, 0.0680761, 0.068197, 0.0665204, 0.0666353, 0.0650396, 
	0.0651491, 0.0636285, 0.0637329, 0.062282, 0.0623817, 0.0609957, 
	0.061091, 0.0597655, 0.0598567, 0.0585877, 0.0586752, 0.0574591, 
	0.0575429, 0.0563764, 0.0564569, 0.0553368, 0.0554142, 0.0543379, 
	0.0544122, 0.053377, 0.0534486, 0.0524522, 0.0525211, 0.0515612, 
	0.0516277, 0.0507023, 0.0507664}.

For $j=\tfrac{3}{2},2,\tfrac{5}{2},\cdots,50,$ the values of 
$\mathfrak{u_{\sfrac{1}{2}}}$ displayed in Fig.~\ref{fig:S2 UR bounds}
are
\\
\texttt{2.36603, 2.36603, 2.32112, 2.32112, 2.28897, 2.28897, 2.26491, 
	2.26491, 2.2461, 2.2461, 2.23089, 2.23089, 2.21825, 2.21825, 2.20753, 
	2.20753, 2.19829, 2.19829, 2.19021, 2.19021, 2.18307, 2.18307, 
	2.1767, 2.1767, 2.17096, 2.17096, 2.16577, 2.16577, 2.16103, 2.16103, 
	2.15668, 2.15668, 2.15268, 2.15268, 2.14898, 2.14898, 2.14553, 
	2.14553, 2.14233, 2.14233, 2.13933, 2.13933, 2.13651, 2.13651, 
	2.13387, 2.13387, 2.13137, 2.13137, 2.12901, 2.12901, 2.12678, 
	2.12678, 2.12466, 2.12466, 2.12265, 2.12265, 2.12073, 2.12073, 
	2.1189, 2.1189, 2.11716, 2.11716, 2.11549, 2.11549, 2.11389, 2.11389, 
	2.11235, 2.11235, 2.11088, 2.11088, 2.10947, 2.10947, 2.10811, 
	2.10811, 2.1068, 2.1068, 2.10553, 2.10553, 2.10432, 2.10432, 2.10314, 
	2.10314, 2.102, 2.102, 2.1009, 2.1009, 2.09984, 2.09984, 2.09881, 
	2.09881, 2.09781, 2.09781, 2.09684, 2.09684, 2.0959, 2.0959, 2.09499, 
	2.09499}.

For $j=\tfrac{3}{2},2,\tfrac{5}{2},\cdots,50,$ the values of 
$\mathfrak{u_2}$ displayed in Fig.~\ref{fig:S2 UR bounds}
are
\\
\texttt{1.75, 1.75, 1.78071, 1.7736, 1.82246, 1.81637, 1.85267, 1.84851, 
	1.87446, 1.87161, 1.89076, 1.88872, 1.90336, 1.90183, 1.91337, 
	1.91219, 1.92153, 1.92058, 1.9283, 1.92752, 1.93399, 1.93334, 
	1.93885, 1.9383, 1.94304, 1.94257, 1.9467, 1.94629, 1.94992, 1.94956, 
	1.95277, 1.95245, 1.95531, 1.95503, 1.9576, 1.95735, 1.95966, 
	1.95944, 1.96153, 1.96133, 1.96324, 1.96305, 1.9648, 1.96463, 
	1.96623, 1.96608, 1.96755, 1.96741, 1.96878, 1.96864, 1.96991, 
	1.96979, 1.97096, 1.97085, 1.97195, 1.97184, 1.97287, 1.97277, 
	1.97373, 1.97363, 1.97453, 1.97445, 1.97529, 1.97521, 1.97601, 
	1.97593, 1.97668, 1.97661, 1.97732, 1.97725, 1.97792, 1.97786, 
	1.9785, 1.97844, 1.97904, 1.97898, 1.97956, 1.9795, 1.98005, 1.98, 
	1.98052, 1.98047, 1.98097, 1.98092, 1.98139, 1.98135, 1.9818, 
	1.98176, 1.98219, 1.98215, 1.98257, 1.98253, 1.98293, 1.98289, 
	1.98327, 1.98324, 1.9836, 1.98357}.

For $j=\tfrac{3}{2},2,\tfrac{5}{2},\cdots,50,$ the values of 
$\mathfrak{u_{max}}$ displayed in Fig.~\ref{fig:S2 UR bounds}
are
\\
\texttt{1.86603, 1.86603, 1.88192, 1.87766, 1.9052, 1.90139, 1.92211, 
	1.91953, 1.93416, 1.93244, 1.94306, 1.94186, 1.94988, 1.94899, 
	1.95525, 1.95458, 1.95959, 1.95906, 1.96317, 1.96274, 1.96616, 
	1.96581, 1.96871, 1.96841, 1.9709, 1.97065, 1.97281, 1.97259, 
	1.97448, 1.97429, 1.97596, 1.97579, 1.97728, 1.97713, 1.97846, 
	1.97832, 1.97952, 1.9794, 1.98049, 1.98038, 1.98136, 1.98127, 
	1.98216, 1.98208, 1.9829, 1.98282, 1.98358, 1.9835, 1.9842, 1.98413, 
	1.98478, 1.98472, 1.98532, 1.98526, 1.98582, 1.98577, 1.98629, 
	1.98624, 1.98673, 1.98668, 1.98714, 1.9871, 1.98753, 1.98749, 
	1.98789, 1.98786, 1.98824, 1.9882, 1.98856, 1.98853, 1.98887, 
	1.98884, 1.98916, 1.98913, 1.98944, 1.98941, 1.9897, 1.98967, 
	1.98995, 1.98992, 1.99019, 1.99016, 1.99042, 1.99039, 1.99063, 
	1.99061, 1.99084, 1.99082, 1.99104, 1.99102, 1.99123, 1.99121, 
	1.99141, 1.99139, 1.99158, 1.99156, 1.99175, 1.99173}.	
\\\vspace{0.0cm}

%-------------------------------------------------------------

For $j=\tfrac{3}{2},2,\tfrac{5}{2},\cdots,50,$ the values of 
$\tfrac{\langle\theta_1,\phi_1 |A_1|\theta_1,\phi_1 \rangle}{a_\textsc{m}}$ given in \eqref{mean value t,f A}
are
\\
\texttt{0.57735, 0.57735,
	0.629941, 0.632993, 0.644427, 0.64715, 0.650684, 
	0.652506, 0.654182, 0.65539, 0.656421, 0.65726, 0.657979, 0.658592, 
	0.659125, 0.659592, 0.660004, 0.660371, 0.660699, 0.660995, 0.661263, 
	0.661507, 0.66173, 0.661934, 0.662122, 0.662296, 0.662457, 0.662606, 
	0.662745, 0.662875, 0.662997, 0.663111, 0.663218, 0.663319, 0.663414, 
	0.663504, 0.663589, 0.66367, 0.663747, 0.663819, 0.663888, 0.663954, 
	0.664017, 0.664077, 0.664134, 0.664189, 0.664242, 0.664292, 0.66434, 
	0.664387, 0.664431, 0.664474, 0.664515, 0.664555, 0.664593, 0.66463, 
	0.664666, 0.6647, 0.664734, 0.664766, 0.664797, 0.664827, 0.664856, 
	0.664884, 0.664912, 0.664938, 0.664964, 0.664989, 0.665013, 0.665037, 
	0.66506, 0.665082, 0.665103, 0.665125, 0.665145, 0.665165, 0.665184, 
	0.665203, 0.665222, 0.66524, 0.665258, 0.665275, 0.665291, 0.665308, 
	0.665324, 0.665339, 0.665355, 0.665369, 0.665384, 0.665398, 0.665412, 
	0.665426, 0.665439, 0.665452, 0.665465, 0.665477, 0.66549, 0.665502}.

For $j=\tfrac{3}{2},2,\tfrac{5}{2},\cdots,50,$ the values of 
$\mathfrak{h}$ showcased in Fig.~\ref{fig:h,u2,umax for A}
are
\\
\texttt{1.38629, 1.38629, 1.38629, 1.38629, 1.38629, 1.38629, 1.38629,
	1.38533, 1.38141, 1.37857, 1.37614, 1.37415, 1.37246, 
	1.371, 1.36974, 1.36863, 1.36765, 1.36678, 1.36599, 1.36529, 1.36465, 
	1.36407, 1.36354, 1.36305, 1.3626, 1.36218, 1.3618, 1.36144, 1.36111, 
	1.3608, 1.36051, 1.36023, 1.35998, 1.35973, 1.35951, 1.35929, 
	1.35909, 1.35889, 1.35871, 1.35854, 1.35837, 1.35821, 1.35806, 
	1.35792, 1.35778, 1.35765, 1.35752, 1.3574, 1.35728, 1.35717, 
	1.35707, 1.35696, 1.35686, 1.35677, 1.35668, 1.35659, 1.3565, 
	1.35642, 1.35634, 1.35626, 1.35619, 1.35612, 1.35605, 1.35598, 
	1.35591, 1.35585, 1.35579, 1.35573, 1.35567, 1.35561, 1.35556, 
	1.3555, 1.35545, 1.3554, 1.35535, 1.3553, 1.35526, 1.35521, 1.35517, 
	1.35512, 1.35508, 1.35504, 1.355, 1.35496, 1.35492, 1.35488, 1.35485, 
	1.35481, 1.35478, 1.35474, 1.35471, 1.35468, 1.35464, 1.35461, 
	1.35458, 1.35455, 1.35452, 1.35449}.

For $j=\tfrac{3}{2},2,\tfrac{5}{2},\cdots,50,$ the values of 
$\mathfrak{u_2}$ showcased in Fig.~\ref{fig:h,u2,umax for A}
are
\\
\texttt{2, 2, 2.09524, 2.10102, 2.12293, 2.12821, 2.13508, 2.13865, 
	2.14193, 2.1443, 2.14633, 2.14799, 2.1494, 2.15061, 2.15167, 2.15259, 
	2.15341, 2.15413, 2.15479, 2.15537, 2.1559, 2.15639, 2.15683, 
	2.15723, 2.15761, 2.15795, 2.15827, 2.15857, 2.15885, 2.15911, 
	2.15935, 2.15957, 2.15979, 2.15999, 2.16018, 2.16036, 2.16053, 
	2.16069, 2.16084, 2.16098, 2.16112, 2.16125, 2.16138, 2.1615, 
	2.16161, 2.16172, 2.16183, 2.16193, 2.16202, 2.16211, 2.1622, 
	2.16229, 2.16237, 2.16245, 2.16253, 2.1626, 2.16267, 2.16274, 
	2.16281, 2.16287, 2.16293, 2.16299, 2.16305, 2.16311, 2.16316, 
	2.16321, 2.16327, 2.16332, 2.16336, 2.16341, 2.16346, 2.1635, 
	2.16354, 2.16359, 2.16363, 2.16367, 2.16371, 2.16374, 2.16378, 
	2.16382, 2.16385, 2.16389, 2.16392, 2.16395, 2.16398, 2.16401, 
	2.16404, 2.16407, 2.1641, 2.16413, 2.16416, 2.16419, 2.16421, 
	2.16424, 2.16427, 2.16429, 2.16431, 2.16434}.

For $j=\tfrac{3}{2},2,\tfrac{5}{2},\cdots,50,$ the values of 
$\mathfrak{u_{max}}$ showcased in Fig.~\ref{fig:h,u2,umax for A}
are
\\
\texttt{2.36603, 2.36603, 2.44491, 2.44949, 2.46664, 2.47073, 2.47603, 
	2.47876, 2.48127, 2.48308, 2.48463, 2.48589, 2.48697, 2.48789, 
	2.48869, 2.48939, 2.49001, 2.49056, 2.49105, 2.49149, 2.49189, 
	2.49226, 2.49259, 2.4929, 2.49318, 2.49344, 2.49368, 2.49391, 
	2.49412, 2.49431, 2.4945, 2.49467, 2.49483, 2.49498, 2.49512, 
	2.49526, 2.49538, 2.49551, 2.49562, 2.49573, 2.49583, 2.49593, 
	2.49603, 2.49612, 2.4962, 2.49628, 2.49636, 2.49644, 2.49651, 
	2.49658, 2.49665, 2.49671, 2.49677, 2.49683, 2.49689, 2.49695, 2.497, 
	2.49705, 2.4971, 2.49715, 2.4972, 2.49724, 2.49728, 2.49733, 2.49737, 
	2.49741, 2.49745, 2.49748, 2.49752, 2.49755, 2.49759, 2.49762, 
	2.49766, 2.49769, 2.49772, 2.49775, 2.49778, 2.49781, 2.49783,
	2.49786, 2.49789, 2.49791, 2.49794, 2.49796, 2.49799, 2.49801, 
	2.49803, 2.49805, 2.49808, 2.4981, 2.49812, 2.49814, 2.49816, 
	2.49818, 2.4982, 2.49822, 2.49823, 2.49825}.
\\\vspace{0.0cm}

%-------------------------------------------------------------

For $j=1,\tfrac{3}{2},\cdots,50,$ the values of 
$\mathfrak{u_{max}}$ exhibited in Fig.~\ref{fig:lmax,Umax}
are
\\
\texttt{2.36603, 2.25971, 2.16889, 2.10727, 2.09897, 2.09021, 2.07744, 
	2.06386, 2.05085, 2.03903, 2.02867, 2.02463, 2.02258, 2.01986, 
	2.01686, 2.01389, 2.01113, 2.00871, 2.00759, 2.00704, 2.00634, 
	2.00558, 2.00484, 2.00416, 2.00357, 2.00321, 2.00301, 2.00278, 
	2.00255, 2.00233, 2.00212, 2.00194, 2.0018, 2.0017, 2.0016, 2.0015, 
	2.00141, 2.00133, 2.00125, 2.00118, 2.00113, 2.00107, 2.00102, 
	2.00097, 2.00093, 2.00089, 2.00085, 2.00081, 2.00078, 2.00075, 
	2.00072, 2.00069, 2.00066, 2.00064, 2.00062, 2.00059, 2.00057, 
	2.00055, 2.00053, 2.00052, 2.0005, 2.00048, 2.00047, 2.00045, 
	2.00044, 2.00043, 2.00041, 2.0004, 2.00039, 2.00038, 2.00037, 
	2.00036, 2.00035, 2.00034, 2.00033, 2.00032, 2.00031, 2.0003, 2.0003, 
	2.00029, 2.00028, 2.00027, 2.00027, 2.00026, 2.00026, 2.00025, 
	2.00024, 2.00024, 2.00023, 2.00023, 2.00022, 2.00022, 2.00021, 
	2.00021, 2.0002, 2.0002, 2.0002, 2.00019, 2.00019}.
\\\vspace{0.0cm}

%-------------------------------------------------------------

For $j=1,\tfrac{3}{2},\cdots,50,$ the values of 
$\mathfrak{h}$ shown in Fig.~\ref{fig:S4 UR bounds}
are
\\
\texttt{0, 1.12467, 0.467583, 0.644239, 0.510273, 0.416216, 0.336635, 0.29168, 
	0.250652, 0.216753, 0.190117, 0.168017, 0.149996, 0.134442, 0.121554, 
	0.110269, 0.100723, 0.0922493, 0.0849666, 0.0784334, 0.0727424, 
	0.0675917, 0.0630542, 0.0589167, 0.0552367, 0.0518593, 0.0488306, 
	0.0460353, 0.0435107, 0.0411692, 0.0390411, 0.0370587, 0.035247, 
	0.0335529, 0.031997, 0.0305369, 0.0291901, 0.0279223, 0.0267482, 
	0.0256398, 0.0246096, 0.0236345, 0.0227253, 0.0218627, 0.021056, 
	0.020289, 0.0195696, 0.0188843, 0.01824, 0.017625, 0.0170455, 
	0.0164914, 0.0159681, 0.015467, 0.0149928, 0.0145381, 0.0141069, 
	0.0136929, 0.0132996, 0.0129215, 0.0125618, 0.0122155, 0.0118855, 
	0.0115675, 0.011264, 0.0109712, 0.0106915, 0.0104213, 0.0101628, 
	0.00991295, 0.00967359, 0.00944199, 0.00921989, 0.0090048, 
	0.00879831, 0.00859819, 0.00840586, 0.00821931, 0.00803986, 
	0.00786568, 0.00769796, 0.00753505, 0.00737805, 0.00722545, 
	0.00707826, 0.0069351, 0.00679691, 0.00666242, 0.00653249, 
	0.00640597, 0.00628366, 0.00616448, 0.00604919, 0.00593679, 
	0.00582798, 0.00572185, 0.00561904, 0.00551871, 0.00542146}.

For $j=1,\tfrac{3}{2},\cdots,50,$ the values of 
$\mathfrak{u_{\sfrac{1}{2}}}$ shown in Fig.~\ref{fig:S4 UR bounds}
are
\\
\texttt{3, 3.73205, 3.43649, 3.53517, 3.46883, 3.41326, 3.37114, 3.33513, 
		3.30647, 3.28143, 3.26077, 3.2424, 3.22684, 3.21281, 3.20069, 
		3.18963, 3.17992, 3.17098, 3.16303, 3.15566, 3.14904, 3.14286, 
		3.13725, 3.13199, 3.12719, 3.12266, 3.1185, 3.11456, 3.11092, 
		3.10746, 3.10425, 3.10119, 3.09834, 3.09561, 3.09306, 3.09061, 
		3.08832, 3.08611, 3.08403, 3.08204, 3.08015, 3.07833, 3.0766, 
		3.07494, 3.07336, 3.07184, 3.07038, 3.06898, 3.06763, 3.06634, 
		3.06509, 3.06389, 3.06274, 3.06162, 3.06054, 3.0595, 3.0585, 3.05753, 
		3.05659, 3.05568, 3.0548, 3.05394, 3.05312, 3.05231, 3.05154, 
		3.05078, 3.05005, 3.04934, 3.04864, 3.04797, 3.04732, 3.04668, 
		3.04606, 3.04545, 3.04487, 3.04429, 3.04373, 3.04319, 3.04266,
		3.04214, 3.04163, 3.04114, 3.04065, 3.04018, 3.03972, 3.03927, 
		3.03883, 3.0384, 3.03798, 3.03757, 3.03716, 3.03677, 3.03638, 
		3.03601, 3.03564, 3.03527, 3.03492, 3.03457, 3.03423}.

For $j=1,\tfrac{3}{2},\cdots,50,$ the values of 
$\mathfrak{u_2}$ shown in Fig.~\ref{fig:S4 UR bounds}
are
\\
\texttt{3, 2.25, 2.76563, 2.65958, 2.74466, 2.80312, 2.85131, 2.87386, 2.89556, 2.91262, 2.9257, 2.93601, 2.94425, 2.95117, 2.95675, 2.96153, 
		2.96548, 2.96892, 2.97182, 2.97438, 2.97656, 2.97851, 2.98021, 
		2.98173, 2.98306, 2.98427, 2.98534, 2.98632, 2.98719, 2.98799, 
		2.98871, 2.98938, 2.98998, 2.99053, 2.99104, 2.99151, 2.99194, 
		2.99235, 2.99272, 2.99306, 2.99338, 2.99369, 2.99396, 2.99423,
		2.99447, 2.9947, 2.99492, 2.99512, 2.99531, 2.99549, 2.99566, 
		2.99582, 2.99597, 2.99612, 2.99625, 2.99638, 2.9965, 2.99662, 
		2.99673, 2.99684, 2.99694, 2.99703, 2.99712, 2.99721, 2.99729, 
		2.99737, 2.99745, 2.99752, 2.99759, 2.99766, 2.99772, 2.99778, 
		2.99784, 2.9979, 2.99795, 2.99801, 2.99806, 2.9981, 2.99815, 2.9982, 
		2.99824, 2.99828, 2.99832, 2.99836, 2.9984, 2.99844, 2.99847, 2.9985, 
		2.99854, 2.99857, 2.9986, 2.99863, 2.99866, 2.99869, 2.99871, 
		2.99874, 2.99876, 2.99879, 2.99881}.

For $j=1,\tfrac{3}{2},\cdots,50,$ the values of 
$\mathfrak{u_{max}}$ shown in Fig.~\ref{fig:S4 UR bounds}
are
\\
\texttt{3, 2.5, 2.875, 2.81354, 2.86259, 2.89639, 2.92288, 2.93487, 2.94638, 2.95534, 2.96215, 2.96749, 2.97174, 2.97528, 2.97814, 2.98058, 2.98259, 2.98434, 2.98581, 2.98711, 2.98821, 2.9892, 2.99005, 
2.99082, 2.99149, 2.9921, 2.99264, 2.99314, 2.99357, 2.99398, 
2.99434, 2.99467, 2.99498, 2.99526, 2.99551, 2.99575, 2.99596, 
2.99617, 2.99635, 2.99653, 2.99669, 2.99684, 2.99698, 2.99711,
2.99723, 2.99735, 2.99745, 2.99756, 2.99765, 2.99774, 2.99783, 
2.99791, 2.99798, 2.99806, 2.99812, 2.99819, 2.99825, 2.99831, 
2.99836, 2.99842, 2.99847, 2.99851, 2.99856, 2.9986, 2.99865,
2.99869, 2.99872, 2.99876, 2.99879, 2.99883, 2.99886, 2.99889,
2.99892, 2.99895, 2.99898, 2.999, 2.99903, 2.99905, 2.99908, 2.9991, 
2.99912, 2.99914, 2.99916, 2.99918, 2.9992, 2.99922, 2.99923, 
2.99925, 2.99927, 2.99928, 2.9993, 2.99931, 2.99933, 2.99934, 
2.99936, 2.99937, 2.99938, 2.99939, 2.99941}.

%\vfill
%\bigskip
%\clearpage

%===========================================

\end{document}